\documentclass[11pt,a4paper]{scrartcl}

\usepackage{hyperref}

\usepackage{CLICdp_LCD}

\usepackage{CLICdp_definitions}
\usepackage{CLICdp_biblatex}

\usepackage{graphicx}
\usepackage{epstopdf}
\usepackage{threeparttable}
\usepackage{units}
\usepackage{xspace}
\usepackage{graphicx}
\usepackage{helvet}
\usepackage{color}
\usepackage{url}
\usepackage{xspace}
\usepackage[point]{rccol}
\usepackage{tikz}
\usetikzlibrary{calc}
\usepackage{multirow}
\usepackage{wrapfig}
\usepackage{upgreek}
\usepackage{tikz}
\usetikzlibrary{calc}
\usepackage{wrapfig}
\usepackage{upgreek}
\usepackage{mathptmx}
\usepackage{helvet}
\usepackage{amsmath}
\usepackage{color}
\usepackage{units}
\usepackage{url}
\usepackage{booktabs}
\usepackage[point]{rccol}
\usepackage{multirow}
\usepackage{graphicx}
\usepackage{caption}
\usepackage{subcaption}
\usepackage{hhline}
\usepackage{todonotes}
\usepackage{comment}

\usepackage{float}
\usepackage[titletoc]{appendix}
\makeatletter
\def\url@myurlfontstyle{%
\@ifundefined{selectfont}{\def\UrlFont{\sf}}{\def\UrlFont{\small\ttfamily}}}
\makeatother

\long\def\symbolfootnote[#1]#2{\begingroup%
\def\thefootnote{\fnsymbol{footnote}}\footnote[#1]{#2}\endgroup}

\title{CLD -  A Detector Concept for the FCC-ee}

\clicdpnote{2019}{001}

\date{\formatdate{25}{11}{2019}}

\addauthor{N.~Bacchetta}{\institute{1}}
\addauthor{J.-J. Blaising}{\institute{2}}
\addauthor{E.~Brondolin}{\institute{1}}
\addauthor{M.~Dam}{\institute{4}}
\addauthor{D.~Dannheim}{\institute{1}}
\addauthor{K.~Elsener}{\institute{1}}
\addauthor{D.~Hynds}{\institute{1}}
\addauthor{P.~Janot}{\institute{1}}
\addauthor{A.M.~Kolano}{\institute{1}}
\addauthor{E.~Leogrande}{\institute{1}}
\addauthor{L.~Linssen}{\institute{1}}
\addauthor{A.~N\"{u}rnberg}{\institute{1}}
\addauthor{E.F.~Perez}{\institute{1}}
\addauthor{M.~Petri\v{c}}{\institute{1}}
\addauthor{P.~Roloff}{\institute{1}}
\addauthor{A.~Sailer}{\institute{1}}
\addauthor{N.~Siegrist}{\institute{1}}
\addauthor{O.~Viazlo}{\institute{1}}
\addauthor{G.G.~Voutsinas}{\institute{1}}
\addauthor{M.A. Weber}{\institute{1}}
\addinstitute{1}{CERN, Geneva, Switzerland}
\addinstitute{2}{LAPP, Annecy, France}
\addinstitute{4}{Niels Bohr Institute, Copenhagen, Denmark}

\abstract{This note gives a conceptual description and illustration of the CLD detector, based on the work for a detector at CLIC. CLD is one of the detectors envisaged at a future 100~km
\epem circular collider (FCC-ee).
The note also contains a brief description of the simulation and reconstruction tools used in the linear collider community, which have been adapted for physics and performance studies of CLD. The detector performance is described
in terms of single particles, particles in jets, jet energy and angular resolution,
and flavour tagging.
The impact of beam-related backgrounds (incoherent \epem pairs and synchrotron radiation photons) on the performance is also discussed.}

\titlecomment{ }

\graphicspath{ {./logos/}{./figures/} }

\begin{document}
\titlepage
\clearpage
\tableofcontents
\clearpage
\newcommand{\radlen}{\ensuremath{X_{0}}\xspace}
\newcommand{\nuclen}{\ensuremath{\lambda_{\mathrm{I}}}\xspace}

\newcommand{\ddhep}{\textsc{DD4hep}\xspace}
\newcommand{\ddg}{\textsc{DDG4}\xspace}
\newcommand{\ddrec}{\textsc{DDRec}\xspace}
\newcommand{\ilcdirac}{\textsc{iLCDirac}\xspace}

\newcommand{\microrad}{\ensuremath{\upmu\mathrm{rad}\xspace}}
\newcommand{\PZgstarToqq}{\ensuremath{\PZ/\gamma^{*}\to \PQq\PAQq}\xspace}

\crefname{section}{chapter}{chapters}
\Crefname{section}{Chapter}{Chapters}
\crefname{subsection}{section}{sections}
\Crefname{subsection}{Section}{Sections}
\crefname{subsubsection}{section}{sections}
\Crefname{subsubsection}{Section}{Sections}

\makeatletter
\def\lr#1{\expandafter\@lr\csname c@#1\endcsname}
\def\@lr#1{%
  \ifnum#1=0%
    isZero%
  \fi
  \ifnum#1=1%
    left%
  \fi
  \ifnum#1=2%
    right%
  \fi
}
\makeatother

\section{Introduction}
\label{sec:Intro}

 The following sections describe a possible future detector for FCC-ee. 
This detector is derived from the most recent CLIC detector model~\cite{CLICdet_note_2017, CLICdet_performance}, 
which features a silicon pixel vertex detector and a silicon tracker, followed by highly granular calorimeters (a silicon-tungsten ECAL and a scintillator-steel HCAL). 
A superconducting solenoid provides a strong magnetic field, and a steel yoke interleaved with resistive plate (RPC) muon chambers closes the field. 

The detector model for FCC-ee is  dubbed `CLD' (CLIC-like detector). 
The overall parameters and all the sub-detectors are described in this note, and the results of full simulation studies illustrate the detector performance for lower level physics observables.

The complex inner region (the Machine-Detector Interface MDI) with the beams crossing at an angle of 30 mrad,
 the final quadrupoles at an L$^{*}$~=~2.2~m, screening and compensating solenoids and the luminosity monitor is described in dedicated chapters of the FCC CDR~\cite{FCC_CDR} and is not touched upon here.
 Instead, a forward region cone with an opening angle of 150~mrad is reserved for the MDI elements in the CLD detector model.

At this stage of the conceptual design, it is assumed that the detector is identical for all the collision energies of FCC-ee, i.e. for operation at the Z~(91.2 GeV), W~(160 GeV), H~(240 GeV) and top~(365 GeV).

This note is structured as follows: In section 2, the overall layout of the CLD detector is described. Section 3 and 4 give details of the vertex and tracking detectors, while section 5 describes the calorimeters.
The magnet system including the muon detectors is described in section 6. An overview of the simulation and reconstruction tools is given in section 7. These full simulation tools are used for an assessment of the CLD detector 
performances, which are also shown in section 7. Appendix A contains a table with the sensor areas of each sub-detector system, the pixel/pad sizes and the resulting total number of channels. 
Appendix B describes a fast-simulation study towards reducing the tracker outer radius, and Appendix C discusses possible ECAL options with fewer layers: both these studies aim at reducing the cost of the ECAL. 
Appendix D shows a comparison between two methods of extracting the jet energy resolution: the RMS90 (commonly used in the linear collider community) and the double-sided Crystal Ball fit (used e.g. by CMS). 
Finally, in Appendix E results of a pilot study on flavour tagging assuming a significantly smaller beam pipe radius are presented.

\section{Overall Dimensions and Parameters}
\label{general}

This section provides information about the general considerations leading to the choice of the main detector parameters. The starting point for the CLD concept was CLICdet~\cite{CLICdet_note_2017, CLICdet_performance}, optimised for a 3~TeV linear \epem collider, and which itself
evolved from the physics and detector studies performed for the CLIC CDR~\cite{cdrvol2}. 

Some important constraints are given by the studies performed for the MDI at FCC-ee~\cite{FCC_CDR}:
\begin{itemize}
\item{the luminosity goal, given the beam blow-up due to the crossing angle of 30 mrad, limits the detector solenoidal field to a maximum of 2 Tesla;}
\item{considerations of synchrotron radiation backgrounds, higher-order mode studies and vacuum requirements define the dimensions of the central beam pipe (inner radius 15~mm, half-length 125~mm);}
\item{experience from the Stanford Linear Collider~\cite{SLC_beam_pipe} indicates that the beam pipe needs to be water-cooled; this is approximated by a 1.2~mm thick Be beam pipe in the simulation model of CLD (0.8 mm for the beam pipe wall thickness, 0.4 mm as the equivalent thickness for the water cooling needed);}
\item{furthermore, a gold layer of 5~\micron thickness is required on the inside of the beam pipe;}
\item{the space inside a 150 mrad polar angular cone\protect\footnotemark is needed for accelerator and MDI elements and can not be used for detectors (other than the luminosity monitor (LumiCal)).}
\end{itemize}
\footnotetext{In a more recent version of the MDI layout, all MDI elements must stay inside a 100 mrad cone - only LumiCal extends to 150 mrad. This implies that the vertex and tracker elements upstream of LumiCal must respect the 150 mrad condition, while the ECAL and HCAL endcap acceptance might be improved in a future layout of the detector, approaching 100 mrad.} 

An additional major constraint stems from the continuous operation of a circular collider like FCC: power-pulsing as foreseen for almost all CLIC detector elements is not possible at FCC.
The impact on cooling needs, material budgets and sampling fractions will depend on technology choices - this issue is discussed for each sub-detector in the corresponding section of this document.
Detailed engineering studies would be needed, but are beyond the scope of a conceptual detector design.
Where possible, approximations based on `best guess' estimates are introduced for the material budget of CLD sub-detectors.

The CLICdet model was adapted for FCC including two major modifications:
\begin{itemize}
\item{the outer radius of the silicon tracker was enlarged from 1.5~m to 2.15~m to compensate for the lower detector solenoid field (2~T instead of 4~T);}
\item{the depth of the hadronic calorimeter was reduced to account for the lower maximum centre-of-mass energy at FCC-ee (5.5~\nuclen instead of 7.5~\nuclen).}
\end{itemize}

\begin{figure}[hbtp] 
         \centering
\includegraphics[width=\linewidth]{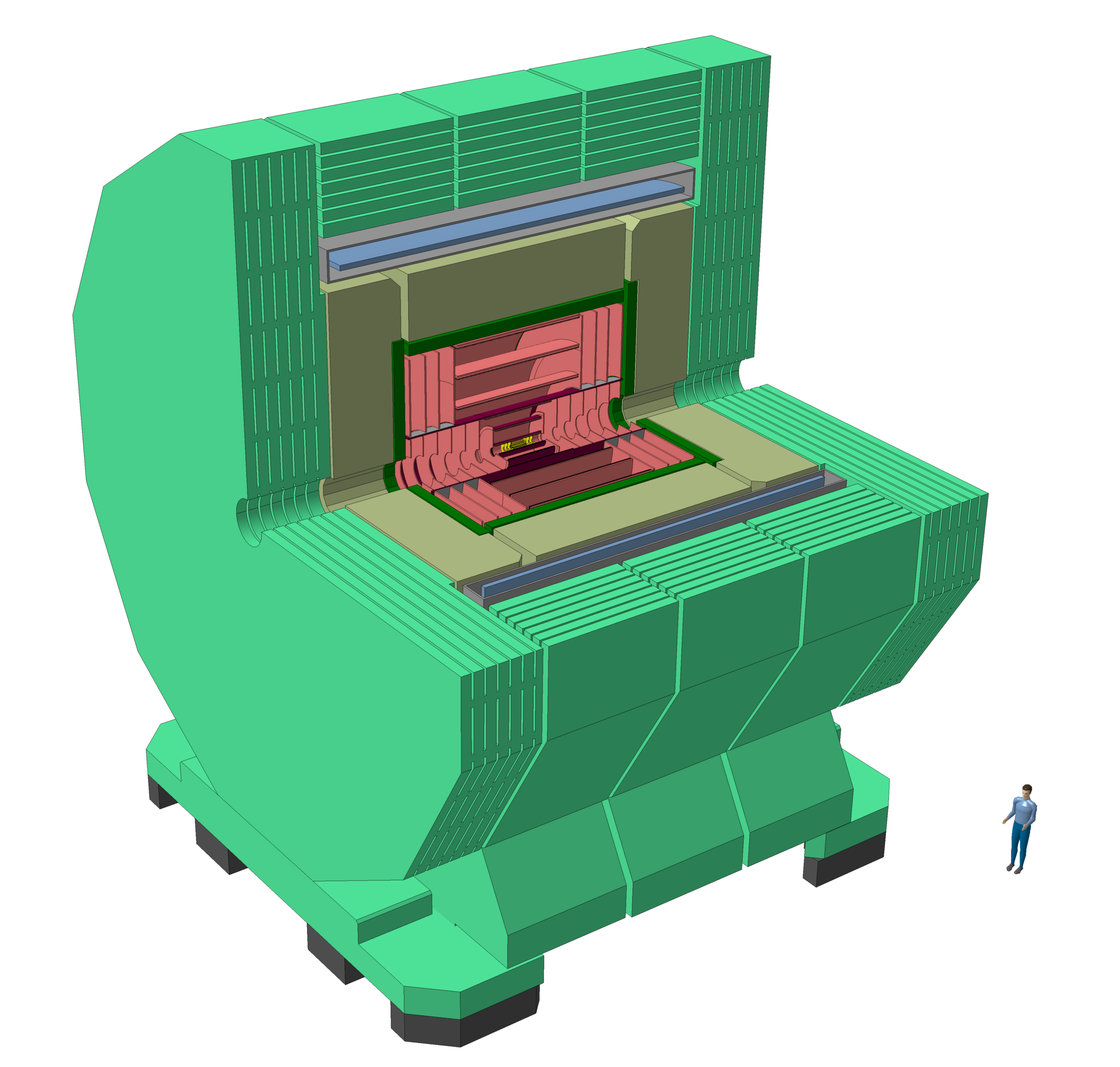}
         \caption{Isometric view of the CLD detector, with one quarter removed.}
         \label{fig.isometric.view}
\end{figure}

\begin{figure}[hbtp] 
         \centering
\includegraphics[scale=0.50]{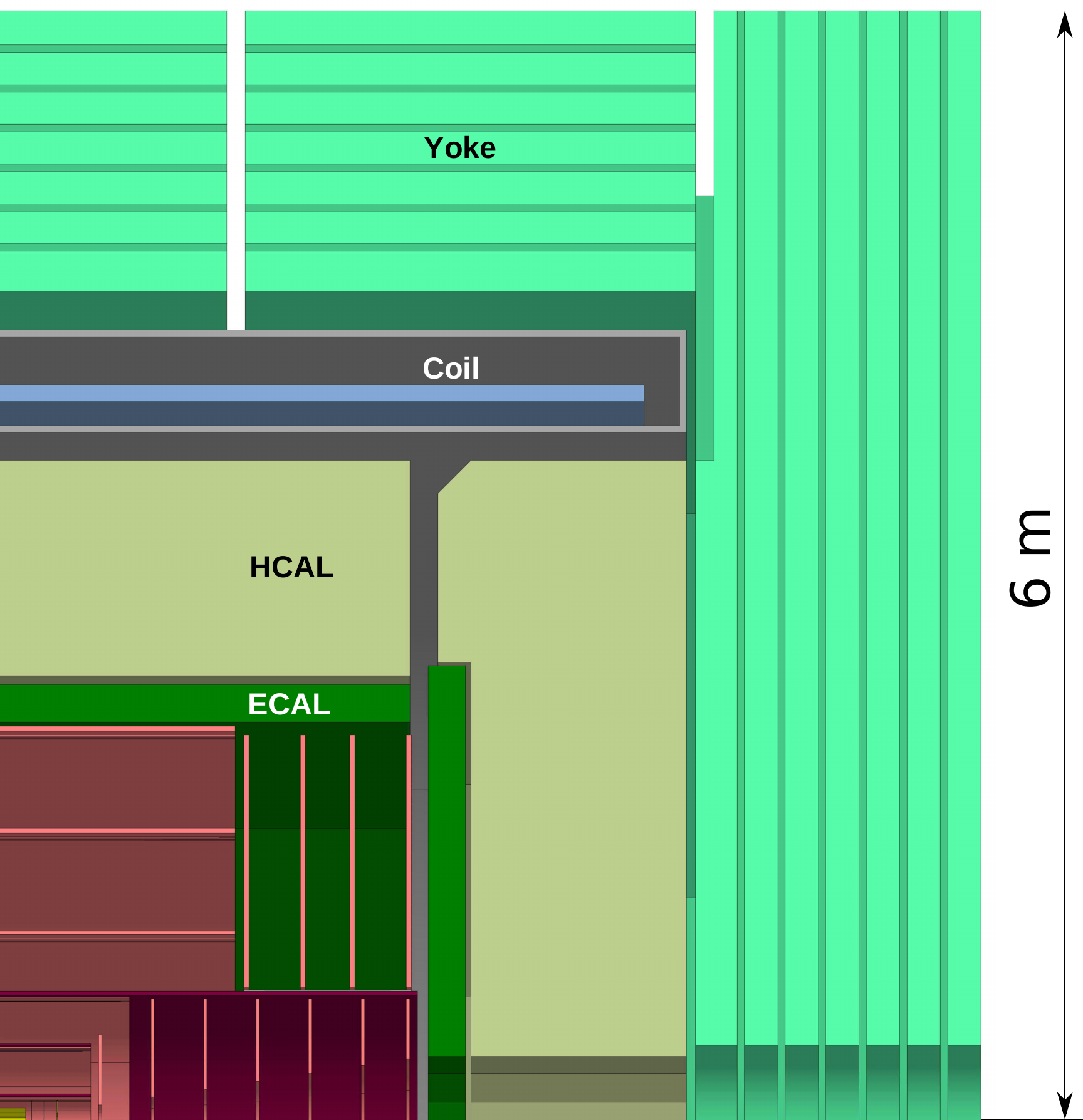}
         \caption{Vertical cross section showing the top right quadrant of CLD. Details of the MDI region are not shown.}
         \label{fig.quarter.view}
\end{figure}

\begin{figure}[hbtp] 
         \centering

\includegraphics[scale=0.25]{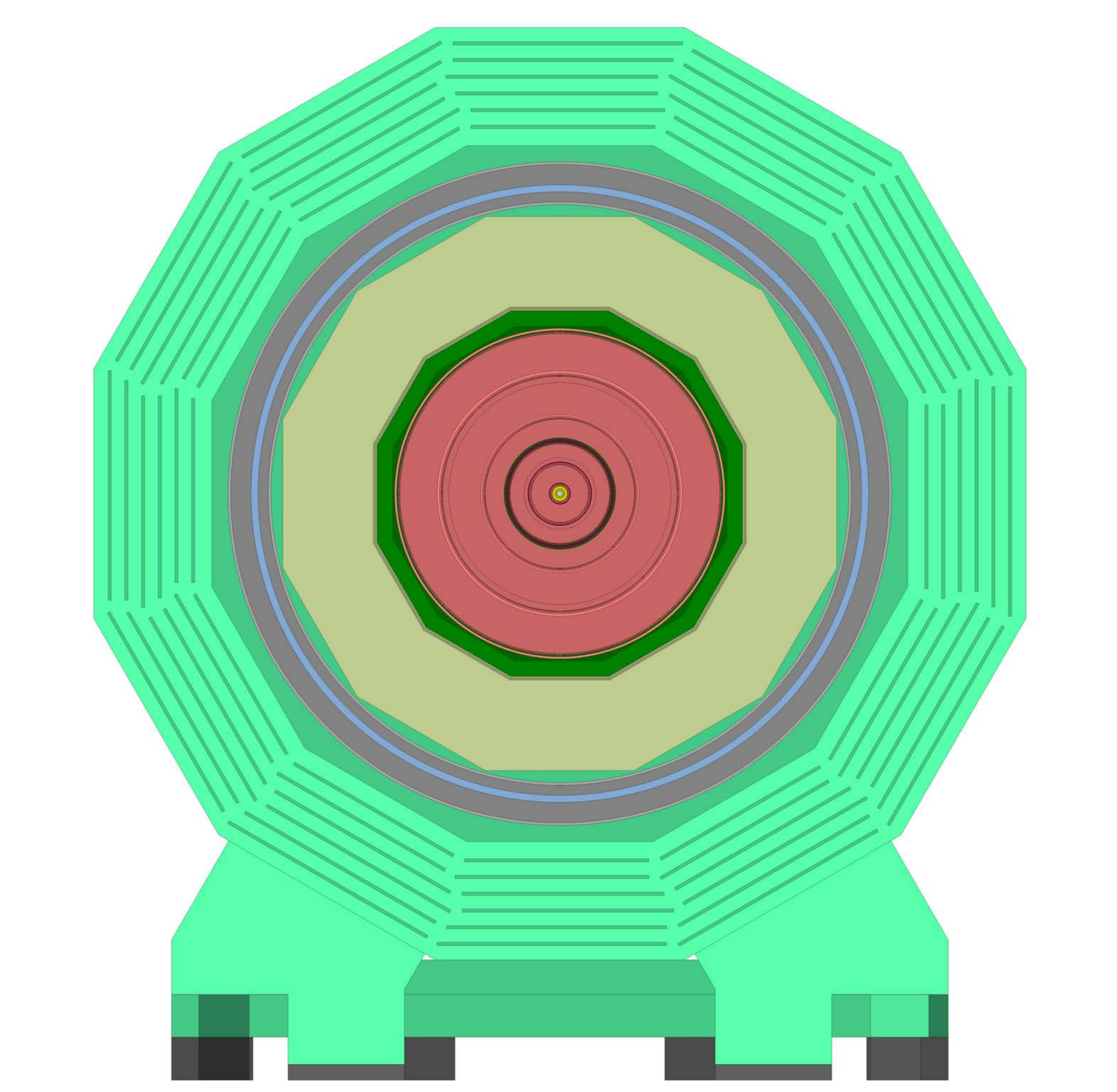}
         \caption{Transverse (XY) cross section of CLD.}
         \label{fig.cross.section}
\end{figure}

A comparison of the main parameters in the CLD and the CLICdet detector models is presented in Table~\ref{table_key_parameters}. 
An illustration of the CLD concept  is given in Figures~\ref{fig.isometric.view},~\ref{fig.quarter.view} and~\ref{fig.cross.section}.

\begin{table}[hbtp]
\caption{\label{table_key_parameters}Comparison of  key parameters of CLD and CLICdet detector models.
The inner radius of the calorimeters is given by the smallest distance of the
calorimeter (dodecagon) to the main detector axis. `HCAL ring' refers to the part of the HCAL endcap surrounding the ECAL endcap.}
\centering
\begin{tabular}{l l l l}
    \toprule
    Concept & CLICdet &  CLD    \\
    \midrule
Vertex inner radius [mm] & 31 & 17.5      \\
Vertex outer radius [mm] & 60 & 58 &       \\
Tracker technology& Silicon & Silicon     \\
Tracker half length [m] & 2.2 &  2.2    \\
Tracker inner radius [m] & 0.127 & 0.127 &   \\
Tracker outer radius [m] & 1.5 & 2.1  \\
Inner tracker support cylinder radius [m] & 0.575 & 0.675 \\
 ECAL absorber  & W & W     \\
    ECAL \radlen & 22 & 22        \\
    ECAL barrel $r_{\min}$ [m] & 1.5 & 2.15         \\
    ECAL barrel $\Delta r$ [mm] & 202 & 202  &    \\
ECAL endcap $z_{\min}$ [m] & 2.31 & 2.31 \\
ECAL endcap $\Delta z$ [mm] & 202  &  202 \\
    HCAL absorber   & Fe  &   Fe   \\
    HCAL \lambdaint & 7.5 & 5.5        \\
    HCAL barrel $r_{\min}$ [m] &1.74  & 2.40       \\
    HCAL barrel $\Delta r$ [mm] & 1590 & 1166      \\
HCAL endcap $z_{\min}$ [m]  & 2.54&2.54 \\
HCAL endcap $z_{\max}$ [m]  & 4.13&3.71 \\
HCAL endcap $r_{\min}$ [mm]  & 250&340 \\
HCAL endcap $r_{\max}$ [m]  & 3.25&3.57 \\
HCAL ring $z_{\min}$ [m]  & 2.36&2.35 \\
HCAL ring $z_{\max}$ [m] & 2.54&2.54 \\
HCAL ring $r_{\min}$ [m] & 1.73&2.48 \\
HCAL ring $r_{\max}$ [m] & 3.25&3.57 \\
Solenoid field [T] & 4 & 2     \\
    Solenoid bore radius [m] & 3.5  & 3.7        \\
    Solenoid length [m] & 8.3 & 7.4        \\
    Overall height [m]  & 12.9 & 12.0     \\
    Overall length [m]  & 11.4 & 10.6    \\
    \bottomrule
\end{tabular}
\label{table.overall}
\end{table}
 
\clearpage

\section{Vertex Detector}
\label{vertex}
\subsection{Overview and Layout}
\begin{figure}[hbtp] 
      \centering
            \includegraphics[scale=0.95]{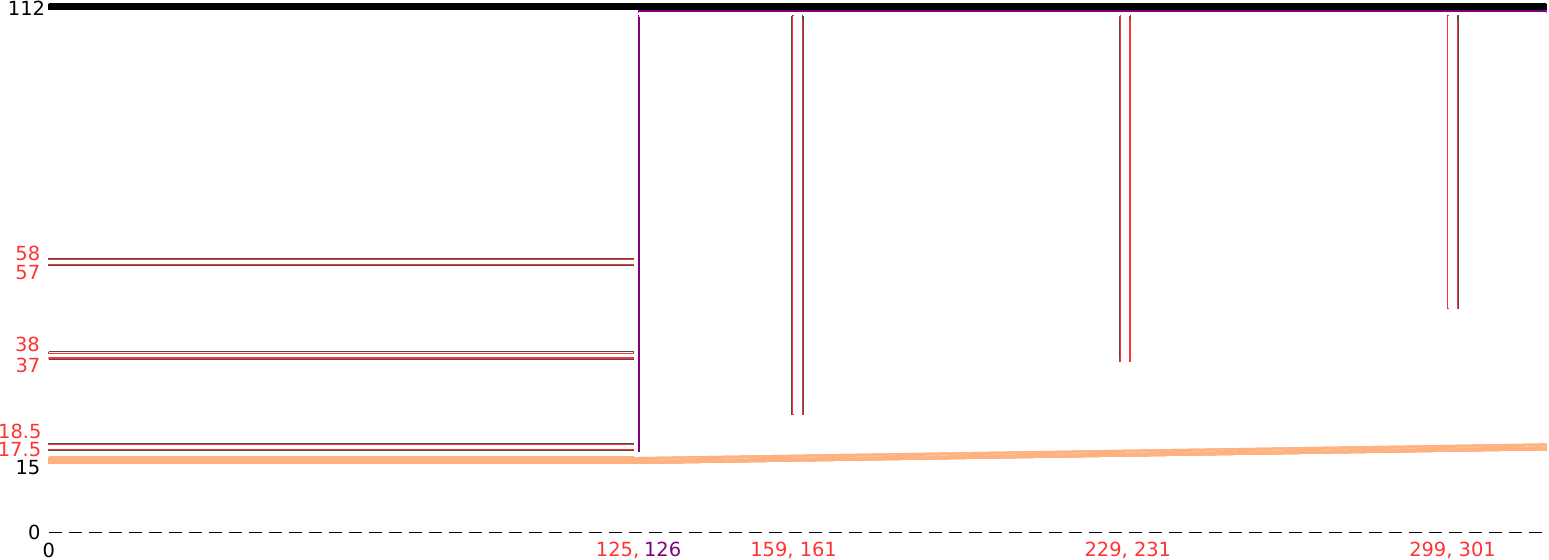}
       \caption{ Sketch of the vertex detector  barrel and forward region (in the ZR plane) of the CLD simulation model. Dimensions are given in mm. 
Red lines indicate sensors, black lines support structure, while magenta lines show cables. The vacuum beam pipe is shown in orange colour.}
       \label{fig.vertex.sketch}
\end{figure}

The vertex detector in the CLD concept, a scaled version of the one in CLICdet, consists of a cylindrical barrel detector closed off in the forward directions by discs.
The layout is based on double layers, i.e. two sensitive layers fixed on a common support structure (which includes cooling circuits).
The barrel consists of three double layers, the forward region is covered by three sets of double-discs on both sides of the barrel. 
An overview of the vertex detector layout is given in Figure~\ref{fig.vertex.sketch}. The total area of the vertex detector sensors is 0.53~m$^2$.

The vertex detector consists of 25~$\times$~25~$\micron^2$  pixels, with a silicon sensor thickness of 50~\micron.
Using pulse height information and charge sharing, a single point resolution of 3~\micron is aimed for.

The inner radius of the innermost vertex barrel layer is determined by the radius and thickness of the central beam pipe, which in turn is given by MDI constraints~\cite{FCC_CDR}.
As a result, the inner edge of the  innermost layer of the vertex barrel is located at R~=~17.5~mm. The location of the additional vertex barrel layers is obtained by scaling-down the layout of the CLICdet vertex detector layout.

\begin{figure}[hbtp]
         \centering
                \includegraphics[width=0.50\textwidth]{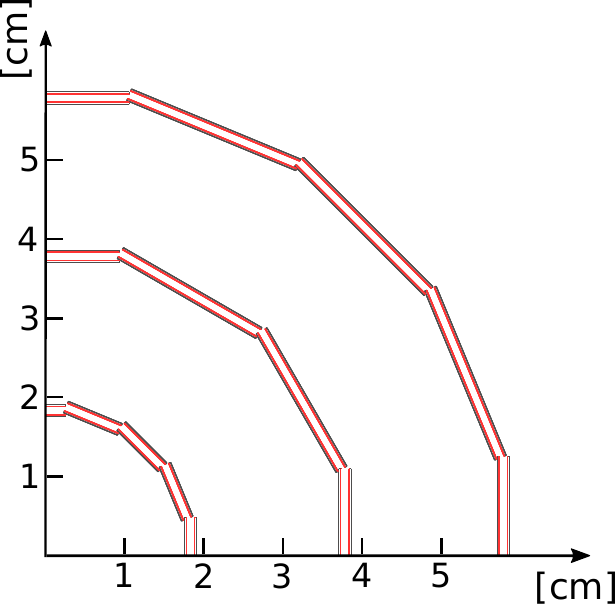}   
 \caption{The double layer arrangement in the vertex barrel detector (XY view).}
\label{fig:vertex.options}
\end{figure}

The overall length of the barrel vertex detector, built from staves, is 250~mm. 
The double layer structure is shown in Figure~\ref{fig:vertex.options}.
 Further details on the dimensions of the vertex detector
barrel layers used in the simulation model are given in Table~\ref{tab.vertex.geometry.for.tracking}. 
Note that the present numbers for the stave widths result from scaling-down the CLICdet layout -- these are the numbers implemented in the simulation model.
 In a forthcoming engineering study, such layout details will have to be revised.

\begin{table}[b!]
   \centering
   \caption{Vertex barrel layout as implemented in the simulations.}
\label{tab.vertex.geometry.for.tracking}
   \begin{tabular}[h!]{c c c c}
\toprule
Barrel layers&	Inner radius [mm]   &  No. of staves & Stave width [mm]\\
\midrule
1 -- 2&17.5 -- 18.5 &  16&  7.3\\
3 -- 4&  37 -- 38&      12 &  20.2\\
5 -- 6&  57 -- 58 &   16  &    23.1\\
\bottomrule
   \end{tabular}
\end{table}
 
\begin{figure}[hbtp] 
      \centering
               \includegraphics[scale=0.6]{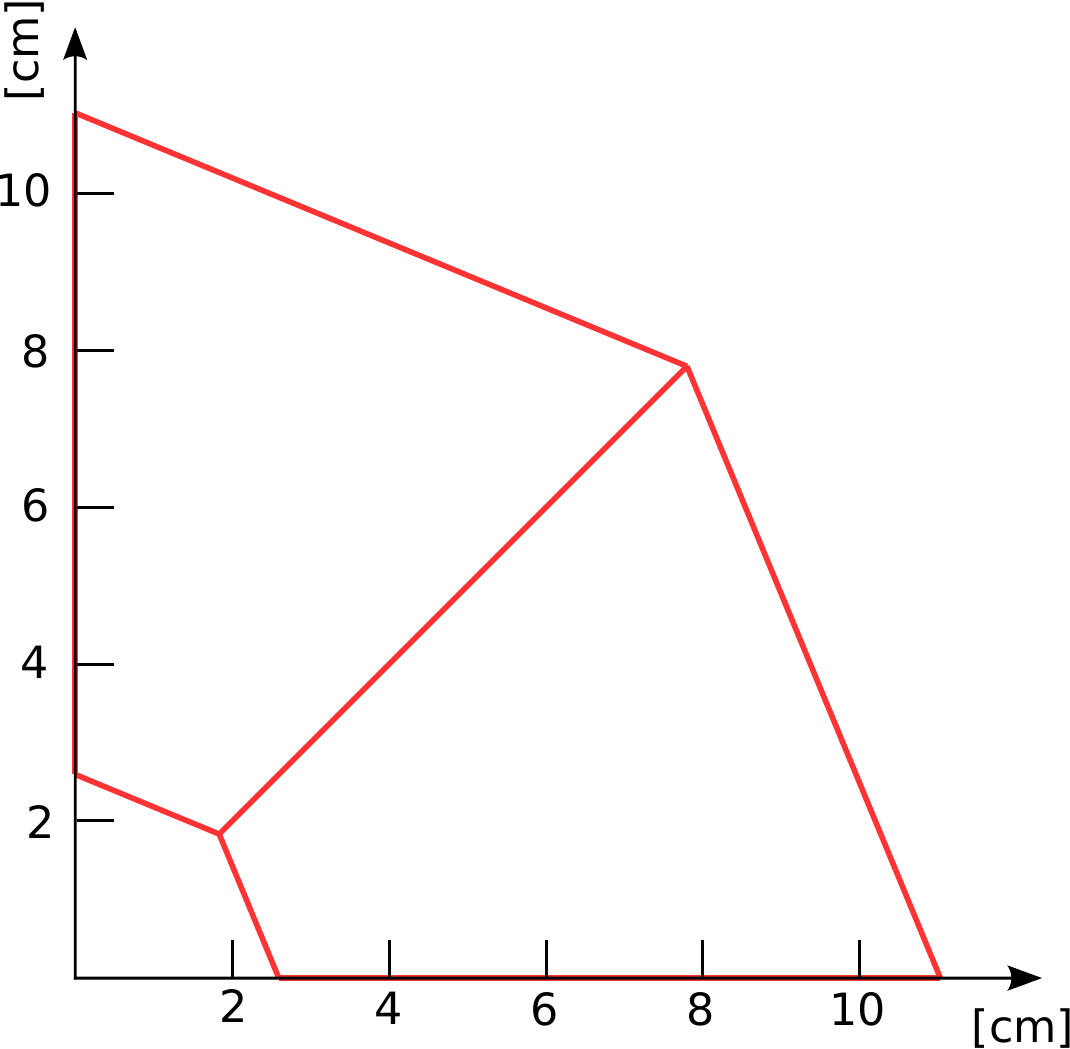}
       \caption{Schematic view of the first vertex forward disc, as implemented in the simulation model (XY view).}
       \label{fig.vertex.petals}
\end{figure}

The vertex detector forward region consists of three discs on each side, each disc is built as a double-layer device.
The discs are located a distance from the IP of 160, 230 and 300~mm, respectively.
They are constructed from 8 trapezoids, approximating a circle. For simplicity the trapezoids are not overlapping in the simulation model. 
The inner radii of the forward discs respect the 150~mrad cone reserved for MDI elements.
The dimensions of the vertex forward discs are given in Table~\ref{tab.vertex.discs}.
Contrary to CLICdet, the forward region vertex detector is built from planar discs\footnote{The design with spirals in CLICdet is motivated by the optimisation of the air flow through the vertex region. 
Air cooling is deemed not to be sufficient for the CLD vertex detector.}.
An example of the vertex petal arrangement as implemented in the simulation is shown in Figure~\ref{fig.vertex.petals}.

\begin{table}[hbtp]
   \centering
   \caption{Dimensions of the vertex discs.}
\label{tab.vertex.discs}
   \begin{tabular}[h!]{c c c c c}
\toprule
Vertex disc&	Inner radius [mm]   &  Outer radius [mm]\\
\midrule
1 & 24&  102\\
2&  34.5 &      102\\
3&  45  &   102\\
\bottomrule
   \end{tabular}
\end{table}

\subsection{Beam-Induced Backgrounds in the Vertex Detector Region}
\label{vtx_background}

Beam-related backgrounds are significant drivers for the vertex and tracker technology choices and the requirements for the read-out of these detectors.
Three types of backgrounds have been studied in detail: incoherent \epem pair production and \gghad production from beam-beam interactions, and background hits from synchrotron 
radiation. The studies using full Monte Carlo simulations and their results are described in detail in section 7.1 of the CDR~\cite{FCC_CDR}.
Examples for incoherent pairs and synchrotron radiation at the lowest and highest energy of FCC-ee operation are given below. 
The rate of \gghad events was found to be negligible (< 0.008 events per bunch crossing at 365 GeV, <0.0007 per bunch crossing at 91.2 GeV).

\begin{figure}[htbp]
  \centering
  \begin{subfigure}{.5\textwidth}
    \centering
    \includegraphics[width=\linewidth]{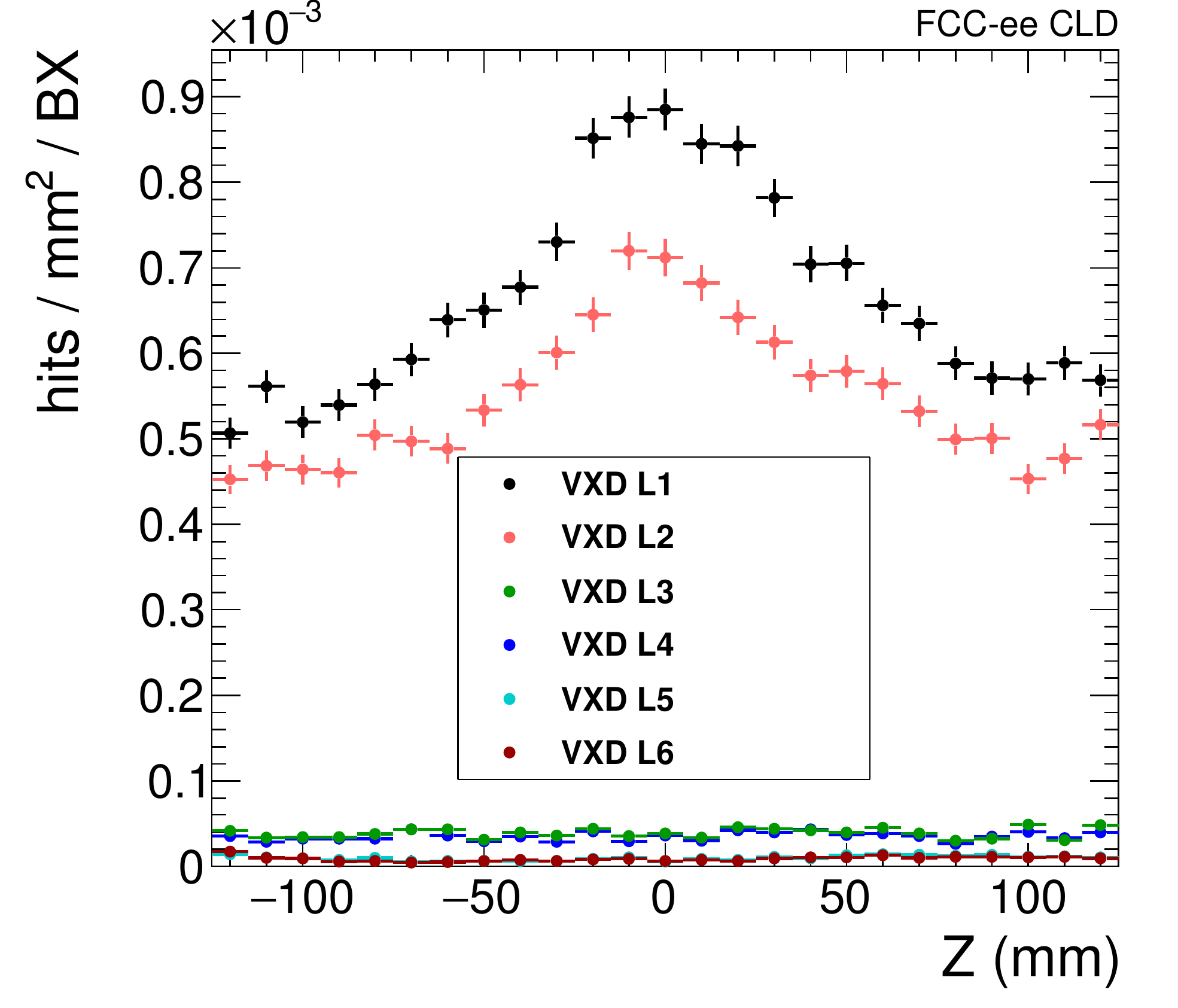}
    \caption{}
    \label{fig:VTXB_pairs_91}
  \end{subfigure}%
  \begin{subfigure}{.5\textwidth}
    \centering
    \includegraphics[width=\linewidth]{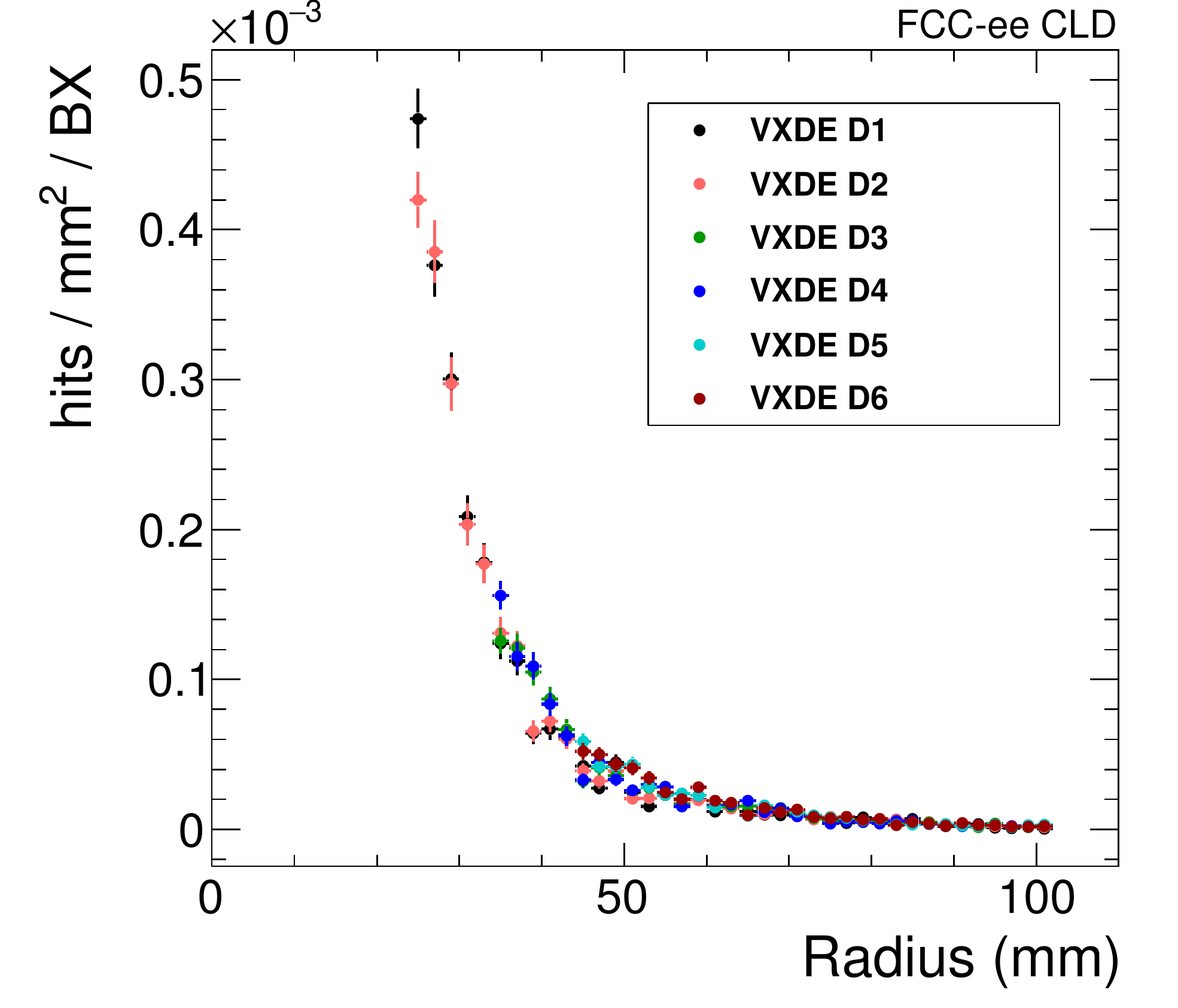}
    \caption{}
    \label{fig:VTXE_Pairs_91}
  \end{subfigure}
  \caption{Hit densities per bunch crossing in the CLD vertex detector barrel layers (a) and  forward discs (b) for particles originating from incoherent pairs. 
Results are shown for operation at 91.2 GeV. VXD L1-L6 indicates the vertex detector barrel layers, while VXDE D1-D6 stands for the vertex detector forward discs.
Vertical error bars indicate the statistical uncertainty, horizontal bars are drawn to show the bin size.
Safety factors for the simulation uncertainties are not included. }
  \label{fig:VTX_pairs_91}
\end{figure}

\begin{figure}[htbp]
  \centering
  \begin{subfigure}{.5\textwidth}
    \centering
    \includegraphics[width=\linewidth]{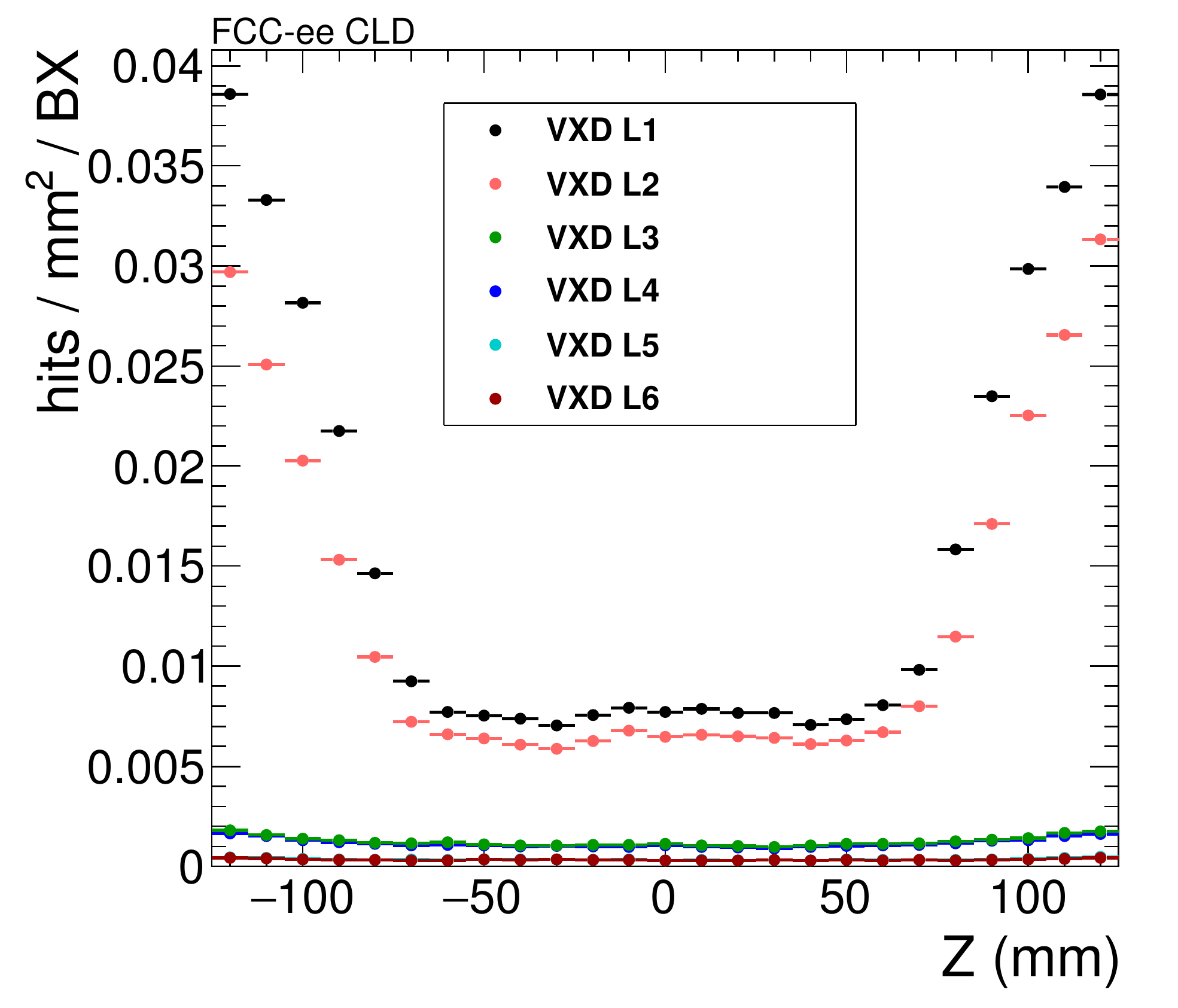}
    \caption{}
    \label{fig:VTXB_pairs_365}
  \end{subfigure}%
  \begin{subfigure}{.5\textwidth}
    \centering
    \includegraphics[width=\linewidth]{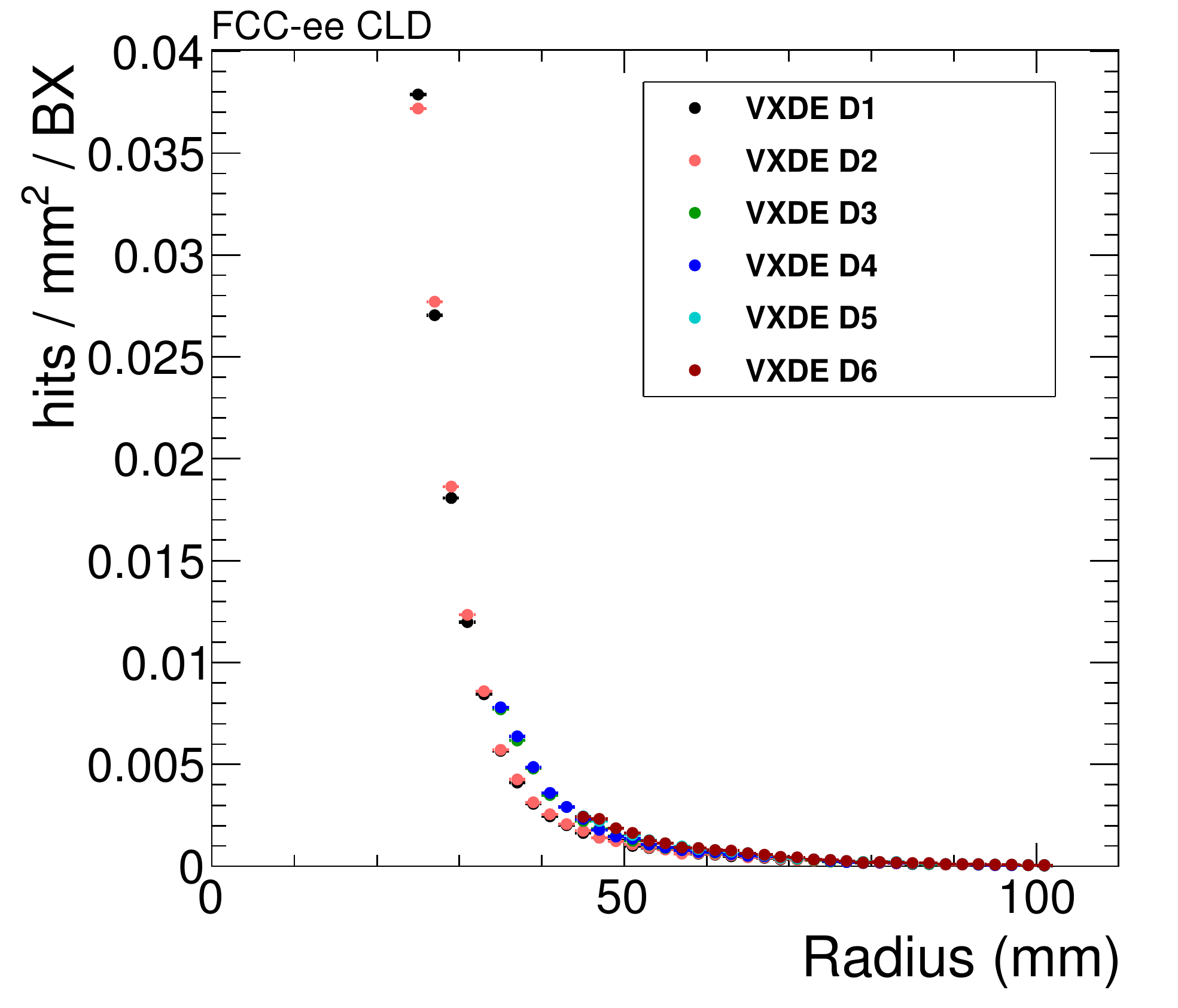}
    \caption{}
    \label{fig:VTXE_Pairs_365}
  \end{subfigure}
  \caption{Hit densities per bunch crossing in the CLD vertex detector barrel layers (a) and  forward discs (b) for particles originating from incoherent \epem pairs, for operation at 365 GeV. 
Vertical error bars indicate the statistical uncertainty, horizontal bars are drawn to show the bin size. Safety factors for the simulation uncertainties are not included. }
  \label{fig:VTX_pairs_365}
\end{figure}

The simulations were performed using the latest version of the CLD detector model and applying a realistic field map (resulting from the detector main solenoid field and the fields of the compensating and screening solenoids).
Figure~\ref{fig:VTX_pairs_91} shows the resulting hit density from incoherent pairs per bunch crossing (BX) for operation at the Z-pole, at 91.2 GeV, while Figure~\ref{fig:VTX_pairs_365} presents the same for the top energy, 365 GeV.
At 91.2 GeV, the result are obtained from simulating 1806 bunch crossings, while at 365 GeV sufficient statistical accuracy was obtained from 371 bunch crossings.
At 365 GeV, backscattering from forward region elements (such as LumiCal) appears to be dominant, leading to more hits at the extremities of the vertex detector barrel layers (Figure~\ref{fig:VTXB_pairs_365}). To the contrary, at 91.2 GeV direct hits (leading to more hits in the centre of the barrel) are dominant (Figure~\ref{fig:VTX_pairs_91}).

\begin{figure}[htbp]
  \centering
  \begin{subfigure}{.5\textwidth}
    \centering
   \includegraphics[width=\linewidth]{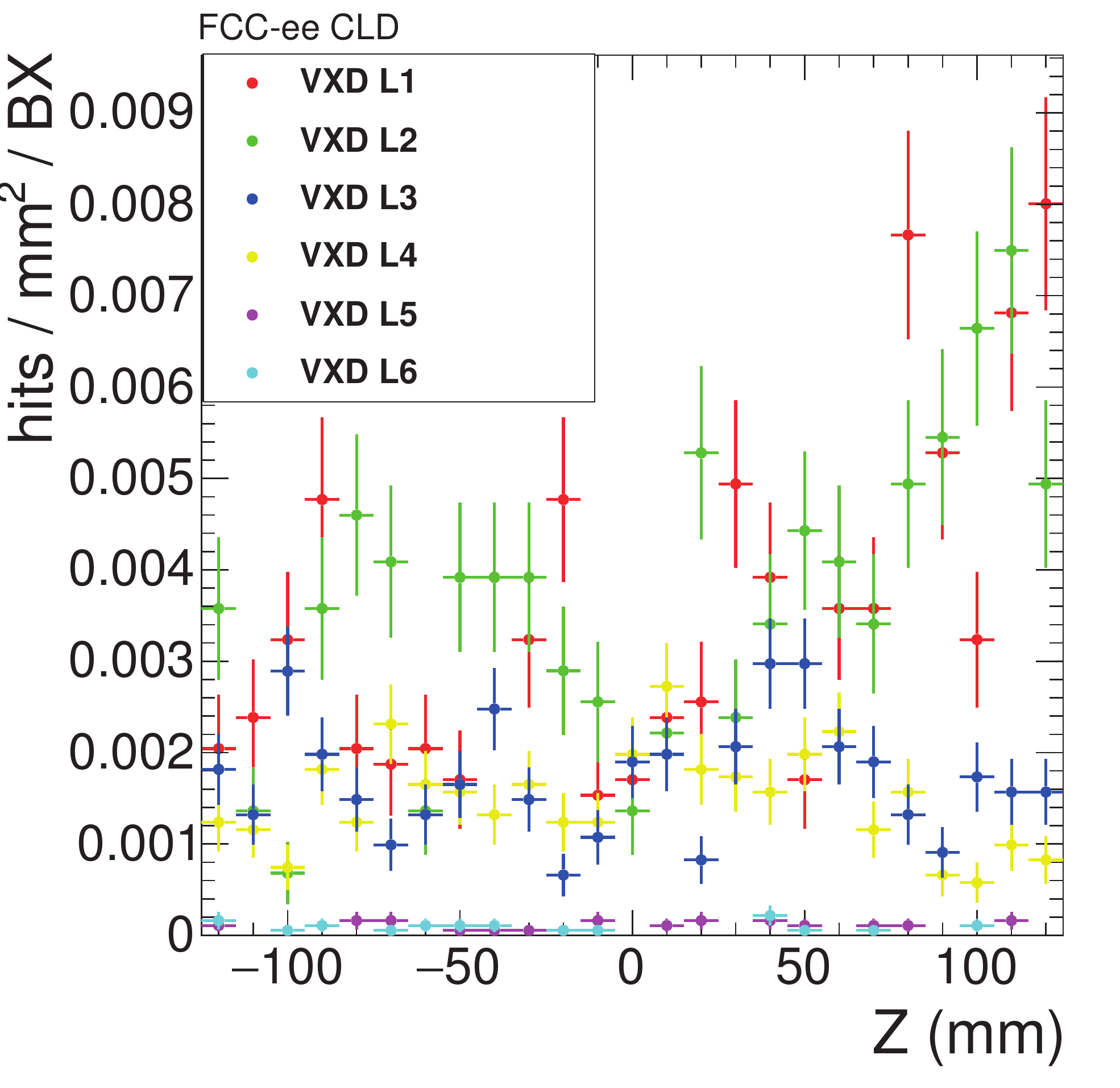}
    \caption{}
    \label{fig:VTXB_sync_365}
  \end{subfigure}%
  \begin{subfigure}{.5\textwidth}
    \centering
    \includegraphics[width=\linewidth]{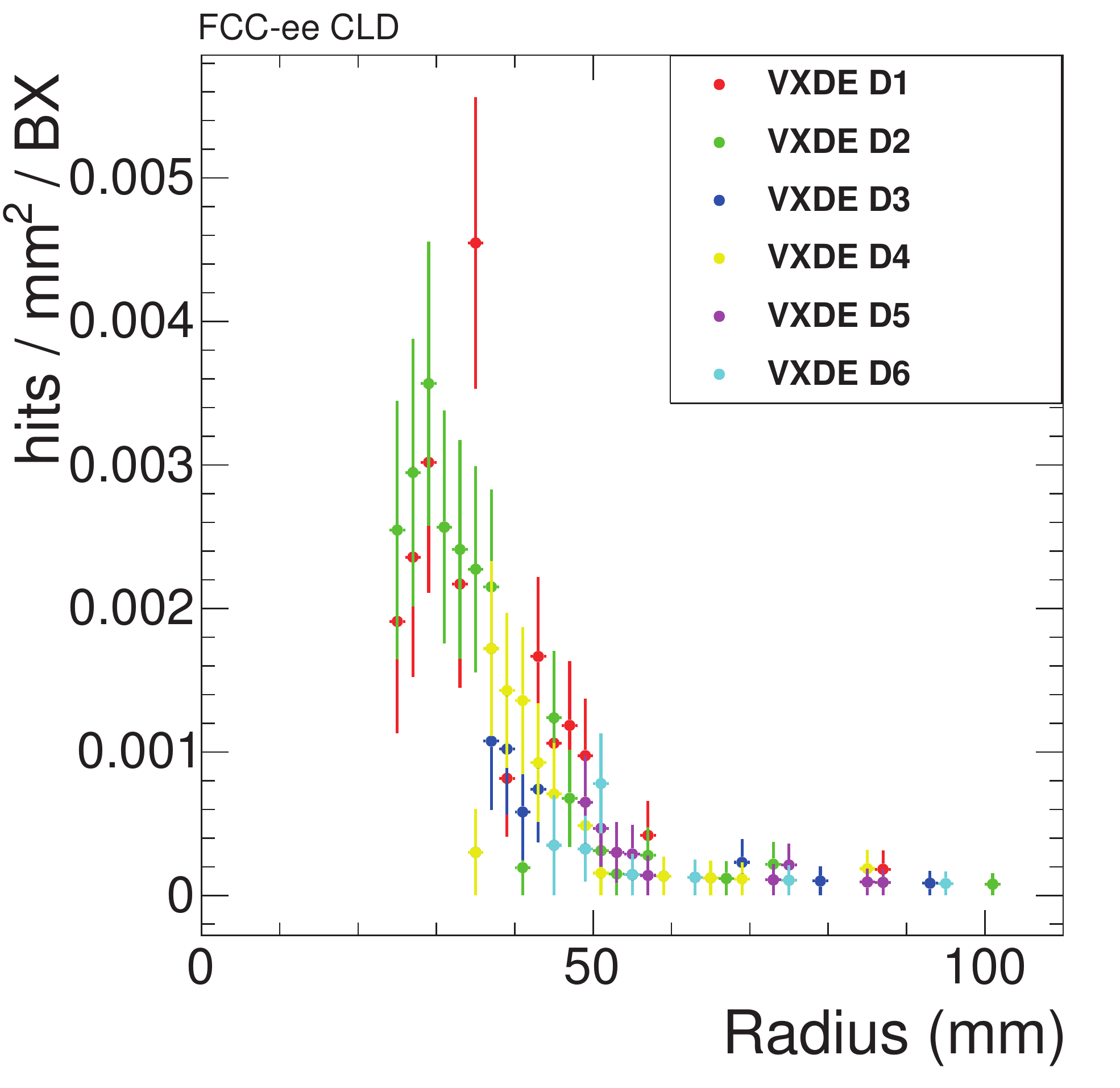}
    \caption{}
    \label{fig:VTXE_sync_365}
  \end{subfigure}
  \caption{As Figure~\ref{fig:VTX_pairs_365} but for hits related to synchrotron radiation photons. }
  \label{fig:VTX_sync_365}
\end{figure}

The corresponding results for hits related to synchrotron radiation photons (averaged over 10 bunch crossings) are shown 
in Figure~\ref{fig:VTX_sync_365}. These  hit rates are about a factor of five lower than the hits from incoherent pairs.
Note that there are no hits from synchrotron radiation observed at 91.2 GeV, even when accumulating data from more than 40000 bunch crossings.

From the hit densities obtained, one can deduce detector occupancies under certain assumptions, in analogy to what is described for CLICdet in~\cite{Nurnberg_Dannheim_2017}.

\begin{equation}
\begin{split}
Occupancy / (readout~window) = Hits / (mm^2 \cdot BX) \times n_{bunches} / (readout~window) \times \\[1em]
 \times (pixel~size) \times (avg.~cluster~size) \times (safety~factor)
\end{split}
\label{eq:occupancy}
\end{equation}

In the following, we are assuming a safety factor of 5 to account for uncertainties in the simulation, and a cluster size of 3 (which will depend on technology choices).
In a scenario where the technology chosen for the ALICE ITS LS2 upgrade~\cite{ALICE_ITS} is assumed, the readout time window\footnote{
Additional studies are required for the case of running at 91.2 GeV: the high rate of Z events (up to 100 kHz) might require some form of time stamping / time of arrival information within the readout window. 
On the other hand, given the rapid evolution of silicon pixel technology, a much shorter readout window with similar powering/cooling requirements is also conceivable.}. would be about 10 $\upmu{}$s.
The present FCC-ee design parameters foresee a bunch spacing of 19.6 ns at 91.2 GeV, and of 3396 ns at 365 GeV, leading to 510 and 3 bunch crossings, respectively,
within the readout window. As a result, the estimated maximal occupancies per readout window (including the safety factor) are expected to be 0.43\% and 0.13\% for operation at 91.2 and 365 GeV, respectively.
It is expected that such occupancies from background hits are acceptable and will not, e.g., impact the performance of the pattern recognition algorithm of the tracking software.

\subsection{Technology Choices, Cooling and Material Budget}
\label{vtx_technology}

The technologies available for vertex pixel sensors, readout electronics, mechanical support structures, and cooling are rapidly evolving. At this stage of the conceptual design for CLD,
the example of the ALICE ITS LS2 upgrade~\cite{ALICE_ITS} is considered an acceptable approximation for the amount of material, to be used in the simulation model.

The ALICE ITS upgrade uses very light-weight structures. The average power dissipation is measured in prototypes to be about 40~mW/cm${^2}$, and water cooling is used for the devices.
Assuming this technology, the total material budget per double layer in the CLD vertex detector is 0.6\%~\radlen. For reference, this corresponds to 50\% more material than in the CLICdet vertex detector
which assumes power pulsing and air cooling.

The low occupancies expected from incoherent pairs and synchrotron radiation (see Section~\ref{vtx_background}), at all energy stages of FCC-ee,
allow one to overlay events from a large number of bunch crossings -- this implies that rather long readout integration times are acceptable for the CLD vertex detector.

A simplified vertex layer layout is implemented in the simulation model. The 50~\micron silicon sensors are separated by a 1~mm air-gap in the barrel and a  2~mm air-gap in the discs. 
On the outside of each sensor, additional material represented by 235~\micron of silicon replaces the combined material of ASIC, support structure, connectivity and cooling. 
The resulting total material budget per double layer corresponds to the thickness in \radlen expected to emerge from the engineering design (and has been achieved in the ALICE ITS upgrade). A summary is given in Table~\ref{table:vertex_material}.
Note that, in analogy to CLICdet, slightly more material is assumed to be needed for the mechanical support of the vertex discs w.r.t. the barrel layers, resulting in a total material budget of 0.7\%~\radlen for the discs.

\begin{table}[h!]
   \centering
   \caption{Vertex detector double layer material budget as implemented in the simulations.}
\label{table:vertex_material}
   \begin{tabular}[h!]{c c c c c c}
\toprule
       Function &Material  & \multicolumn{2}{c}{Barrel} & \multicolumn{2}{c}{Discs} \\
       \cmidrule(lr){3-4} \cmidrule(l){5-6}
&	  &  Thickness  & Material budget & Thickness & Material budget \\
       &	  &  [\micron{}] &  \radlen~[\%{}] &  [\micron{}] &   \radlen~[\%{}] \\
\midrule
ASIC, support etc.& Silicon  &   235 &  0.259& 280& 0.298\\
Sensor& Silicon &  50&  0.053 & 50&  0.053 \\
Gap&  Air  &   1000  &   0.001 &2000 &0.001\\
Sensor& Silicon &  50&  0.053 & 50&  0.053 \\
ASIC, support etc.& Silicon  &   235 &  0.259& 280& 0.298\\
\midrule
total & & &0.625 & &0.703 \\
\bottomrule
   \end{tabular}
\end{table}

\clearpage

\section{Tracking System}
\label{tracker}
\subsection{Overview and Layout}

\begin{figure}[b!] 
         \centering
                 \includegraphics[scale=0.75]{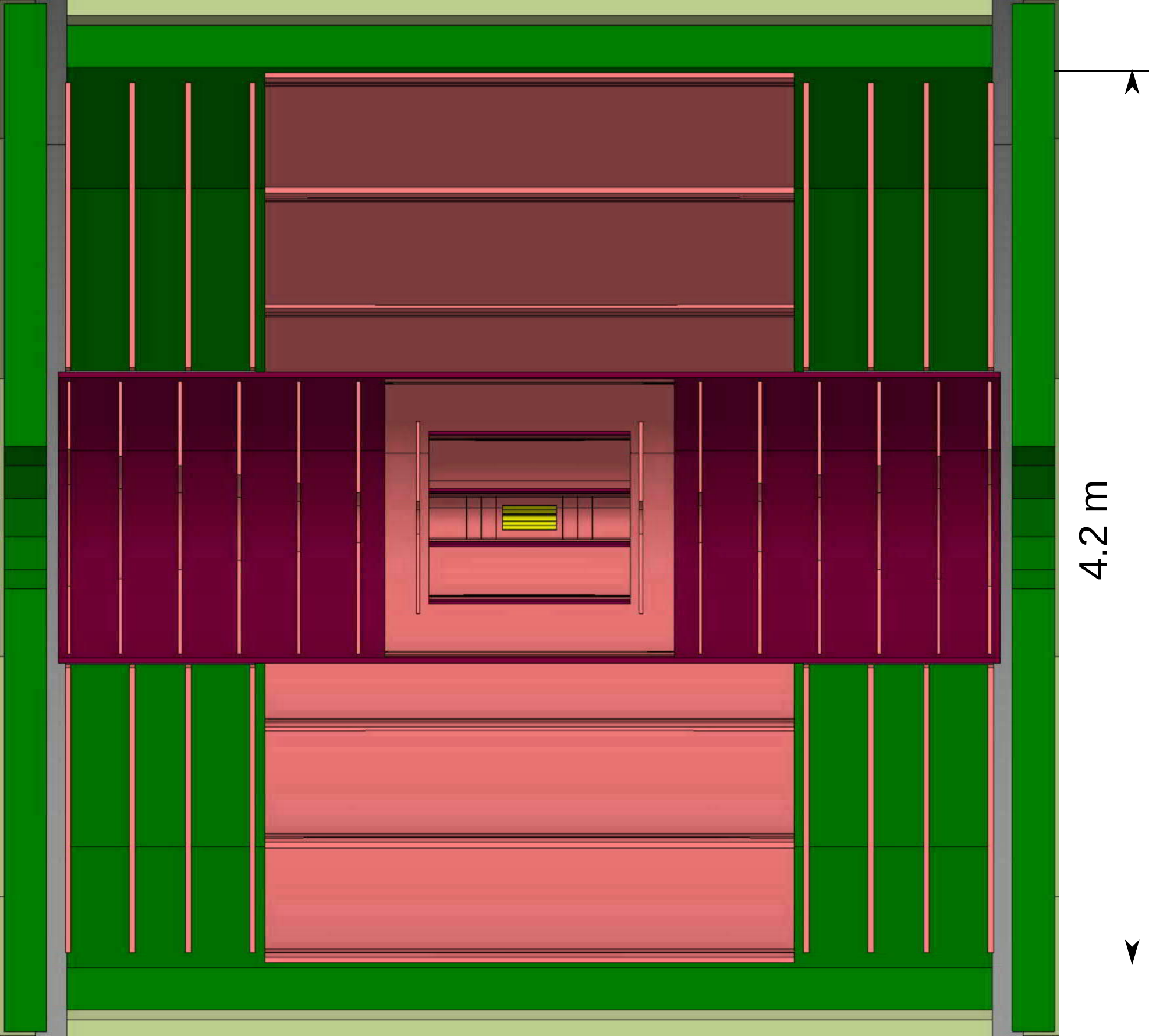}
         \caption{Overall layout of the CLD tracking system: the vertex barrel detector is shown in yellow, the tracking layers in lighter red. The area in darker red illustrates the main support tube for the inner tracking region and the vertex detector. The surrounding ECAL is shown in green. The tracking system covers polar angles larger than 150 mrad. }
         \label{fig.tracker}
\end{figure}

In analogy to CLICdet, the CLD concept features an all-silicon tracker. 
Engineering and maintenance considerations led to the concept of a main support tube
for the inner tracker region (including the vertex detector). 
The inner tracker consists of three barrel layers and seven forward discs.
The outer tracker completes the system with an additional three barrel layers and four discs.
The overall layout of the silicon tracker in CLD is shown in Figure~\ref{fig.tracker}. 

The tracking volume has a half-length of 2.2~m and a maximum radius of 2.1~m. 
This radius allows to achieve a similar momentum resolution in the CLD tracking system
with a 2~T magnetic field as in the CLICdet tracker with a 4~T field and a radius of 1.5~m.
The main support tube has an inner and outer radius of  0.686 and 0.690~m, respectively, and a half-length of 2.3~m.
The layout respects the 150~mrad cone reserved for beam- and MDI-equipment.
The overall geometrical parameters of the tracker are given in Table~\ref{tab.tracker.barrel} and~\ref{tab.tracker.endcap} for the barrel and discs, respectively. 

\begin{table}[hbtp]
\caption{Main parameters of the tracker barrel layout, radius R and half-length L/2. ITB and OTB denote inner and outer 
tracker barrel, respectively.}
\label{tab.tracker.barrel}
\centering
\begin{tabular}{c c c c}
    \toprule
    Layer No. &   Name &  R [mm] & L/2 [mm]\\
    \midrule
1 & ITB1 &127 & 482       \\    
2 & ITB2& 400& 482  \\
3 & ITB3& 670& 692 \\
\midrule
4 & OTB1& 1000& 1264  \\
5 & OTB2& 1568&  1264      \\
6 & OTB3& 2136& 1264 \\
    \bottomrule
\end{tabular}
\end{table}
\begin{table}[hbtp]
\centering
\caption{Main parameters of the tracker discs. ITD and OTD denote inner and outer tracker discs, respectively.}
\label{tab.tracker.endcap}
\begin{tabular}{c c c c c c}
    \toprule
    Disc No. & Name & Z [mm] &$R_{\mathrm{in}}$ [mm]& $R_{\mathrm{out}}$ [mm]\\
    \midrule
1 &ITD1& 524 & 79.5  & 457       \\    
2 &ITD2& 808& 123.5 & 652  \\
3 &ITD3& 1093 & 165 & 663 \\
4 &ITD4& 1377 & 207.5  & 660.5  \\
5 &ITD5& 1661& 249.5 & 657     \\
6 &ITD6& 1946 & 293 & 640   \\
7 &ITD7& 2190 & 330 & 647      \\
\midrule
8 &OTD1& 1310 & 718 & 2080    \\
9 &OTD2& 1617 & 718 & 2080      \\
10 &OTD3& 1883 & 718 & 2080      \\
11&OTD4& 2190 & 718 & 2080     \\
    \bottomrule
\end{tabular}

\end{table}

\begin{figure}[hbtp]
   \centering
   \includegraphics[width=.6\linewidth]{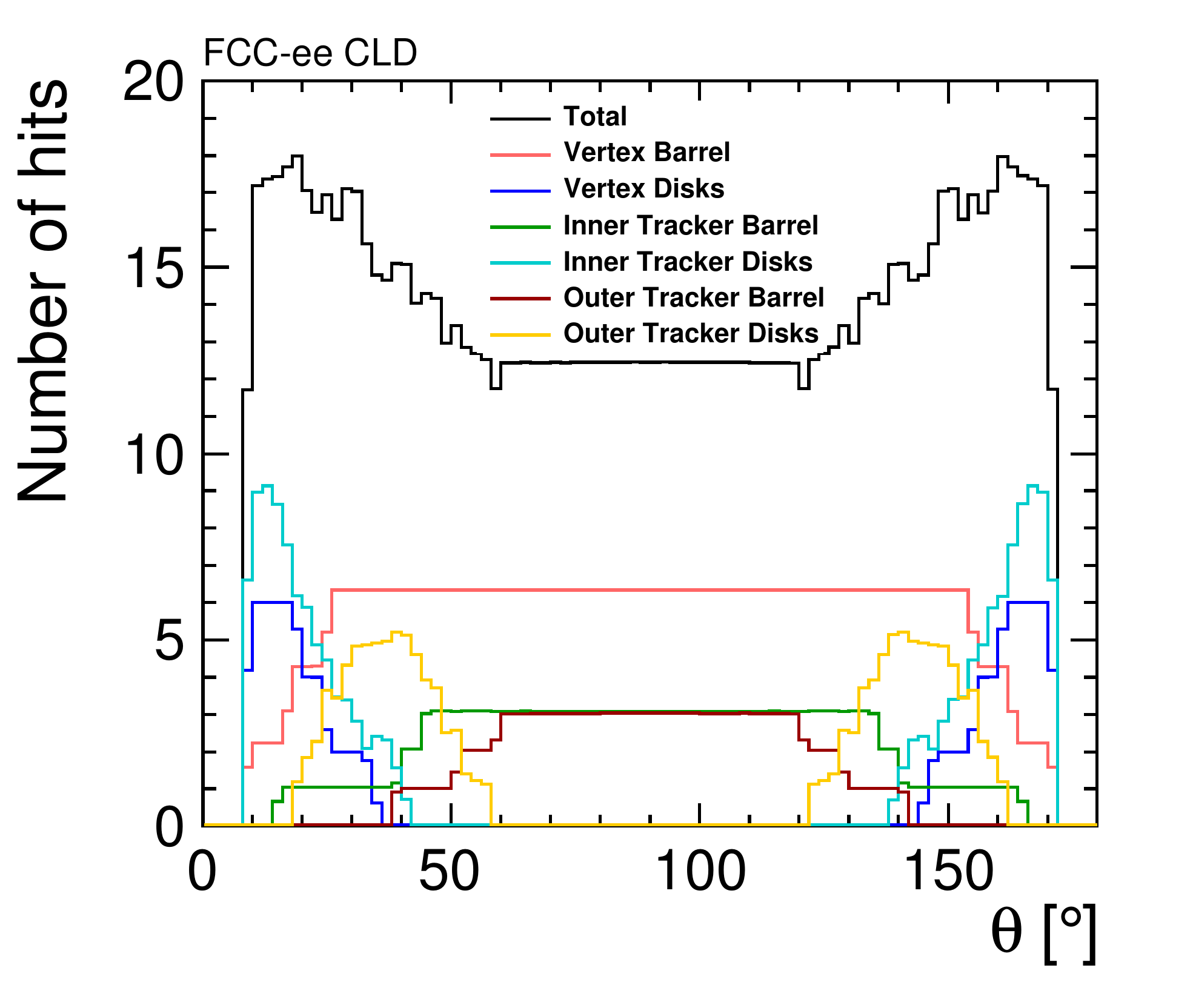}
   \caption{The coverage of the tracking systems as a function of the polar angle $\theta$. Shown is the mean number of hits (averaged over the azimuthal angle $\phi$) created by Geantinos\protect\footnotemark in full simulation. At least eight hits are measured for all tracks with a polar angle down to 8.5\degrees.}
   \label{fig:trackingcoverage}
\end{figure}

The pixel vertex detector and the silicon tracker are treated as one unified tracking system in simulation and reconstruction.
The number of expected hits in CLD as a function of polar angle $\theta$ is shown in Figure~\ref{fig:trackingcoverage}.

\footnotetext{Geantino is an artificial particle from \geant{}~which is used as a geometrical probe.}

Preliminary engineering studies have been performed for CLICdet to define the support structures and cooling systems needed for the tracker barrel layers and discs.
For the outer tracker barrel support, these studies were completed by building and testing a prototype~\cite{Fernando_outer_tracker_barrel}. 
At the present level of a conceptual design, the same concepts and material thicknesses are used for CLD.
The material budget needed in addition to the 200~\micron thick
layer of silicon  (sensors plus ASICs or monolithic structure) is estimated from CLICdet studies~\cite{sroka_klempt_2016}. 
In addition to the cylindrical main support tube, two carbon fibre structures (`interlink structures') are needed, 
to mount the inner and outer tracker barrel layers and to route connections. Preliminary sketches of these two interlink structures exist~\cite{sroka_klempt_2016}.

The building blocks from which the tracker detection layers are constructed, are modules of sensor plus ASIC. 
They are glued on one side to multi-layer carbon fibre structures acting as supports and containing the cooling.
On the other side, these modules are glued to the elements needed for connectivity.
In the tracker discs, 
modules are arranged into petals, which in turn are assembled into the full discs. 
In the inner and outer barrel, the silicon sensor size for all modules is 30~$\times$~30 mm$^2$. In the present simulation model, in the barrel an overlap between modules of 0.1\,mm is implemented in azimuthal direction 
-- there is no overlap along the detector axis.
The outer tracker discs are assembled from the same type of modules, 30~$\times$~30 mm$^2$, while the inner tracker discs are made of modules with 15~$\times$~15 mm$^2$ sensors.
In all tracker discs, a considerable overlap between petals is foreseen while modules inside the petals have no overlap.
Details of the present ideas on module support, overlaps and other engineering issues can be found in~\cite{sroka_klempt_2016} for the case of CLICdet -- the same design principles are followed for CLD.

\begin{figure}[hbtp] 
         \centering
                 \includegraphics[scale=0.8]{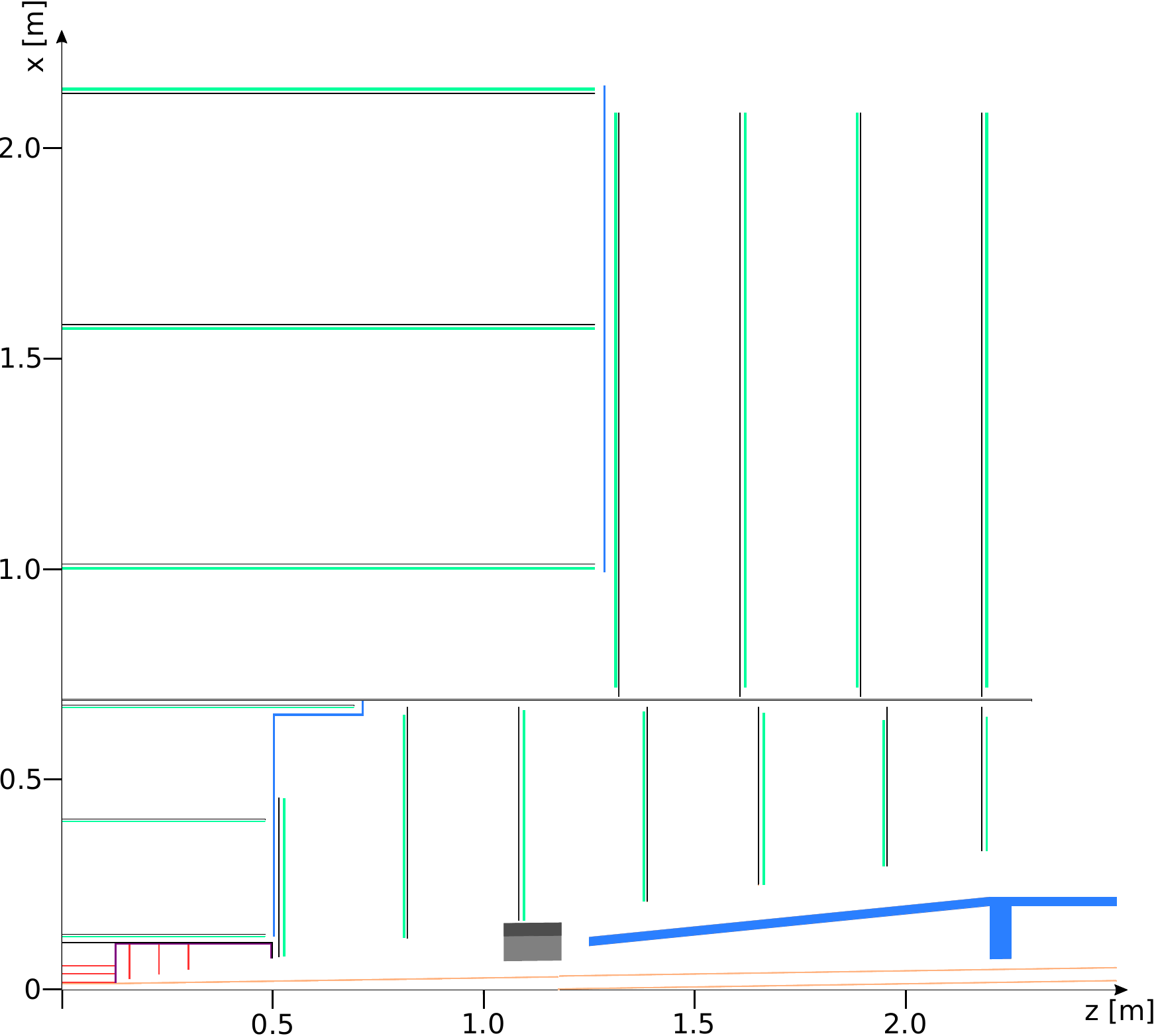}
         \caption{XZ-view of the tracker as implemented in the simulation model. 
The black lines indicate the tracker support structures including cooling and cables, and the main support tube. 
The green lines represent the tracker sensor layers.
The blue lines show the carbon-fibre interlink structures. 
The vertex detector is shown in the centre (in red). 
Cables going outwards from the vertex detector are represented in magenta. 
The LumiCal is represented by the small grey box near z~=~1.2~m. 
The conical and cylindrical structures (in blue) downstream of LumiCal represent material (steel) of the compensating and shielding solenoids. 
Also indicated is the vacuum tube (in orange). }
         \label{fig.tracker1}
\end{figure}

\begin{figure}[hbtp] 
         \centering
                 \includegraphics[scale=0.5]{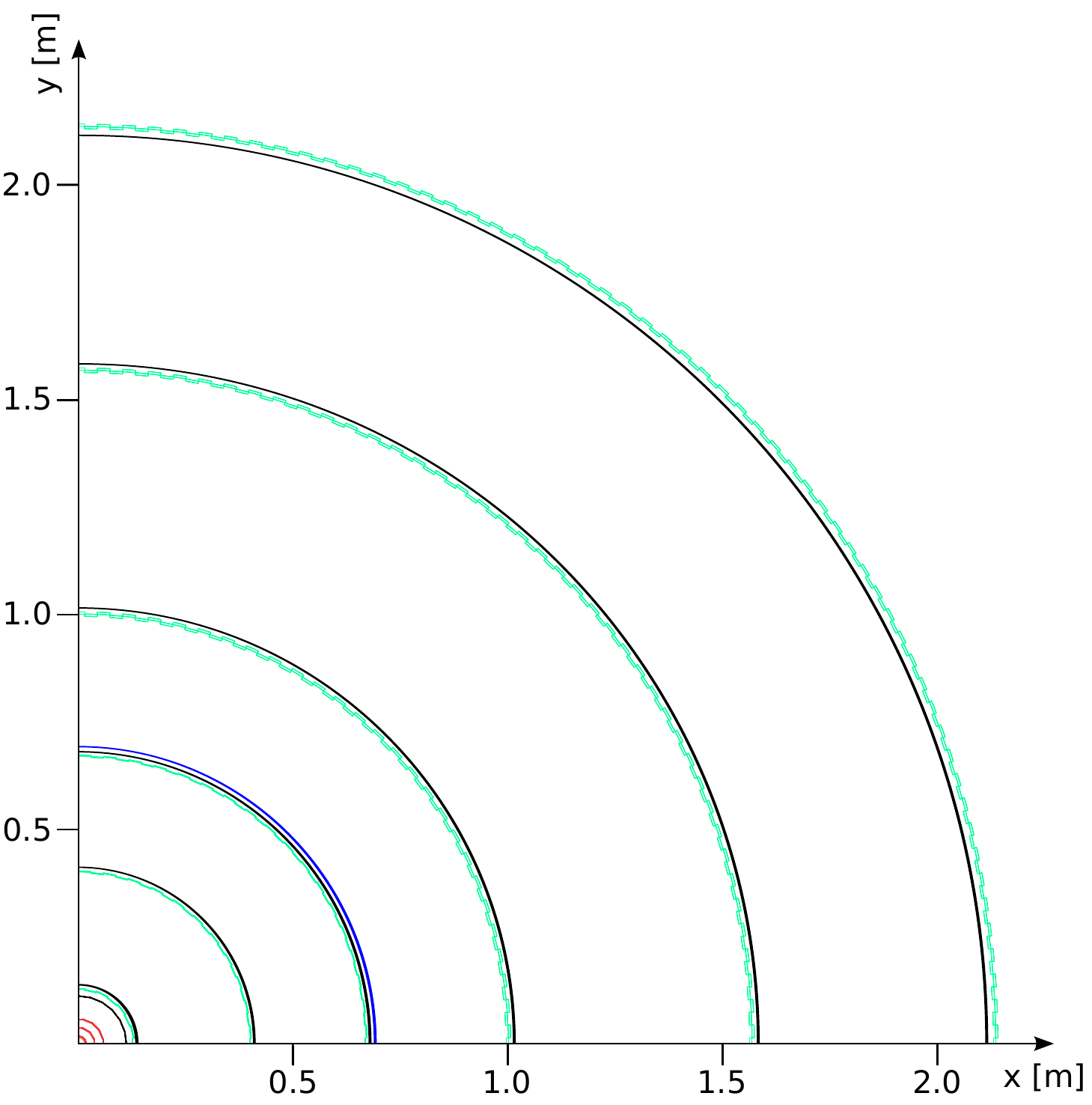}
         \caption{XY-view of the tracker barrel layers as implemented in the simulation model. The black lines indicate the support structures including cooling and cables, the green lines represent the tracker sensor layers. 
The blue line shows the main support tube. The vertex detector (in red) is shown in the centre.}
         \label{fig.tracker2}
\end{figure}

\begin{figure}[hbtp] 
         \centering
                 \includegraphics[scale=0.20]{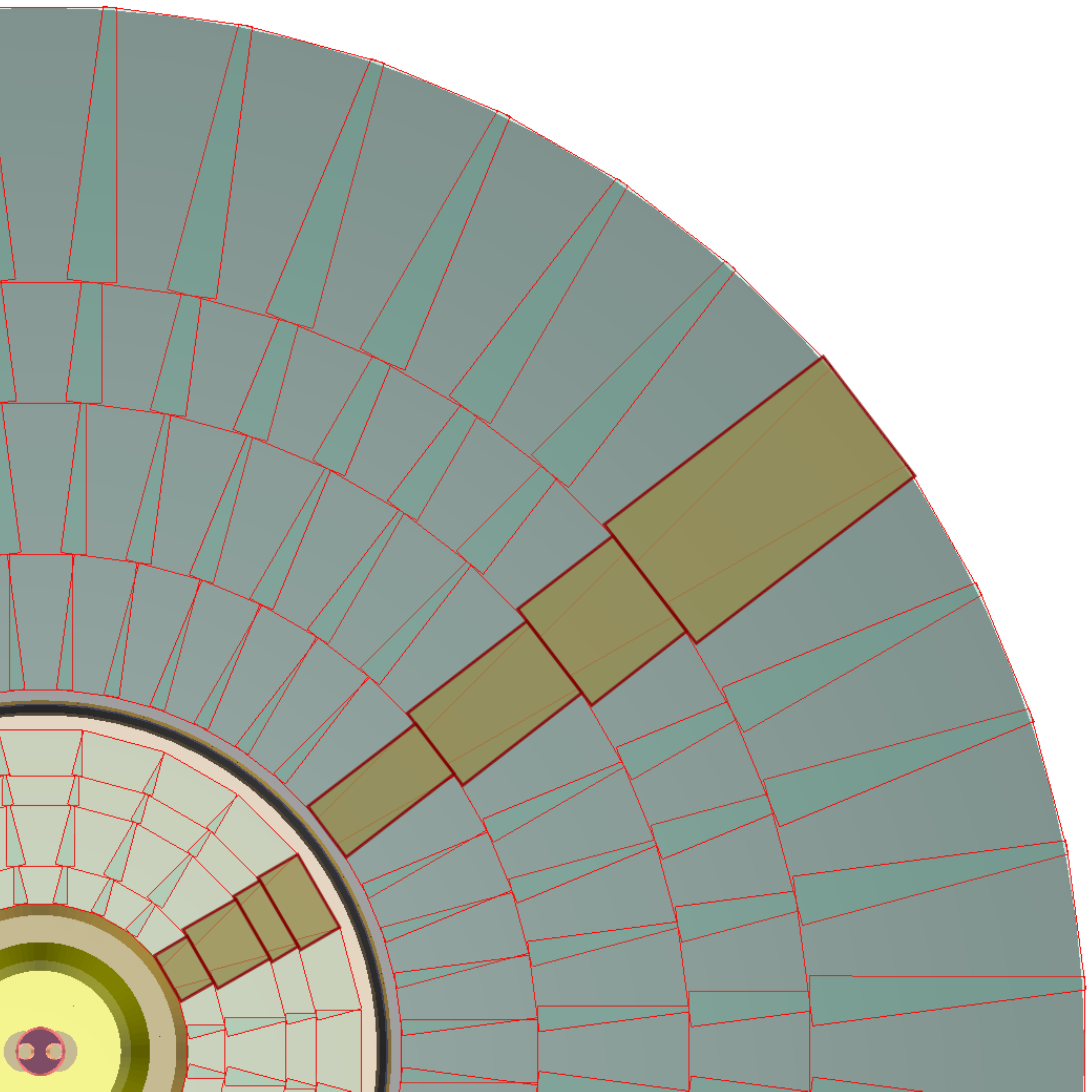}
         \caption{XY-illustration of the tracker disc implementation in the simulation model: the overlap between petals is visible as darker, smaller wedges. The petals are constructed from three or four rectangular volumes (example highlighted in yellow/brown), 
a simplification with respect to the engineering layout in~\cite{sroka_klempt_2016}. This graph is a view from z~=~1800~mm in downstream direction, thus showing the ITD6 and OTD3 inner and outer tracker discs.}
         \label{fig:forward_petals}
\end{figure}

This preliminary engineering model is implemented in the
simulation model of the tracker, with emphasis on the correct total material budget per layer (in units of \radlen).
The current implementation is shown in Figures~\ref{fig.tracker1} and~\ref{fig.tracker2}. 
Simplifications with respect to the engineering model include the use of larger rectangular surfaces instead of the small modules in the tracker disc petals, as shown in Figure~\ref{fig:forward_petals}.

The total material budget for the different tracker layers is listed in Tables~\ref{tab.tracker.barrel.mat} and ~\ref{tab.tracker.disk.mat}. 
The material budget for the modules (sensor and electronics) plus cooling and connectivity, 
is estimated to be 1.09\%~\radlen per layer for ITBs and 1.15-1.28\%~\radlen for OTBs. 
This material budget does not include the tracker support structures, made of carbon fibre components,
which differ from layer to layer and amount to 0.13\%~\radlen to 0.37\%~\radlen per layer. The CLD simulation model
includes this additional material.

\begin{figure}[hbtp]
   \centering
   \includegraphics[width=.71\linewidth]{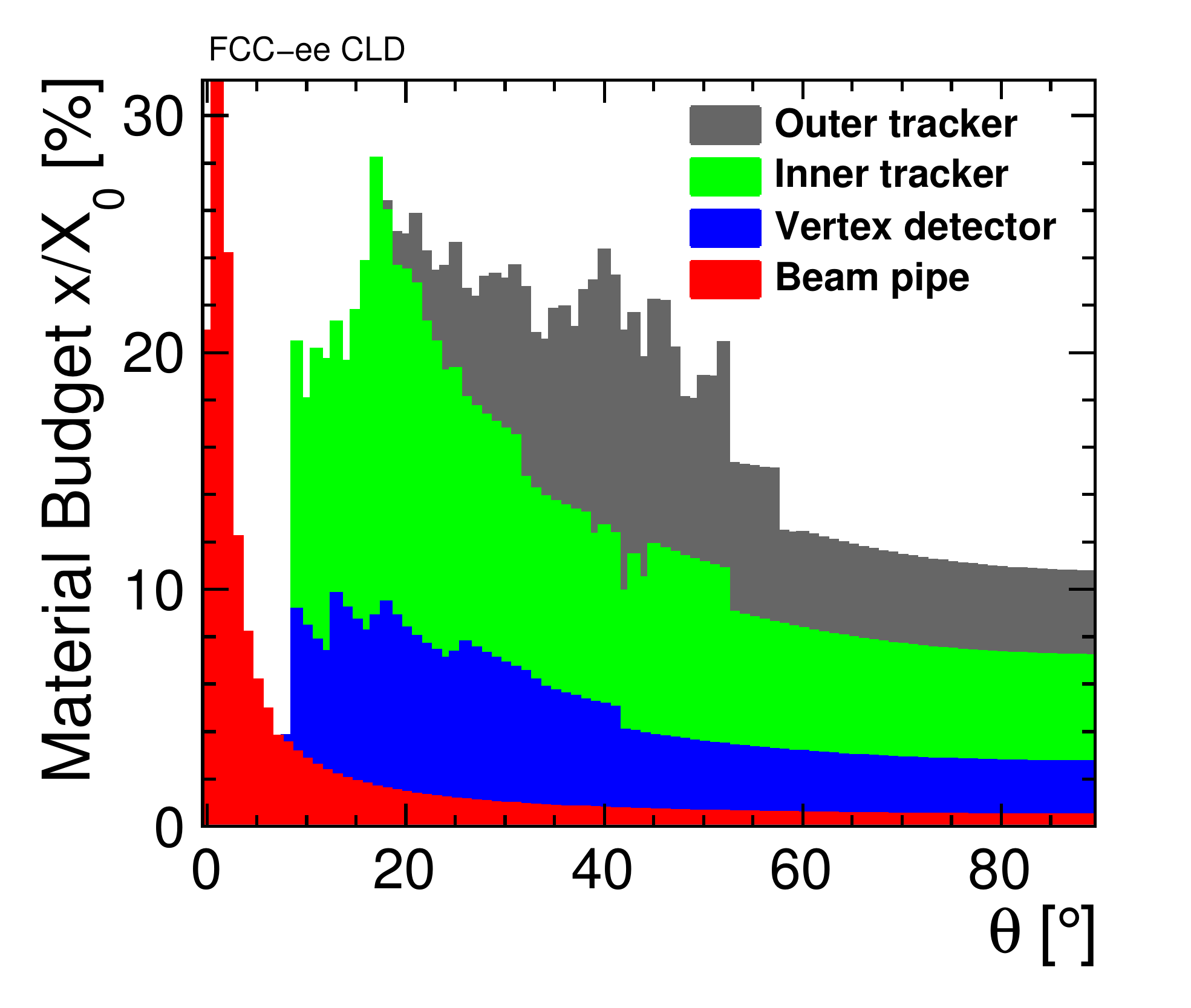}
   \caption{Stacked material budget of the different regions inside the tracking system, as a function of the polar angle and averaged over all azimuthal angles. Contributions from sensitive layers, cables, supports and cooling are included in the respective regions.
}
   \label{fig:material_budget}
\end{figure}

Details on the different contributions to the total material budget can be found in~\cite{sroka_klempt_2016}. 
In the CLD simulation model, the inner and outer interlink structures are inserted with 0.6\%~\radlen, approximately accounting for the foreseen graphite structure plus cables.
An additional support structure, for the vertex barrel layers and discs, is also inserted with  0.6\%~\radlen.
The main support tube, in its preliminary design, amounts to 1.25\%~\radlen.
The total material budget considering all elements up to the calorimeters is shown in Figure~\ref{fig:material_budget}.

\begin{table}
\parbox{.45\linewidth}{
\centering
\caption{Material budget of the tracker barrel layers -- total per barrel layer, as implemented in the simulation.}
\label{tab.tracker.barrel.mat}
\begin{tabular}{l l}
    \toprule
    Layer Name &  \radlen [\%] \\
    \midrule

ITB1 -- 3 & 1.09 \\    
\midrule
OTB1 & 1.28 \\
OTB2 -- 3 & 1.15 \\
    \bottomrule
\end{tabular}
}
\hfill
\parbox{.45\linewidth}{
\centering
\caption{Material budget of the tracker discs -- total per disc, as implemented in the simulation. The range given reflects the overlap between modules.
}
\label{tab.tracker.disk.mat}
\begin{tabular}{l l}
    \toprule
    Disc Name &  \radlen [\%] \\
    \midrule
ITD1 & 1.34 -- 1.87 \\    
ITD2 & 1.28 -- 2.13 \\
ITD3 & 1.39 -- 2.03 \\
ITD4 & 1.39 -- 1.76 \\
ITD5 & 1.39 -- 1.79 \\
ITD6 & 1.41 -- 1.75 \\
ITD7 & 1.34 -- 1.68 \\
\midrule
OTD1 -- 4 & 1.37 -- 1.91\\
    \bottomrule
\end{tabular}
}
\end{table}

\clearpage
\subsection{Beam-Induced Backgrounds in the Tracking Region}
\label{tracker_bg}

\begin{figure}[htbp]
  \centering
  \begin{subfigure}{.5\textwidth}
    \centering
    \includegraphics[width=\linewidth]{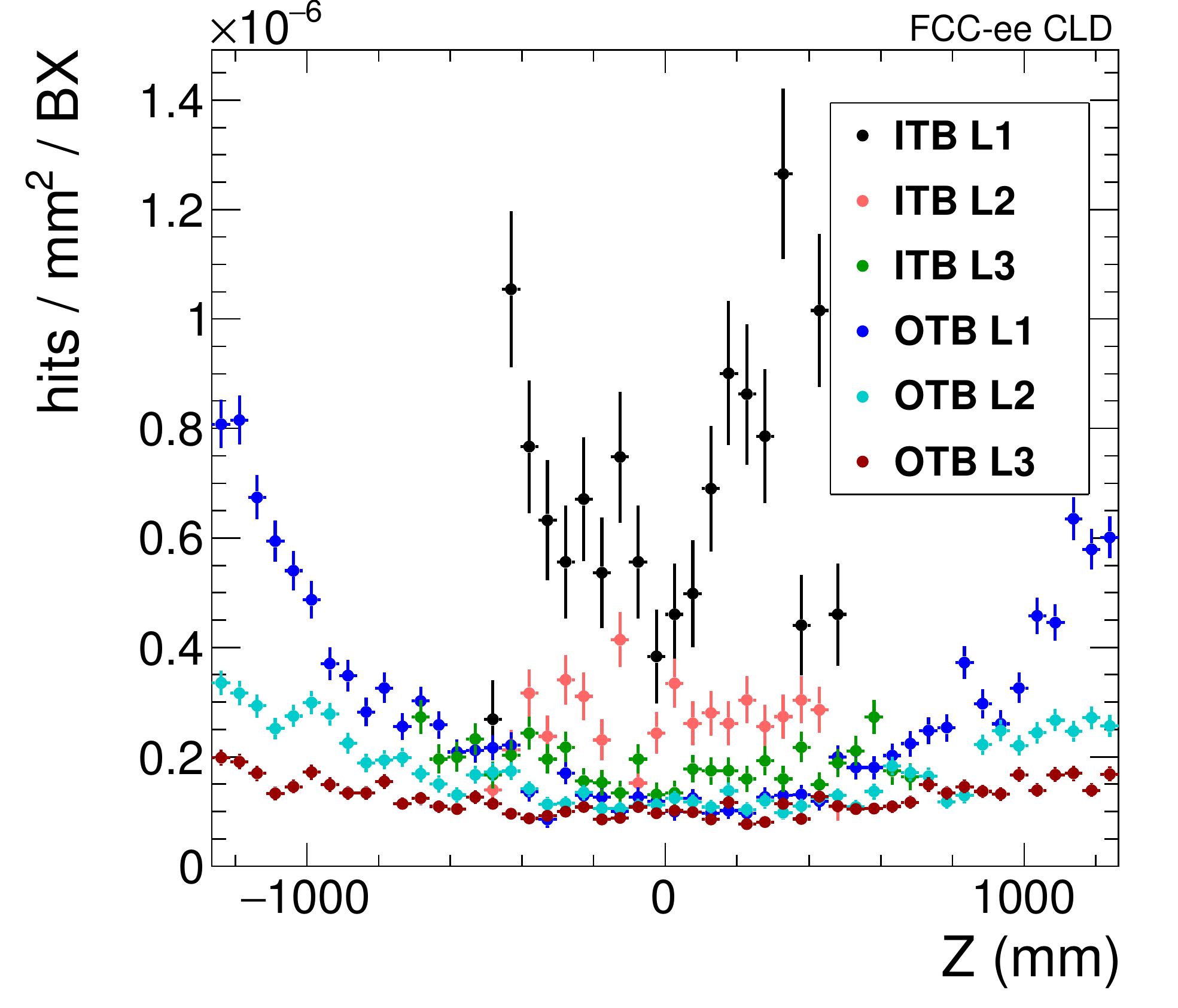}
    \caption{}
    \label{fig:TRB_pairs_91}
  \end{subfigure}%
  \begin{subfigure}{.5\textwidth}
    \centering
   \includegraphics[width=\linewidth]{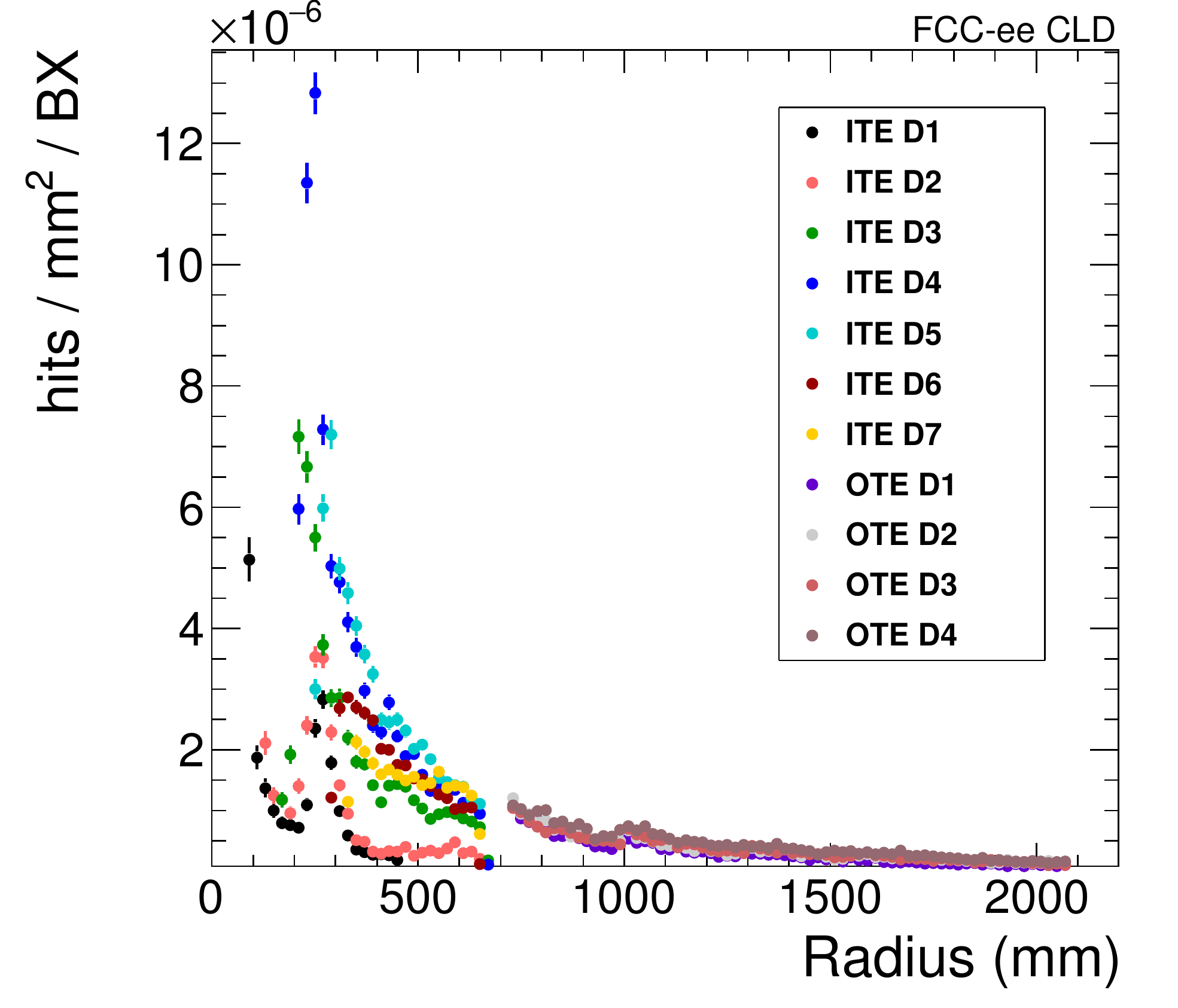}
    \caption{}
    \label{fig:TRE_pairs_91}
  \end{subfigure}
  \caption{Hit densities in the CLD tracking detector barrel layers (a) and  discs (b) for particles originating from incoherent pairs.
 Results are shown for operation at 91.2 GeV. ITB and OTB stand for inner and outer tracker barrel layers, ITE and OTE for inner and outer
tracker forward discs, respectively.
Vertical error bars show the statistical uncertainty, horizontal bars indicate the bin size.
Safety factors for the simulation uncertainties  are not included. }
  \label{fig:TR_pairs_91}
\end{figure}

\begin{figure}[b!]
  \centering
  \begin{subfigure}{.5\textwidth}
    \centering
    \includegraphics[width=\linewidth]{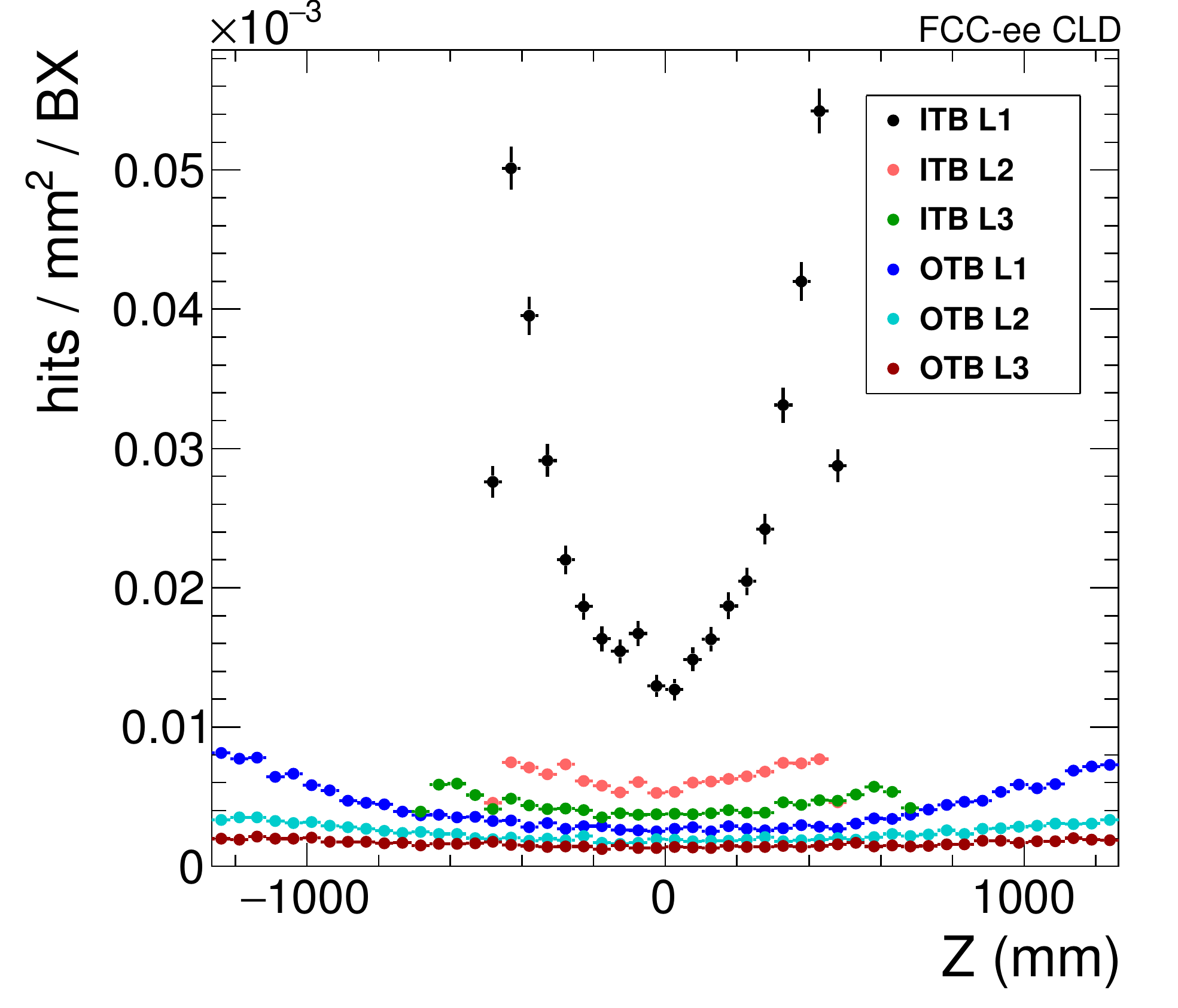}
    \caption{}
    \label{fig:TRB_pairs_365}
  \end{subfigure}%
  \begin{subfigure}{.5\textwidth}
    \centering
   \includegraphics[width=\linewidth]{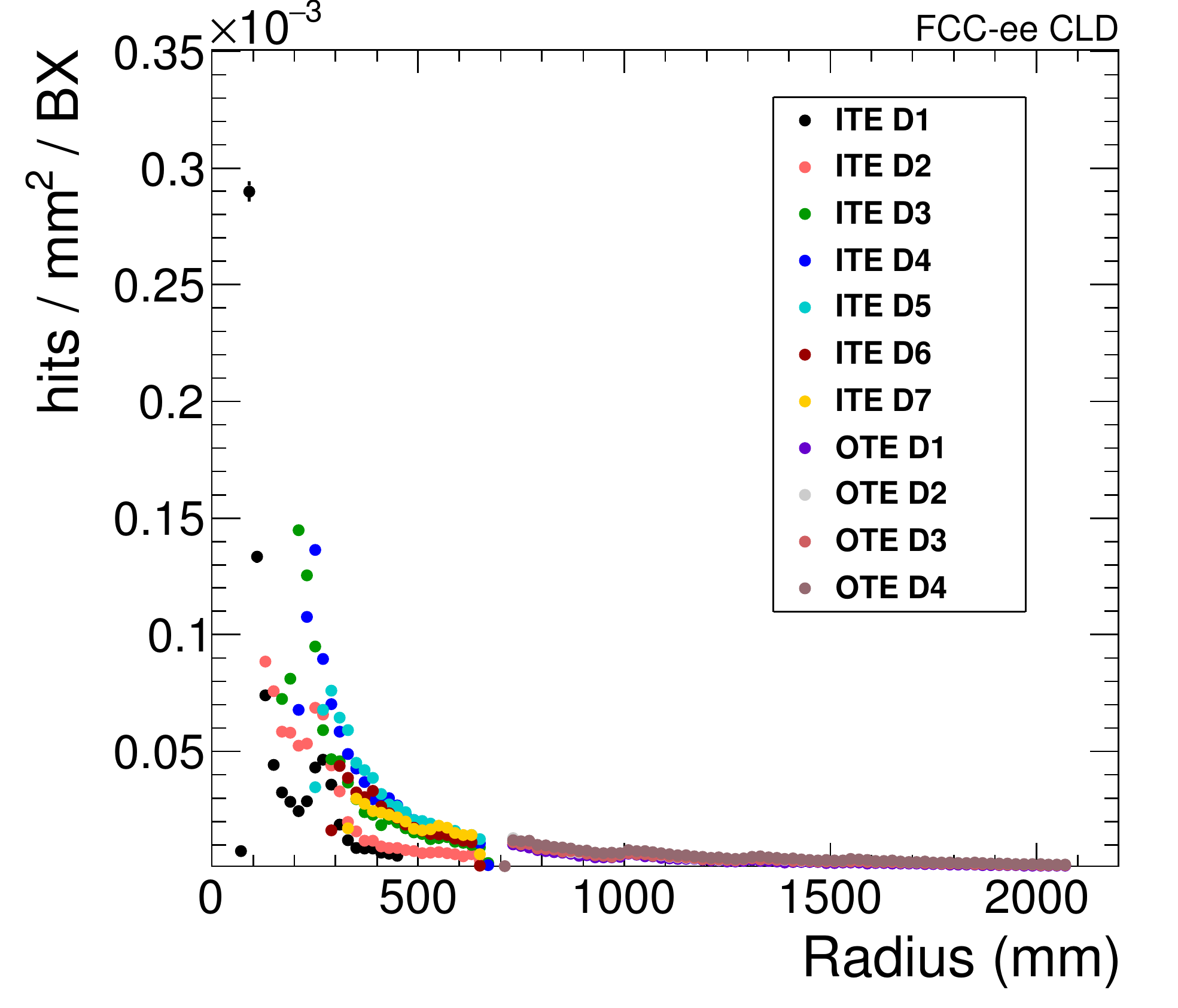}
    \caption{}
    \label{fig:TRE_pairs_365}
  \end{subfigure}
  \caption{Hit densities in the CLD tracking detector barrel layers (a) and  discs (b) for particles originating from incoherent pairs, for operation at 365 GeV. 
Vertical error bars show the statistical uncertainty, horizontal bars indicate the bin size.
Safety factors for the simulation uncertainties  are not included. }
  \label{fig:TR_pairs_365}
\end{figure}

\begin{figure}[b!]
  \centering
  \begin{subfigure}{.5\textwidth}
    \centering
    \includegraphics[width=\linewidth]{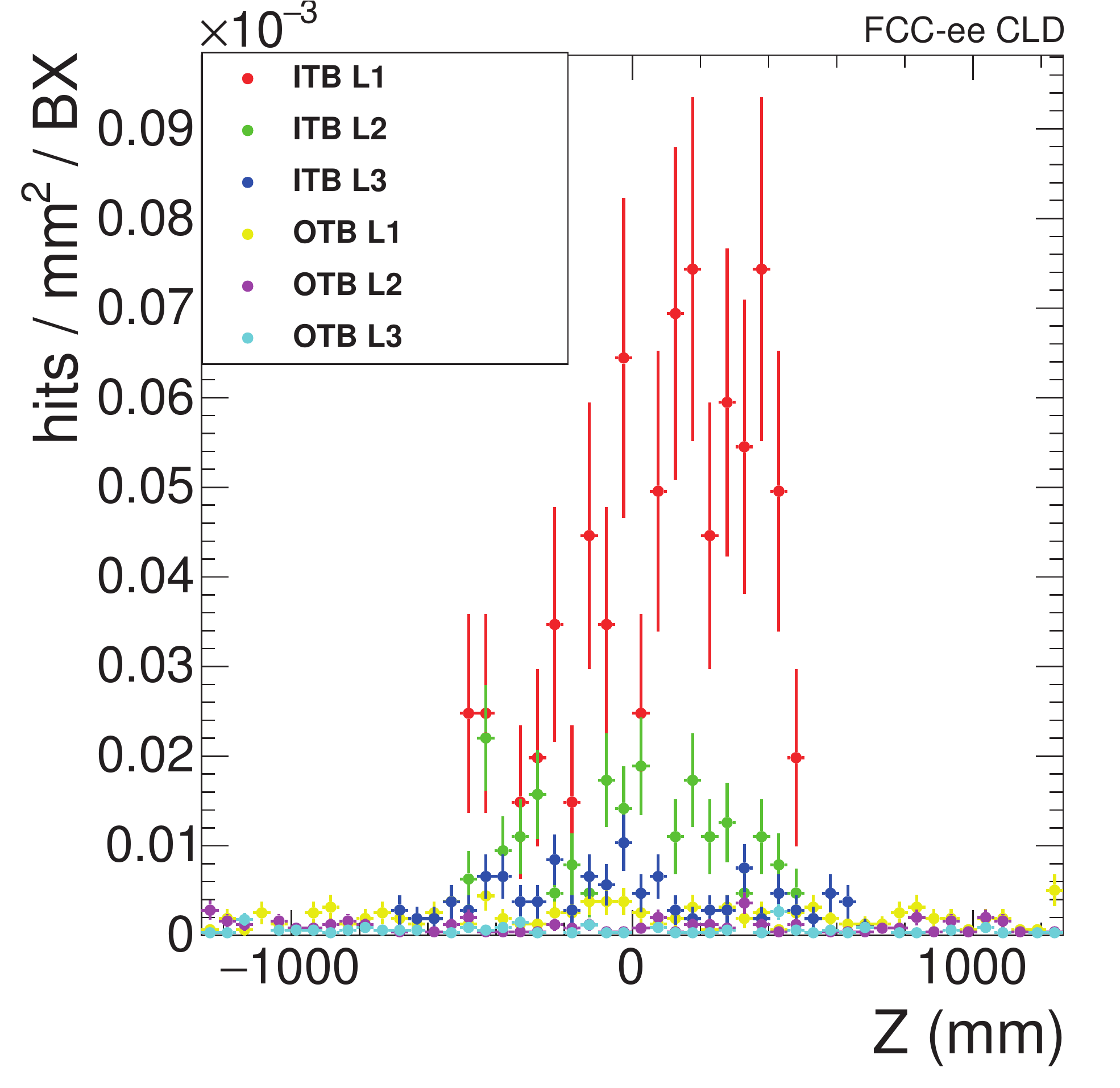}
    \caption{}
    \label{fig:TRB_sync_365}
  \end{subfigure}%
  \begin{subfigure}{.5\textwidth}
    \centering
    \includegraphics[width=\linewidth]{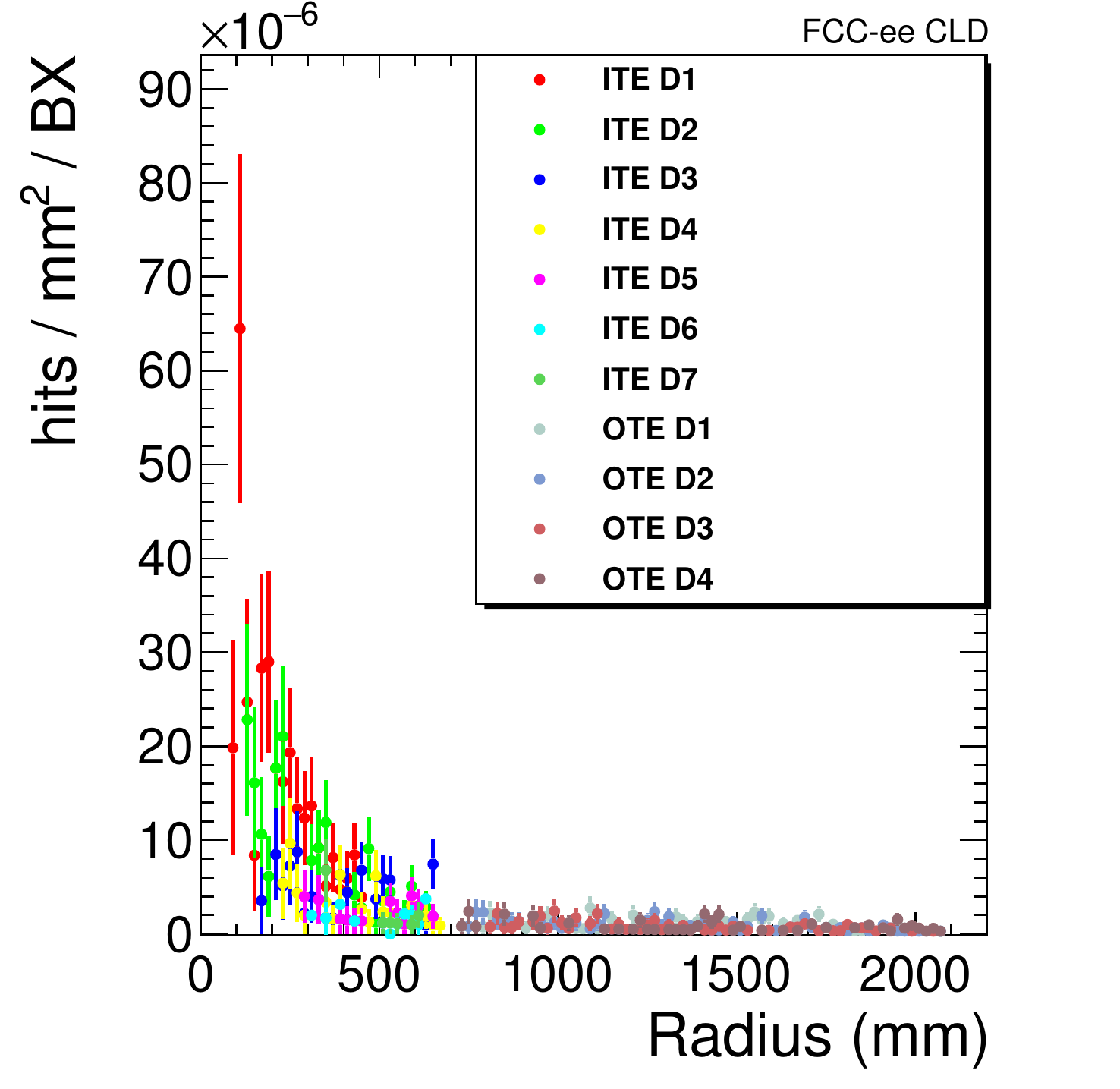}
    \caption{}
    \label{fig:TRE_sync_365}
  \end{subfigure}
  \caption{As Figure~\ref{fig:TR_pairs_365} but for hits related to synchrotron radiation photons. }
  \label{fig:TR_sync_365}
\end{figure}

The results of full detector simulation studies showing background hits from incoherent pairs in the tracker region are given in Figures~\ref{fig:TR_pairs_91} 
for operation at the Z pole, 91.2 GeV. Note that at this collision energy, no hits are observed originating from synchrotron radiation photons.
The  results for 365 GeV are given in Figures~\ref{fig:TR_pairs_365} for pairs and~\ref{fig:TR_sync_365} for synchrotron radiation. 
From these hit densities, in analogy to Section~\ref{vtx_background}, the expected peak occupancies can be obtained. 
Similarly to CLICdet, for the CLD tracker we assume that very small strip or pixel detectors with a maximum cell size of
50~\micron width and 0.3~mm length will be used. In a technical design phase, the layout details will have to be refined. 
Assuming a cluster size of 3, a safety factor of 5 and a readout window of 10 $\upmu{}$s, the occupancy at 91.2 GeV operation will be less than 1\%. 
For the same assumptions, the occupancy at 365 GeV from pairs and synchrotron radiation combined is expected to be below 0.15\%.

Note that the highest occupancies in the tracker detectors are found in a rather small region of the first two inner tracker discs. Similarly to what is described for CLICdet in~\cite{Nurnberg_Dannheim_2017},
to further reduce the occupancy these regions can be replaced by pixel detectors.

\vspace{1cm}
\subsection{Technology Choices and Cooling}

The sensors of the ALICE ITS upgrade tracker technology~\cite{ALICE_ITS} appear to be a suitable choice for the CLD tracker. 
The integration time window of the chip of 10 $\upmu{}$s does not appear to give rise to too high occupancies,
and the power dissipation of 40 mW/cm$^2$ is low.
A leak-less de-mineralised water cooling system is used.
The average material budget in the ALICE ITS outer tracker is 0.8\% \radlen per layer, with twelve peaks of 1.2\% and 1.4\% \radlen in the azimuthal distribution.
The material budget assumed for the CLD tracking detector layers varies from 1\% to 2\% \radlen.

Since the tracker outer radius (2.1~m) and surface area in CLD (195 m$^2$) is much larger than the one of the ALICE ITS upgrade (0.4~m and 9.4 m$^2$), a number of engineering issues will have to be investigated. 
Not least, the total heat load in the tracker will be around 180 kW and an adequate cooling infrastructure to reach the tracker elements, 
which are distributed over a large volume, will be needed.

\clearpage

\section{Calorimetry}
\label{calo}

\subsection{Introduction}

Extensive studies in the context of ILC and CLIC have revealed that high granularity particle flow calorimetry appears to be a promising option to reach the required jet energy resolution of 3--4\%. Such a performance is necessary to allow the distinction e.g. of W and Z bosons on an event-by-event basis.

In contrast to a purely calorimetric measurement, particle flow calorimetry requires the reconstruction of the four-vectors of all visible particles in an event. The momenta of charged particles
(about 60\% of the jet energy) are measured in the tracking detectors. Photons (about 30\% of the jet energy) and neutral hadrons are measured in the electromagnetic and hadronic calorimeter, respectively. An overview of particle flow and the PandoraPFA software can be found in~\cite{Thomson:2009rp, Pandora_2012, Marshall:2015rfaPandoraSDK, Marshall:2012ryPandoraPFA}. Experimental tests of particle flow calorimetry are described in detail in~\cite{sefkow_review}.
A recent report provides updates on results obtained by the CALICE collaboration~\cite{DET_RD_for_CLIC}.

A precise measurement of hit times in the calorimeters can be used for the association to a bunch crossing and for background rejection. Following the studies in CALICE and for CLICdet, with calorimeters very similar to the ones in CLD,
a hit time resolution of a few nanoseconds should be achievable. As shown in section~\ref{jet_energy_res}, the beam-induced backgrounds appear to have a limited impact on the performance, even without using timing cuts - these might be found to be
needed in future studies of physics processes.

\subsection{Electromagnetic Calorimeter}
\label{ecal}
 
The segmentation of the ECAL has to be sufficient to resolve energy depositions from nearby particles in high energy jets. Studies performed in the context of the ILC and CLIC suggest a calorimeter transverse segmentation of  5~$\times$~5~mm$^2$.~\footnote{For reasons dating back to the former MOKKA drivers, the software implementation of the ECAL shows cells of  5.1~$\times$~5.1~mm$^2$.}
The technology chosen as baseline option for the detectors at the linear colliders is a silicon-tungsten sandwich structure. To limit the leakage beyond the ECAL, a total depth of around 22-23~\radlen is chosen. 

To investigate the ECAL performance for different longitudinal sampling options, a series of full simulation studies was performed for the CLICdet study~\cite{CLICdet_note_2017}.
As a result, a longitudinal segmentation with 40 identical Si-W layers (using 1.9~mm thick W plates) was found to give the best photon energy resolution over a wide energy range.
This detector design is also implemented in the CLD simulation model.

The overall dimensions of the ECAL are given in Table~\ref{tab:ECAL_layout}. Note that in a forthcoming version of the CLD design the forward acceptance (i.e. the parameter ECAL endcap $r_{\min}$) can be reduced
according to the latest MDI layout (a cone of 100 mrad instead of 150 mrad cone should be reserved for accelerator elements).

\begin{figure}[hbtp] 
         \centering
                 \includegraphics[scale=0.5]{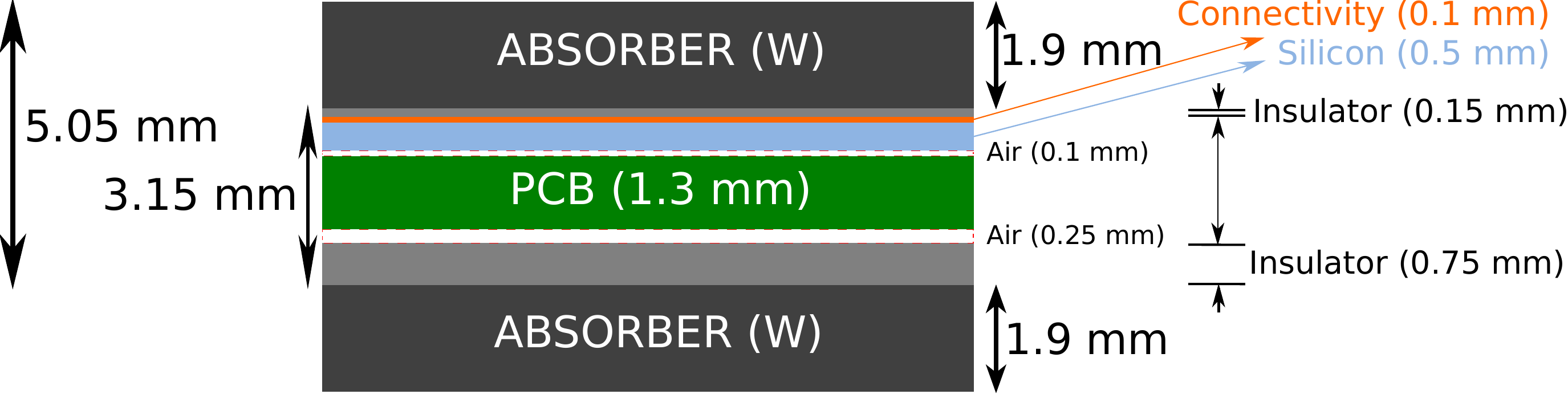}
         \caption{Schematic drawing of the ECAL segmentation as implemented in the simulation model.}
         \label{fig.ecal.layer.structure}
\end{figure}

\begin{figure}[hbtp] 
         \centering
\includegraphics[width=0.85\textwidth]{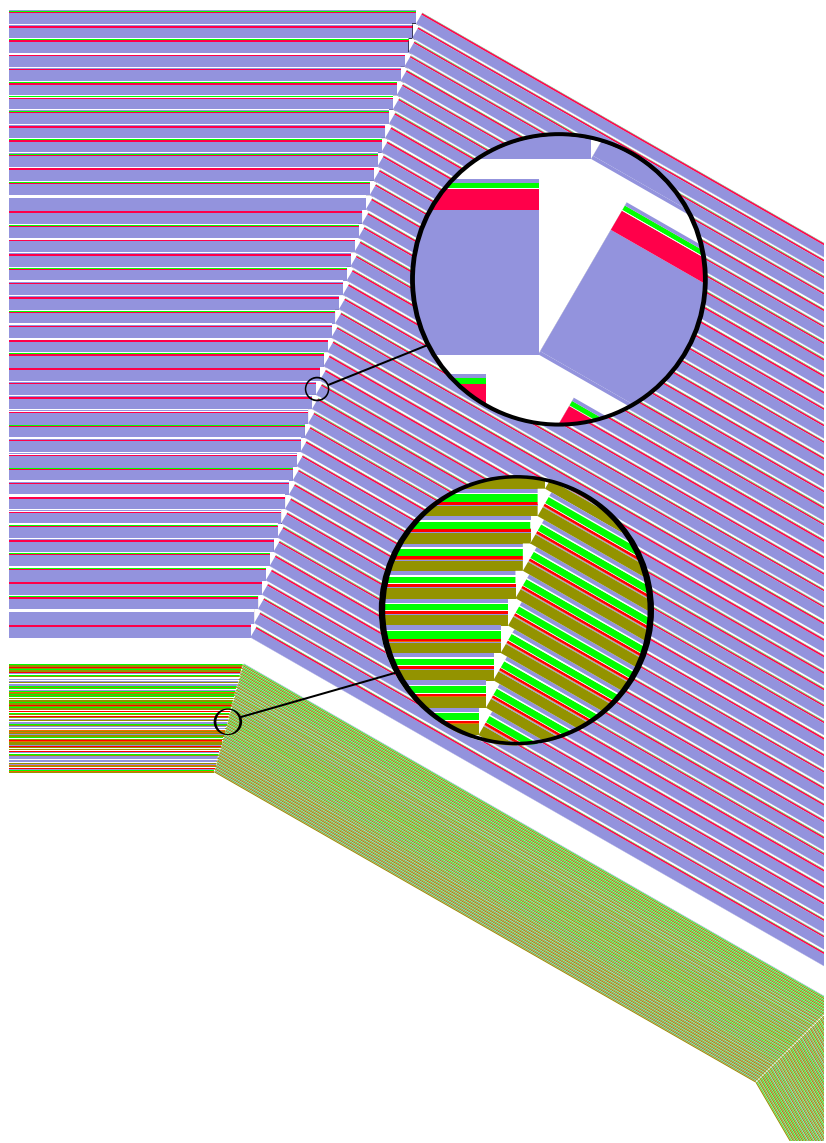}

         \caption{Implementation of ECAL and HCAL in the simulation model (the reader may need to zoom in to see all the details). The region of a junction between two sectors (of the dodecagon) in the barrel region is shown. In the ECAL, the olive-green regions indicate the tungsten layers, while the red regions symbolise the silicon sensors. Purple layers are G10 insulator, green is PCB and connectivity, and white is air. In the HCAL, blue regions indicate the steel layers (with thin steel sheets for the cassette), red stands for the scintillator, while green and white are PCB and air as in the ECAL. }
         \label{fig:ecal_impl}
\end{figure}

The detailed ECAL layer stack as implemented in the simulation model is shown in Figure~\ref{fig.ecal.layer.structure} and is given in Table~\ref{tab:ECAL}.
A distance of 3.15~mm between W plates is chosen to accommodate sensors and readout, in analogy to CLICdet and ILD at ILC.
Note that the ECAL starts with an absorber layer, followed by a sensor/electronics layer, and so on. The last element in the ECAL stack is a sensor/electronics layer.
A section of the ECAL barrel as implemented in the simulations is shown in Figure~\ref{fig:ecal_impl}.

\begin{table}[hbtp]
   \centering
   \caption{ECAL layout as implemented in the simulation model (dimensions in mm). Further details on channel numbers and silicon sensor areas are given in Appendix~\ref{sec:Appendix_I}.}
   \label{tab:ECAL_layout}
   \begin{tabular}{l r }\toprule
 ECAL barrel $r_{\min}$ & 2150 \\
 ECAL barrel $r_{\max}$  & 2352 \\
ECAL barrel $z_{\max}$  & 2210 \\
ECAL endcap $z_{\min}$  & 2307 \\
ECAL endcap $z_{\max}$  & 2509 \\
ECAL endcap $r_{\min}$  &340 \\
ECAL endcap $r_{\max}$  & 2455 \\
     \bottomrule
   \end{tabular}
\end{table}

\begin{table}[hbtp]
   \centering
   \caption{Parameters for the ECAL layer stack as implemented in the simulation model, with a total of 40 SiW layers.}
   \label{tab:ECAL}
   \begin{tabular}{l c c}\toprule
                Function &  Material  &  Layer thickness [mm] \\
\midrule
Absorber & tungsten alloy & 1.90 \\
\midrule
Insulator & G10 & 0.15 \\
Connectivity &  mixed (86\% Cu) & 0.10 \\
Sensor & silicon & 0.50 \\
Space & air & 0.10 \\
PCB & mixed (82\% Cu) & 1.30 \\
Space & air & 0.25 \\
Insulator & G10 & 0.75 \\
\midrule
Total between W plates & & 3.15 \\
\midrule
Total SiW layer& & 5.05 \\
     \bottomrule
   \end{tabular}
\end{table}

\clearpage

\clearpage

\subsection{Hadronic Calorimeter}
\label{hcal}

\begin{figure}[hbtp] 
         \centering
                 \includegraphics[scale=0.45]{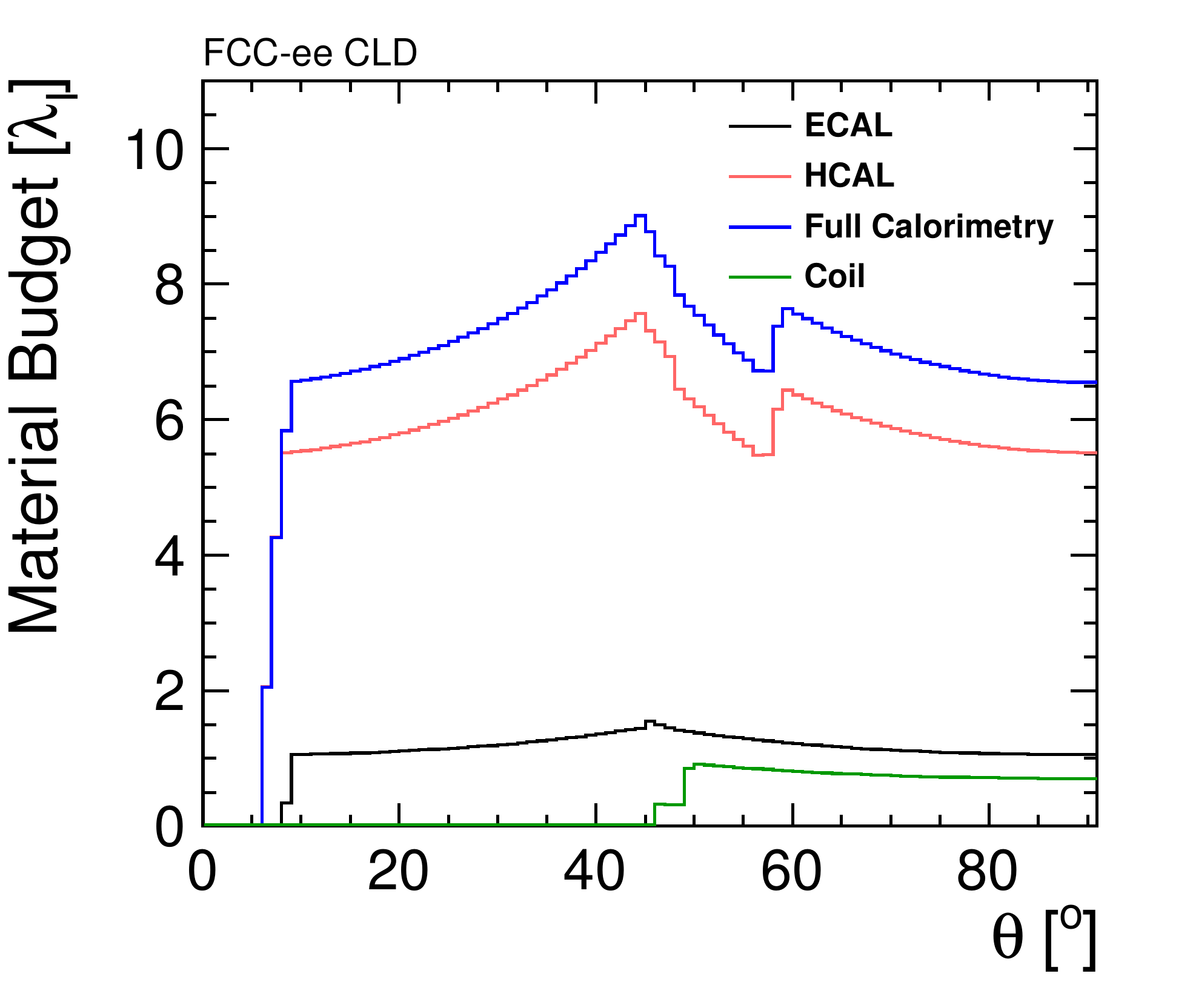}
         \caption{Nuclear interaction lengths \nuclen in the calorimeters as a function of the polar angle $\theta$. The interaction length corresponding to the material of the superconducting coil is shown for completeness.}
         \label{fig.hcal.lambda}
\end{figure}

Detailed optimisation studies have been performed for the HCAL foreseen in the detectors at ILC and CLIC.
Details of recent studies for CLICdet are described in~\cite{thesis_steven_green} and~\cite{hcal_note_2015}.
The proposed hadronic calorimeter of CLD has a structure and granularity as the one in CLICdet. 
It consists of steel absorber plates, each of them 19 mm thick,  interleaved with scintillator tiles,
similar to the CALICE calorimeter design for the ILD detector at ILC~\cite{ilc_det_tdr}.
The gap for the sensitive layers and their cassette is 7.5~mm. 
The polystyrene scintillator in the cassette is 3~mm thick with a
tile size of $30\times30~\mathrm{mm}^{2}$. Analogue readout of the tiles with SiPMs is envisaged.  
The HCAL consists of 44 layers and thus is around 5.5~\nuclen deep, which brings the
combined thickness of ECAL and HCAL to 6.5~\nuclen (see \autoref{fig.hcal.lambda}).
In the studies performed for the ILD detector at ILC (500 GeV), this depth of the calorimetry for hadrons was found to be sufficient.
The overall dimensions of the HCAL are summarised in Table~\ref{tab:HCAL_layout}. 
In the simulations, the part of the HCAL endcap which surrounds the ECAL endcap (see Figure~\ref{fig.quarter.view})
 is treated as a separate entity called the "HCAL ring". 
Note that, in analogy to the ECAL, in a forthcoming version of the CLD design the forward acceptance can be improved (i.e. the parameter HCAL endcap $r_{\min}$ reduced).

\begin{figure}[hbtp] 
         \centering
                 \includegraphics[scale=0.5]{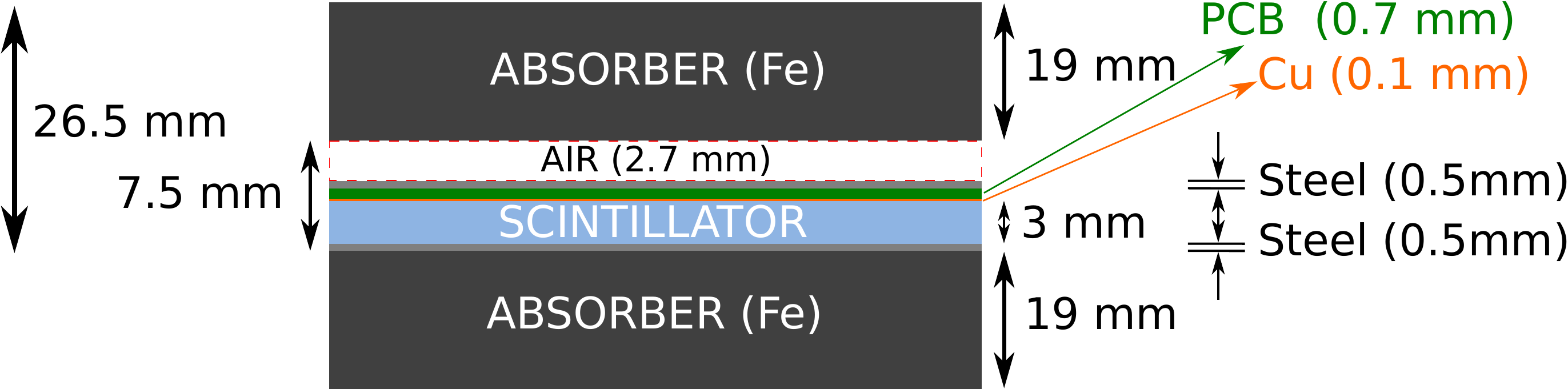}
         \caption{Schematic drawing of HCAL segmentation as implemented in the simulation model.}
         \label{fig.hcal.layer.structure}
\end{figure}

The detailed HCAL layer stack as implemented in the simulation model is shown in Figure~\ref{fig.hcal.layer.structure} and is given in Table~\ref{tab:HCAL}.

A section of the HCAL barrel as implemented 
in the simulations is shown in Figure~\ref{fig:ecal_impl}.

\begin{table}[hbtp]
   \centering
   \caption{HCAL overall layout  as implemented in the simulation model (dimensions in mm). Further details on channel numbers and silicon sensor areas are given in Appendix~\ref{sec:Appendix_I}.}
   \label{tab:HCAL_layout}
   \begin{tabular}{l r }\toprule
HCAL barrel $r_{\min}$  & 2400 \\
HCAL barrel $r_{\max}$  & 3566 \\
HCAL barrel $z_{\max}$  & 2210 \\
HCAL endcap $z_{\min}$  & 2539 \\
HCAL endcap $z_{\max}$  & 3705 \\
HCAL endcap $r_{\min}$  & 340 \\
HCAL endcap $r_{\max}$  & 3566 \\
HCAL ring $z_{\min}$  & 2353.5 \\
HCAL ring $z_{\max}$ & 2539 \\
HCAL ring $r_{\min}$ & 2475 \\
HCAL ring $r_{\max}$  & 3566 \\
     \bottomrule
   \end{tabular}
\end{table}
\begin{table}[hbtp]
   \centering
   \caption{Parameters for the HCAL layer stack as implemented in the simulation model, with a total of 44 Fe-Scintillator layers.}
   \label{tab:HCAL}
   \begin{tabular}{l c c}\toprule
                Function &  Material  &  Layer thickness [mm] \\
\midrule
Absorber & steel & 19 \\
\midrule
Space & air & 2.7\\
Cassette & steel & 0.5 \\
PCB & mixed & 0.7 \\
Conductor & Cu & 0.1 \\
Scintillator & polystyrene & 3 \\
Cassette & steel & 0.5 \\
\midrule
Total between steel plates & & 7.5 \\
\midrule
Total Fe-scintillator layer& & 26.5 \\
     \bottomrule
   \end{tabular}
\end{table}

\clearpage
\subsection{Beam-Induced Backgrounds in the Calorimeter Region}
\label{calorimeter_bg}

\begin{figure}[htbp]
  \centering
  \begin{subfigure}{.5\textwidth}
    \centering
    \includegraphics[width=\linewidth]{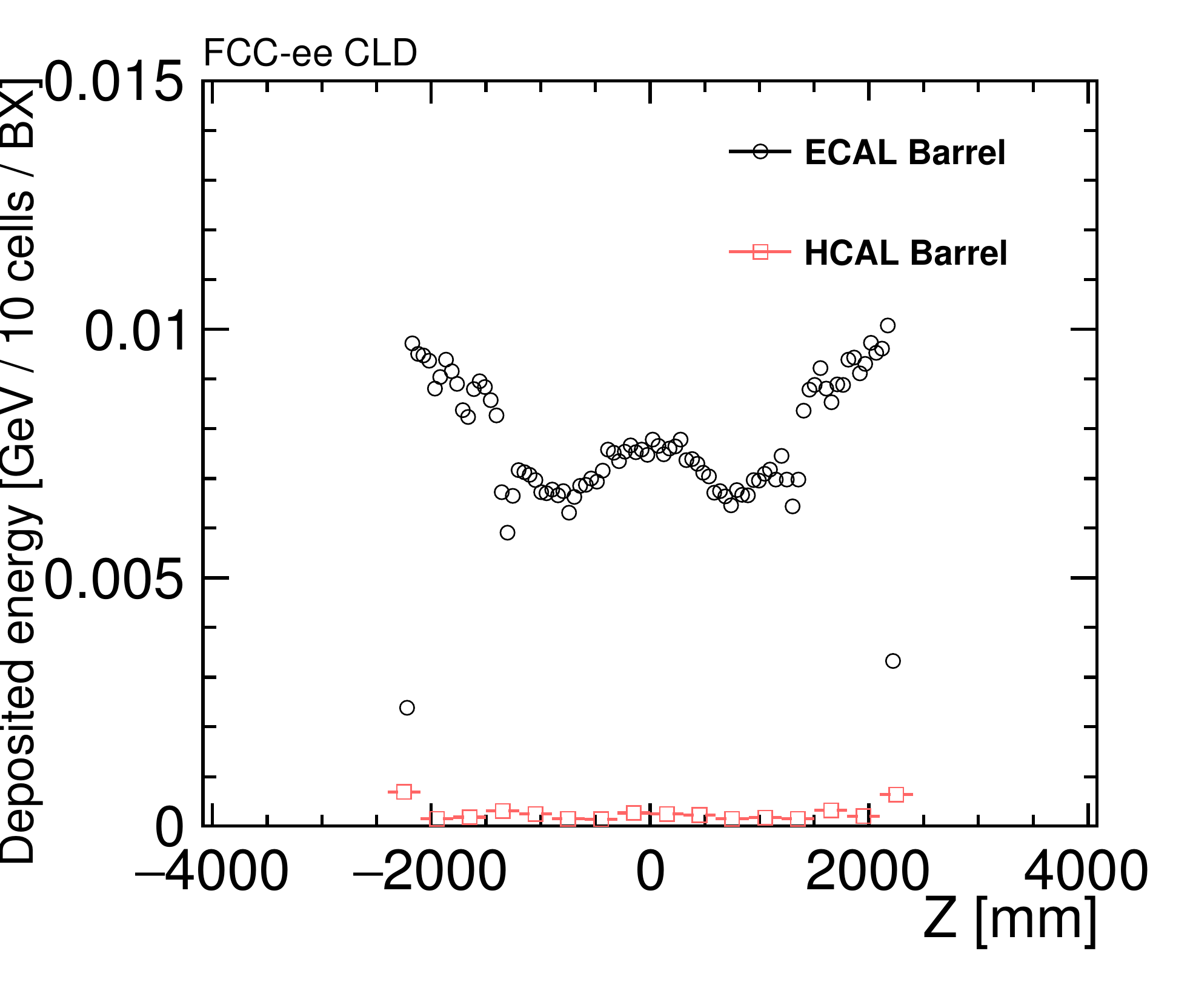}
    \caption{}
    \label{fig:caloBarrel_bg_pairs_91}
  \end{subfigure}%
  \begin{subfigure}{.5\textwidth}
    \centering
   \includegraphics[width=\linewidth]{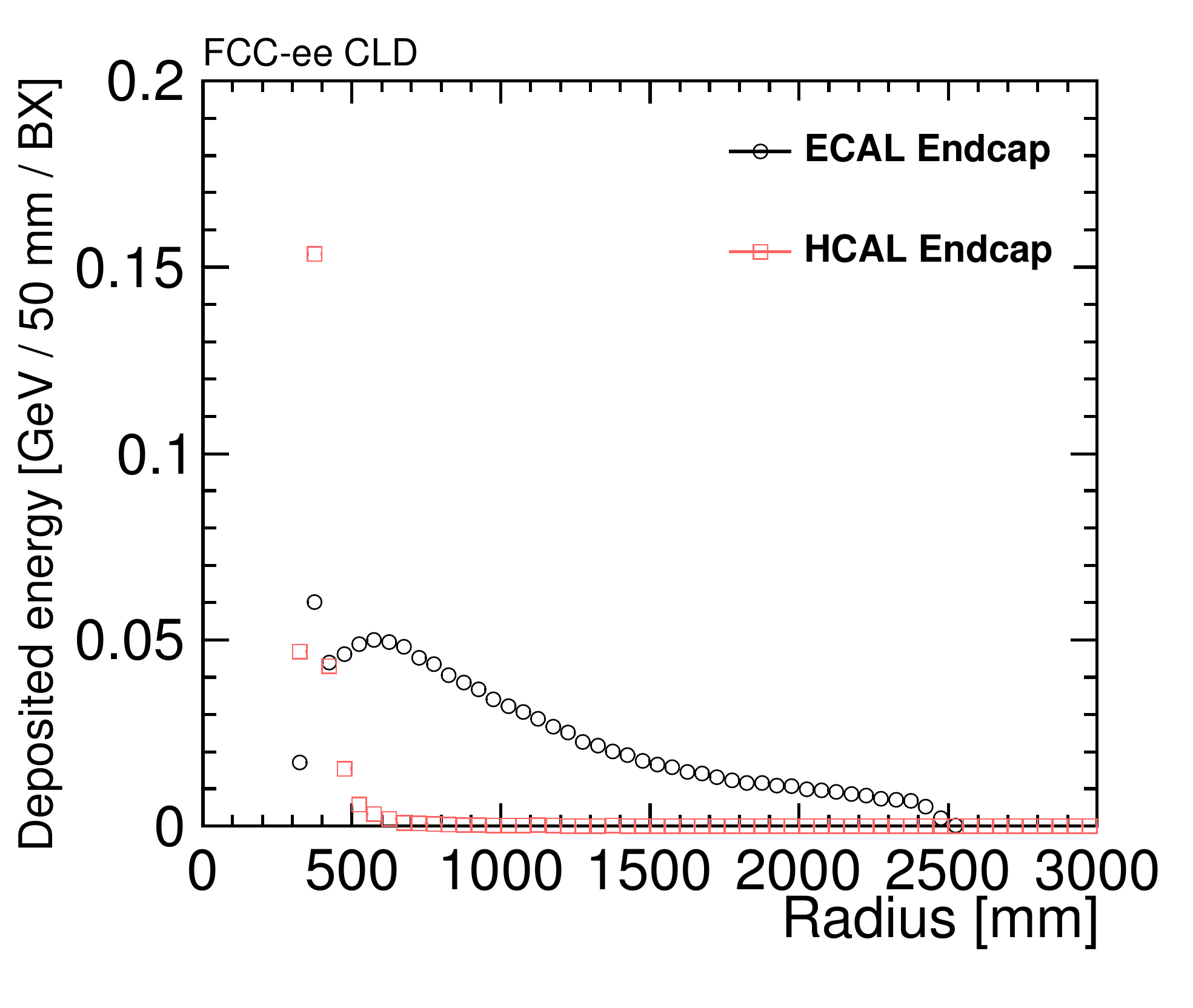}
    \caption{}
    \label{fig:caloEndcap_bg_pairs_91}
  \end{subfigure}
  \caption{Calibrated deposited energy in the CLD calorimeter barrel part (a) and endcap (b) for particles originating from incoherent pairs.
 Results are shown for operation at 91.2 GeV. 
Horizontal error bars indicate the bin size. Statistical errors are smaller than the size of the symbols in the figure.
Note that in (a) the bin size is 10 cells, i.e. 51~mm for ECAL and 300 mm for HCAL.
Safety factors for  simulation uncertainties are not included. }
  \label{fig:calo_bg_pairs_91}
\end{figure}

\begin{figure}[htbp]
  \centering
  \begin{subfigure}{.5\textwidth}
    \centering
    \includegraphics[width=\linewidth]{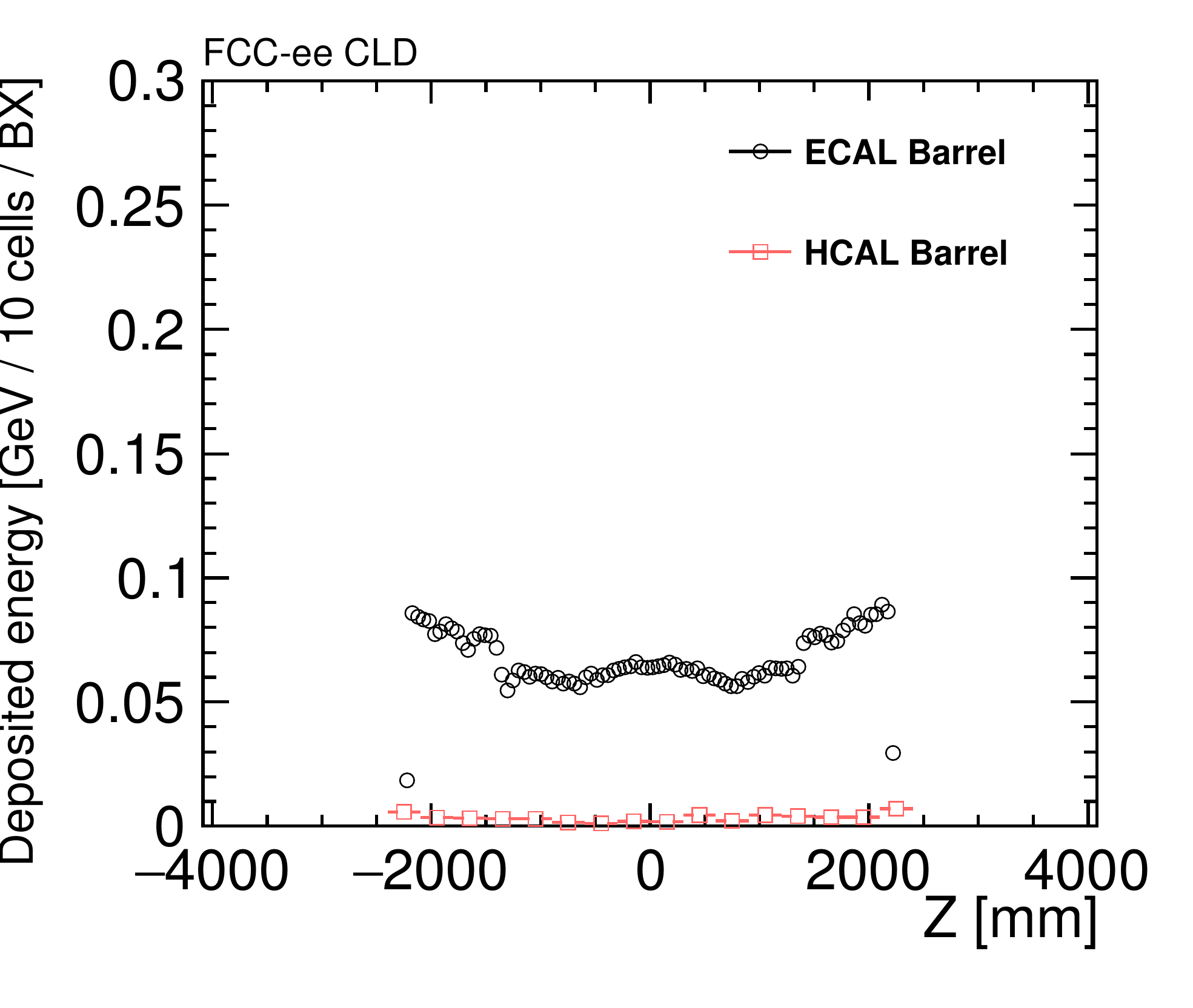}
    \caption{}
    \label{fig:caloBarrel_bg_pairs_365}
  \end{subfigure}%
  \begin{subfigure}{.5\textwidth}
    \centering
   \includegraphics[width=\linewidth]{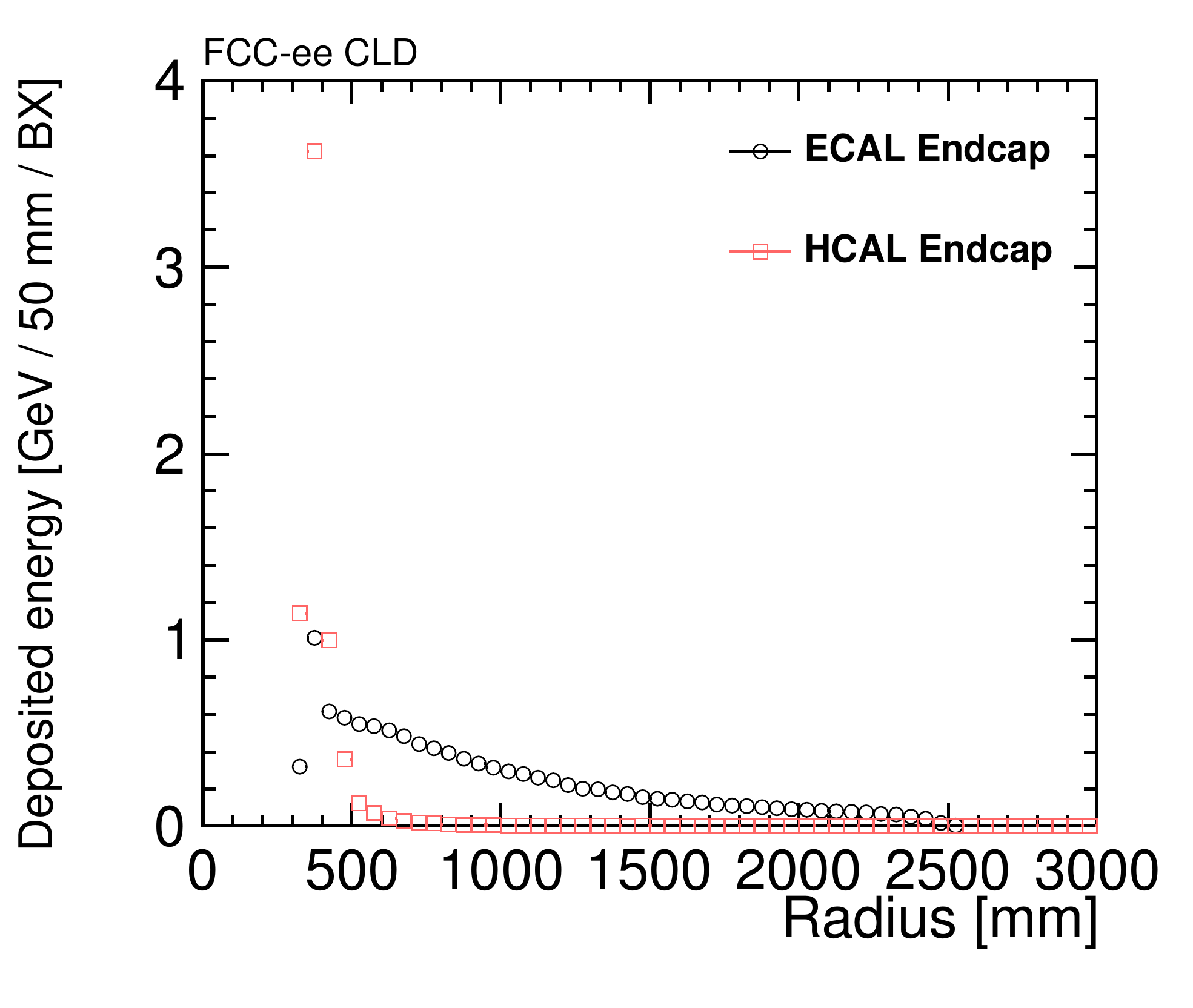}
    \caption{}
    \label{fig:caloEndcap_bg_pairs_365}
  \end{subfigure}
  \caption{Calibrated deposited energy in the CLD calorimeter barrel part (a) and endcap (b) for particles originating from incoherent pairs.
 Results are shown for operation at 365 GeV.
Horizontal error bars indicate the bin size. Statistical errors are smaller than the size of the symbols in the figure.
Note that in (a) the bin size is 10 cells, i.e. 51~mm for ECAL and 300 mm for HCAL.
Safety factors for simulation uncertainties are not included. }
  \label{fig:calo_bg_pairs_365}
\end{figure}

Full detector simulation studies similar to the ones described in Chapters~\ref{vtx_background} and~\ref{tracker_bg} have been performed to estimate the effect of the incoherent \epem pair background in the calorimeters. 
The total deposited energy per bunch crossing scaled with the calibration constants has been studied as a function of the longitudinal position in the barrel and as a function of the radius in the endcap.
The energy from incoherent pairs deposited in the ECAL and HCAL are given in Figures~\ref{fig:calo_bg_pairs_91} and~\ref{fig:calo_bg_pairs_365} for operation at 91.2 and 365 GeV, respectively. 
The largest amount of energy is deposited in the forward region in the calorimeter endcaps close to the beam-pipe. This correlates with the corresponding observation of hit densities in the Vertex and Tracker detectors. 

\begin{figure}[htbp]
  \centering
  \begin{subfigure}{.5\textwidth}
    \centering
    \includegraphics[width=\linewidth]{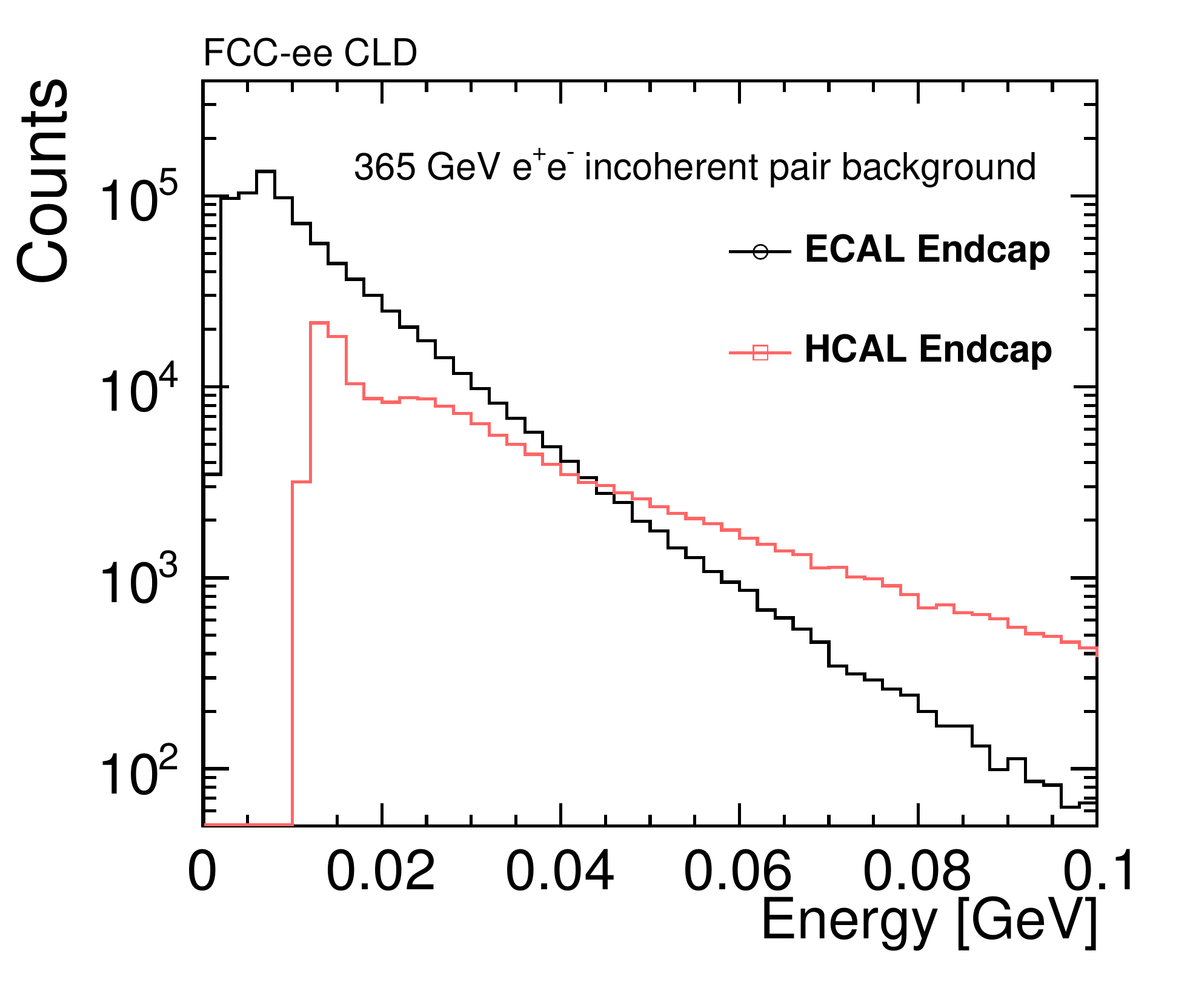}
    \caption{}
    \label{fig:caloDepositedEnergy_bg_pairs_365}
  \end{subfigure}%
  \begin{subfigure}{.5\textwidth}
    \centering
   \includegraphics[width=\linewidth]{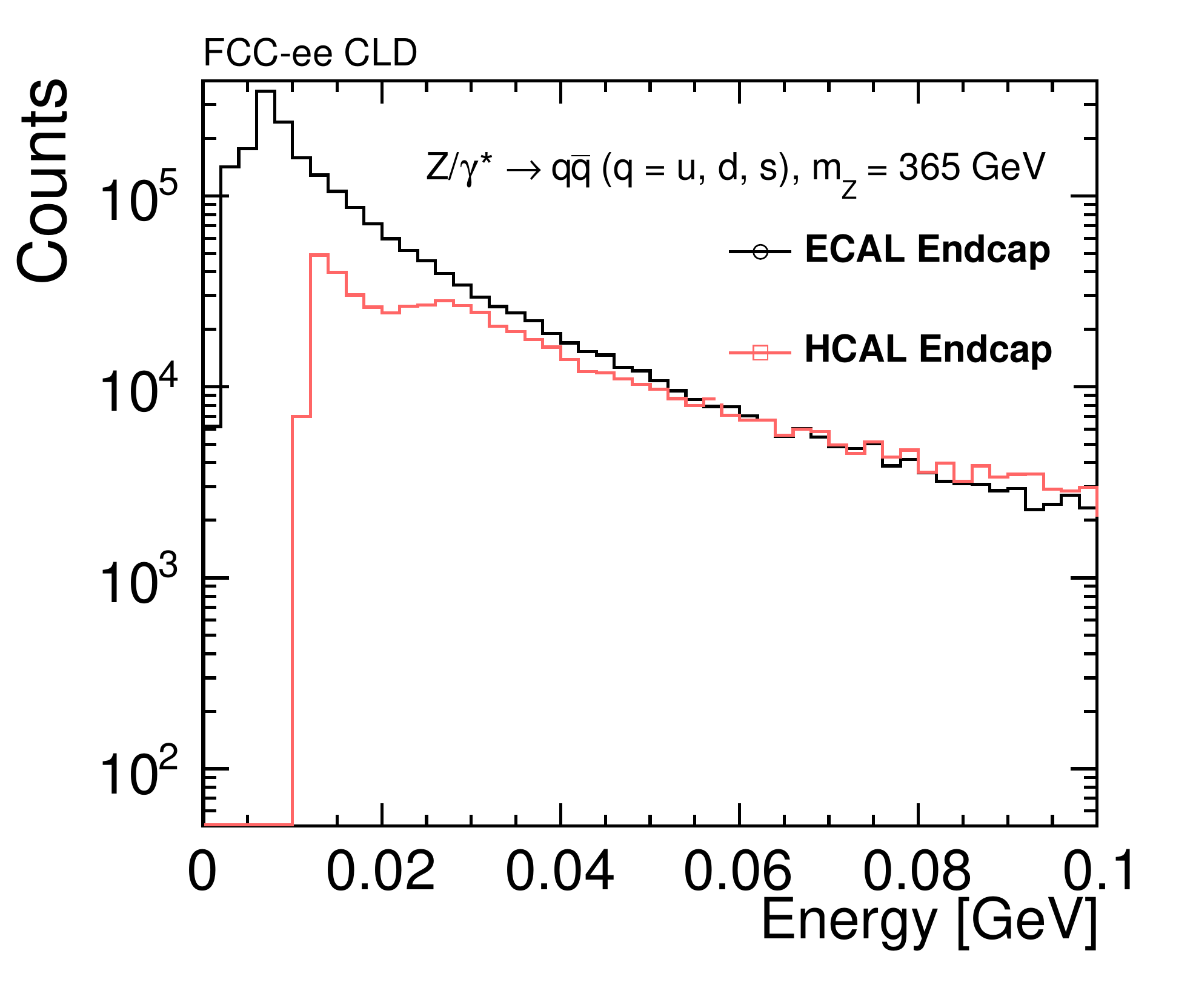}
    \caption{}
    \label{fig:caloDepositedEnergy_zuds_365}
  \end{subfigure}
  \caption{
	  Calibrated deposited energy per cell for particles originating from incoherent pairs at 365 GeV (a) and for particles from Z$\rightarrow q\bar{q}$, $q=(u,d,s)$ event (b).
Safety factors for the simulation uncertainties are not included. }
	  \label{fig:caloDepositedEnergy_bg_pairs_vs_zuds_365}
\end{figure}

The distributions of deposited energy per cell from incoherent pair background and for physics events from Z$\rightarrow q\bar{q}$, $q=(u,d,s)$ at 365 GeV operation are shown in Figure~\ref{fig:caloDepositedEnergy_bg_pairs_vs_zuds_365}. 
Particles from incoherent pair background are overall softer compared to the ones from physics events, however, since the amount of deposited energy per cell does not depend strongly on the 
momentum of the particle one   observes comparable distributions for both cases. As a result, discrimination of pair background by selecting a threshold in the energy deposit per cell does not appear feasible - this is, however, not a problem,
since the impact from pair background is found to be minor.

\subsection{Technology Choices and Cooling}

Presently, the technology assumed for the simulation model of the CLD ECAL silicon-tungsten sampling calorimeter is identical to the solution pursued by ILD/CALICE.
In this layout, a thin copper sheet in contact with the distributed ASICs (via a thermally conducting grease) allows one to remove the heat. A leak-less
water cooling system is foreseen to be connected at the outer end of each module~\cite{Cooling_Katia_talk, ECAL_cooling, Calo_cooling_AIDA, Grondin_2}. 
With power pulsing, the total heat load from the 77 million
channels of the ECAL (barrel plus endcaps) would only be 4.6 kW.

Without power pulsing and using the same technology a heat load 50 to 100 times higher is expected. This will impose the use of a different cooling scheme,
and might lead to a different ECAL design -- possibly inspired by the solution chosen for the CMS HGCal project~\cite{CMS_HGCAL}.

The ILD/CALICE technology has also been assumed for the CLD HCAL simulation model. The steel absorber plates in this calorimeter are 19 mm thick.
Cooling as foreseen in ILD consists of conducting the heat to the edges of the modules using these steel plates, where water cooling manifolds are connected
to them. In the technology chosen for the ILD HCAL, and using power pulsing, the average power dissipated is found to be 40 $\mathrm{\mu}$W per channel.

CALICE HCAL prototype layers with 4 $\times$ 144 channels have operated continuously (no power pulsing) and a preliminary version of the cooling system has been tested successfully~\cite{Calo_cooling_AIDA}. Such a number of channels operated continuously corresponds to the heat dissipation of a fully equipped ILD layer with power pulsing.

For the HCAL of CLD, with much higher power dissipation, the cooling system will have to be redesigned. Modifying the absorber layers by adding copper plates might be a solution.
This will lead to a different longitudinal sampling in the HCAL and will impact the performance of the calorimeter. 

Detailed engineering studies and further simulations will be needed to assess the different design options for ECAL and HCAL -- this type of work goes beyond the scope of this study.

\clearpage

\section{Magnet System}
\label{Magnet}
\subsection{Superconducting Solenoid}
\label{solenoid}

The solenoid magnetic field of the FCC-ee detectors is 2 T, which is limited by MDI constraints. Design details of the superconducting solenoid are described in~\cite{FCC_CDR}.

In the simulation model, the magnetic field in CLD is 2~T throughout the volume inside the superconducting coil.
The field in the yoke barrel is 1~T, pointing in the opposite direction with respect to the inner field. 
The simulation model currently assumes no field in the yoke endcap nor outside the yoke.

The solenoid of CLD
is implemented in the simulation model with parameters as shown in Table~\ref{tab:coil}. 
The material budget of the solenoid corresponds to about 0.7~\nuclen, as indicated in Figure~\ref{fig.hcal.lambda}.

\begin{table}[hbtp]
  \caption{Description of the solenoid elements (coil and cryostat) as implemented in the simulation model. For all elements the material, the longitudinal extent in one half of
the detector $z_{\mathrm{min/max}}$ and the radial extent $r_{\mathrm{min/max}}$ are given.}
  \label{tab:coil}
  \centering 
  \begin{tabular}{l l l l l l l}
    \toprule
  Element &  Material  & $z_{\min}$ [mm] & $z_{\max}$ [mm] & $r_{\min}$ [mm] & $r_{\max}$ [mm]   \\\midrule
Inner Barrel &    Steel     &                0   &            3705  &          3719  &           3759   \\
Coil &	    Aluminium &                0   &            3467  &          3885  &           3975    \\
Outer Barrel &	    Steel     &                0   &            3705  &          4232  &           4272    \\
 End plates &          Steel     &           3665   &           3705 &         3759   &           4232  \\
	    \bottomrule
	  \end{tabular}
	\end{table}

\subsection{Yoke and Muon Detectors}
\label{yoke}

\begin{figure}[hbtp] 
         \centering
                 \includegraphics[scale=0.28]{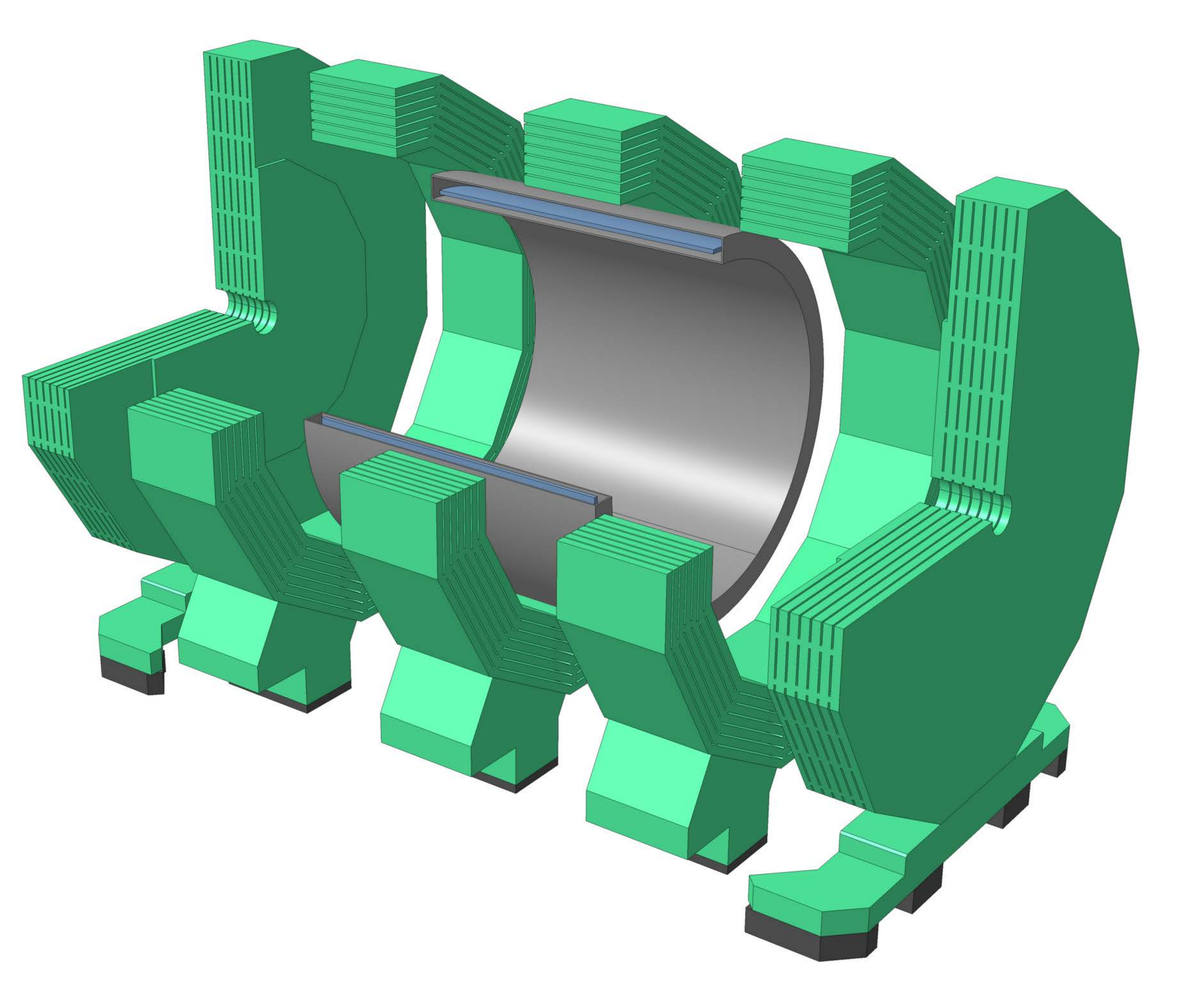}
         \caption{Segmentation of the iron return yoke of CLD into endcaps and three barrel rings.}
         \label{fig.yoke_sections}
\end{figure}

The iron return yoke is structured into three rings in the barrel region and the two endcaps, as shown in Figure~\ref{fig.yoke_sections}.
The thickness of the yoke is reduced w.r.t.  CLICdet, in correspondence to the lower solenoid field (2~T vs. 4~T).

\begin{figure}[hbtp] 
         \centering
                 \includegraphics[scale=0.40]{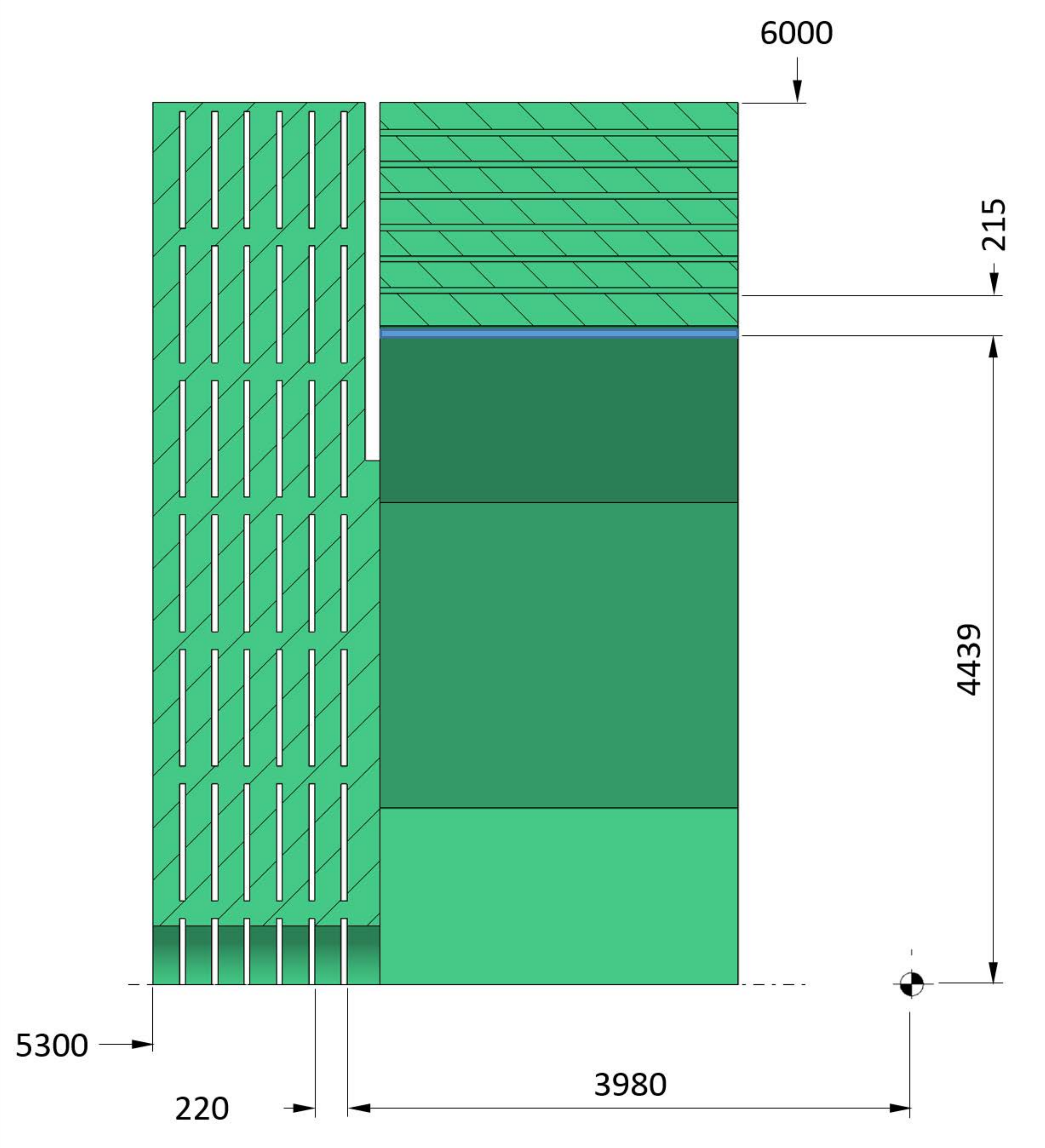}
         \caption{Schematic cross section of the muon system layout in the yoke of CLD. Dimensions are in mm. The staggering of the layers is not visible in this cross section. }
         \label{fig.muon_system}
\end{figure}

A muon identification system with 6 layers as in CLICdet is implemented.
An additional 7th layer is inserted in the barrel as close as possible to the coil.
This layer may serve as tail catcher for hadron showers. 
The muon system layout in CLD is shown in Figure~\ref{fig.muon_system}. 

The muon detection layers are proposed to be built as RPCs with cells of $30\times30~\mathrm{mm}^{2}$.
Alternatively, crossed scintillator bars could
be envisaged. The free space between yoke steel layers
is 40 mm, which is considered sufficient given present-day technologies for building RPCs. In analogy to CMS and CLICdet, the yoke layers and thus the muon detectors are staggered to avoid gaps (see Figure~\ref{fig.cross.section}).

\clearpage

\section{Physics Performance}
\label{physics}

\subsection{Simulation and Reconstruction}
\label{sec:simreco}

The detector simulation and reconstruction software tools used for the results
presented in the following are developed together with the linear collider community. The
\ddhep~\cite{frank13:dd4hep} detector simulation and geometry framework was
developed in the AIDA and AIDA-2020  projects~\cite{AIDA}. Larger simulation and reconstruction samples
were produced with the \ilcdirac{} grid production
tool~\cite{dirac08,ilcdirac13}. 
The software packages of iLCSoft-2019-07-09 have been used throughout this study 
with the CLD geometry version FCCee\_o1\_v04, unless otherwise specified.

\subsubsection{Event Generation}
\label{sec:event-generation}

The detector performance is studied with single particles or simple event
topologies. The individual particles are used to probe the track reconstruction
and the particle ID.\@ The reconstruction of particles inside jets is tested
through \PZgstarToqq events decaying into pairs of u, d, or s quarks at different centre-of-mass energies. These events were created with
\whizard{}~\cite{Kilian:2007gr,Moretti:2001zz}. To study the track reconstruction and particle ID
in complex events, and for the flavour tagging studies, \uu{}, \dd{}, \ssbar{},
\cc{}, \bb{}, and \ttbar{} events created with
\whizard{} are used. In all cases, parton showering,
hadronisation, and fragmentation is performed in \pythia{}~6.4~\cite{PYTHIAmanual} with the
fragmentation parameters tuned to the OPAL data taken at
LEP~\cite[Appendix B]{cdrvol2}. 
The generation of the beam-related background events (dominated by incoherent pairs and synchrotron radiation photons) are described elsewhere~\cite{FCC_CDR}.

\subsubsection{Detector Simulation}
\label{sec:detector-simulation}

The CLD detector geometry is described with the \ddhep{}
software framework, and simulated in \geant{}~\cite{Agostinelli2003,
  Allison2006, Allison2016186} via the
\ddg{}~\cite{frank15:ddg4} package of \ddhep{}. The \geant{} simulations are
performed with the \texttt{FTFP\_BERT} physics list of \geant{} version 10.02p02.
In the CLD simulations, the tracker is assumed to be built using the higher granularity "enlongated pixel" technology.

\subsubsection{Event Reconstruction}
\label{sec:event-reconstruction}

The reconstruction software is implemented in the linear collider
\marlin{}-framework~\cite{MarlinLCCD}, the reconstruction algorithms take
advantage of the geometry information provided via the \ddrec{}~\cite{sailer17:ddrec}
data structures and surfaces. 
If the effect of beam-induced background is to be studied, 
the reconstruction starts with the
overlay of background events via the \emph{Overlay Timing} processor~\cite{LCD:overlay},
which also selects only the energy deposits inside appropriate timing windows around the physics event.
In
the next step, the hit positions in the tracking detectors are smeared with
Gaussian distributions according to the single point resolutions per layer. The calorimeter hits are scaled with the calibration
constants obtained from the reconstruction of mono-energetic 10~GeV photons and 50~GeV~\PKzL{}.

\begin{figure}[htbp]
  \centering
  \includegraphics[width=0.65\linewidth]{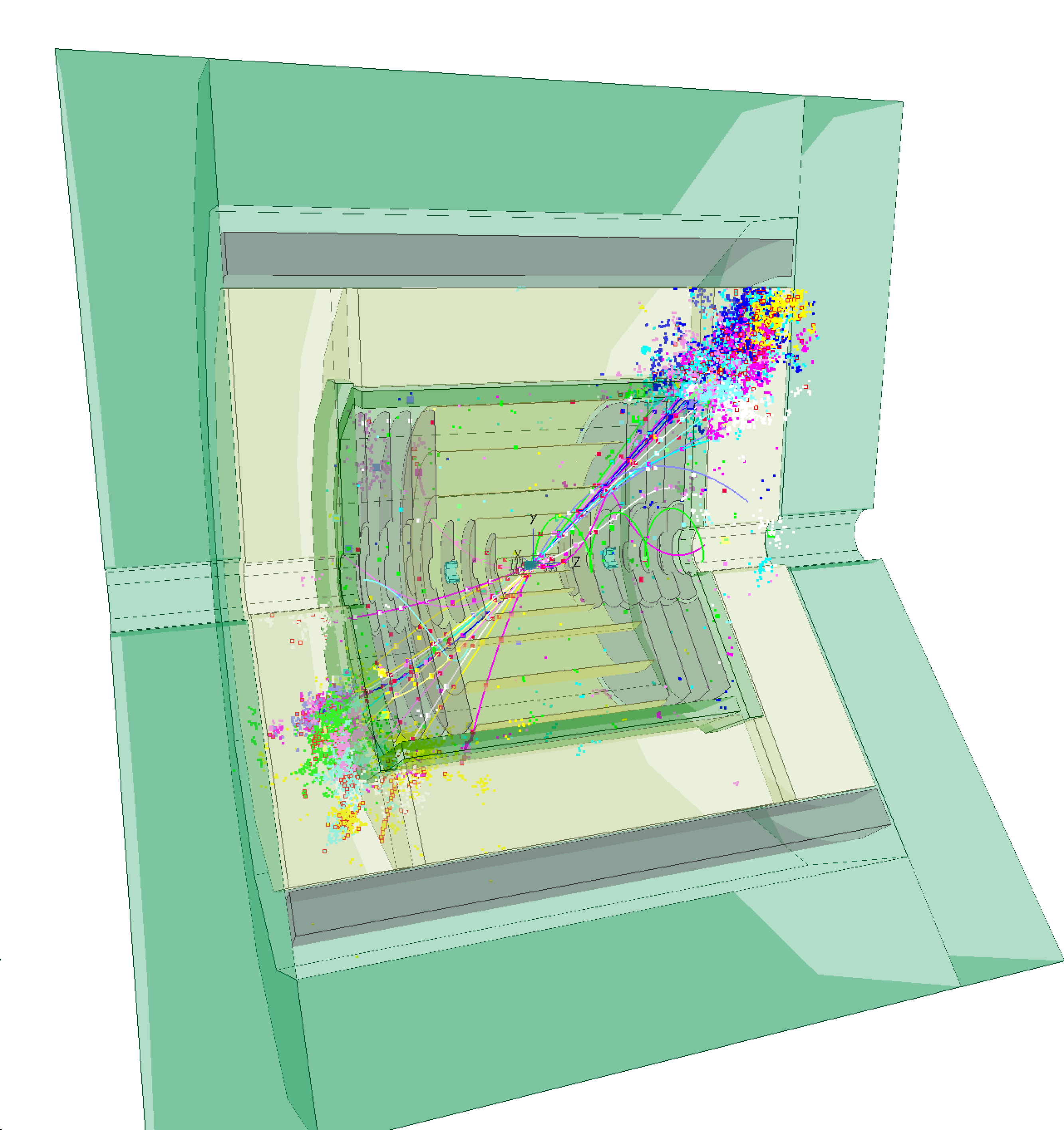}
\caption{Event display in the CLD detector for a \PZgstarToqq event, with $m_{\PZ}$ = 365 GeV.}
\label{fig:event_display}
\end{figure}

As an illustration of the CLD detector model and event simulation, an event display is shown in Figure~\ref{fig:event_display}.

\paragraph{Tracking}

The tracking algorithm used in reconstruction at CLD is referred to as
\emph{ConformalTracking}~\cite{Brondolin:2019awm}. 
In modern pattern recognition algorithms, the use of cellular networks has
proven to be a powerful tool, providing robustness against missing hits and the
addition of noise to the system~\cite{Emilia_proceedings}. For a detector with solenoid field and barrel
plus endcap configuration, cellular automata (CA) may be applied to provide
 efficient track finding. Several aspects of CA algorithms may however
impact performance negatively: producing many possible hit combinations requires
a fit to be performed on a large number of track candidates. This may be costly
in processing time. Methods to reduce combinatorics at this stage may, in turn, compromise
on the final track finding performance. One way around such issues is the
additional application of conformal mapping. 

\begin{figure}[htbp]
  \centering
  \includegraphics[width=0.5\linewidth]{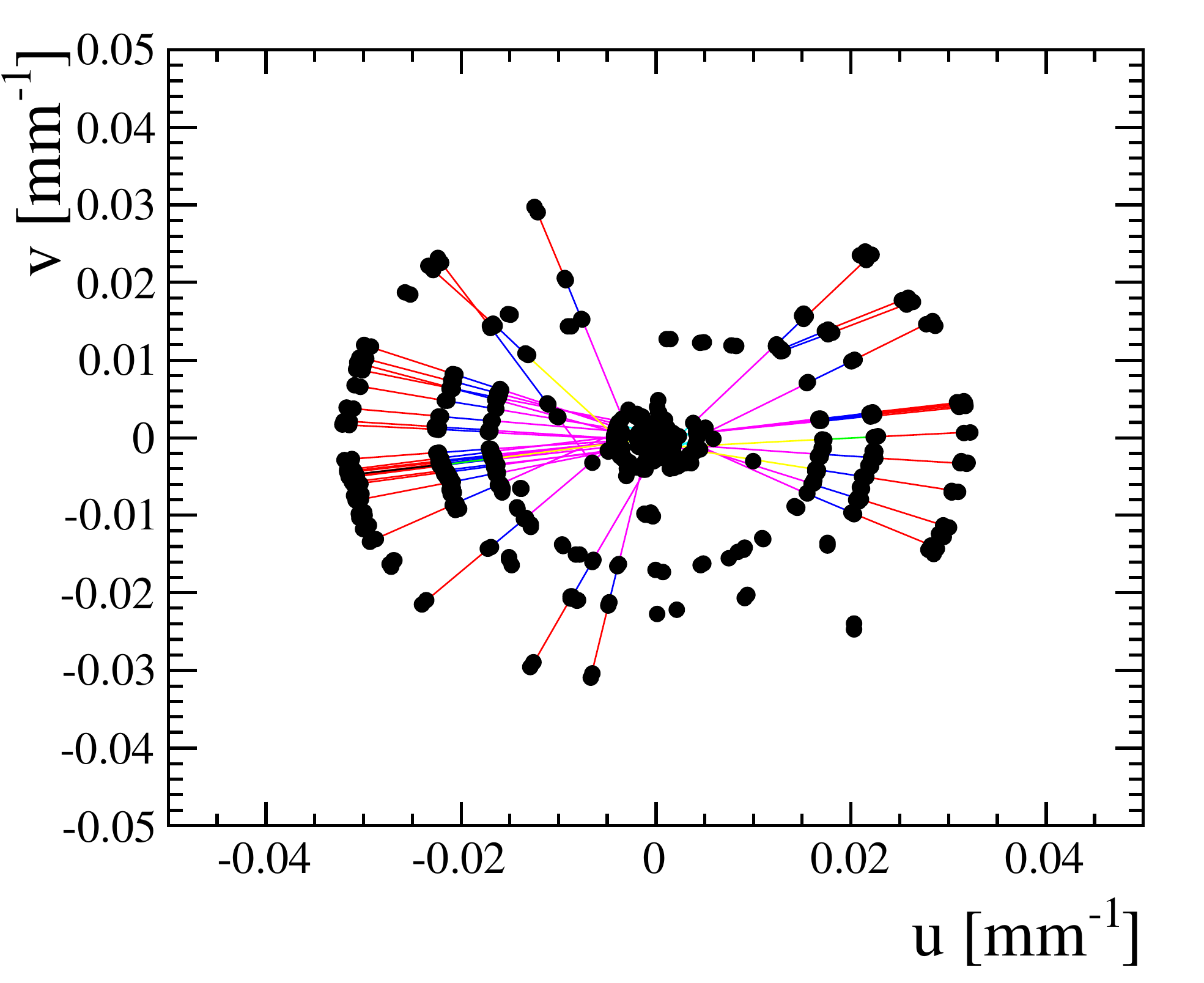}
\caption{Cellular automaton in conformal space.}
\label{fig:confMapping}
\end{figure}

Conformal mapping is a geometry transform which has the effect of mapping
circles passing through the origin of a set of axes (in this case the global
$xy$ plane) onto straight lines in a new $uv$ co-ordinate system. By performing
such a transform on an $xy$ projection of the detector (where the $xy$ plane is
the bending plane of the solenoid), the pattern recognition can be reduced to a
straight line search in two dimensions. Cellular automata can then be applied
in this 2-D space, with the use of a simple linear fit to differentiate between track
candidates. \Cref{fig:confMapping} shows an example of a cellular automaton in conformal space.

To make this approach flexible to changes in the geometry (or for
application to other detector systems), all hits in conformal space are treated
identically, regardless of sub-detector and layer. Cells between hits are
produced within a given spatial search window, employing kd-trees for fast
neighbour lookup~\cite{Emilia_proceedings}. This provides additional robustness against missing hits in
any given detection layer. A second 2D linear fit in the $sz$ parameterisation of
the helix is also implemented, to recover the lost information resulting from
the 2D projection onto the $xy$ plane and reduce the number of ``ghost'' tracks.
A minimum number of 4 hits is required to reconstruct a track.

For displaced tracks, which do not comply with the requirement of passing through the origin of the global $xy$ plane, 
second-order corrections are applied to the transformation equations. Additionally, a strategy change has been proven necessary, in terms of:
\begin{itemize}
\item broader angles in the search for nearest neighbours
\item minimum number of 5 hits to reconstruct a displaced track
\item inverted order, from tracker to vertex hits
\end{itemize}

A summary of the full pattern recognition chain is given in~\cref{tab:iterations}, 
including the values of the parameters used in every step of the track finding.
For more details concerning the definition of the cuts, the reader is referred to~\cite{Brondolin:2019awm}.

\begin{table}
  \centering
  \caption{Overview of the configuration for the different steps of the pattern recognition chain. On the right, some of the parameters of relevance for the cellular automaton as used for CLD are shown:
  the maximum angle between cells $\alpha_{\textnormal{max}}$, the maximum cell length $l_{\textnormal{max}}$, 
  the minimum number of hits on track $N^{\textnormal{hits}}_{\textnormal{min}}$, the maximum $\chi^{2}_{\textnormal{max}}$ for valid track candidates, and $p_{\textnormal{T, cut}}$ used to discriminate between the two variations of the algorithm of track extension.}
  \label{tab:iterations}
  \begin{tabular}{cllccccc}\toprule
  \multirow{3}{*}{\textbf{Step}} & \multirow{3}{*}{\textbf{Algorithm}} & \multirow{3}{*}{\textbf{Hit collection}} & \multicolumn{5}{c}{\textbf{Parameters}}\\
  \cmidrule{4-8}
  & & & $\alpha_{\textnormal{max}}$ & $l_{\textnormal{max}}$ & $N^{\textnormal{hits}}_{\textnormal{min}}$ & $\chi^{2}_{\textnormal{max}}$ & $p_{\textnormal{T, cut}}$\\
  & & & [\SI{}{rad}] & [\SI{}{\mm}$^{-1}$] & - & -  & [\SI{}{\GeV}]\\
  \midrule
  0 & Building & Vertex Barrel & 0.01 & 0.03 & 4 & 100 & - \\ 
  1 & Extension & Vertex Endcap & 0.01 & 0.03 & 4 & 100 & 10 \\ 
  2 & Building & Vertex & 0.05 & 0.03 & 4 & 100 & - \\ 
  3 & Building & Vertex & 0.1 & 0.03 & 4 & 2000 & - \\ 
  4 & Extension & Tracker & 0.1 & 0.03 & 4 & 2000 & 1 \\ 
  5 & Building & Vertex \& Tracker & 0.1 & 0.015 & 5 & 1000 & - \\ 
  \bottomrule
  \end{tabular}
\end{table}

The tracks found by the pattern recognition in conformal space are then fitted in global space with a Kalman filter method.
The performance studies presented in this note assume a homogeneous magnetic field of 2 T.

\paragraph{Particle Flow Clustering}

The calorimeter clusters are reconstructed in the particle flow approach by
\pandora~\cite{Marshall:2015rfaPandoraSDK,Thomson:2009rp,Marshall:2012ryPandoraPFA}. \pandora{} uses the reconstructed tracks and
calorimeter hits as input to reconstruct all visible particles. The procedure is optimised
to achieve the best jet energy resolution. This may not be the ideal procedure for isolated particles,
which can benefit from a dedicated treatment. The output of the particle flow
reconstruction are \emph{particle flow objects} (PFOs).

\subsubsection{Treatment of  Background}
\label{sec:treatment-background}

The largest impact on the detector performance from beam-induced backgrounds comes in the form of incoherent 
\epem pairs and photons from synchrotron radiation (see section 7.1 of the CDR~\cite{FCC_CDR}).
When studying the detector performance degradation due to these backgrounds, the following number of bunch crossings 
are overlaid to the physics event\footnote{The number of incoherent \epem pair particles  within the detector acceptance, per bunch crossing, is found to be approximately 6 at 91.2 GeV and 290 at 365 GeV c.m. energy (see Table 7.1 in the CDR [3]).}, placed at bunch crossing number 1:
\begin{itemize}
\item{at 91.2 GeV centre-of-mass energy: 20 bunch crossings}
\item{at 365 GeV centre-of-mass energy: 3 bunch crossings}
\end{itemize}
All hits inside the time window are then passed forward to the reconstruction.

Given the different bunch spacing at the two energies (19.6 ns at 91.2 GeV, 3396 ns at 365 GeV), the number of overlaid bunch crossings corresponds to a detector integration time of, respectively, 400 ns and 10$\upmu{}$s. 
The latter is in accordance with the assumed vertex and tracker readout time (see Section~\ref{vtx_technology}).
The overlay of only 20 background events (400 ns integration time at 91.2 GeV), on the other hand, is currently imposed by a  limitation from software/computing.

\subsection{Performance of Lower Level Physics Observables}
\label{observables}
 
\subsubsection{Single Particle Performances}
\paragraph{Position, Angular and Momentum Resolutions}

The results in this section demonstrate the combined performance of the tracking system (vertex and tracker sub-detectors) with those of the tracking algorithm, described in Section~\ref{sec:event-reconstruction}. 

To identify heavy-flavour quark states and tau-leptons with high efficiency, a precise measurement of the impact parameter and of the charge of the tracks originating from the secondary vertex is required. 
Monte Carlo simulations for linear collider experiments~\cite{cdrvol2} show that these goals can be met with a constant term in the transverse impact-parameter resolution of $a \simeq$ 5~$\micron$ and a multiple-scattering term of $b \simeq$ 15~$\micron$, 
using the canonical parameterisation
\begin{equation}
\sigma(d_{0}) = \sqrt{a^2+b^2 \cdot \mathrm{GeV^2}/(p^2 \sin^3(\theta))}.
\label{eq:d0}
\end{equation}

\begin{figure}
\centering
\begin{subfigure}{.5\textwidth}
  \centering
  \includegraphics[width=\linewidth]{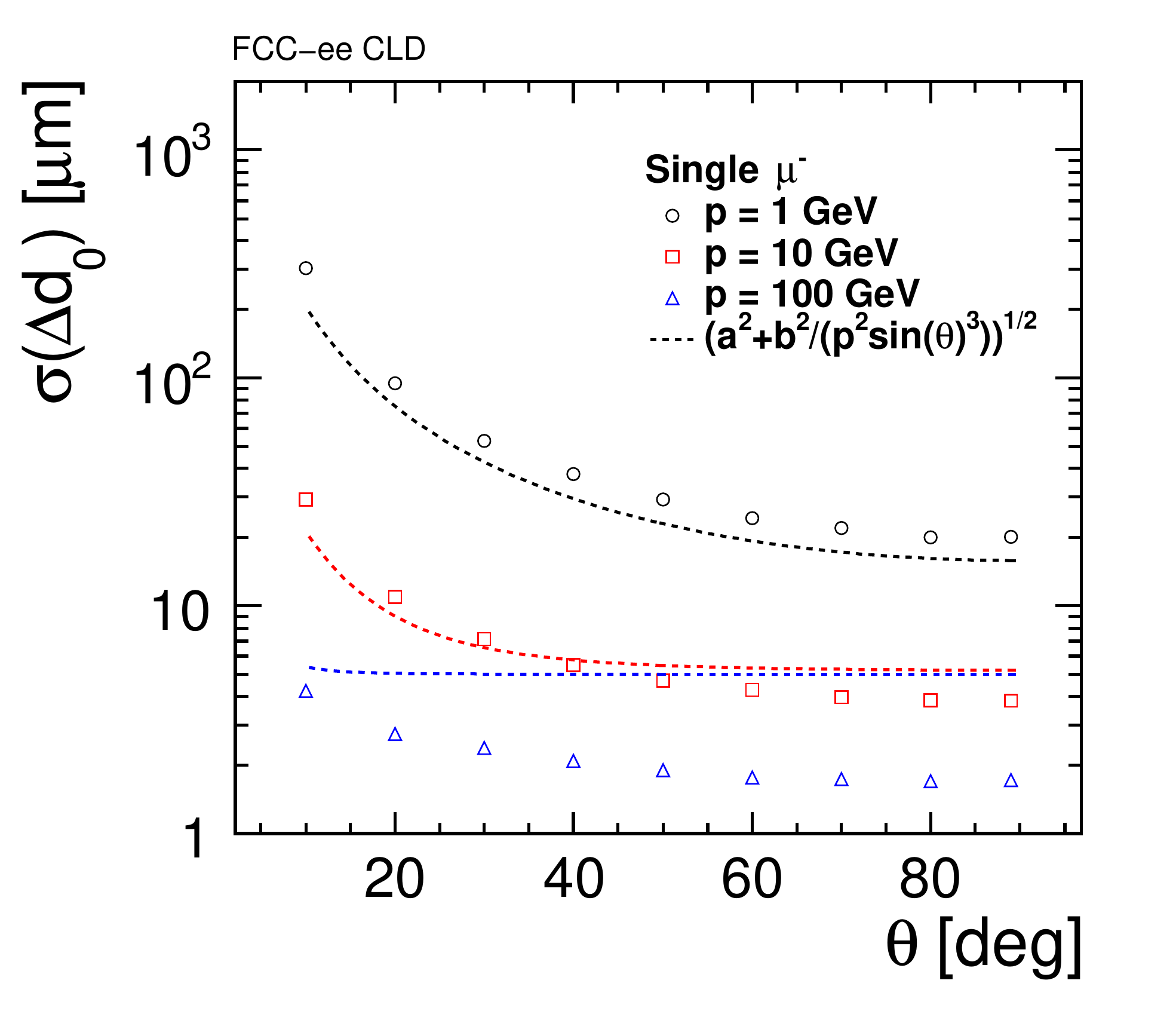}
  \caption{$d_{0}$ resolution}
  \label{fig:d0res}
\end{subfigure}%
\begin{subfigure}{.5\textwidth}
  \centering
  \includegraphics[width=\linewidth]{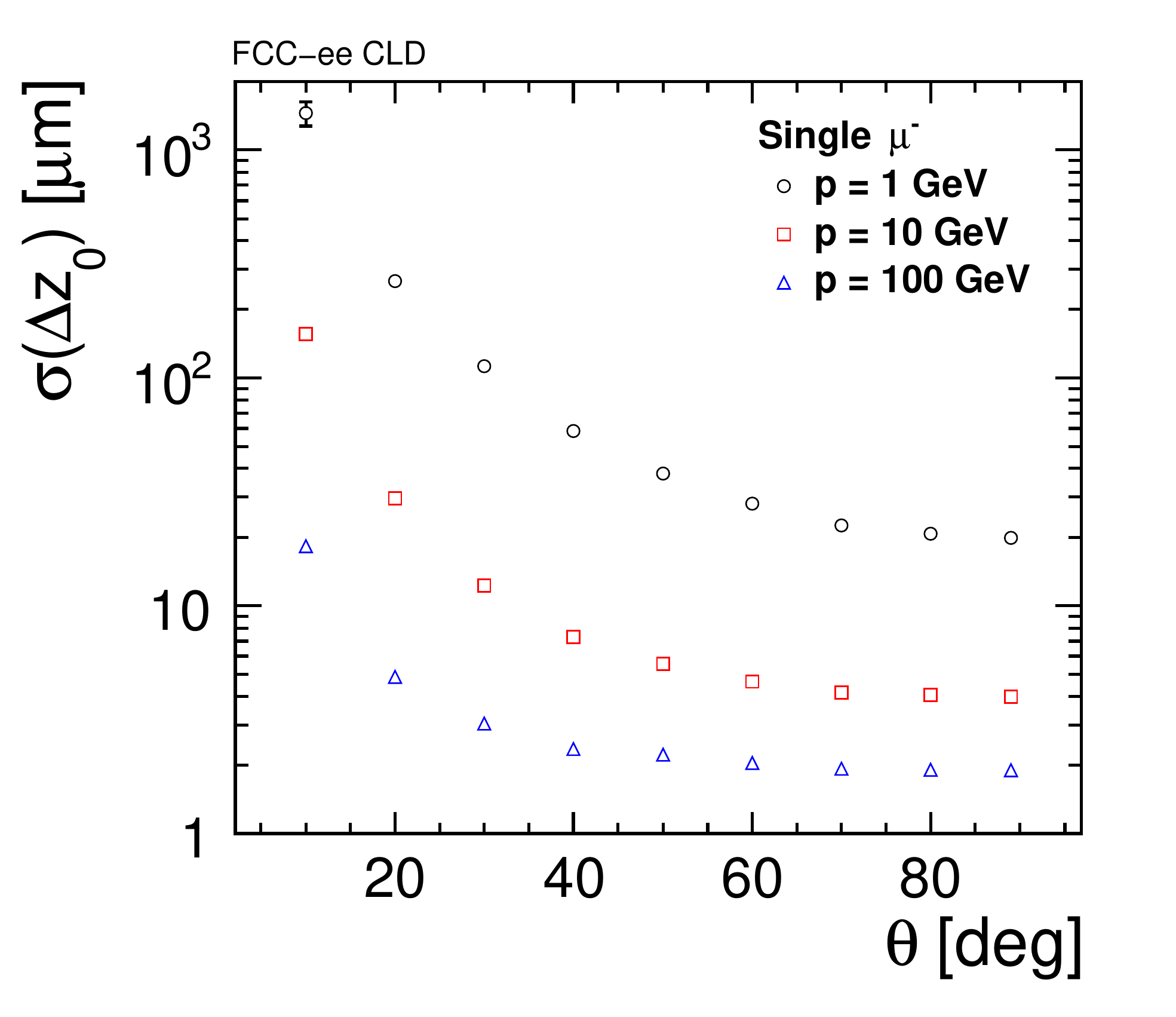}
  \caption{$z_{0}$ resolution}
  \label{fig:z0res}
\end{subfigure}
\caption{Impact-parameter resolutions, obtained with the baseline vertex detector layout for CLD, (a) in the transverse and (b) in the longitudinal plane, for muon tracks with momenta of 1, 10 and 100 GeV. Also shown is the parameterisation of Equation~\ref{eq:d0} with the target values $a$ = 5~$\micron$ and $b$ = 15~$\micron$.}
\label{fig:impact_parameter_res}
\end{figure}

Figure~\ref{fig:impact_parameter_res} shows the impact-parameter resolutions obtained for CLD, for isolated muon tracks with momenta of 1, 10 and 100 GeV
originating from the nominal interaction point. 
Each data point corresponds to 10\,000 muons at fixed energy and polar angle.
For each dataset, the resolution is calculated as the width of a Gaussian fit of the residual distributions, i.e. the difference between the reconstructed and simulated parameters per track.
In Figure~\ref{fig:d0res}, superimposed to the data points for the transverse impact parameter resolution are the curves obtained with Equation~\ref{eq:d0} for the different energies. High-energy muons show a resolution well below the high-momentum limit of 5~$\micron$ at all polar angles, while for 10~GeV muons this is achieved only for central tracks above 30\degrees{}. The data points for 1~GeV muons are systematically above the parameterisation, by 15--35\%.
The achieved longitudinal impact-parameter resolution, shown in Figure~\ref{fig:z0res}, for muons at all energies and polar angles is smaller than the longitudinal bunch length of 1.5 mm at the highest collision energy.
Note that while the $d_0$ resolution at high energies is determined by the layout of the vertex detector and the single point resolution, the $z_0$ resolution is in addition influenced by the polar angle resolution~\cite{Grefe_thesis}.

\begin{figure}
\centering
\begin{subfigure}{.5\textwidth}
  \centering
  \includegraphics[width=\linewidth]{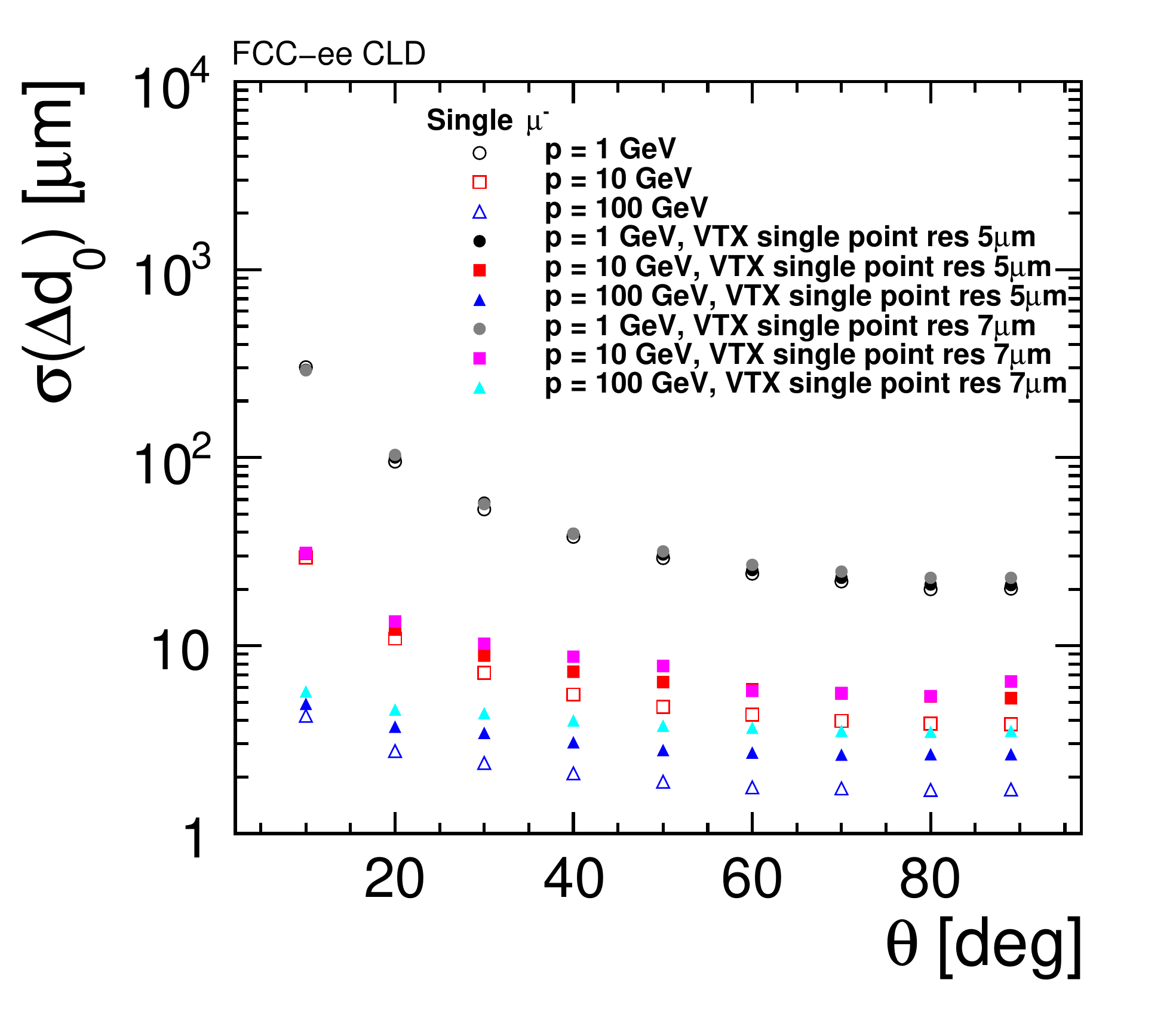}
  \caption{$d_{0}$ resolution}
  \label{fig:d0res_spr}
\end{subfigure}%
\begin{subfigure}{.5\textwidth}
  \centering
  \includegraphics[width=\linewidth]{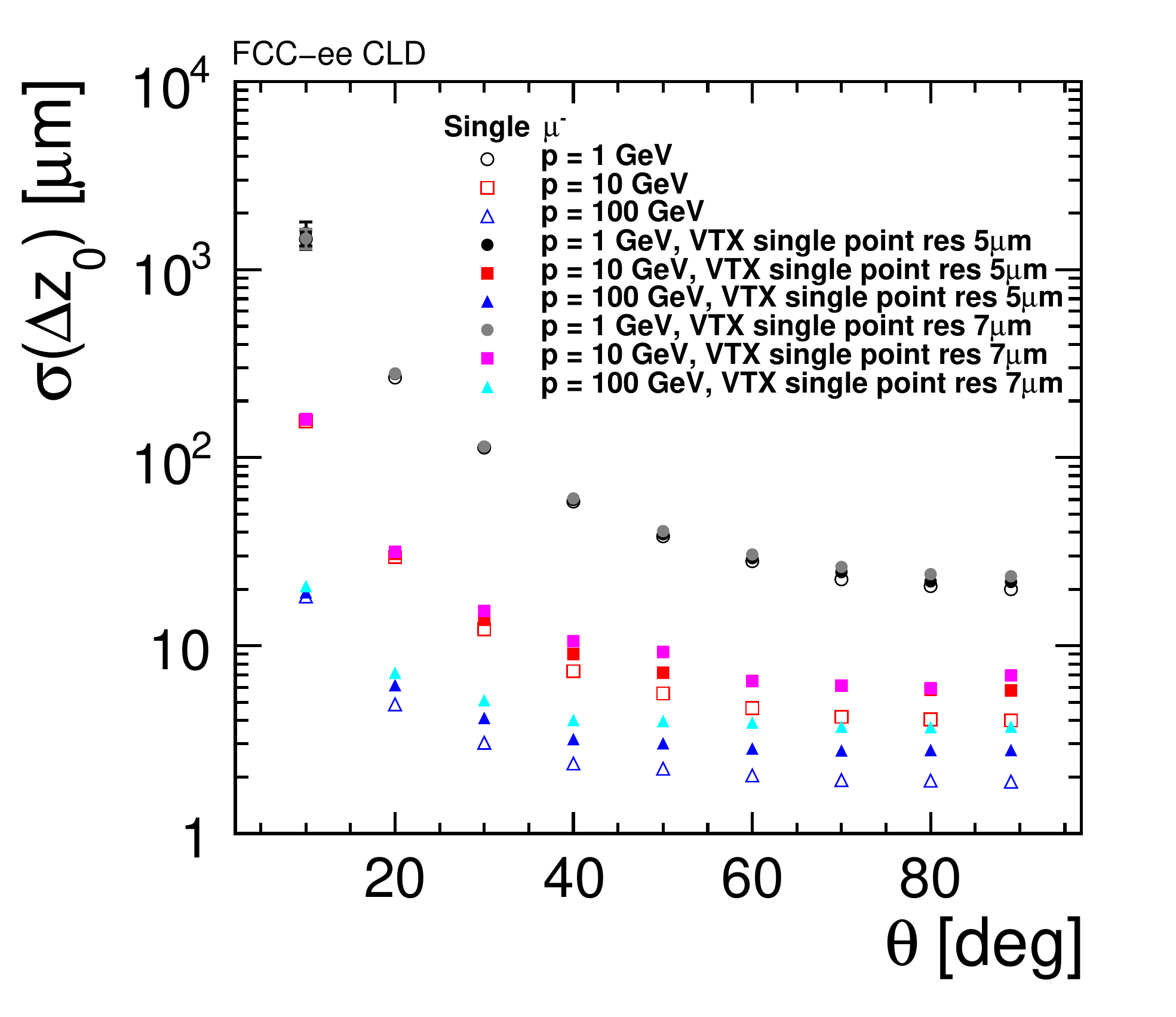}
  \caption{$z_{0}$ resolution}
  \label{fig:z0res_spr}
\end{subfigure}
\caption{Impact-parameter resolution as a function of polar angle (a) in the transverse and (b) in the longitudinal plane, for muon tracks with momenta of 1, 10 and 100 GeV. 
Empty markers refer to the baseline vertex detector with single point resolution of 3~$\micron$. 
Two other sets of curves are plotted, which correspond to a single point resolution in the vertex detector of 5~$\micron$ and 7~$\micron$ respectively.}
\label{fig:impact_parameter_res_spr}
\end{figure}

The dependence of the impact-parameter resolution on the pixel technology has been studied by varying the single point resolution for the vertex layers from the baseline value of 3~$\micron$ 
to 5~$\micron$ and 7~$\micron$. 
The resulting resolutions are shown in Figure~\ref{fig:impact_parameter_res_spr}. 
The single point resolution dominates at higher energies, especially in the barrel region, where a change from 3 to 5~$\micron$ results in an increase 
by approximately 50$\%$ for both $d_{0}$ and $z_{0}$ resolutions. However, even in the worst scenario of 7~$\micron$ single point resolution, 
the $d_{0}$ resolution for 100~GeV tracks does not exceed the target value for the high-momentum limit of $a \simeq$ 5~$\micron$. For the 10~GeV tracks, on the other hand, the resolution is at the limit.
For 1~GeV muons, for which multiple scattering dominates, the effect of a single point resolution variation from 3 to 5~$\micron$ reaches up to 6$\%$. 

The detector performance depends on the spatial resolution in the silicon detectors, which is defined by the pixel sizes. 
The default single point resolutions in the simulation model are:
\begin{itemize}
\item vertex barrel and discs: 3~$\micron$~$\times$~3~$\micron$
\item inner tracker barrel and discs: 7~$\micron$~$\times$~90~$\micron$
\begin{itemize}
\item except first inner tracker discs: 5~$\micron$~$\times$~5~$\micron$
\end{itemize}
\item outer tracker barrel and discs: 7~$\micron$~$\times$~90~$\micron$
\end{itemize}

\begin{figure}
\centering
\begin{subfigure}{.5\textwidth}
  \centering
  \includegraphics[width=\linewidth]{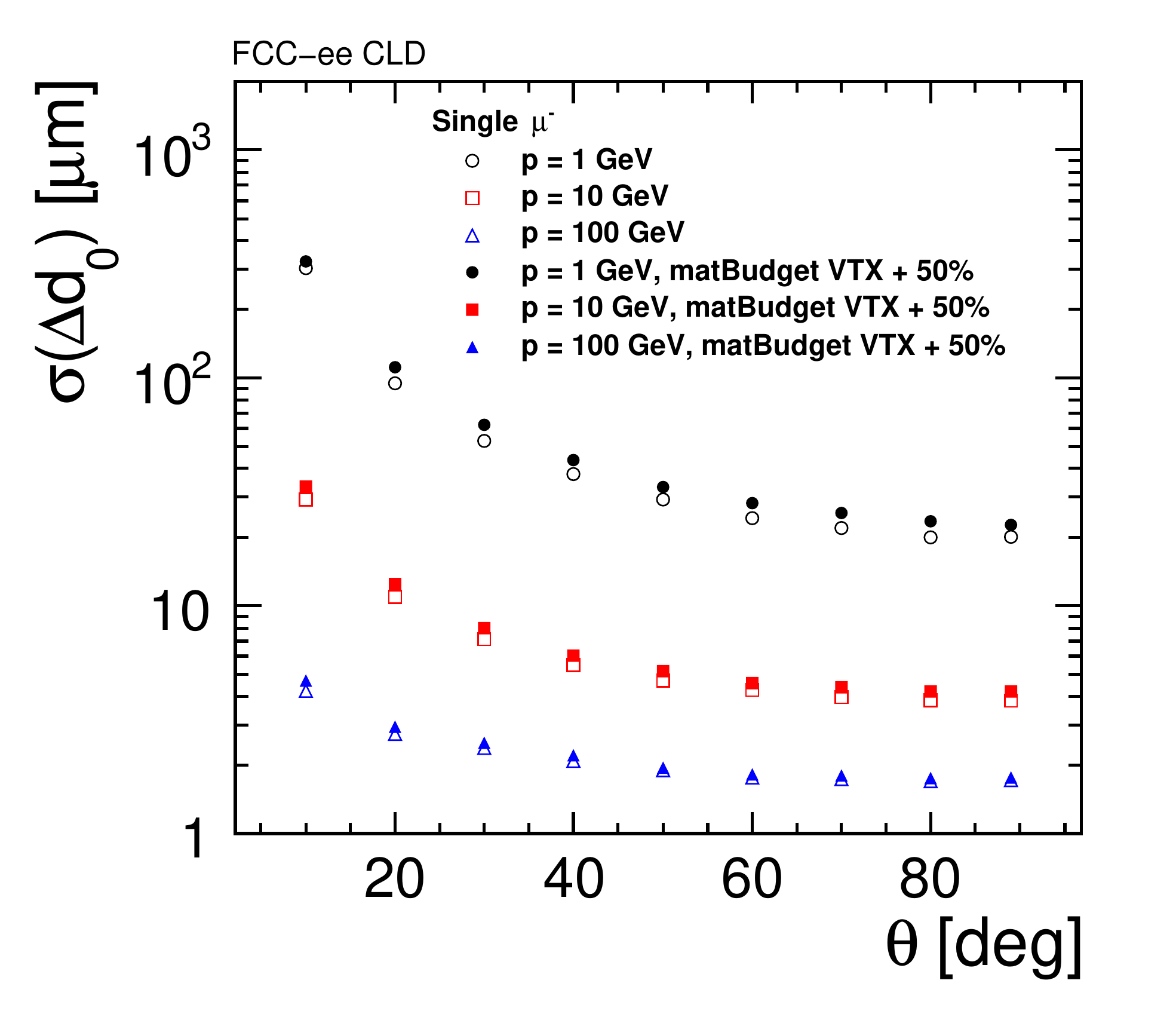}
  \caption{$d_{0}$ resolution}
  \label{fig:d0res_matBudgetVTX}
\end{subfigure}%
\begin{subfigure}{.5\textwidth}
  \centering
  \includegraphics[width=\linewidth]{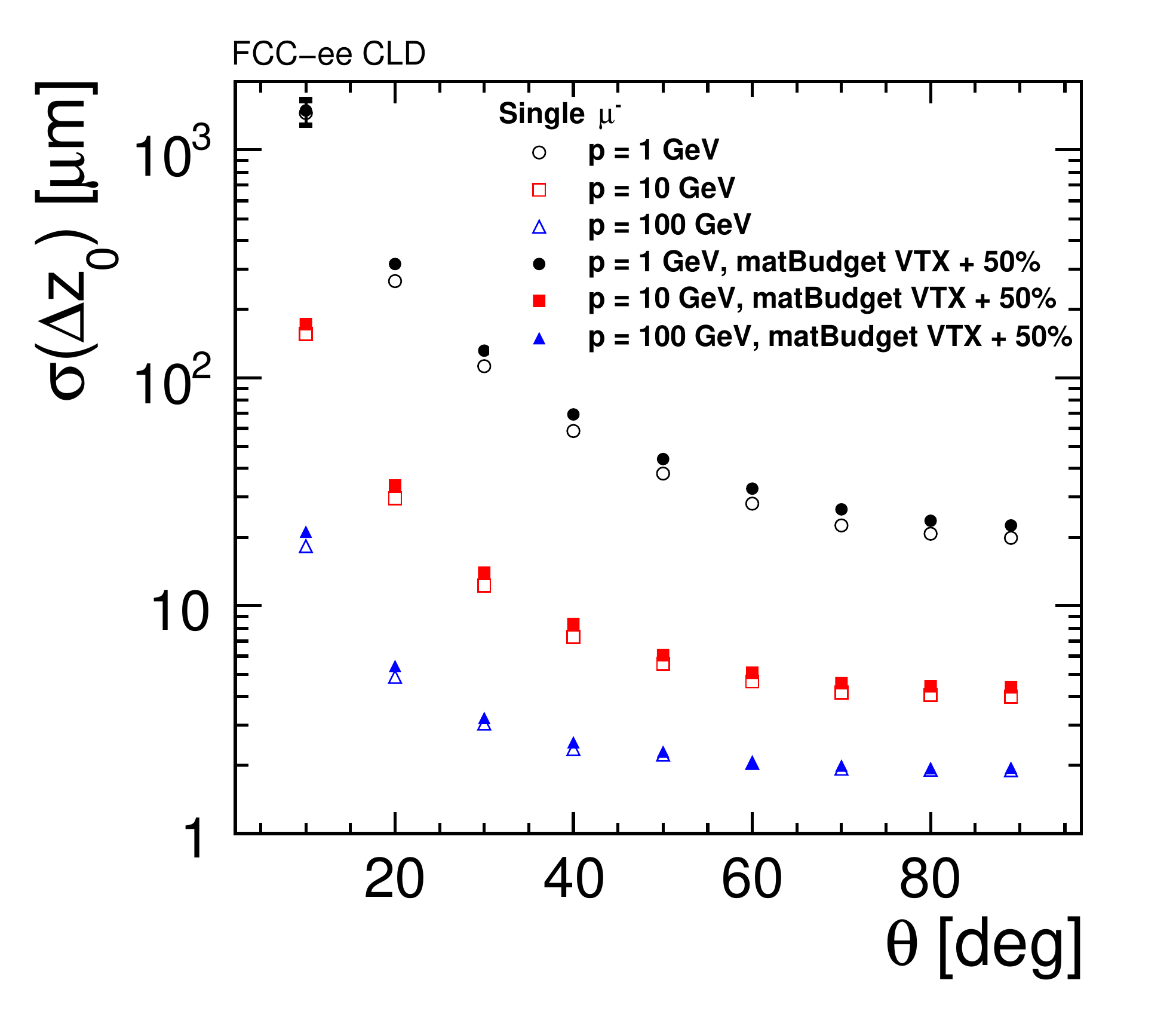}
  \caption{$z_{0}$ resolution}
  \label{fig:z0res_matBudgetVTX}
\end{subfigure}
\caption{Impact parameter resolution as a function of polar angle (a) in the transverse and (b) in the longitudinal plane, for muon tracks with momenta of 1, 10 and 100 GeV. Empty markers refer to the default design of the CLD vertex detector. 
Results shown with full markers are obtained with a detector model with the material budget in the vertex detector layers increased by 50$\%$ with respect to CLD.}
\label{fig:impact_parameter_res_matBudgetVTX}
\end{figure}

So far no detailed cooling studies have been performed. Adopting the mechanics and cooling of the ALICE ITS upgrade concept, as described in Section~\ref{vtx_technology}, has been a first attempt in including material to account for water cooling in the vertex region.
In order to test the sensitivity of the performance to assumptions about cooling, supports and cabling, the material budget of the CLD vertex detector layers has been further increased by 50\%. 
Results for the impact-parameter resolution with this additional material are shown in Figure~\ref{fig:impact_parameter_res_matBudgetVTX}, together with the default values for CLD. 
As expected, low-momentum tracks are the most affected by the change in the material budget, while the effect is negligible for tracks of 100~GeV. 
However, it is encouraging to notice that the biggest variation, on 1~GeV tracks, amounts to only 10$\%$.

\begin{figure}
\centering
\begin{subfigure}{.5\textwidth}
  \centering
  \includegraphics[width=\linewidth]{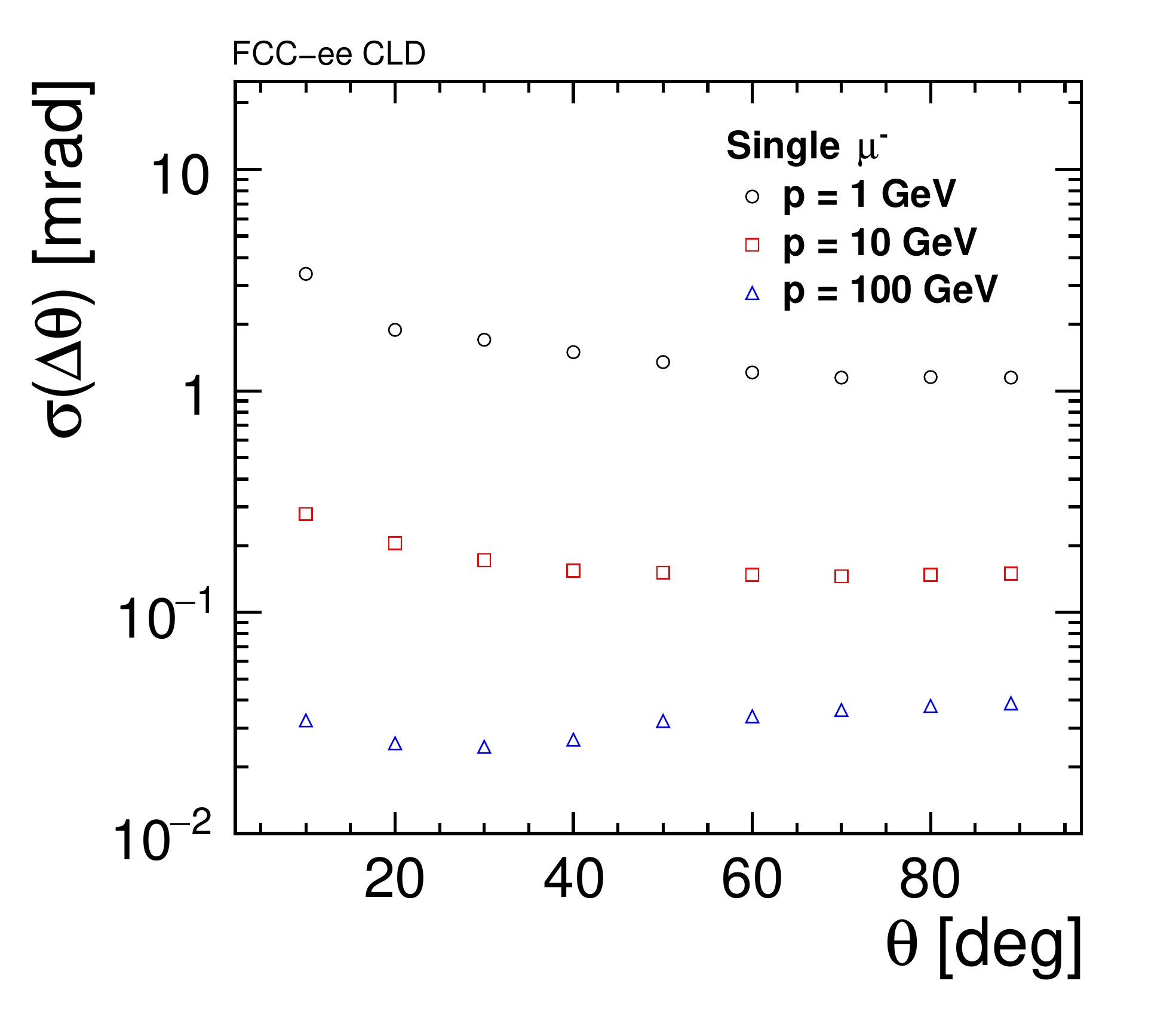}
  \caption{$\theta$ resolution}
  \label{fig:ThetaResolution}
\end{subfigure}%
\begin{subfigure}{.5\textwidth}
  \centering
  \includegraphics[width=\linewidth]{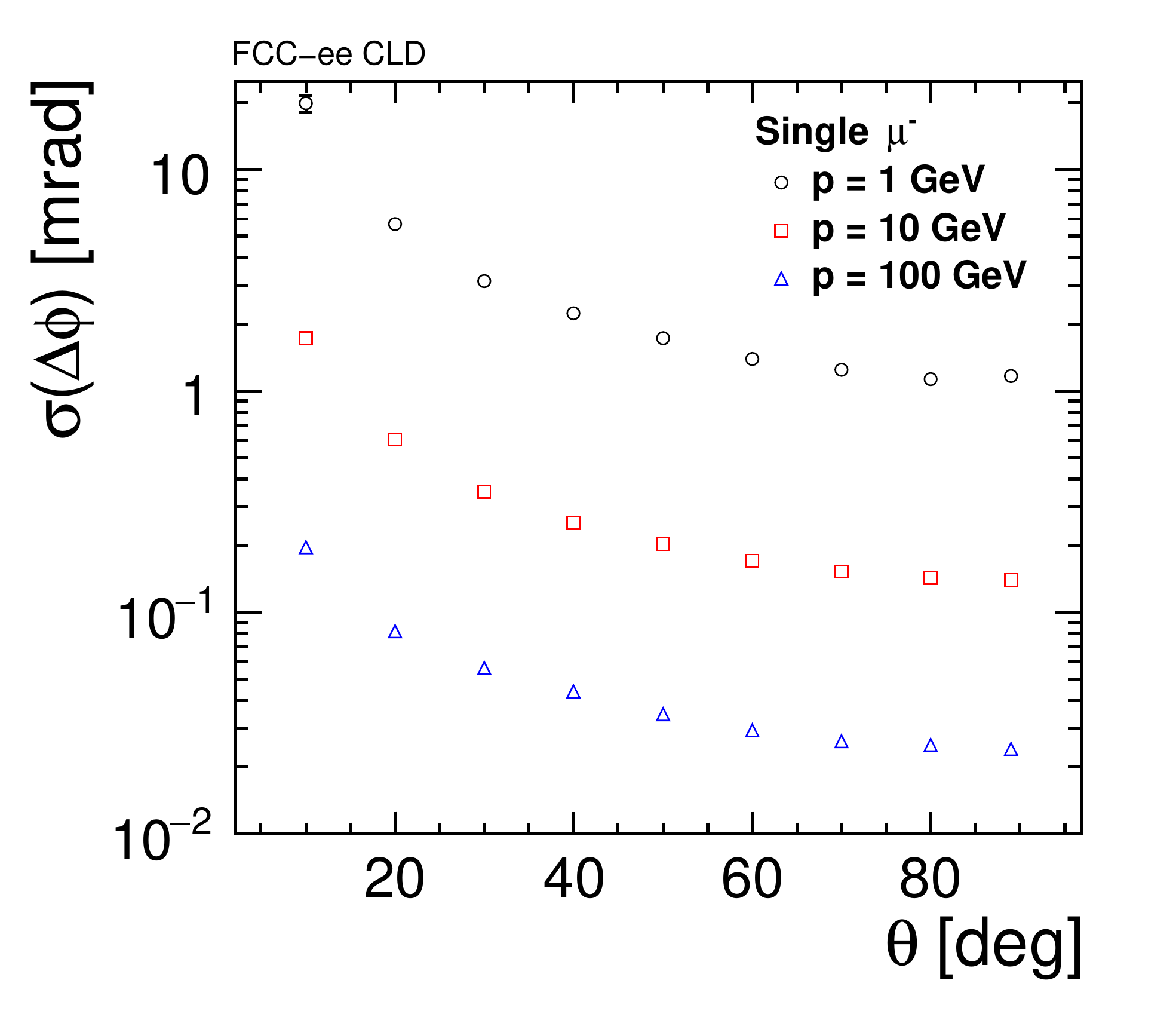}
  \caption{$\phi$ resolution}
  \label{fig:PhiResolution}
\end{subfigure}
\caption{(a) Polar and (b) azimuthal angular resolution as a function of polar angle for muon tracks with momenta of 1, 10 and 100 GeV.}
\label{fig:angular_res}
\end{figure}

Figure~\ref{fig:angular_res} shows the polar angular resolution (left) and the azimuthal angular resolution (right), both as a function of the polar angle $\theta$, for muon tracks of 1, 10 and 100 GeV. 
Both resolutions improve while moving from the forward to the transition region and then level up in the barrel. 
The only exception is the trend of the $\theta$ resolution for high energy muons, as it increases in the barrel region, where the single point resolution becomes dominant.
For the high energy muons, the $\phi$ resolution reaches a minimum of 0.023 mrad in the barrel and the $\theta$ resolution reaches the same value in the transition region.

\begin{figure}
\centering
\begin{subfigure}{.5\textwidth}
  \centering
  \includegraphics[width=\linewidth]{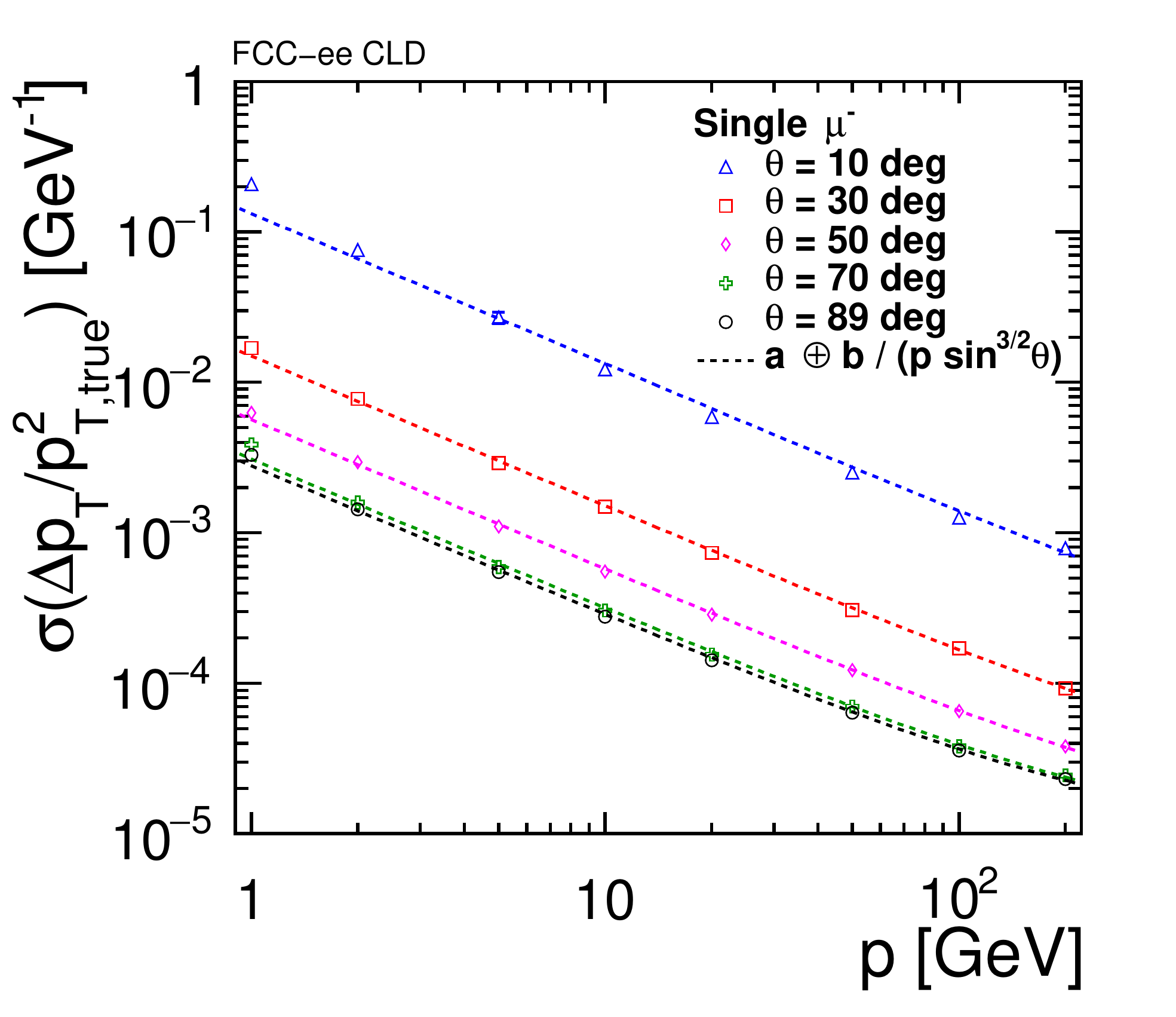}
  \caption{vs $p$}
  \label{fig:MomRes_vs_p}
\end{subfigure}%
\begin{subfigure}{.5\textwidth}
  \centering
  \includegraphics[width=\linewidth]{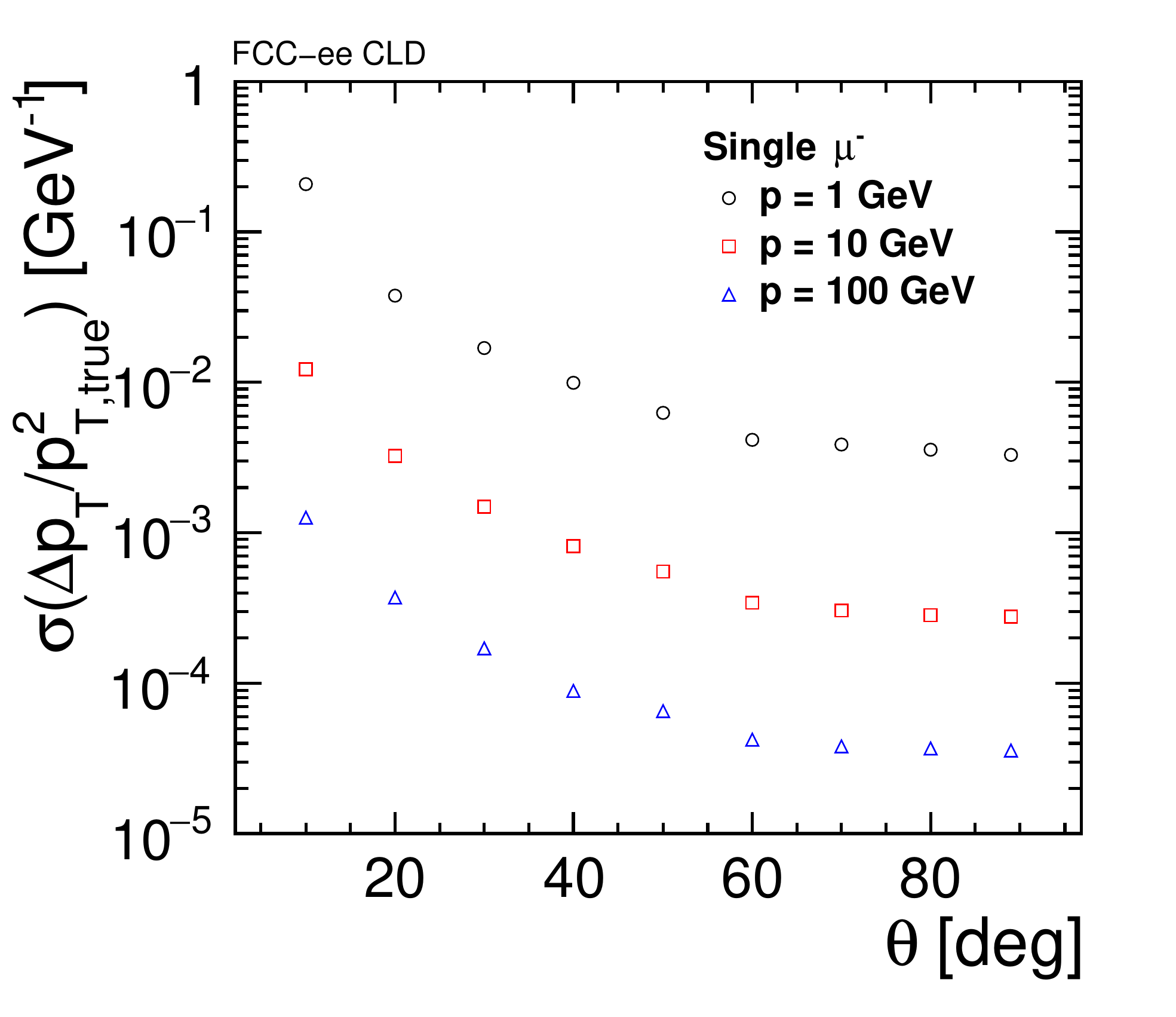}
  \caption{vs $\theta$}
  \label{fig:MomRes_vs_theta}
\end{subfigure}
\caption{Transverse momentum resolution for single muons (a) as a function of momentum at fixed $\theta$~=~10\degrees{}, 30\degrees{}, 50\degrees{}, 70\degrees{} and 89\degrees{} 
and (b) as a function of polar angle at fixed momentum $p$~=~1, 10 and 100 GeV. The dashed lines show a fit to the parameterisation given in Equation~\ref{eq:momRes}. The fitted parameters are given in Table~\ref{tab:parameters}.}
\label{fig:mom_res}
\end{figure}

The \pT resolution $\sigma(\Delta p_{\mathrm{T}}/p_\mathrm{{T}}^2)$ for single muons is determined from a single Gaussian fit of the distribution $(p_\mathrm{{T,MC}}-p_\mathrm{{T,rec}})/p_\mathrm{{T,MC}}^2$ and is shown in Figure~\ref{fig:mom_res} as a function of the momentum $p$ and of the polar angle $\theta$. Dashed lines in Figure~\ref{fig:MomRes_vs_p} correspond to the fit of the data points according to the parameterisation:
\begin{equation}
\sigma(\Delta p_\mathrm{{T}}/p_\mathrm{{T}}^2) = a \oplus \frac{b}{p \sin^{3/2}\theta}
\label{eq:momRes}
\end{equation}
where parameter~$a$ represents the contribution from the curvature measurement and parameter~$b$ is the multiple-scattering contribution. The values of these parameters for the different curves are summarised in Table~\ref{tab:parameters}. 
A \pT~resolution of 3.5$\cdot 10^{-5} \mathrm{GeV}^{-1}$ is achieved for 100\,GeV tracks in the barrel. 
For low-momentum tracks, the data points slightly deviate from the parameterisation due to the multiple scattering becoming dominant. 

\begin{table}[b!]
  \begin{center}
    \caption{Fit parameters from Equation~\ref{eq:momRes}, for the fitted curves shown in Figure~\ref{fig:MomRes_vs_p}. }
      \centering
        \begin{tabular}{c l c l c l}
	\toprule
$\theta$ & a & b \\
	\midrule
	    10\degrees{}   & $7.4 \cdot 10^{-5}$ & 0.010 \\
	    30\degrees{}   & $1.8 \cdot 10^{-5}$ & 0.005 \\
             50\degrees{}   & $9.2 \cdot 10^{-6}$ & 0.004 \\
             70\degrees{}   & $8.3 \cdot 10^{-6}$ & 0.003 \\
             89\degrees{}   & $8.6 \cdot 10^{-6}$ & 0.003 \\
\bottomrule 
        \end{tabular}

\label{tab:parameters}
\end{center}
\end{table}

\begin{figure}
\centering
\begin{subfigure}{.5\textwidth}
  \centering
  \includegraphics[width=\linewidth]{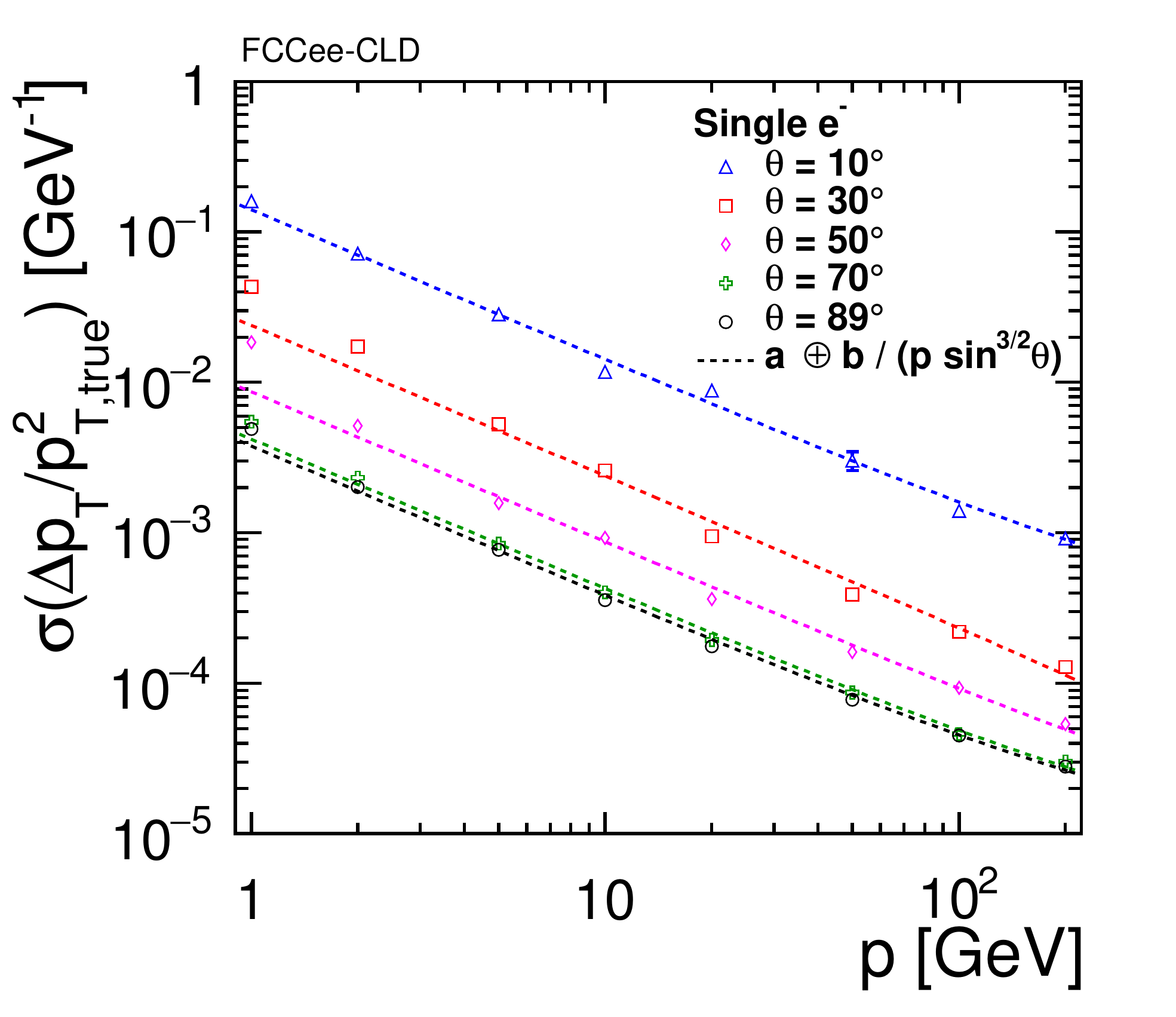}
  \caption{electrons}
  \label{fig:MomRes_vs_p_electrons}
\end{subfigure}%
\begin{subfigure}{.5\textwidth}
  \centering
  \includegraphics[width=\linewidth]{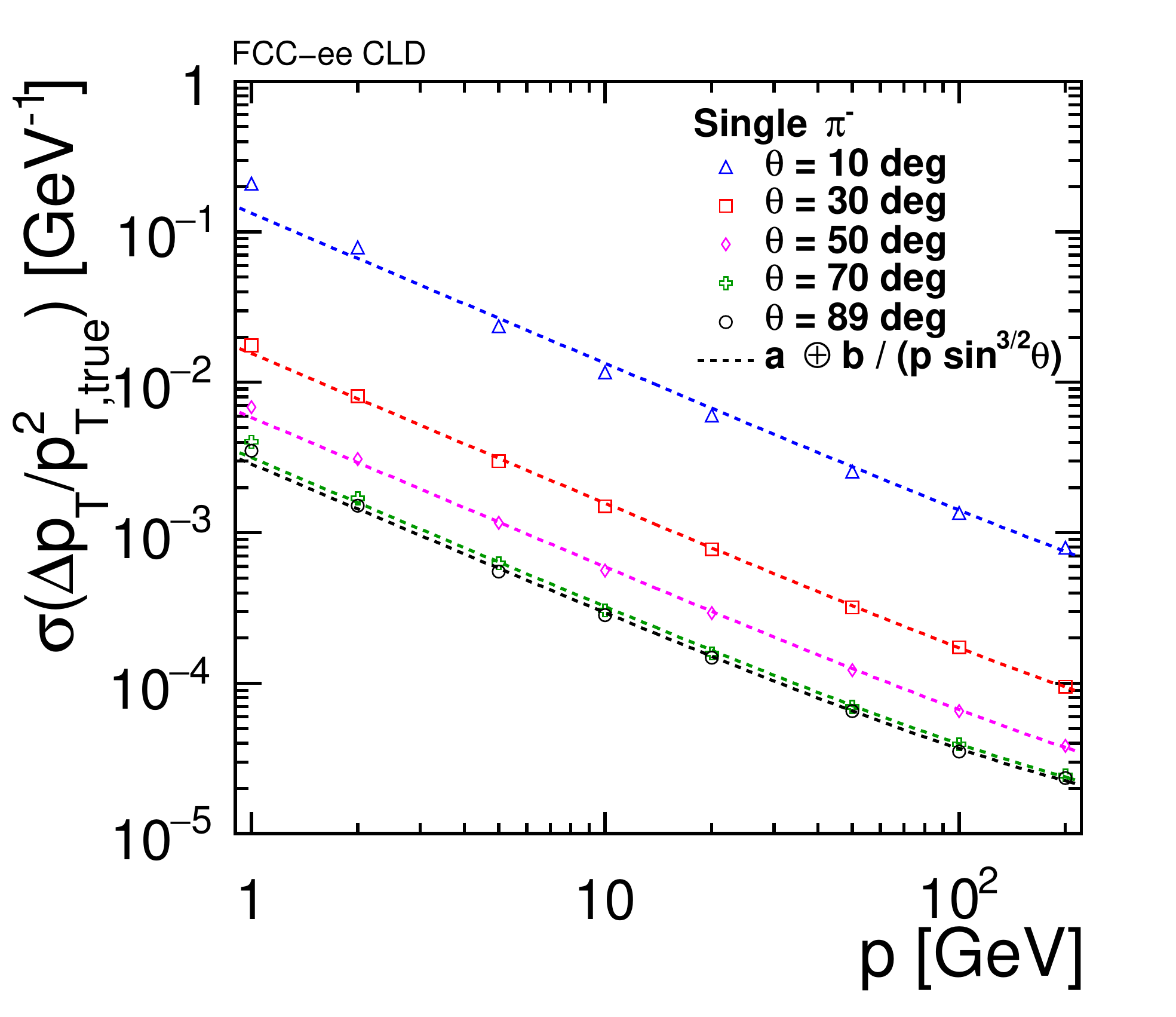}
  \caption{pions}
  \label{fig:MomRes_vs_p_pions}
\end{subfigure}
\caption{Transverse momentum resolution as a function of momentum at fixed $\theta$ = 10\degrees{}, 30\degrees{}, 50\degrees{}, 70\degrees{} and 89\degrees{} (a) for single electrons and (b) for single pions. 
The dashed lines show a fit to the parameterisation given in Equation~\ref{eq:momRes}.}
\label{fig:mom_res_pdg}
\end{figure}

Similarly, the momentum resolution for isolated electron and pion tracks was studied and is shown in Figure~\ref{fig:mom_res_pdg}, to test the tracking performance for other particles than muons. The residual distributions are fitted with a Gaussian in the 3$\sigma$ interval around 0, thus neglecting the electrons in the low-energy tail of the momentum distribution that have irradiated high-energyBremsstrahlung photons.
The resulting performances at high energies are very similar to those for isolated muon tracks.

\paragraph{Tracking Efficiency}

Tracking efficiency is defined as the fraction of the reconstructable Monte Carlo particles that have been reconstructed. 
A particle is considered reconstructable if it is stable at generator level (genStatus = 1)~\footnote{Particles with lifetimes $\tau$ such that $c\tau$ $\ge$ 10 mm.}, if \pT > 100~MeV, $|\cos(\theta)|$ < 0.99 and if it has at least 4 unique hits (i.e. hits which do not occur on the same sub-detector layer). 

\begin{figure}[htbp]
  \centering
 \includegraphics[width=0.623\linewidth]{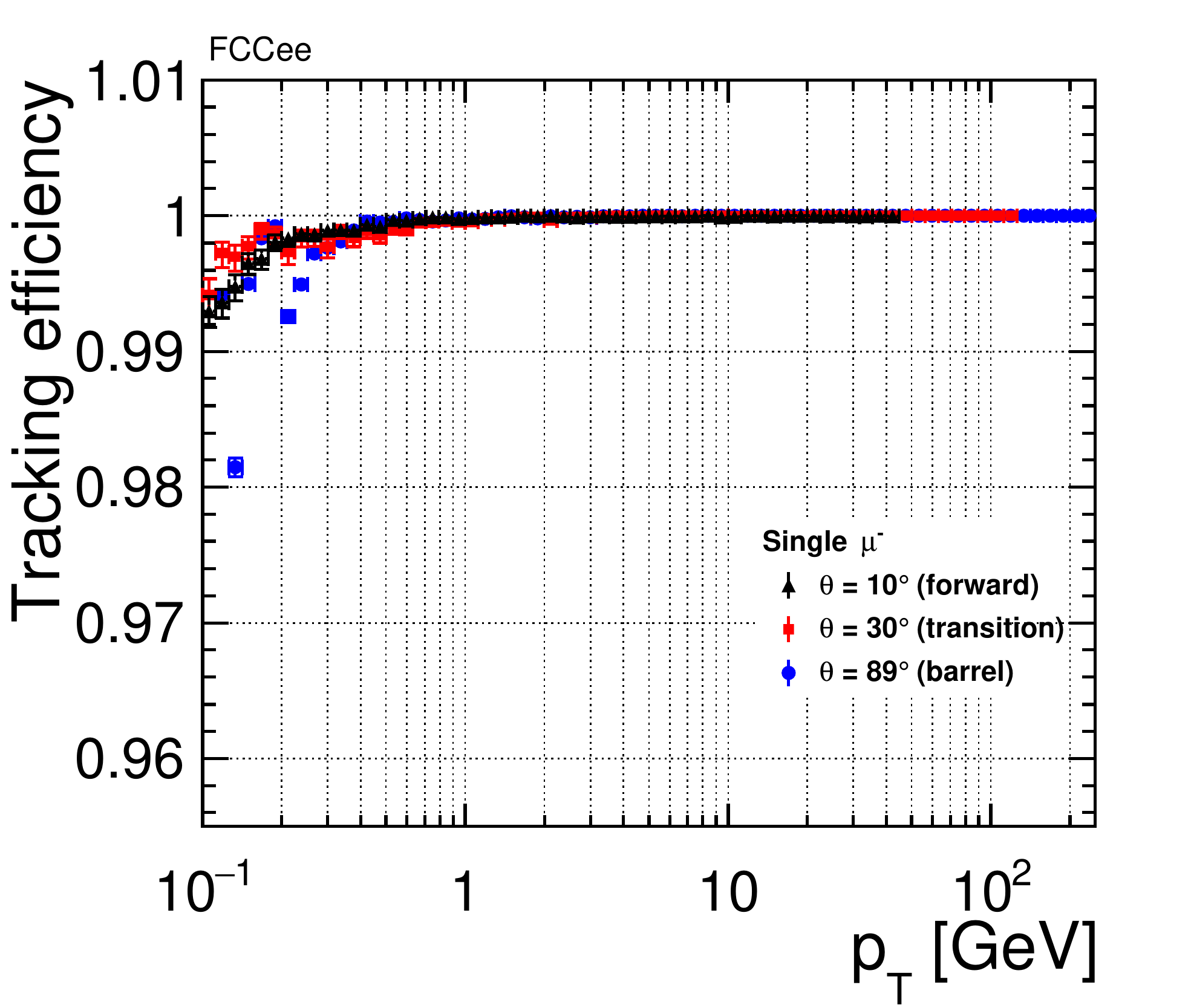}
         \caption{Tracking efficiency as a function of \pT for muons with energies up to 250 GeV, at polar angles $\theta$ = 10\degrees{}, 30\degrees{} and 89\degrees{}.
}
   \label{fig:eff_vs_pT}
\end{figure}

The efficiency for isolated muon tracks, shown in Figure~\ref{fig:eff_vs_pT}, has been computed by reconstructing 2 million muons simulated at polar angles $\theta$ = 10\degrees{}, 30\degrees{}, 89\degrees{} 
and with a power-law energy distribution (maximum energy 250 GeV) to favour statistics at low-\pT. The tracking is fully efficient for single tracks with transverse momentum greater than 400~MeV.

\begin{figure}
\centering
\begin{subfigure}{.5\textwidth}
  \centering
  \includegraphics[width=\linewidth]{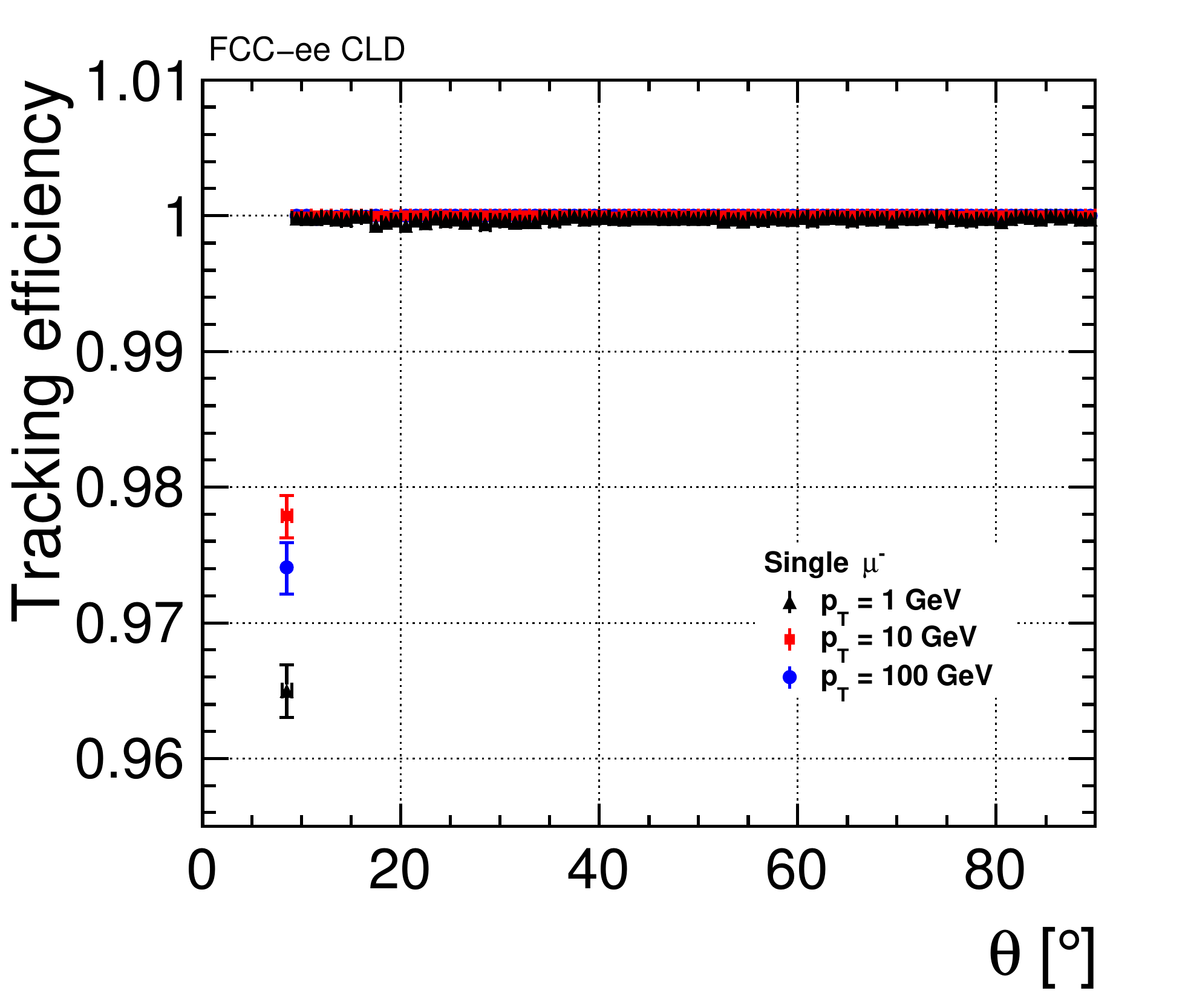}
  \caption{vs $\theta$}
  \label{fig:eff_vs_theta}
\end{subfigure}%
\begin{subfigure}{.5\textwidth}
  \centering
  \includegraphics[width=\linewidth]{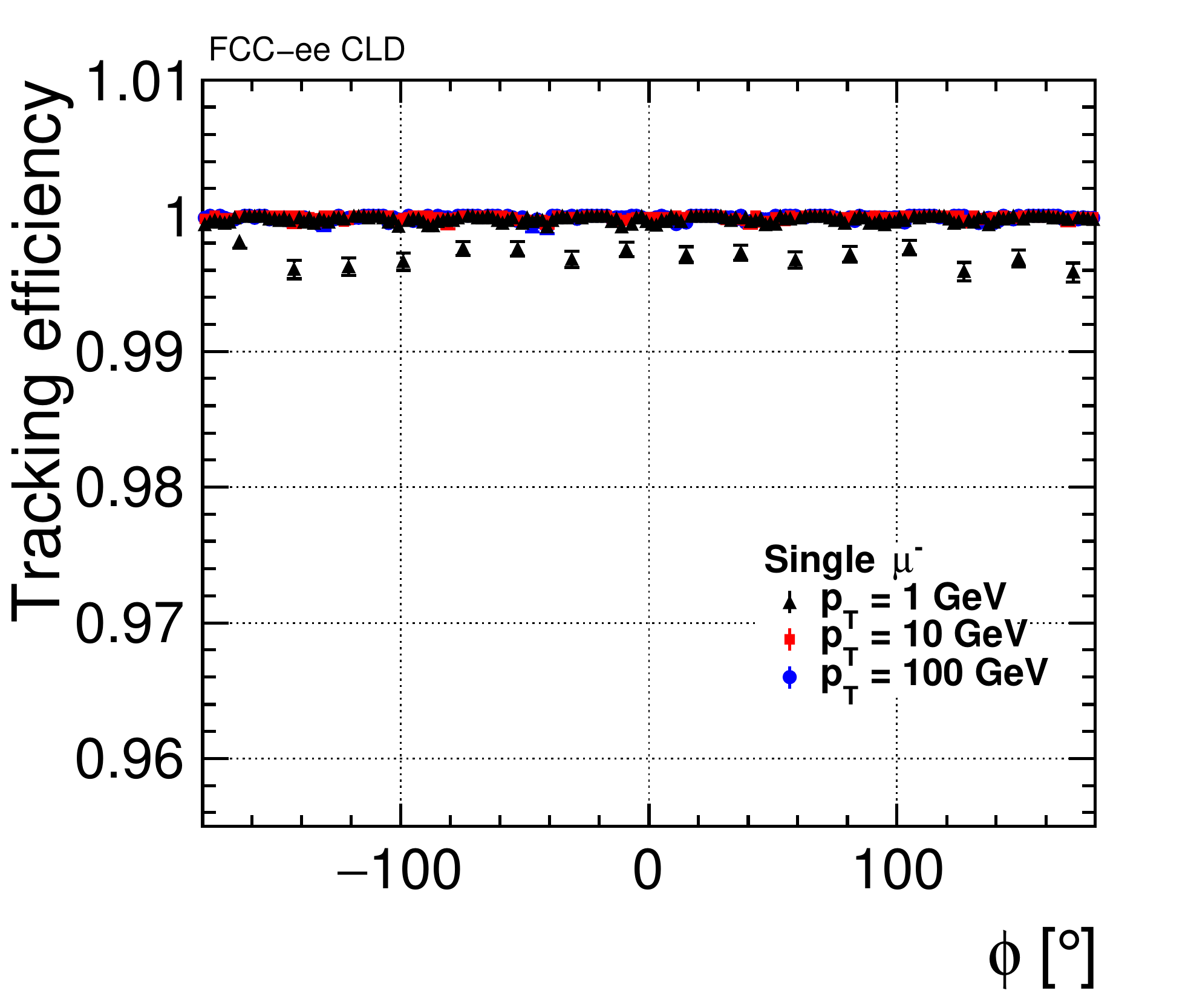}
  \caption{vs $\phi$}
  \label{fig:eff_vs_phi}
\end{subfigure}
\caption{Tracking efficiency as a function of (a) polar and (b) azimuthal angle for muons with momenta of 1, 10 and 100 GeV.}
\label{fig:eff_vs_angle}
\end{figure}

Figure~\ref{fig:eff_vs_angle} shows the same efficiency as a function of polar angle (a) and azimuthal angle (b). An efficiency drop for all energies is observed only in the most forward bin (8\degrees{}). The oscillation pattern shown in Figure~\ref{fig:eff_vs_phi} reflects the position of overlaps between modules of the layers.

\begin{figure}
\centering
\begin{subfigure}{.5\textwidth}
  \centering
  \includegraphics[width=\linewidth]{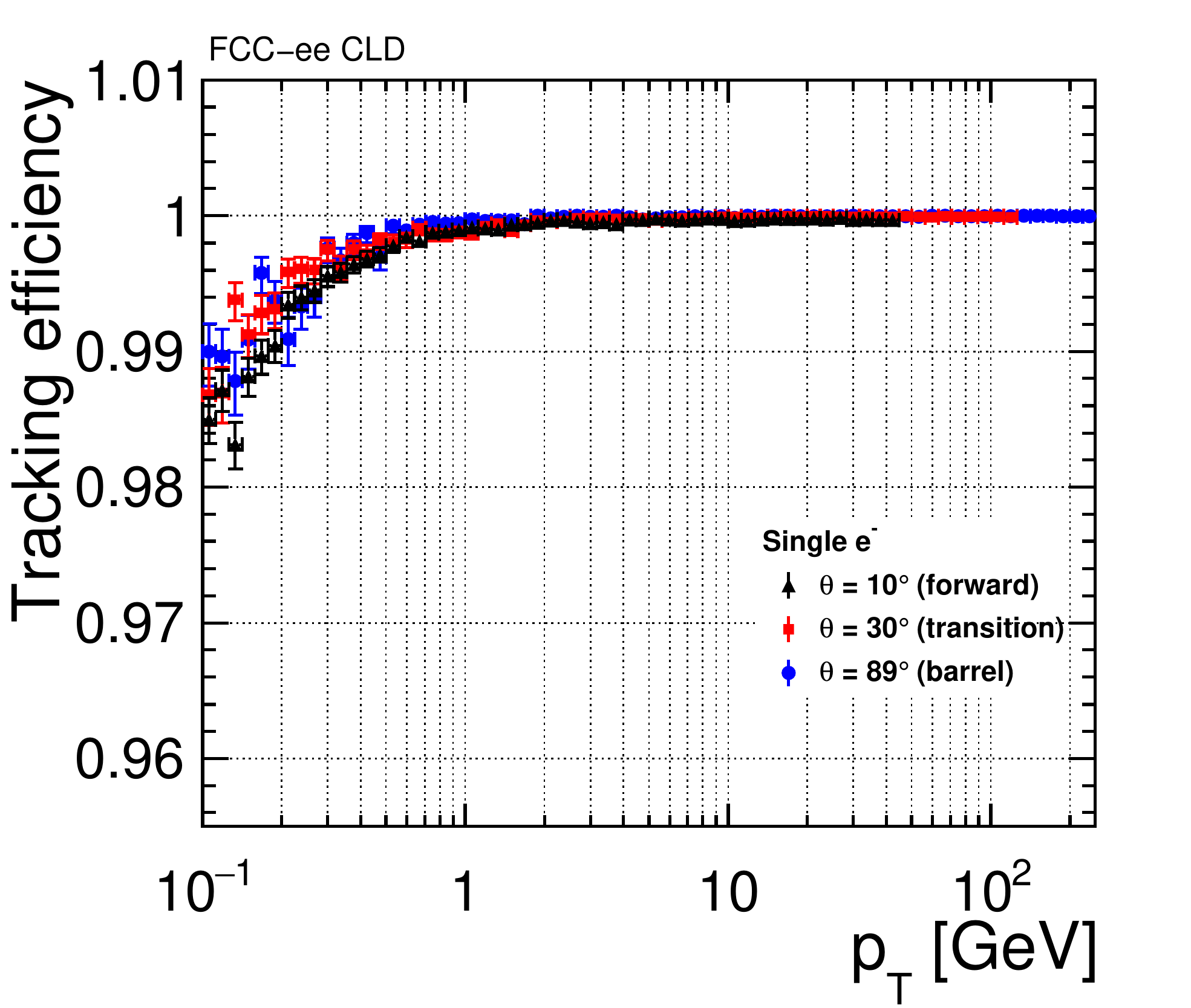}
  \caption{electrons}
  \label{fig:electrons_eff_vs_pT}
\end{subfigure}%
\begin{subfigure}{.5\textwidth}
  \centering
  \includegraphics[width=\linewidth]{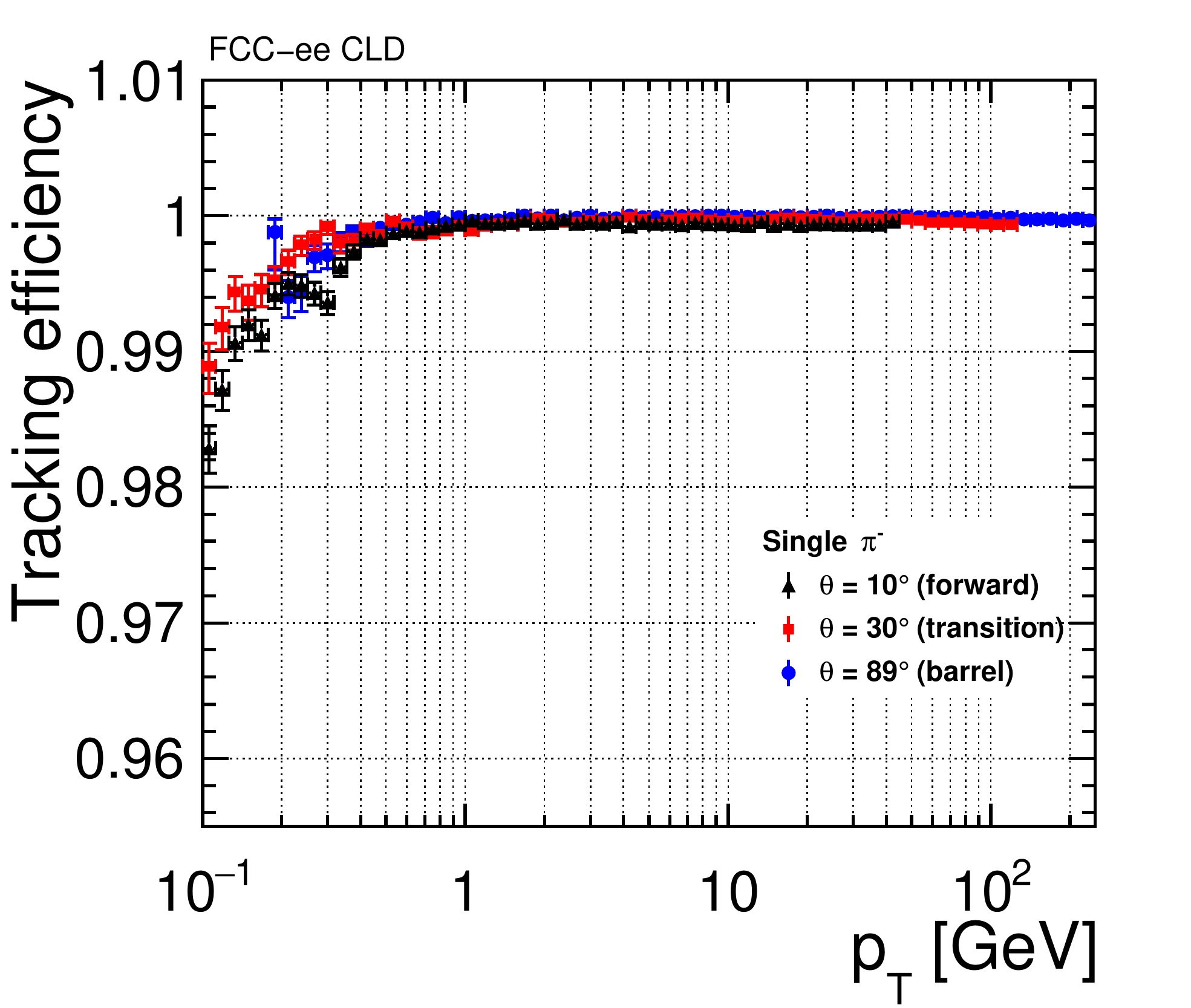}
  \caption{pions}
  \label{fig:pions_eff_vs_pT}
\end{subfigure}
\caption{Tracking efficiency vs $p_{T}$, (a) for electrons and (b) for pions.}
\label{fig:eff_pions_elect}
\end{figure}

Similarly, the tracking efficiency for 2 million isolated electrons and pions simulated at polar angles $\theta$ = 10\degrees{}, 30\degrees{} and 89\degrees{} and with a power-law energy distribution, is shown in Figure~\ref{fig:eff_pions_elect}. 

Electrons in the investigated momentum range lose energy mainly through bremsstrahlung.  
For electrons at all angles, the efficiency reaches 100\% above 1~GeV. 
At low transverse momentum, at any of the probed angles, the efficiency does not drop below a minimum of 98\%. 

The trend for pions does not differ from that of electrons at the same angle. The efficiency is 100\% down to 600~MeV, and it drops slightly, but remains above 98\%, for lower-$\pT$ tracks. 

\begin{figure}[htbp]
  \centering
 \includegraphics[width=0.623\linewidth]{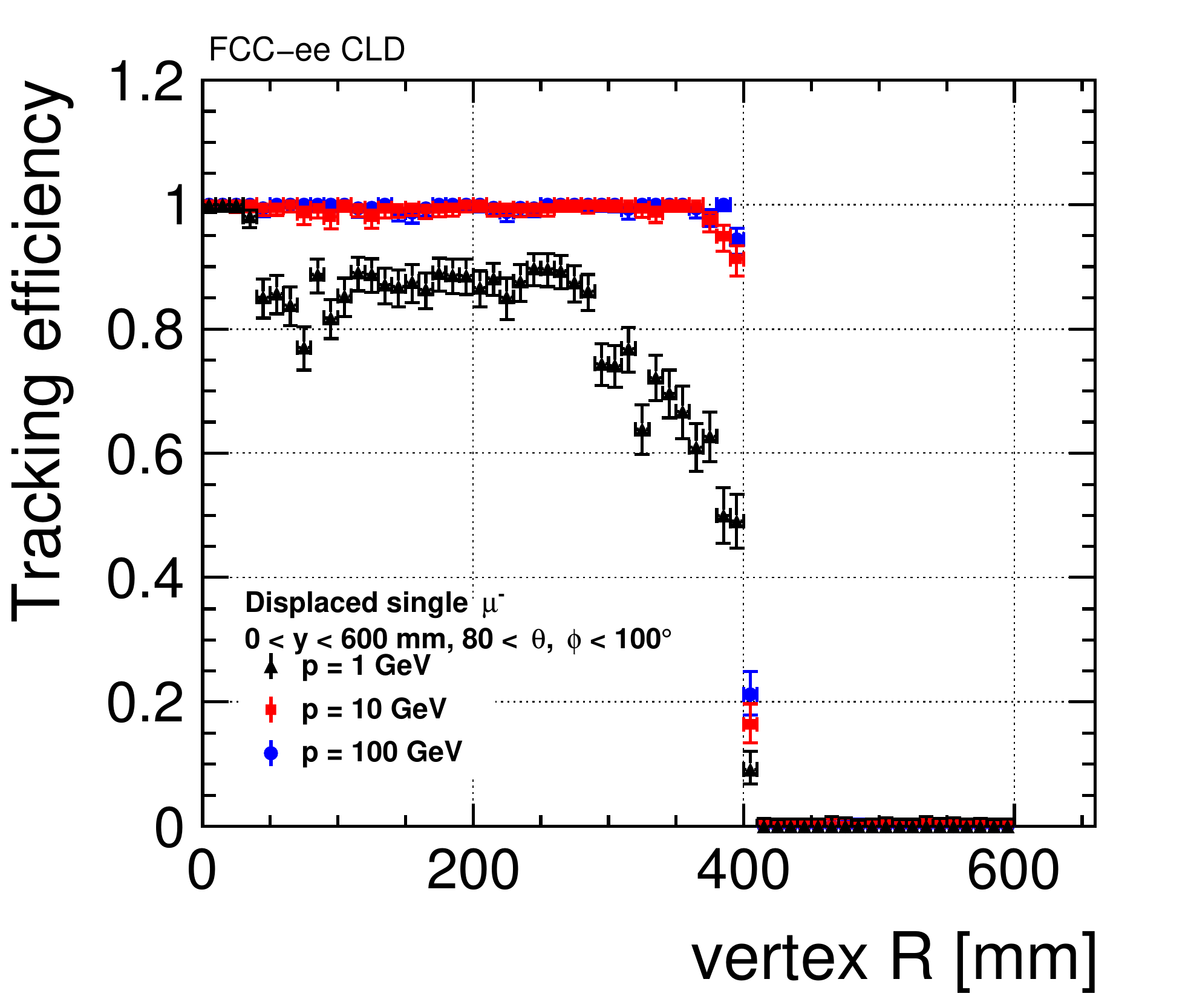}
         \caption{Tracking efficiency as a function of production vertex radius for muons with momenta of 1, 10 and 100\,GeV, uniformly generated in a range 0 < $y$ < 600 mm and 80\degrees{} < $\theta$,$\phi$ < 100\degrees{}.}
   \label{fig:eff_displaced_vs_vertexR}
\end{figure}

To probe the tracking performance for displaced tracks, 10$^4$ single muons have been simulated, requiring their production vertex to be within 0 cm < $y$ < 60 cm and their angular distribution in a 10\degrees{} cone around the $y$ axis, i.e. 80\degrees{} < $\theta,\phi$ < 100\degrees{}. This is done so that particles are produced in the barrel region only and they traverse roughly the same amount of material. The efficiency as a function of production vertex radius, i.e. $\sqrt{x^2+y^2}$, is shown in Figure~\ref{fig:eff_displaced_vs_vertexR} for muons with momenta of 1, 10 and 100 GeV. For 1 GeV muons the efficiency is around 100\%, except for those that are produced after the first two vertex barrel double layers (radius R > 38~mm), for which the efficiency drops by 15\%.
Due to energy loss while traversing the detector layers, some particles have not enough left-over momentum to leave the required minimum number of hits. For higher-energy muons, instead, the efficiency is constantly 100\% over most of the probed production vertex range. Regardless of the energy, an abrupt fall-off is observed for all tracks with a production radius of 400 mm or more. This is an effect of the reconstruction cuts, since for displaced tracks a minimum number of 5 hits is required to make the track, while only 4 sensitive layers are traversed by tracks starting beyond R = 400 mm.

\paragraph{Particle Reconstruction and Identification}

Particles are reconstructed and identified using the \pandora{} Particle Flow Analysis Toolkit~\cite{Marshall:2015rfaPandoraSDK}. 
The particle flow reconstruction algorithms of Pandora have been studied extensively  in full \geant{} simulations of the ILD and the CLIC\_ILD detector concept~\cite{Marshall:2012ryPandoraPFA} as well as of CLICdet. 
Particle flow aims to reconstruct each visible particle in the event using information from all sub-detectors. The high granularity of calorimeters is essential in achieving the desired precision measurements. 
Electrons are identified using clusters largely contained within ECAL and matched with a track. Muons are determined from a track and a matched cluster compatible with a minimum ionizing particle signature in ECAL and HCAL, 
plus corresponding hits in the muon system. A hadronic cluster in ECAL and HCAL matched to a track is used in reconstructing charged hadrons. Hadronic clusters without a corresponding track are interpreted as neutrons,
 and photons are reconstructed from an electromagnetic cluster in ECAL\@. In jets typically 60\% of the energy originates from charged hadrons and 30\% from photons. The remaining 10\% of the jet energy are mainly carried by neutral hadrons.

\begin{figure}[pb]
  \centering
  \begin{subfigure}{.5\textwidth}
    \centering
    \includegraphics[width=\linewidth]{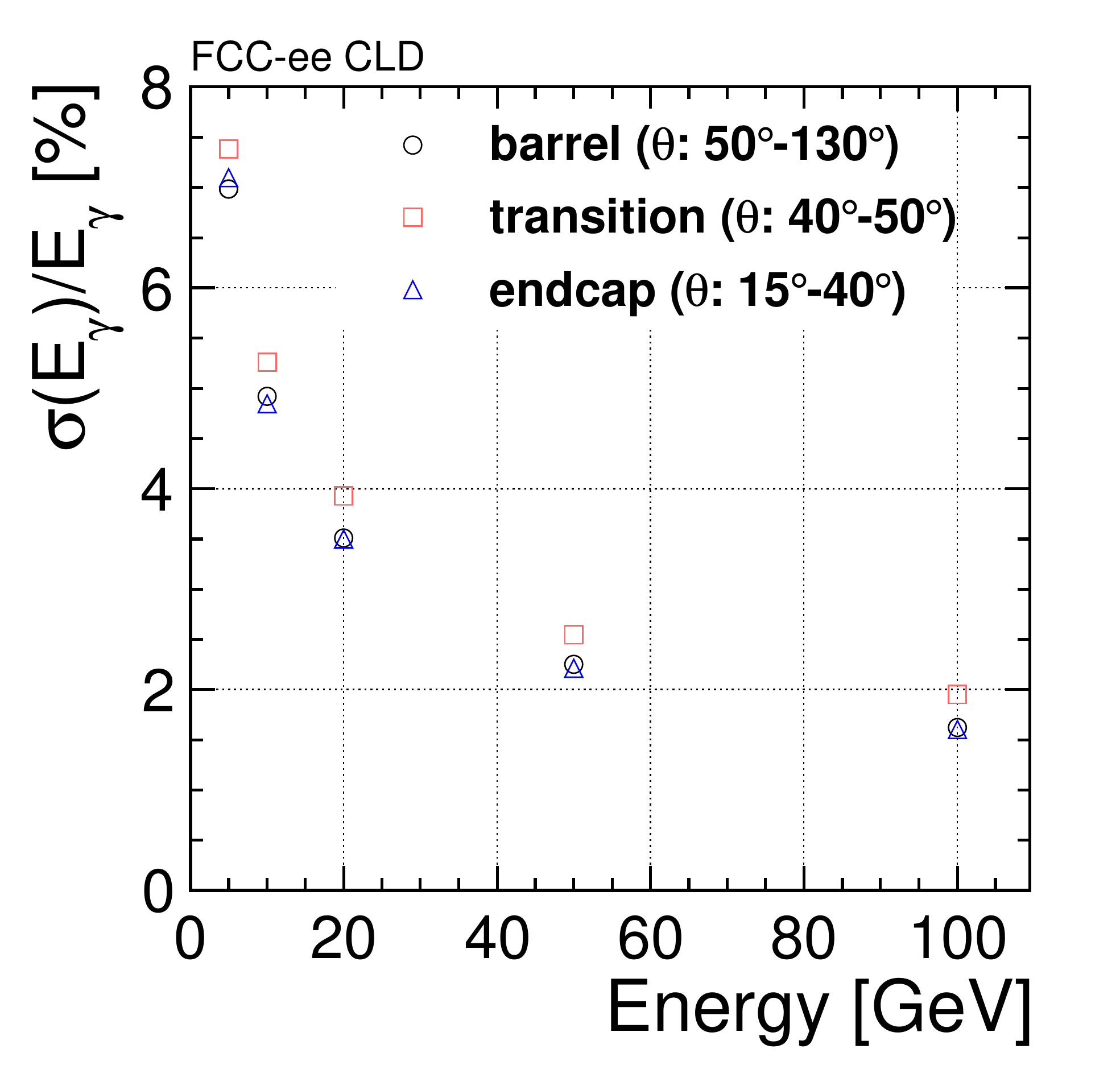}
    \caption{}
    \label{fig:photonResolutionVsEnergy}
  \end{subfigure}%
  \begin{subfigure}{.5\textwidth}
    \centering
   \includegraphics[width=\linewidth]{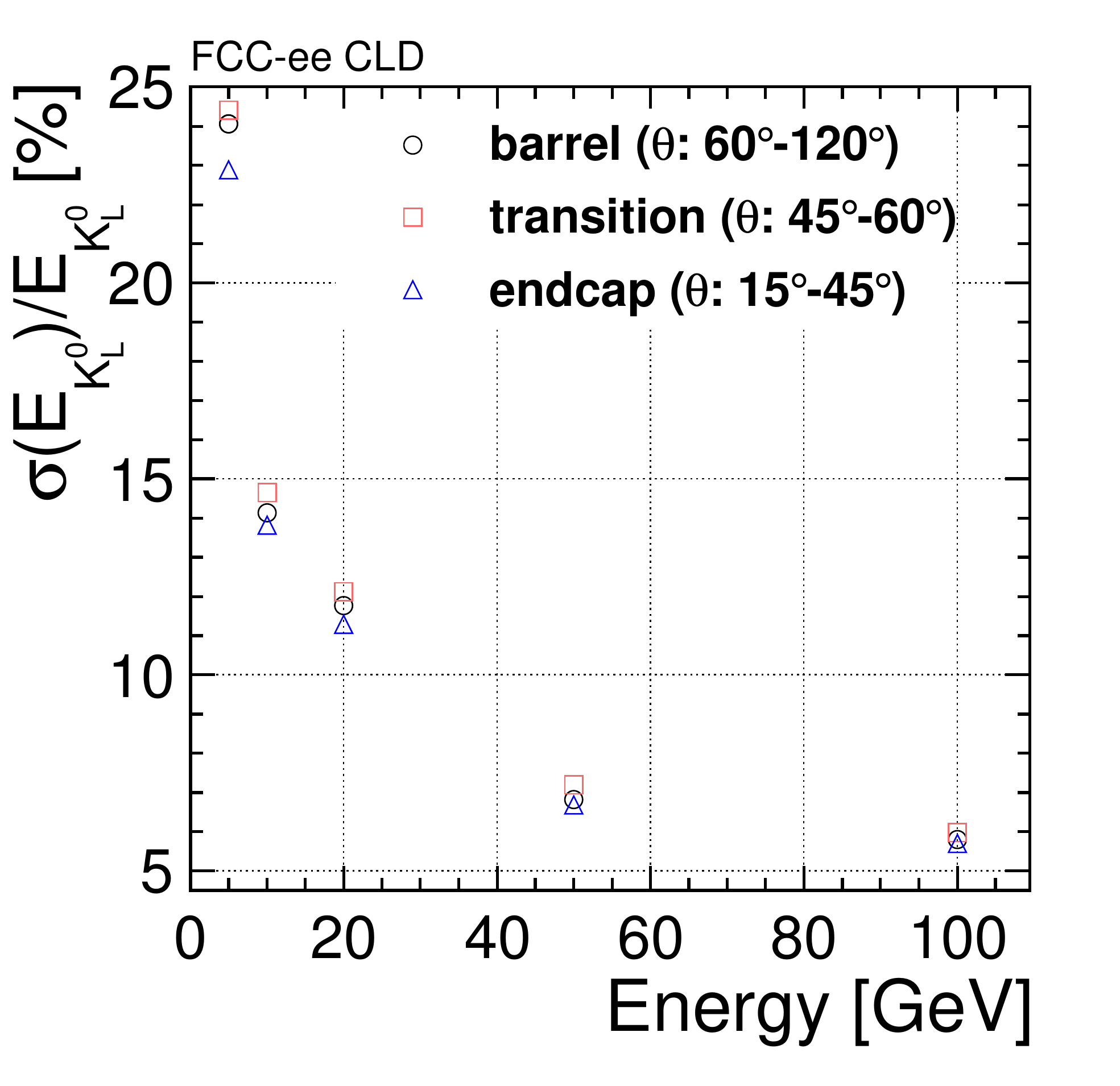}
    \caption{}
    \label{fig:K0LResolutionVsEnergy}
  \end{subfigure}
  \caption{(a) Photon energy resolution and (b) neutral hadron resolutions of \PKzL's  as a function of energy. Results are shown for the barrel region, transition region and endcap.}
\end{figure}

The performance of the Pandora reconstruction algorithms is studied in single particle events at several energies, generated as flat distributions in $\cos\theta$. The ECAL energy resolution is studied using single photon events. At each energy point in three different regions (barrel, endcap, and transition region) the photon energy response distribution is iteratively fitted with a Gaussian within a range $\pm3\sigma$. The $\sigma$ of the Gaussian is a measure for the energy resolution in ECAL\@. 
The energy dependence of the photon energy resolution of CLD is shown in Figure~\ref{fig:photonResolutionVsEnergy}, for the three detector regions. The stochastic term is $15\%/\sqrt{E}$, determined from a two parameter fit within the energy range of 5 and 100 GeV\@.

For hadrons the HCAL hits are re-weighted using the software compensation technique implemented within Pandora, and developed by the CALICE collaboration~\cite{CALICE_sc_2012, Tran:2017tgrSoftwareCompensation}. In the non-compensating calorimeters of CLD the detector response for electromagnetic sub-showers is typically larger than for hadronic showers. On average the electromagnetic component of the shower has larger hit energy densities. The weights applied depend on the hit energy density and the unweighted energy of the calorimeter cluster, where hits with larger hit energy densities receive smaller weights. In a dedicated calibration procedure within Pandora, software compensation weights are determined using single neutron and \PKzL{} events over a wide range of energy points. At each energy point equal statistics is required, i.e.\ using the same number of events for neutrons and \PKzL{}. Only events with one cluster fully contained within ECAL plus HCAL are used in this calibration. Software compensation improves the energy resolution of hadronic clusters. 
The resulting energy resolution of neutral hadrons is shown for \PKzL as a function of energy in Figure~\ref{fig:K0LResolutionVsEnergy}.

\begin{figure}[t!]
  \centering
  \begin{subfigure}{.5\textwidth}
    \centering
   \includegraphics[width=\linewidth]{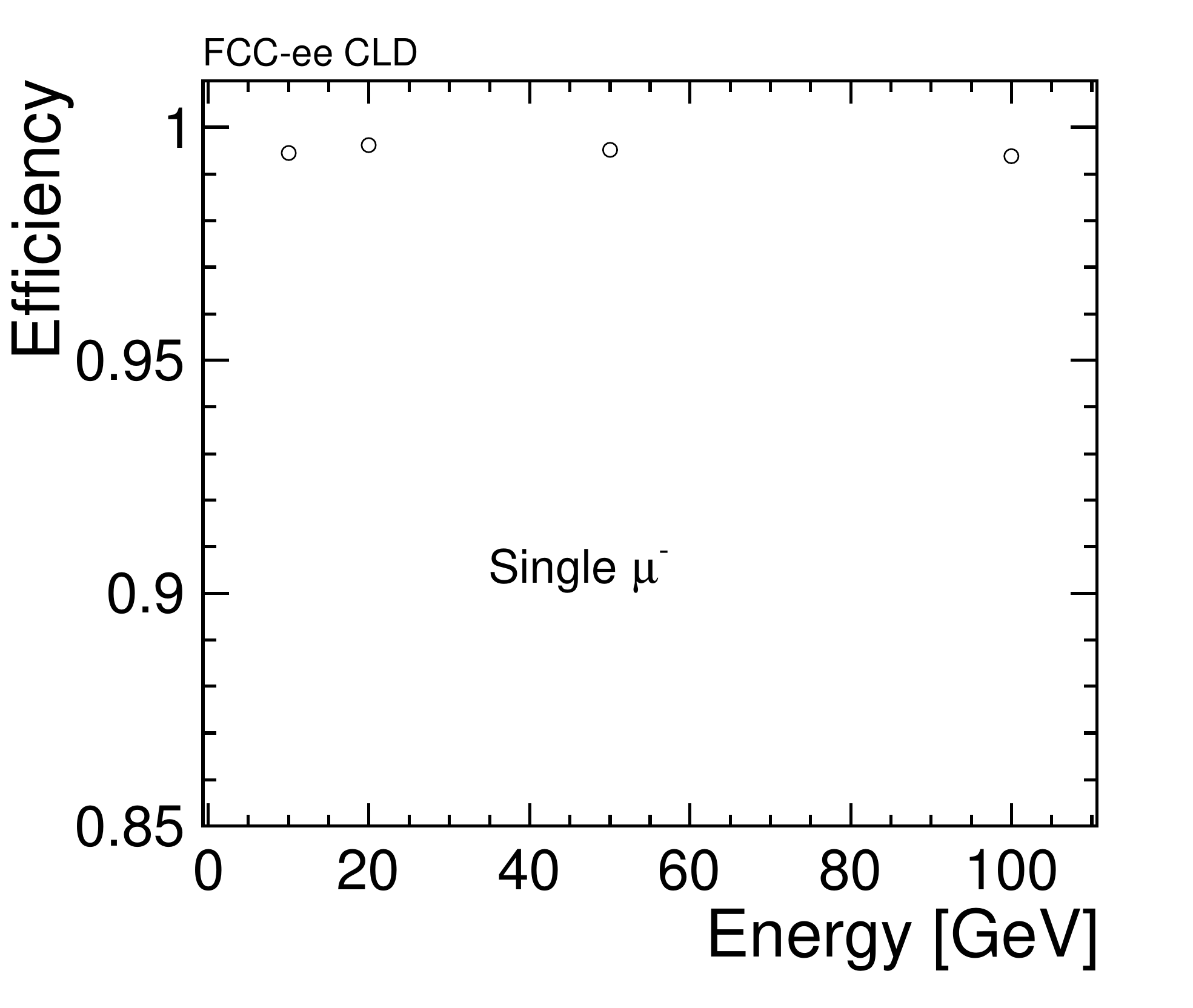}
    \caption{}
    \label{fig:particleGun_muIDEffVsE}
  \end{subfigure}%
  \begin{subfigure}{.5\textwidth}
    \centering
   \includegraphics[width=\linewidth]{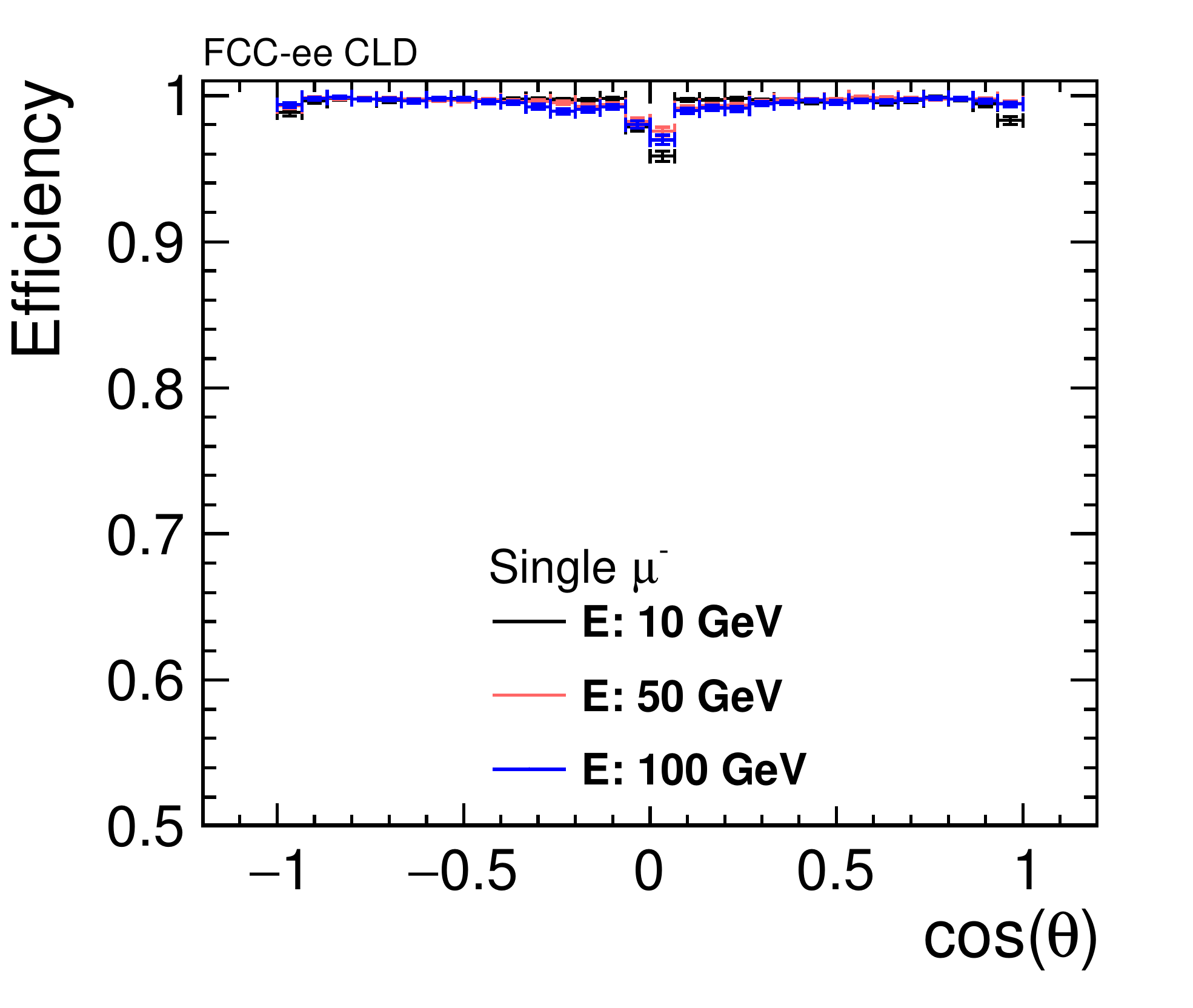}
    \caption{}
    \label{fig:particleGun_muIDEffVsTheta}
  \end{subfigure}
  \caption{Particle identification efficiency for muons (a) as a function of energy and (b) as a function of $\cos\theta$ for three different energies. Error bars (statistical errors) are sometimes smaller than the dots.}
\end{figure}

\begin{figure}[t!]
  \centering
  \begin{subfigure}{.5\textwidth}
    \centering
   \includegraphics[width=\linewidth]{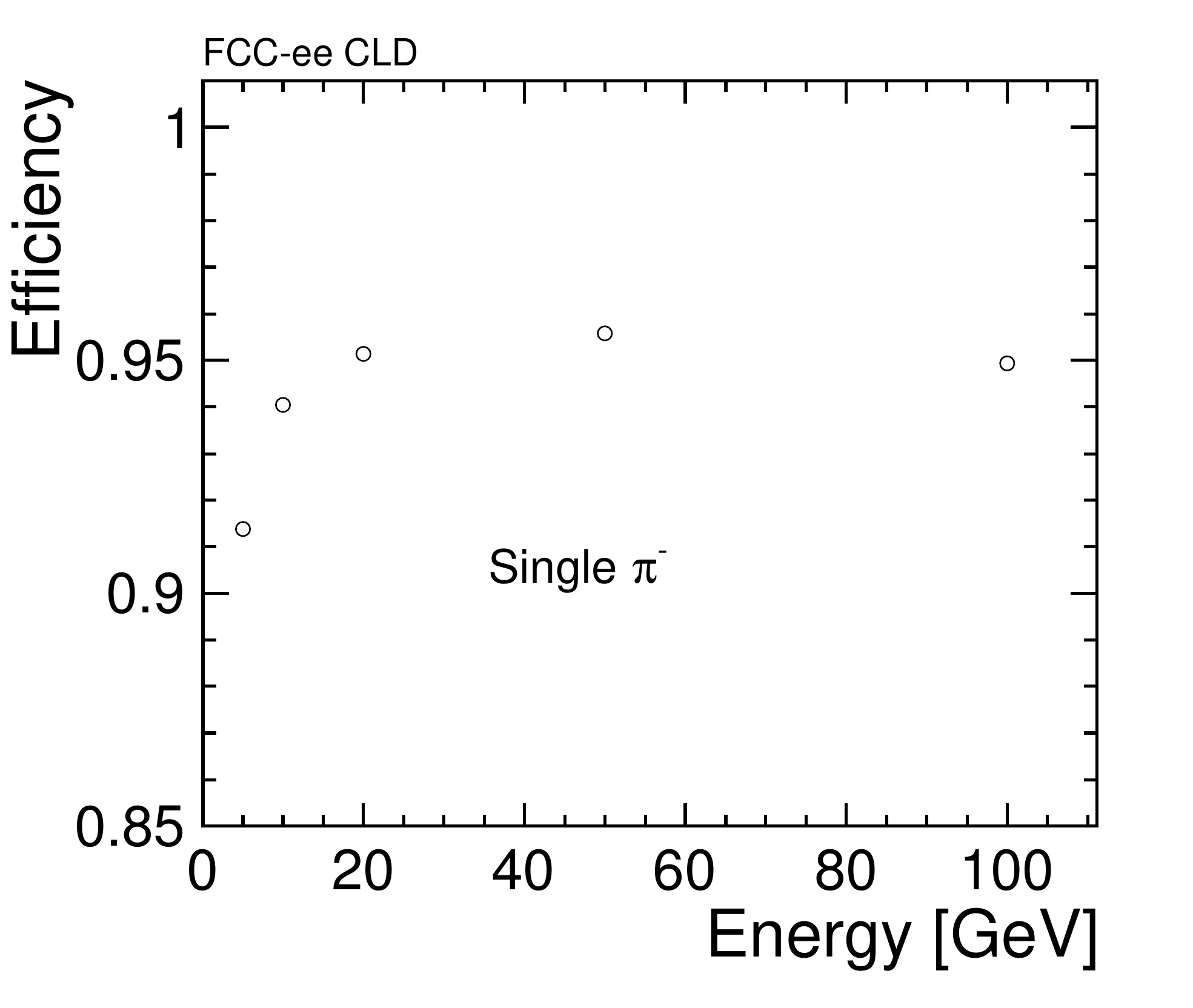}
    \caption{}
    \label{fig:particleGun_piIDEffVsE}
  \end{subfigure}%
  \begin{subfigure}{.5\textwidth}
    \centering
   \includegraphics[width=\linewidth]{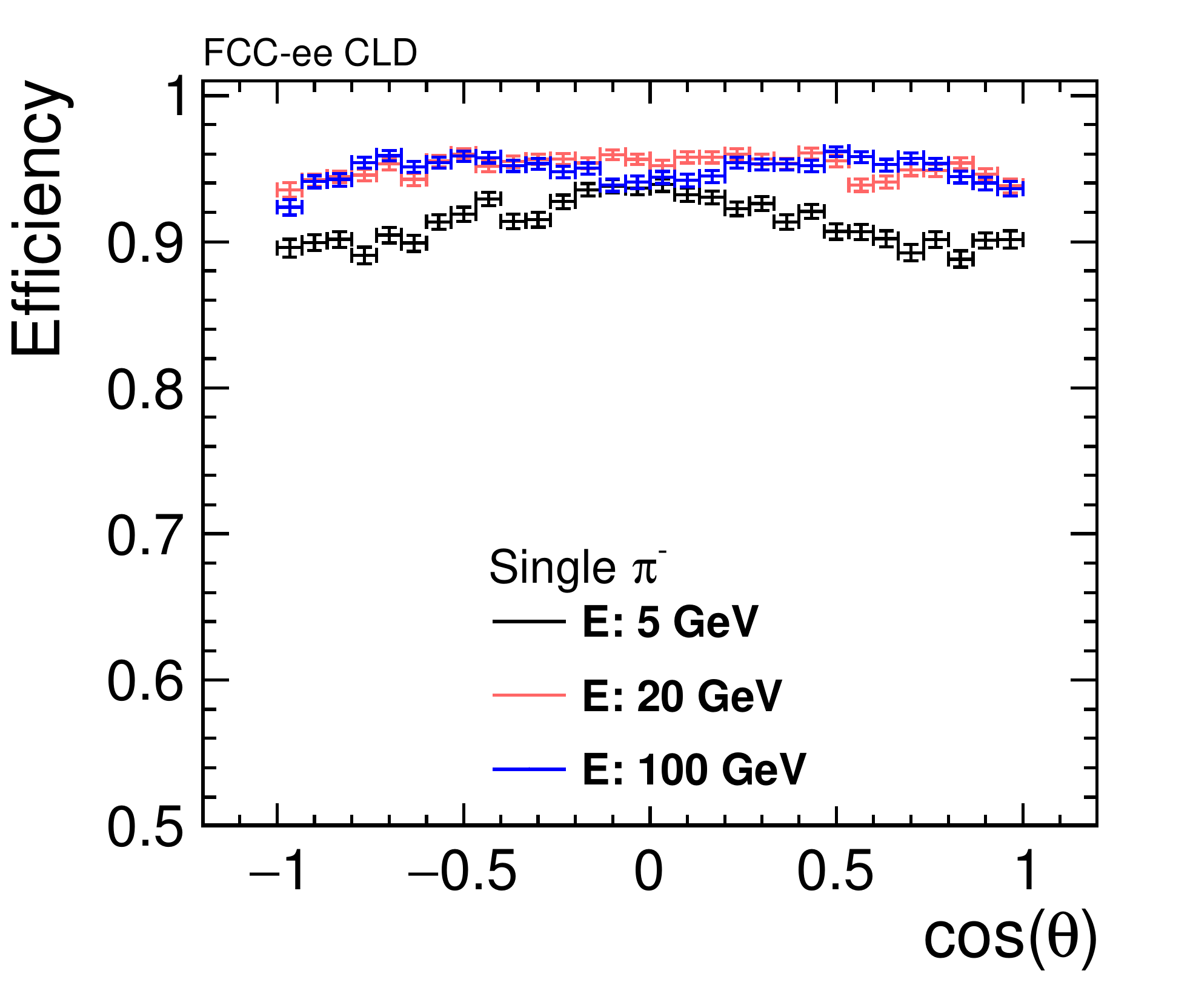}
    \caption{}
    \label{fig:particleGun_piIDEffVsTheta}
  \end{subfigure}
  \caption{Particle identification efficiency for pions (a) as a function of energy and (b) as a function of $\cos\theta$ for three different energies. Error bars (statistical errors) are sometimes smaller than the dots.}
\end{figure}

The particle identification efficiency of Pandora particle flow algorithms is studied in single particle events separately for muons, electrons, photons, and pions. The events are produced as flat distributions in $\cos\theta$. 
The reconstructed particle is required to be of the same type as the `true' particle.
It has to satisfy angular matching criteria $|\phi_{reconstructed}-\phi_{true}|< 2$ mrad and $|\theta_{reconstructed}-\theta_{true}|< 1$ mrad. 
The reconstructed transverse momentum of charged particles has to be within 5$\%$ of the transverse momentum of the `true' particle.
The particle identification efficiency is studied as a function of energy and $\cos\theta$. 
The result for muons is illustrated in Figure~\ref{fig:particleGun_muIDEffVsE} as a function of energy and in Figure~\ref{fig:particleGun_muIDEffVsTheta} as a function of $\cos\theta$. 
The efficiency is above 99\% for all energies, and generally flat as a function of $\cos\theta$. The small dip around 90\degrees, which is also observed in CLICdet, is the subject of further investigations.

The efficiency of pion identification is about 90\% at low energies, 95--96\% at energies from 20 GeV up to 100 GeV (Figure~\ref{fig:particleGun_piIDEffVsE}), and flat as a function of $\cos\theta$ at high energies (Figure~\ref{fig:particleGun_piIDEffVsTheta}).
The pion inefficiency is mainly due to mis-identification of the particle type by Pandora. At high energies, the most common case is mis-identification of a pion as a muon. This happens when the shower starts late in the calorimeter and some particles from the shower reach the muon chambers. At low energies, the pion is mostly mis-reconstructed as an electron. The shower can start in the ECAL which can lead to a significant amount of energy being deposited in it. 45\% of 5 GeV pions in the barrel region (60\degrees $< \theta <$ 120\degrees) have most of their energy deposited in the ECAL instead of the HCAL. This is the case only for 30\% of 20 GeV pions. This can lead to confusion of the particle identification algorithm at low pion energies.
This example demonstrates that further tuning of the Pandora algorithms will be required, e.g. to improve the particle identification efficiency for lower energy pions.

While for muons and pions the energy is accurately reconstructed, for electrons the reconstructed energy has, due to bremsstrahlung, a long tail towards lower values compared to the true energy. A simple bremsstrahlung recovery algorithm is applied which uses close by photons (within $|\phi_{reconstructed}-\phi_{true}|< 20$ mrad and $|\theta_{reconstructed}-\theta_{true}|< 1$ mrad) to dress the electron momentum by summing their four momenta. Additionally, reconstructed electrons, which were dressed with photons, are required to satisfy a looser energy matching requirement (reconstructed energy by the calorimeter has to be within 5~$\sigma_\mathrm{{E}}$) since part of the their energy has been measured by the calorimeter only.

\begin{figure}[htbp]
  \centering
  \begin{subfigure}{.5\textwidth}
    \centering
    \includegraphics[width=\linewidth]{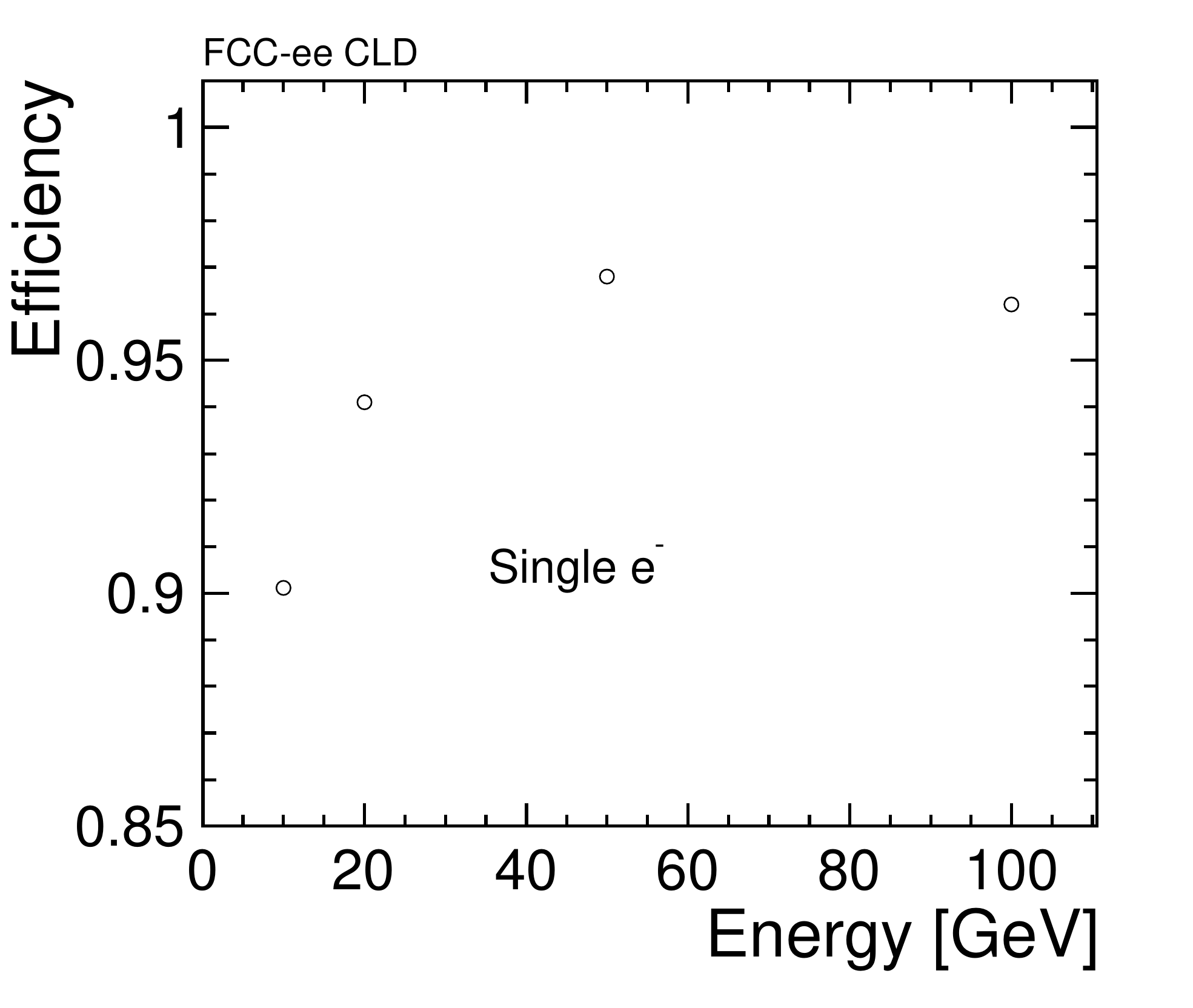}
    \caption{}
    \label{fig:particleGun_elIDEffVsE}
  \end{subfigure}%
  \begin{subfigure}{.5\textwidth}
    \centering
    \includegraphics[width=\linewidth]{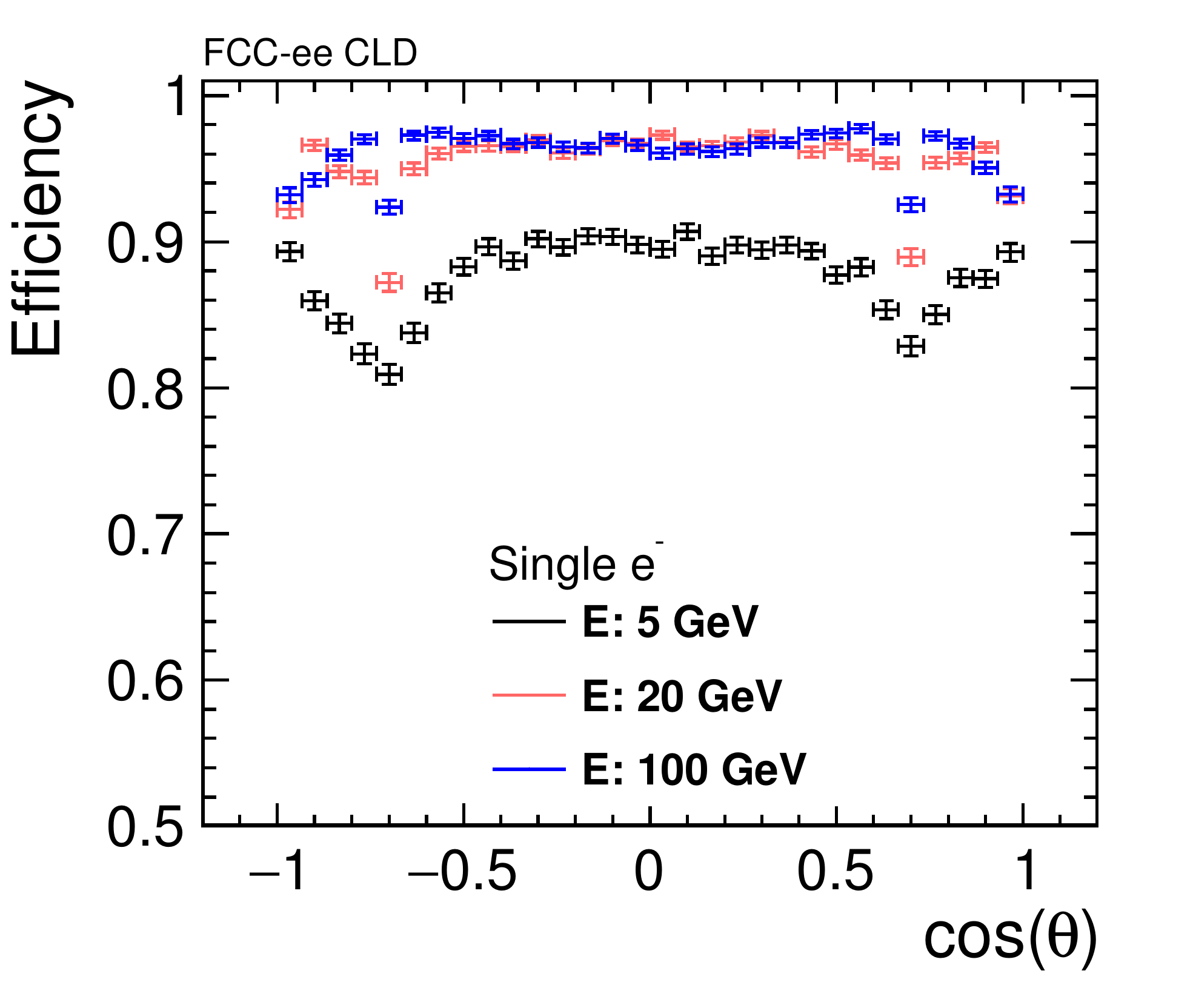}
    \caption{}
    \label{fig:particleGun_elIDEffVsTheta}
  \end{subfigure}
  \caption{Particle identification efficiency for electrons (a) as a function of energy and (b) as a function of $\cos\theta$ for three different energies. Error bars (statistical errors) are sometimes smaller than the dots.}
\end{figure}

The electron identification efficiency as a function of energy is shown in Figure~\ref{fig:particleGun_elIDEffVsE}.
For energies of about 20 GeV and higher, the efficiencies reach 95\%.
The efficiency in the endcap and barrel are similar, as can be seen in Figure~\ref{fig:particleGun_elIDEffVsTheta}.
 In the transition region the efficiency is 5\%--10\% lower than in the barrel or endcaps individually. For reasons which are still under investigation, in this region PandoraPFA mis-identifies a fraction of the electrons as pions.

\begin{figure}[htbp]
  \centering
  \begin{subfigure}{.5\textwidth}
    \centering
    \includegraphics[width=\linewidth]{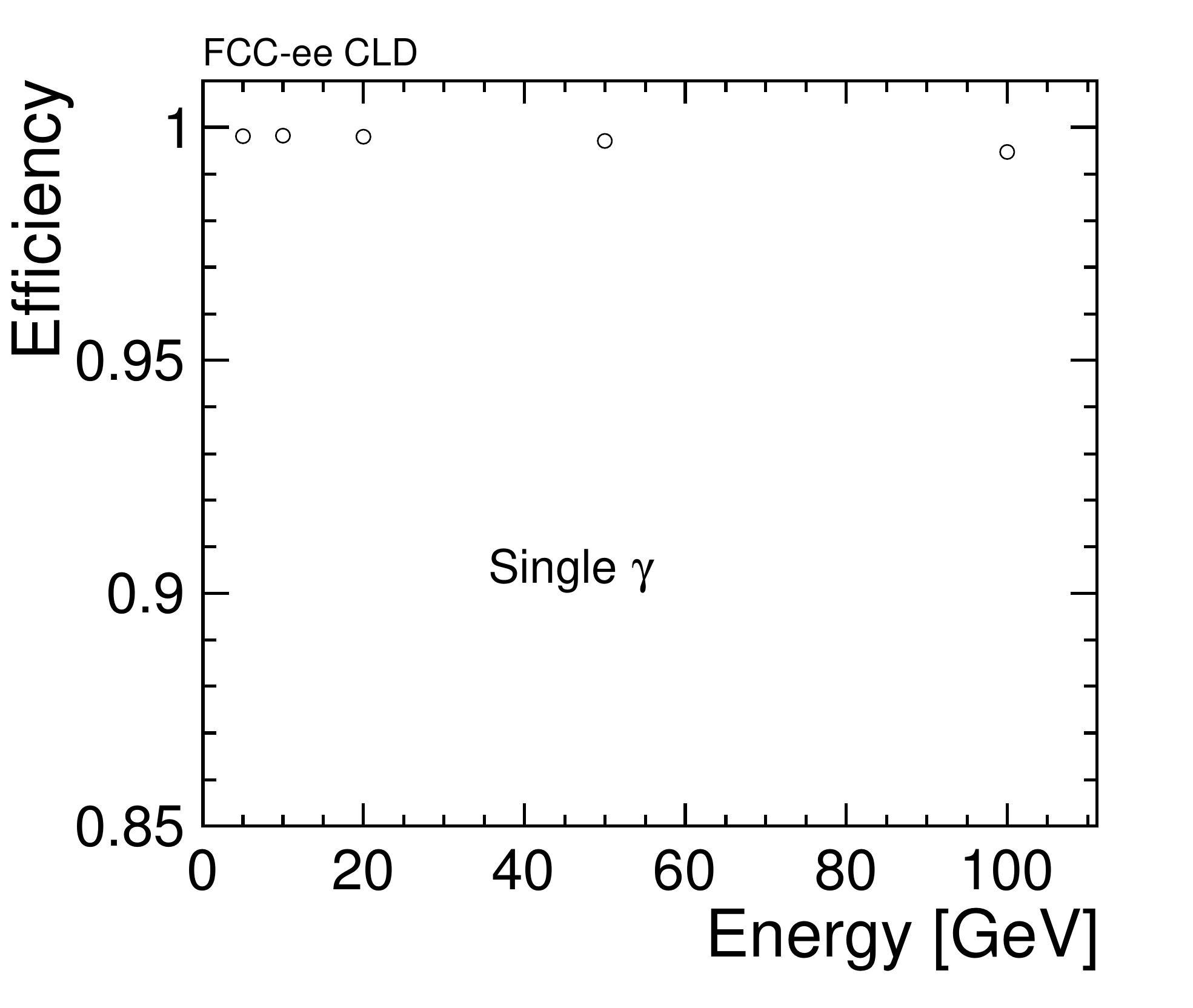}
    \caption{unconverted photons}
    \label{fig:particleGun_phUnconvIDEffVsE}
  \end{subfigure}%
  \begin{subfigure}{.5\textwidth}
    \centering
	  \includegraphics[width=\linewidth]{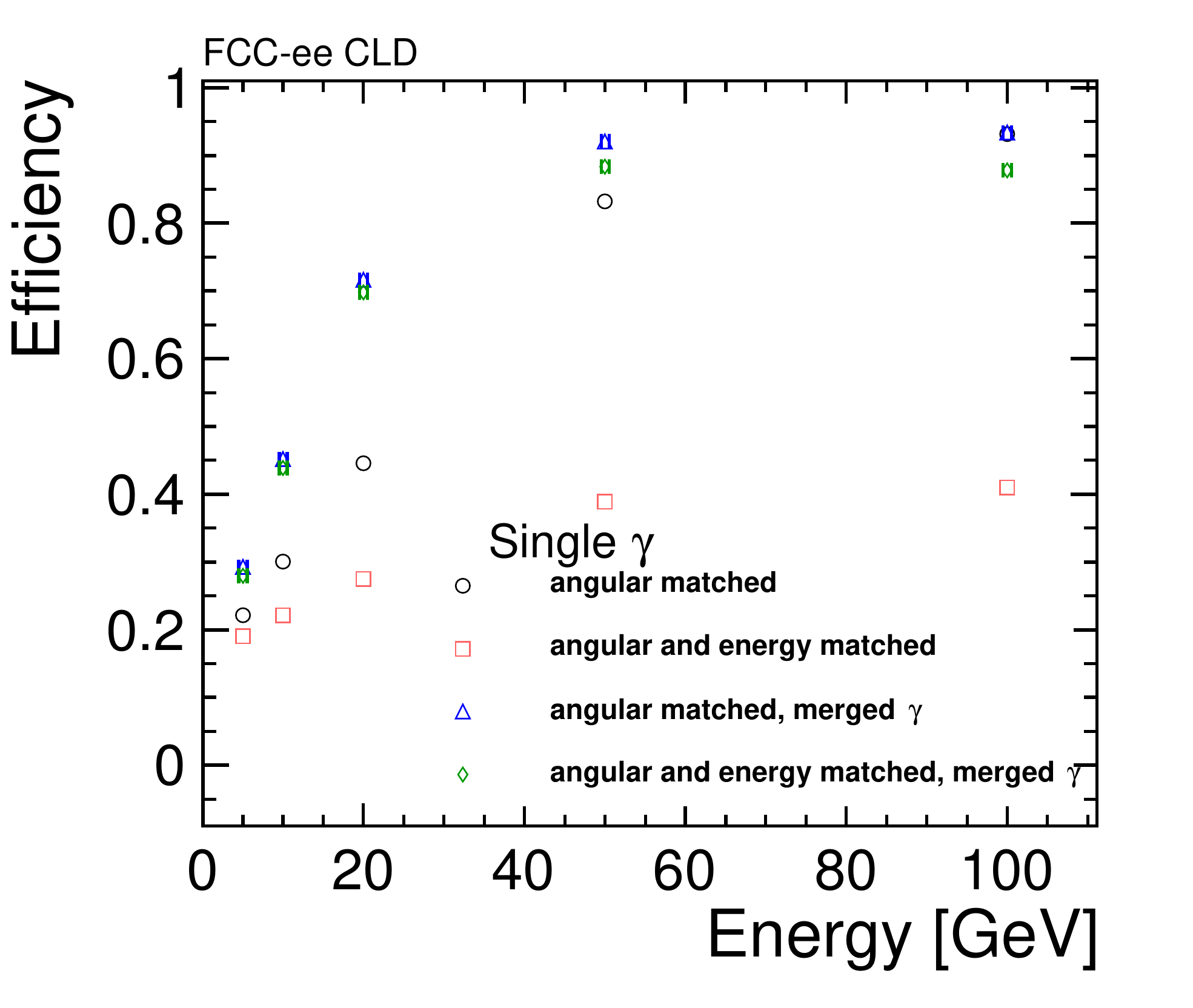}
    \caption{converted photons}
    \label{fig:particleGun_phConvIDEffVsTheta}
  \end{subfigure}
  \caption{Particle identification efficiency as a function of energy (a) for unconverted and (b) for converted photons. Error bars (statistical errors) are often smaller than the dots.
In the case of converted photons, the efficiency is shown when requiring either angular, or angular and energy matching. 
In the case of converted photons both criteria are, additionally, applied after merging leading photon candidates (i.e.\ their electromagnetic clusters) into one new photon candidate.}
   \label{fig:photon_id_eff}
\end{figure}

For single photons, the signatures for unconverted and converted photons are considered separately. 
The fraction of converted photons is around 12\% overall, for all energy points. 
This fraction increases from around 8\% at 90\degrees{} to around 20\% for very forward polar angles.
The particle identification efficiency for unconverted and converted photons is shown in Figure~\ref{fig:photon_id_eff}.
Reconstructed photons are required to satisfy the same spatial matching criteria as charged particles and their reconstructed energy has to be within 5~$\sigma_\mathrm{{E}}$ (based on the two parameter fit on the resolution curves shown in Figure~\ref{fig:photonResolutionVsEnergy}).
For unconverted photons the identification efficiency is above 99\% for all energies (Figure~\ref{fig:particleGun_phUnconvIDEffVsE}). 

For converted photons, requiring only angular matching results in a high identification efficiency at high energies, reaching 83\% at 50 GeV\@.
Adding the energy matching criterion to the leading photon in the event leads to a strongly reduced efficiency, as expected (see the red squares in Figure~\ref{fig:particleGun_phConvIDEffVsTheta}).
In many conversion events Pandora PFA running in its default configuration reconstructs two photons. 
Merging both reconstructed candidate clusters, if they are within $|\phi_{reconstructed}-\phi_{true}|< 20$ mrad and $|\theta_{reconstructed}-\theta_{true}|< 1$ mrad, and applying the identification criteria on the merged candidate,
 significantly improves the efficiency for the angular and energy matched case (see the blue triangles in Figure~\ref{fig:particleGun_phConvIDEffVsTheta}).

 Around 60\% of all conversions occur before reaching the last 4 layers of the tracker. 
The tracking algorithm requires at least four hits in the tracker. Work has started on a specific conversion algorithm in Pandora which should improve identification of converted photons particularly at low energies.  

\subsubsection{Performances for Complex Events}

\paragraph{Tracking Efficiency}

The tracking efficiency for particles in jets has been studied in samples at the lowest (91 GeV) and highest (365 GeV) centre-of-mass energy at which the FCC-ee is designed to operate.
The tracking performance for the same physics samples has been analysed also with the overlay of incoherent
\epem pairs. 

In the following, results will be presented for 10\,000 \PZgstarToqq events of 365 GeV mass, with and without background overlay. 
A comparison with \PZgstarToqq  events of 91 GeV mass is also shown. Finally, tracking efficiency and fake rate are studied for tracks in \bb{} events at 365 GeV.

Efficiency is defined as the fraction of reconstructable Monte Carlo particles which have been reconstructed as \emph{pure} tracks. A track is considered pure if most of its hits ($\ge75\%$) belong to the same Monte Carlo particle. The definition of a reconstructable particle is the same as given for the single particle efficiency.

\begin{figure}[htbp]
  \centering
  \includegraphics[width=0.623\linewidth]{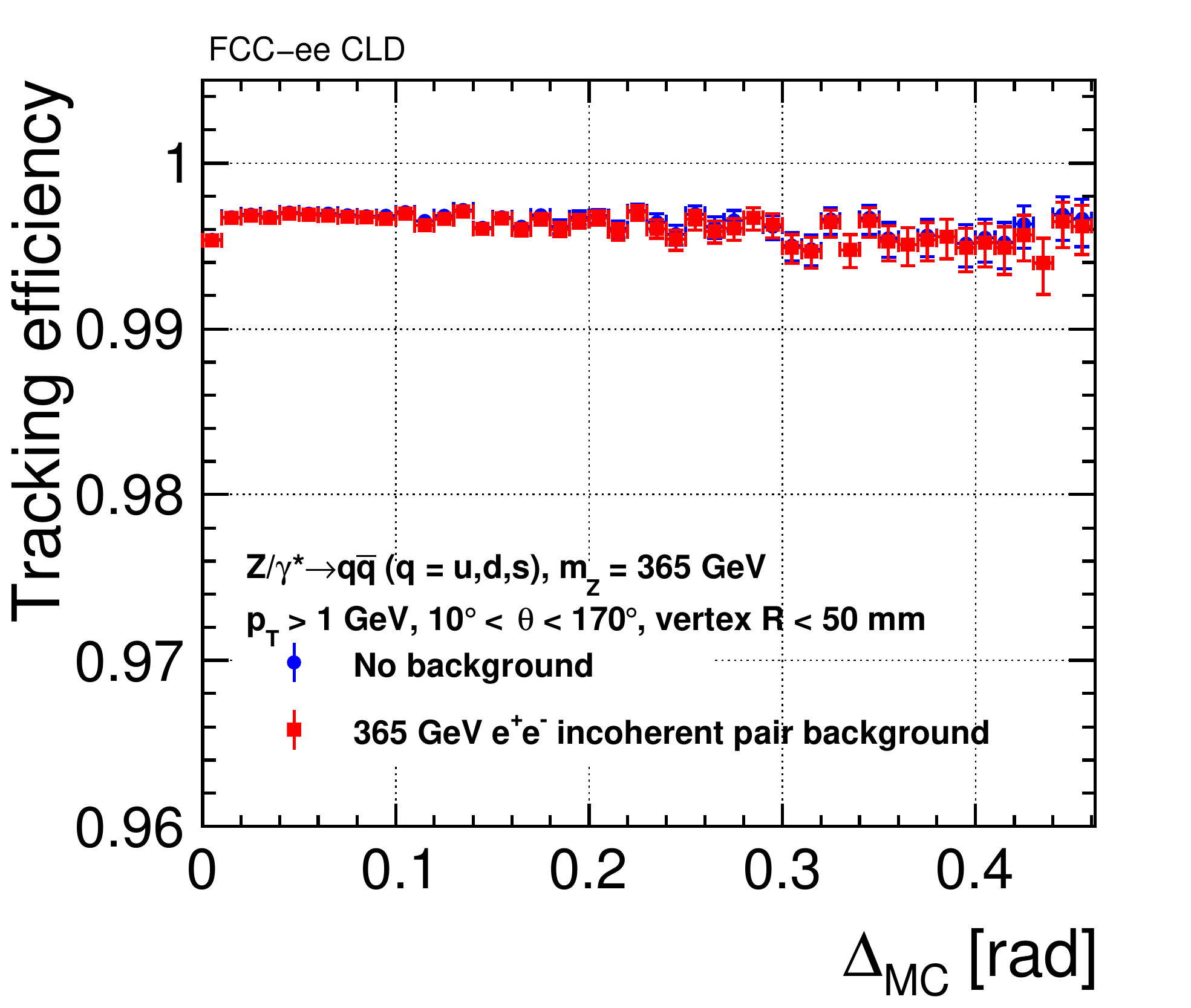}
         \caption{Tracking efficiency as a function of particle proximity $\Delta_{\mathrm MC}$ for \PZgstarToqq events, with $m_{\PZ}$ = 365 GeV, with and without background overlay.}
   \label{fig:Zuds365GeV_eff_dist}
\end{figure}

In jet events, the vicinity of other particles may affect the performance of the pattern recognition in assigning the right hits to the proper track. 
Therefore, the tracking efficiency in \PZgstarToqq events of 365~GeV mass has been monitored as a function of the particle proximity. 
The latter is defined as the smallest distance between the Monte Carlo particle associated to the track and any other Monte Carlo particle, 
$\Delta_{\mathrm MC} = \sqrt{{(\Delta\eta)}^{2}+{(\Delta\phi)}^{2}}$, where $\eta$ is the pseudorapidity. The efficiency is shown in Figure~\ref{fig:Zuds365GeV_eff_dist}, 
in which the following cuts are applied: $\pT > 1~\mathrm{GeV}$, $10\degrees{} < \theta < 170\degrees{}$ and production radius smaller than 50 mm. 
Results with and without background overlay are comparable within statistical errors. 
Being a minimum cut on particle proximity necessary, 
the value of 0.02 rad is chosen and applied in all following tracking efficiency results. 

\begin{figure}[htbp]
  \centering
  \includegraphics[width=0.623\linewidth]{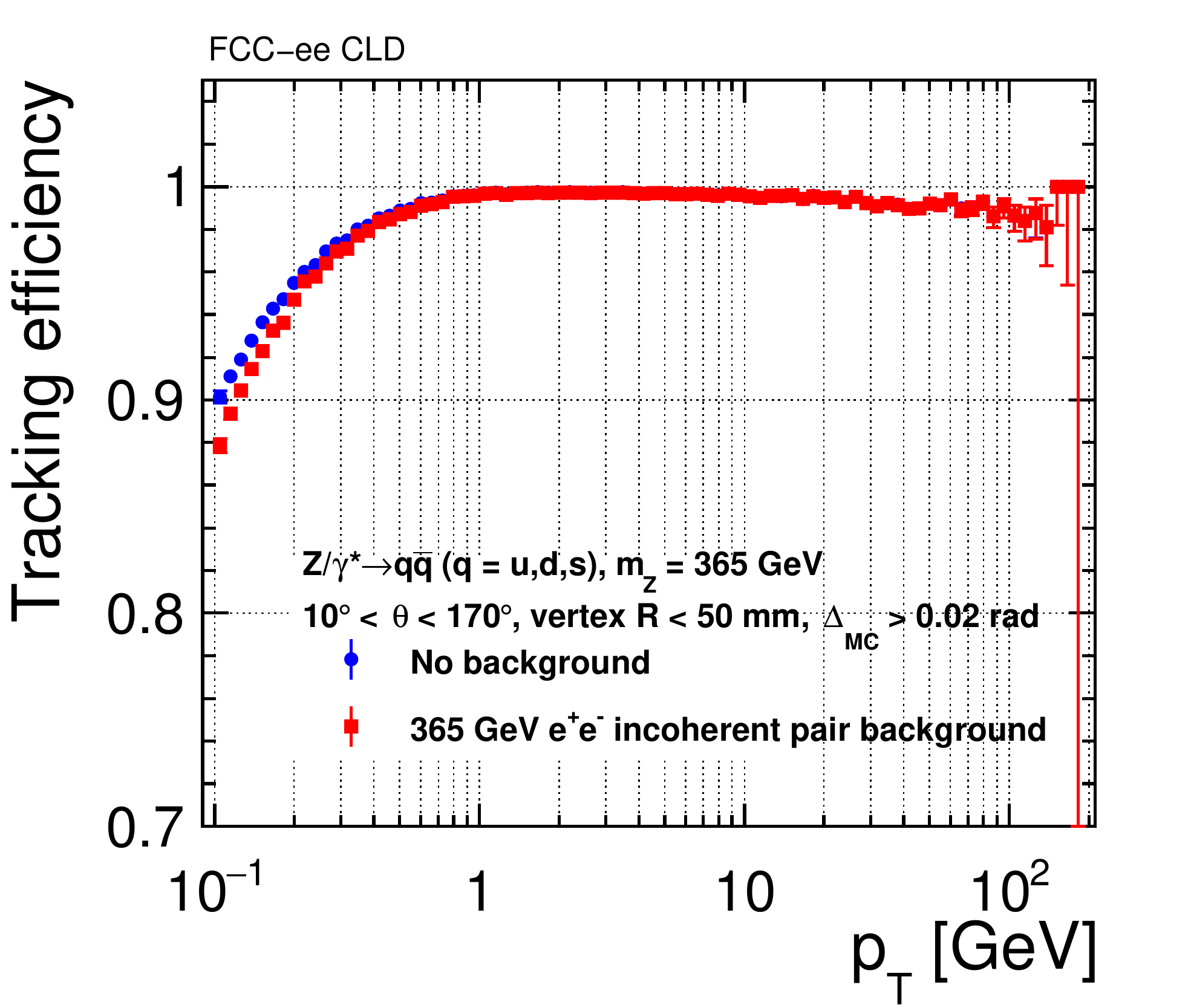}
\caption{Tracking efficiency as a function of \pT{} for \PZgstarToqq events, with $m_{\PZ}$ = 365 GeV, with and without background overlay.}
\label{fig:Zuds365GeV_eff_pt}
\end{figure}

Figure~\ref{fig:Zuds365GeV_eff_pt} shows the tracking efficiency in \PZgstarToqq events of 365~GeV mass as a function of transverse momentum, with and without background overlay. The following cuts are applied for each particle in this plot: $10\degrees{} < \theta < 170\degrees{}$, particle proximity $\Delta_{\mathrm MC}$ larger than 0.02 rad and production radius smaller than 50 mm. Above 1~GeV, the tracking is fully efficient, while below 1~GeV it goes down to a minimum of 90\% and 88\%, without and with background respectively. The effect of background, slightly visible in the low-$\pT$ region, is otherwise negligible.

\begin{figure}[htbp]
\centering
\begin{subfigure}{.5\textwidth}
  \centering
  \includegraphics[width=\linewidth]{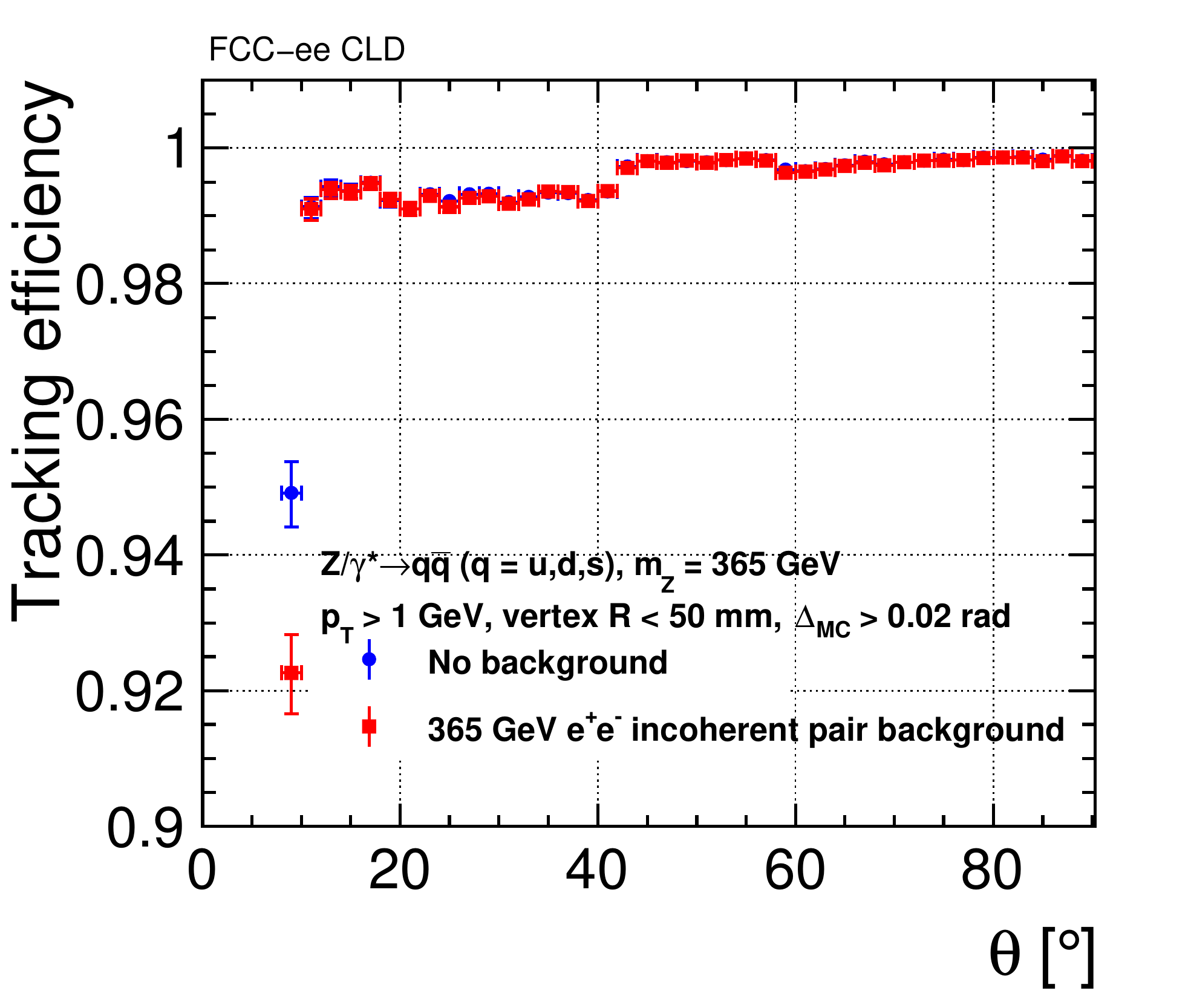}
  \caption{vs $\theta$}
  \label{fig:Zuds365GeV_eff_theta}
\end{subfigure}%
\begin{subfigure}{.5\textwidth}
  \centering
  \includegraphics[width=\linewidth]{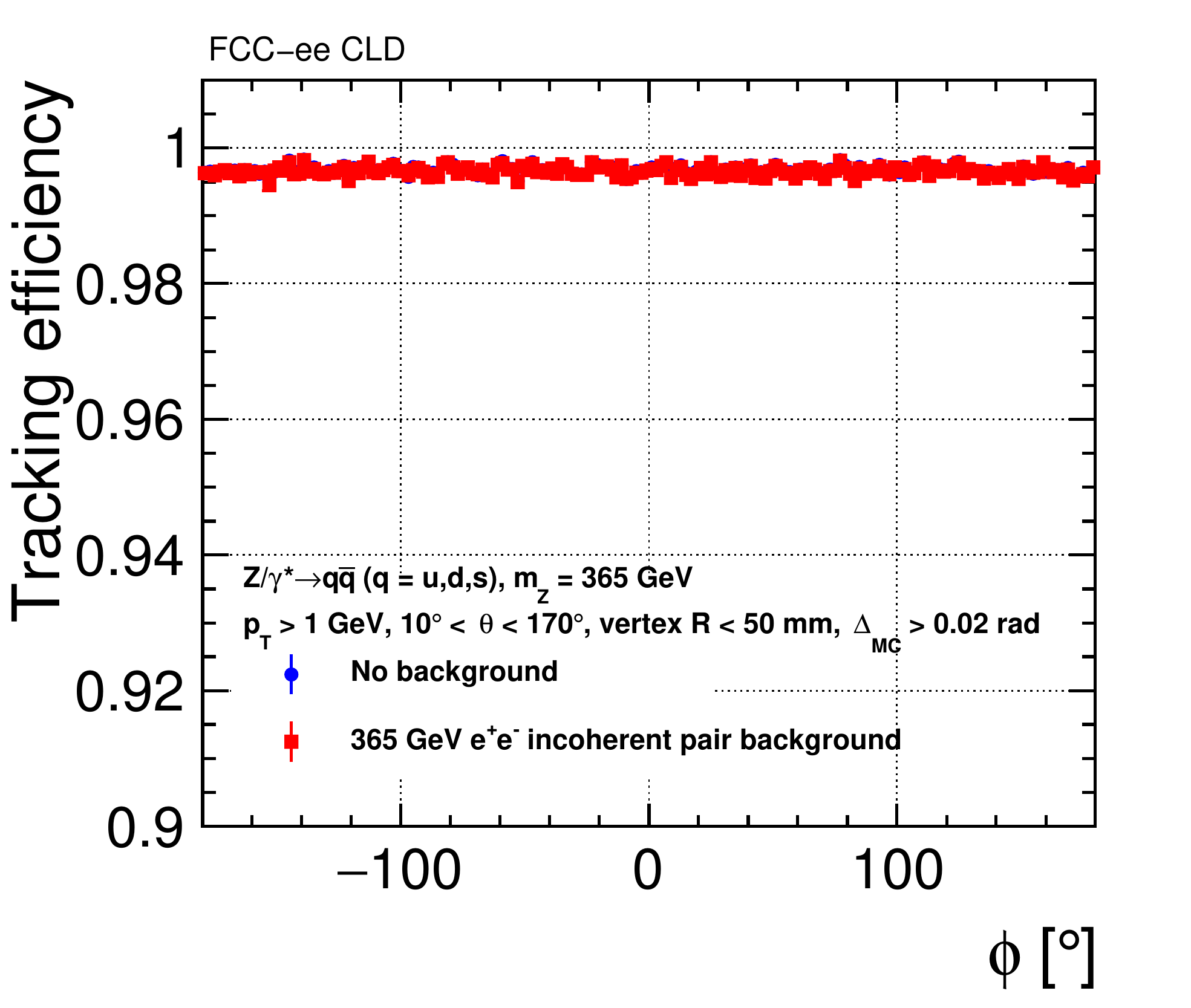}
  \caption{vs $\phi$}
  \label{fig:Zuds365GeV_eff_phi}
\end{subfigure}
\caption{Tracking efficiency as a function of (a) polar and (b) azimuthal angle for \PZgstarToqq events, with $m_{\PZ}$ = 365 GeV, with and without background overlay.}
\label{fig:Zuds365GeV_eff_angle}
\end{figure}

For the same events, the tracking efficiency is shown in Figure~\ref{fig:Zuds365GeV_eff_angle} as a function of polar (left) and azimuthal angle (right). The following cuts are applied in this plot: $\pT > 1~\mathrm{GeV}$, particle proximity $\Delta_{\mathrm MC}$ larger than 0.02 rad and production radius smaller than 50 mm. The tracking efficiency approaches 100\% for very central tracks and is around 99\% towards the forward region. For very forward tracks, i.e. below 10\degrees{}, a maximum efficiency loss of less than 5\% and 8\% is observed, without and with background, within statistical uncertainties. The dependence on the azimuthal angle is flat and does not show any particular feature. In both figures, the impact of incoherent pairs is observed to be small.

\begin{figure}[htbp]
  \centering
  \includegraphics[width=0.623\linewidth]{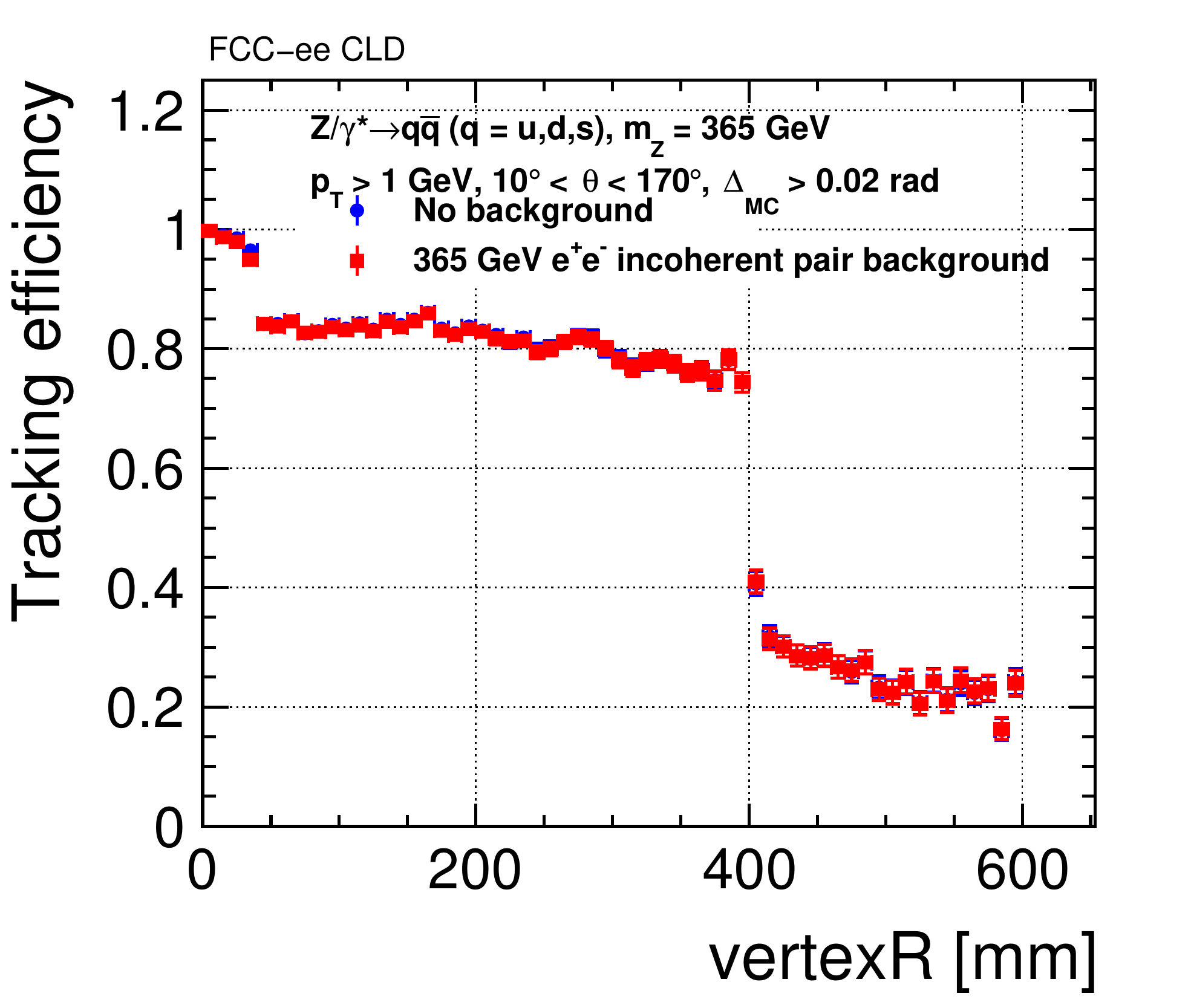}
\caption{Tracking efficiency as a function of the production vertex radius for \PZgstarToqq events, with $m_{\PZ}$ = 365 GeV, with and without background overlay.}
\label{fig:Zuds365GeV_eff_vertexR}
\end{figure}

Finally, the tracking efficiency as a function of the production vertex radius is shown in Figure~\ref{fig:Zuds365GeV_eff_vertexR}. In this plot, the following cuts are applied: $\pT > 1~\mathrm{GeV}$, $10\degrees{} < \theta < 170\degrees{}$ and particle proximity $\Delta_{\mathrm MC}$ larger than 0.02 rad. The trend reflects the same behaviour observed for single displaced low-momentum muons in Figure~\ref{fig:eff_displaced_vs_vertexR}, since the low-energy component of the particle spectrum dominates. The effect of background from incoherent pairs is negligible.

\begin{figure}[htbp]
  \centering
  \includegraphics[width=0.623\linewidth]{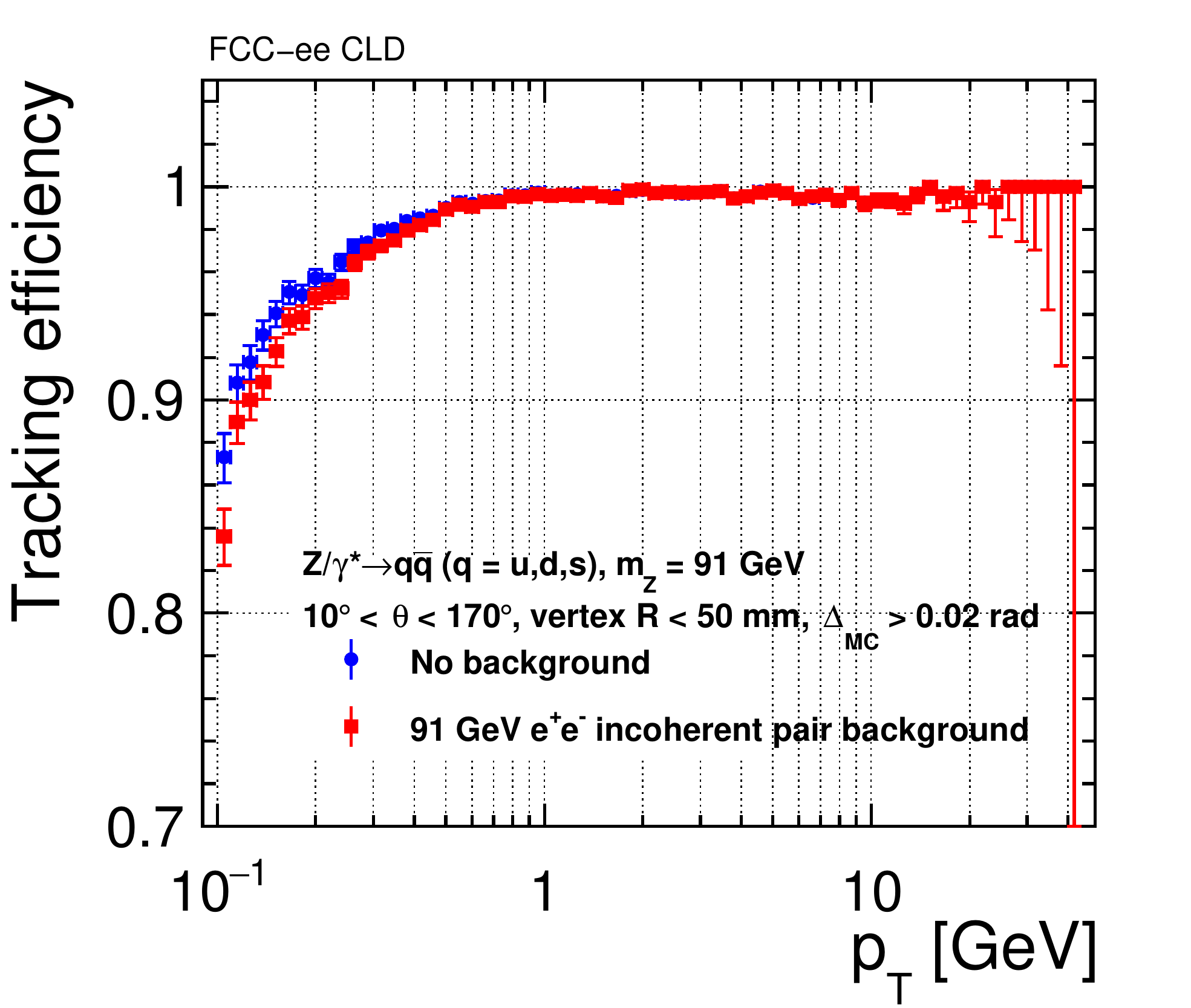}
\caption{Tracking efficiency as a function of \pT{} for \PZgstarToqq events, with $m_{\PZ}$ = 91 GeV, with and without background overlay.}
\label{fig:Zuds91GeV_eff_pt}
\end{figure}

Figure~\ref{fig:Zuds91GeV_eff_pt} shows the equivalent of Figure~\ref{fig:Zuds365GeV_eff_pt} for \PZgstarToqq events of 91 GeV mass, with and without background overlay. The tracking performs equally well at both centre-of-mass energies.

The tracking performance has been completed by studying \bb{} events at 365 GeV, in terms of efficiency and fake rate, the latter being the fraction of \emph{impure} tracks among all reconstructed tracks. 
A track is considered impure if less than 75\% of its hits belong to the same Monte Carlo particle.
Unlike the tracking efficiency, the fake rate is shown as a function of the reconstructed observables, e.g. \pT{} and $\theta$, as obtained from the track parameters. The cuts introduced in the plots are also applied to reconstructed observables.

\begin{figure}[htbp]
\centering
\begin{subfigure}{.5\textwidth}
  \centering
  \includegraphics[width=\linewidth]{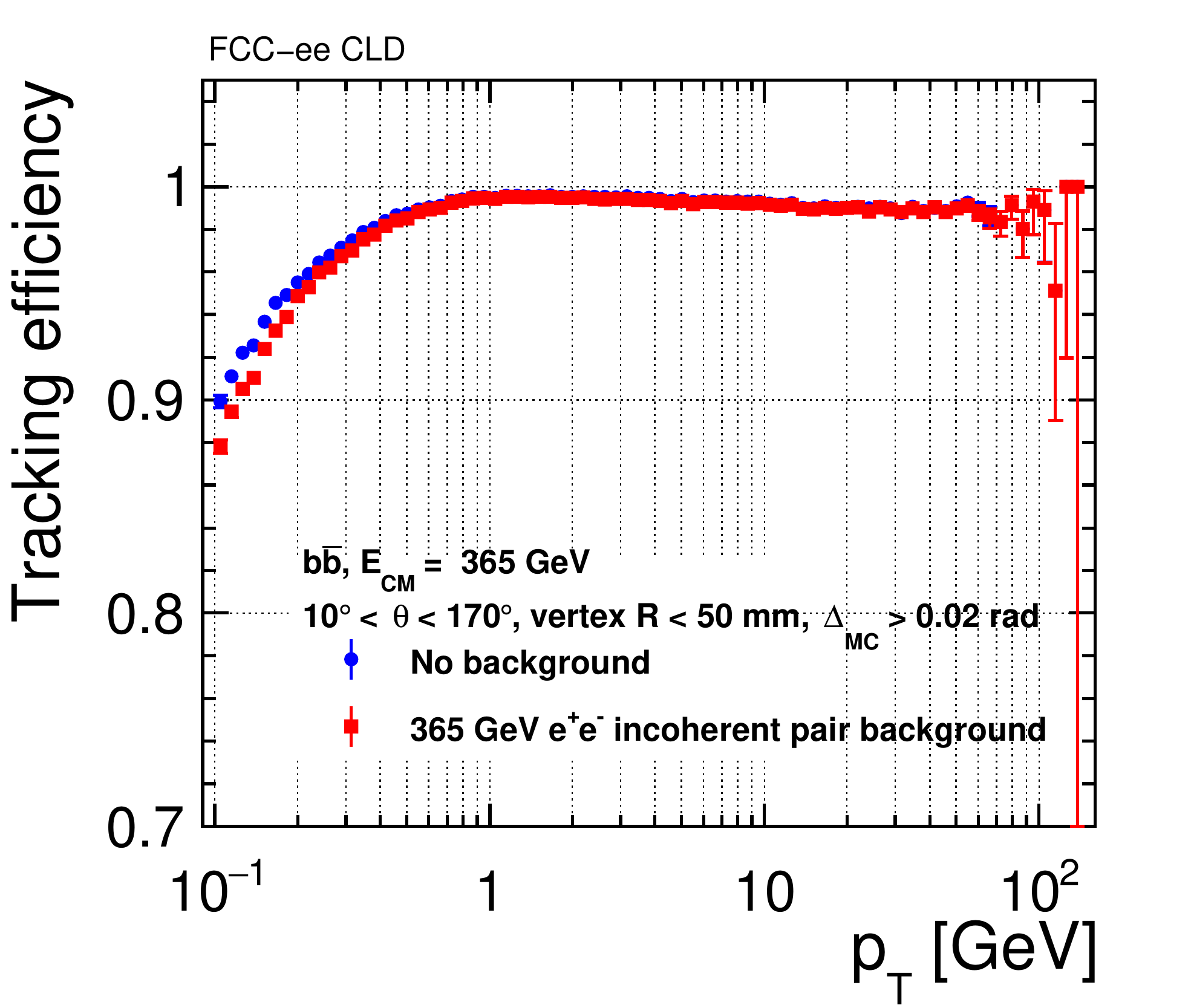}
  \caption{efficiency}
  \label{fig:bbbar365GeV_eff_pt}
\end{subfigure}%
\begin{subfigure}{.5\textwidth}
  \centering
  \includegraphics[width=\linewidth]{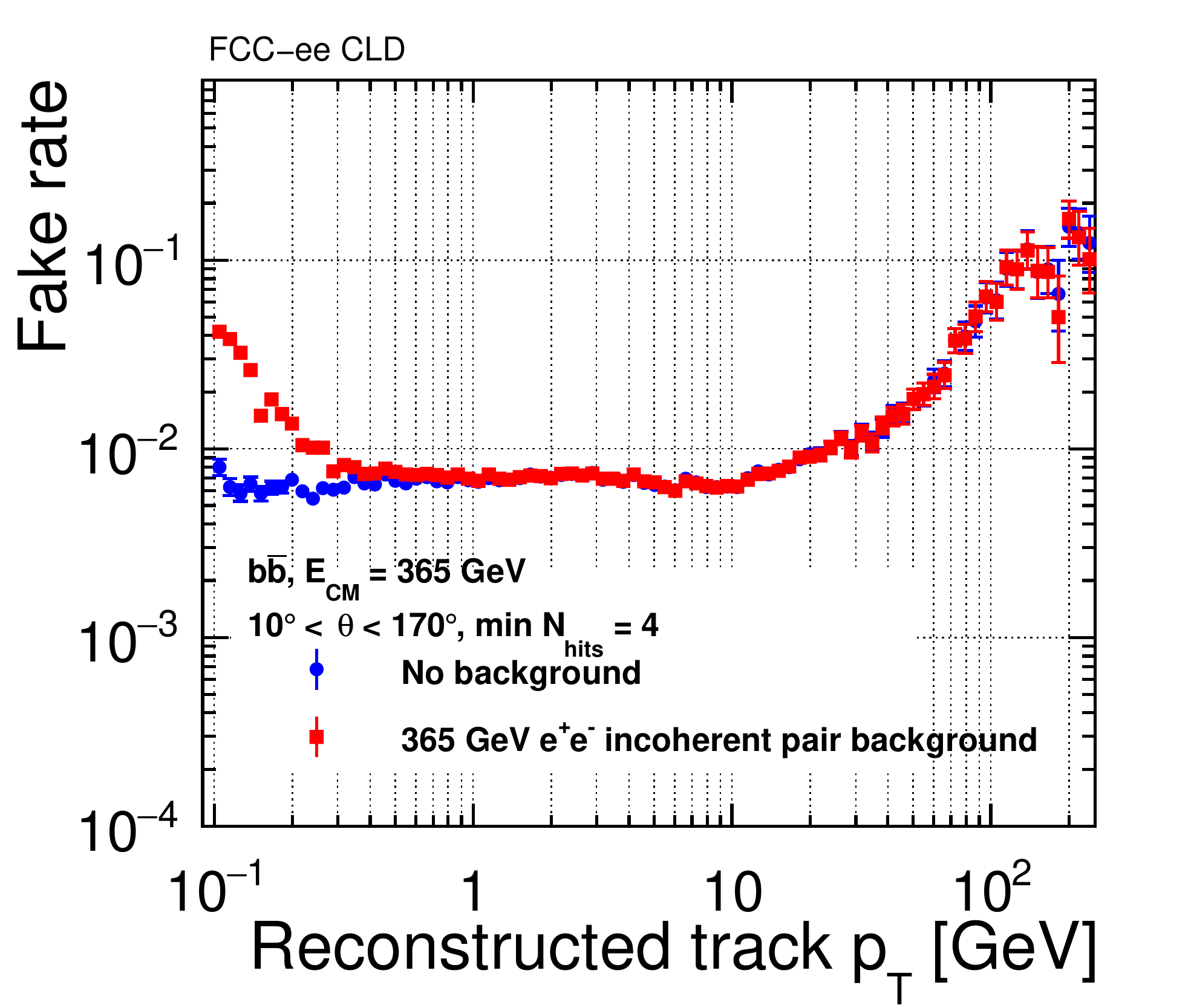}
  \caption{fake rate}
  \label{fig:bbbar365GeV_fake_pt}
\end{subfigure}
\caption{(a) Tracking efficiency and (b) fake rate as a function of transverse momentum for \bb{} events at 365 GeV, with and without background overlay.}
\label{fig:bbbar365GeV_pt}
\end{figure}

Figure~\ref{fig:bbbar365GeV_pt} shows efficiency (left) and fake rate (right) as a function of transverse momentum. In Figure~\ref{fig:bbbar365GeV_eff_pt}, the following cuts are applied for each particle: $10\degrees{} < \theta < 170\degrees{}$, particle proximity $\Delta_{\mathrm MC}$ larger than 0.02 rad and production radius smaller than 50 mm. The low-$\pT$ region is comparable with the one for \PZgstarToqq events. However, the efficiency for \bb{} events reaches a maximum of 100\% around 1~GeV and then decreases progressively at higher transverse momentum down to a minimum of 98\%, as tracks are more straight and confusion in selecting hits from close-by particles arises. Figure~\ref{fig:bbbar365GeV_fake_pt} shows the fake rate as a function of reconstructed \pT{}, with the following cuts applied: $10\degrees{} < \theta < 170\degrees{}$ and minimum number of hits on track equal to 4.
The fake rate is higher in the low-$\pT$ region with background overlaid, due to the intrinsic difficulty to reconstruct low-energetic particles. It reaches a minimum between 1 GeV and 10 GeV, and increases again for higher momentum, for the same reason of confusion in pattern recognition described above. The maximum fake rate reached is around 10\%. Both efficiency and fake rate depend visibly on the background overlay only in the low-$\pT$ region.

\begin{figure}[htbp]
\centering
\begin{subfigure}{.5\textwidth}
  \centering
  \includegraphics[width=\linewidth]{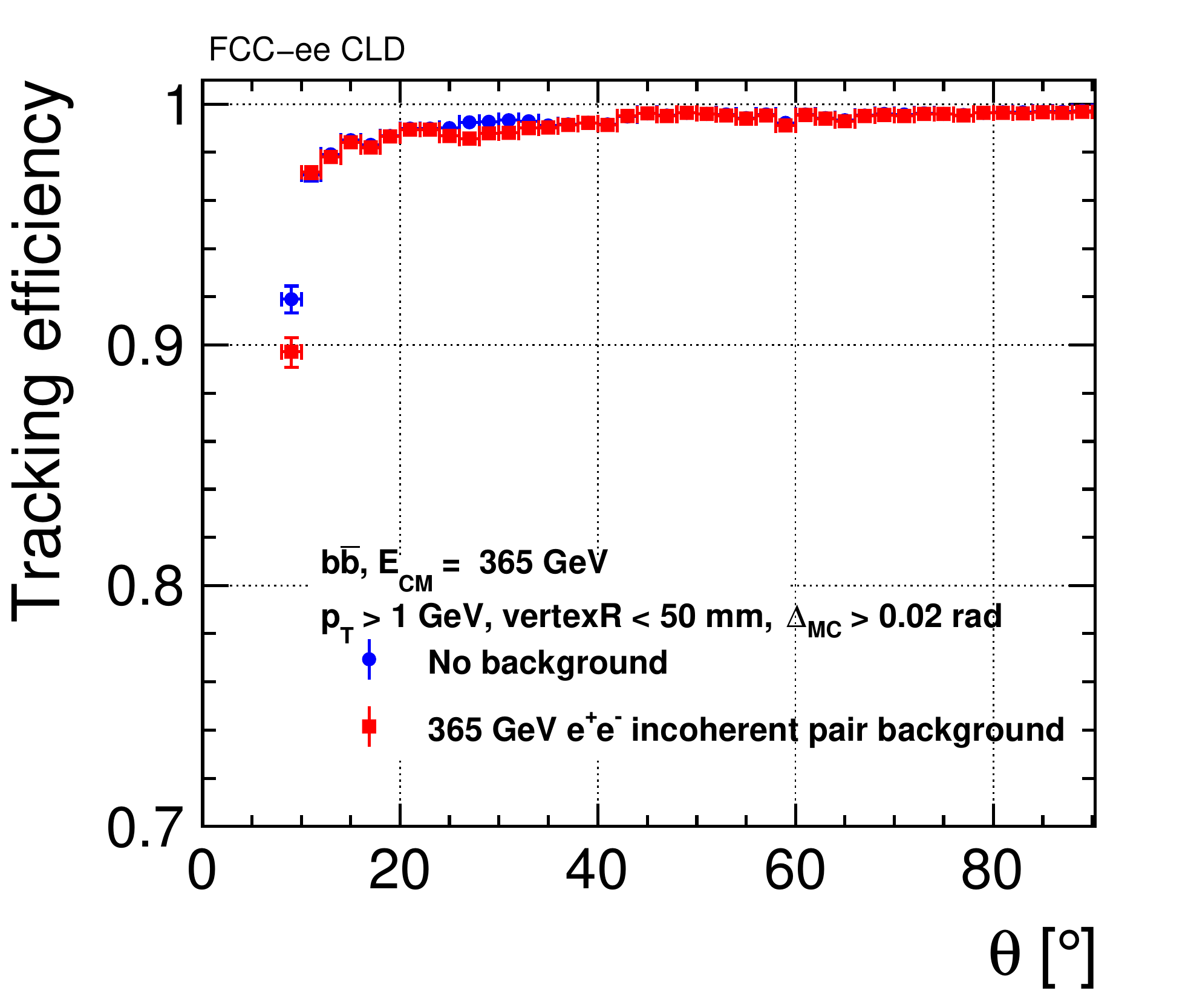}
  \caption{efficiency}
  \label{fig:bbbar365GeV_eff_theta}
\end{subfigure}%
\begin{subfigure}{.5\textwidth}
  \centering
  \includegraphics[width=\linewidth]{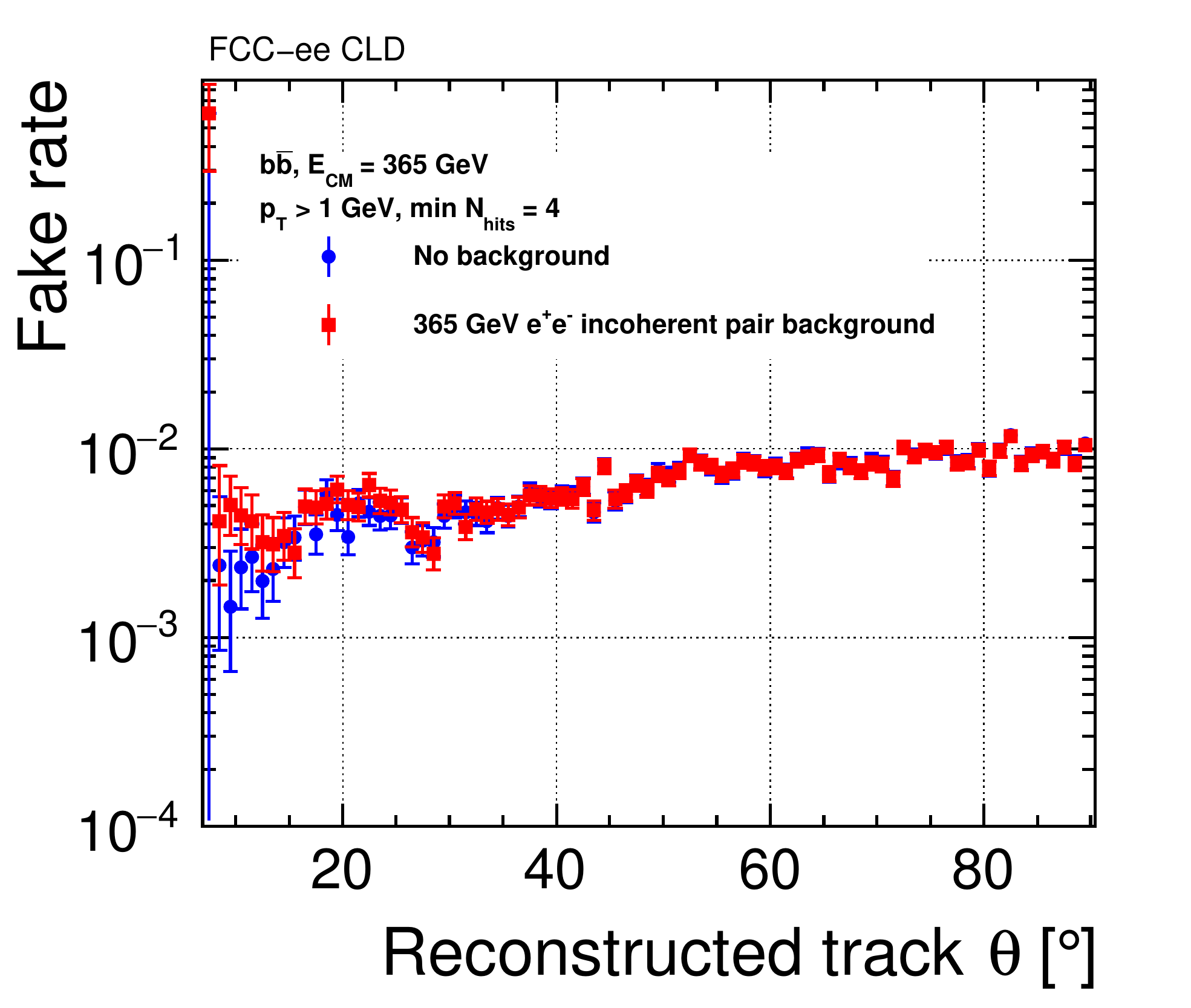}
  \caption{fake rate}
  \label{fig:bbbar365GeV_fake_theta}
\end{subfigure}
\caption{(a) Tracking efficiency and (b) fake rate as a function of $\theta$ for \bb{} events at 365 GeV, with and without background overlay.}
\label{fig:bbbar365GeV_theta}
\end{figure}

Figure~\ref{fig:bbbar365GeV_theta} shows the same efficiency (left) and fake rate (right) as a function of polar angle. The following cuts are applied in Figure~\ref{fig:bbbar365GeV_eff_theta}: $\pT > 1~\mathrm{GeV}$, particle proximity $\Delta_{\mathrm MC}$ larger than 0.02 rad and production radius smaller than 50 mm. The efficiency reaches a value above 99.5\% in the central region, and decreases progressively down to a minimum of 90\% at $\theta$ smaller than 10\degrees.
The following cuts are applied in Figure~\ref{fig:bbbar365GeV_fake_theta}: $\pT > 1~\mathrm{GeV}$ and minimum number of hits on track equal to 4. The fake rate reaches a maximum of 1\% in the central region and decreases in the forward region, except for a peak in the most forward $\theta$ bin (down to 8\degrees), with low statistical significance.

\begin{figure}[htbp]
\centering
  \centering
  \includegraphics[width=0.623\linewidth]{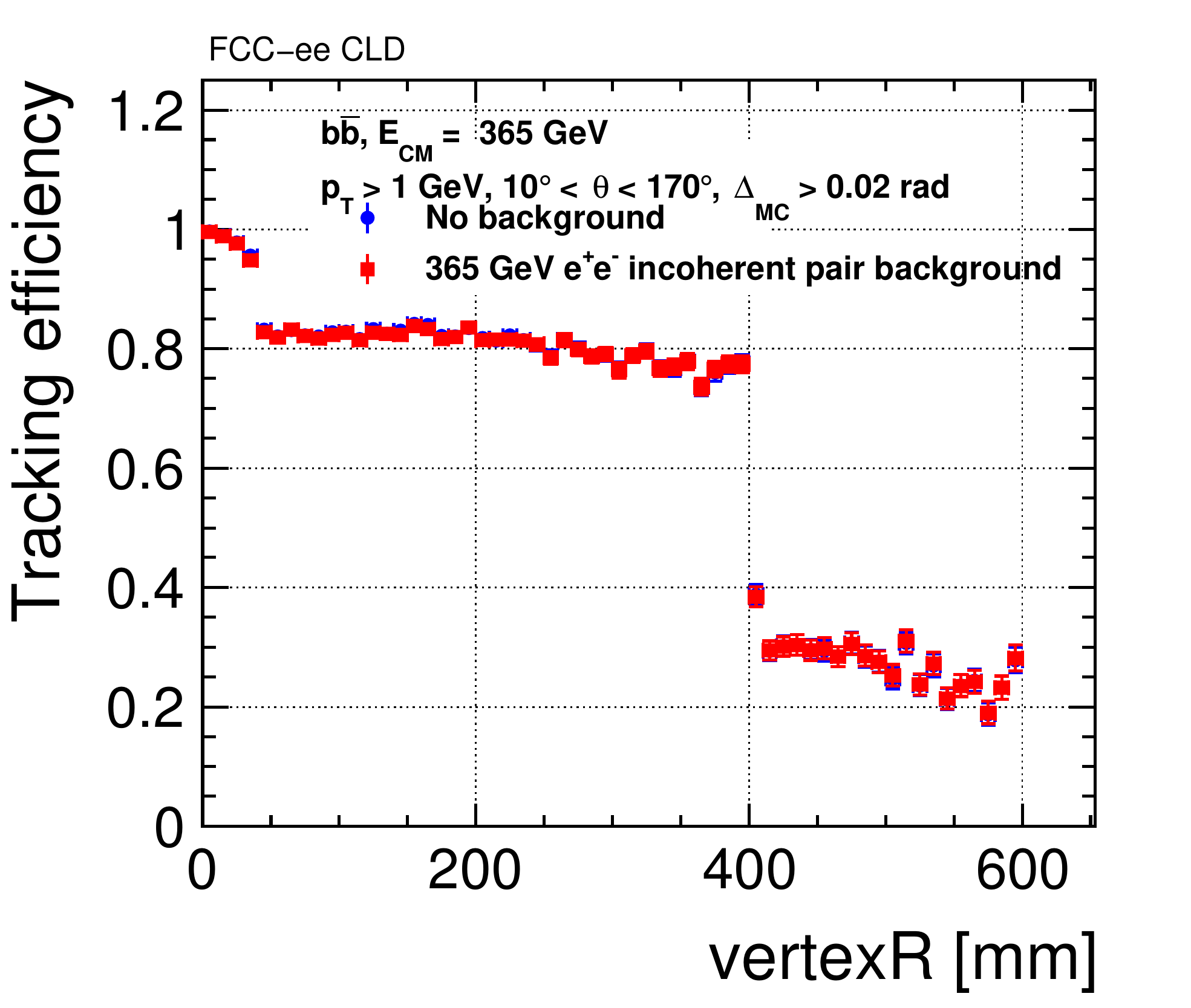}
\caption{Tracking efficiency as a function of production vertex radius for \bb{} events at 365 GeV, with and without background overlay.}
\label{fig:bbbar365GeV_vertexR}
\end{figure}

Finally, in Figure~\ref{fig:bbbar365GeV_vertexR} the efficiency is plotted as a function of production vertex radius. The following cuts are applied: $\pT > 1~\mathrm{GeV}$, $10\degrees{} < \theta < 170\degrees{}$ and particle proximity $\Delta_{\mathrm MC}$ larger than 0.02 rad. The efficiency trend is comparable with the results previously discussed for low-momentum muons and \PZgstarToqq  events (Figures~\ref{fig:eff_displaced_vs_vertexR} and~\ref{fig:Zuds365GeV_eff_vertexR}). The effect of the background on the tracking performances also in these events is found to be negligible.

\paragraph{Lepton Identification}

Lepton identification efficiencies for muons and electrons have been studied for CLICdet in complex \ttbar{} samples at 3~TeV.
 In that study direct leptons from W decays were considered. Muons were identified with more than 98\% efficiency at all energies, and overlay of background had no impact on the result.
The electron identification efficiency was found to be higher than 90\% for electrons with more than 20 GeV energy~\cite{CLICdet_performance}.

\subsubsection{Jet Performance}
\label{jet_energy_res}

A precise jet energy measurement,  using highly granular calorimeters and Particle Flow algorithms,  allows differentiating between different jet topologies, e.g.\ between jets originating from W and Z boson decays.
The jet performance in CLD is studied in di-jet events using  \PZgstarToqq decays at two centre-of-mass energies. 
The Pandora particle flow algorithms~\cite{Marshall:2015rfaPandoraSDK,Marshall:2012ryPandoraPFA} are used to reconstruct each particle, 
combining information from tracks, calorimeter clusters and hits in the muon system. 
Software compensation is applied to clusters of reconstructed hadrons in HCAL to improve their energy measurement, 
using local energy density information provided by the high granularity of the calorimeter system~\cite{Tran:2017tgrSoftwareCompensation}.

In the most direct approach, the jet energy resolution can be determined by comparing the energy sum of all reconstructed particles with the sum of all stable particles (excluding neutrinos) on MC truth level~\cite{Buttar:2008jxParticleLevel}. This method is used in the first part of this section. Alternatively, as shown in the second part, a jet clustering algorithm can be used.

$\mathrm{RMS}_{90}$ is used as a measure for the jet energy resolution. $\mathrm{RMS}_{90}$ is defined as the RMS in the smallest range of the reconstructed energy containing 90\% of the events~\cite{Marshall:2012ryPandoraPFA}\footnote{$\mathrm{RMS}_{90}$(E$_j$) = $\mathrm{RMS}_{90}$($\Sigma$E$_i$)/$\sqrt2$}. 
This provides a good measure for the resolution of the bulk of events, while it is relatively insensitive to the presence of tails. 
As an alternative method, fitting of the jet energy response with a double-sided Crystal Ball function~\cite{Oreglia:1980cs} has been investigated (see Appendix~D), and the results compared with $\mathrm{RMS}_{90}$. 
While both methods give comparable results, $\mathrm{RMS}_{90}$ is found to be more conservative.

\begin{figure}[htbp]
  \centering
  \includegraphics[width=0.5\linewidth]{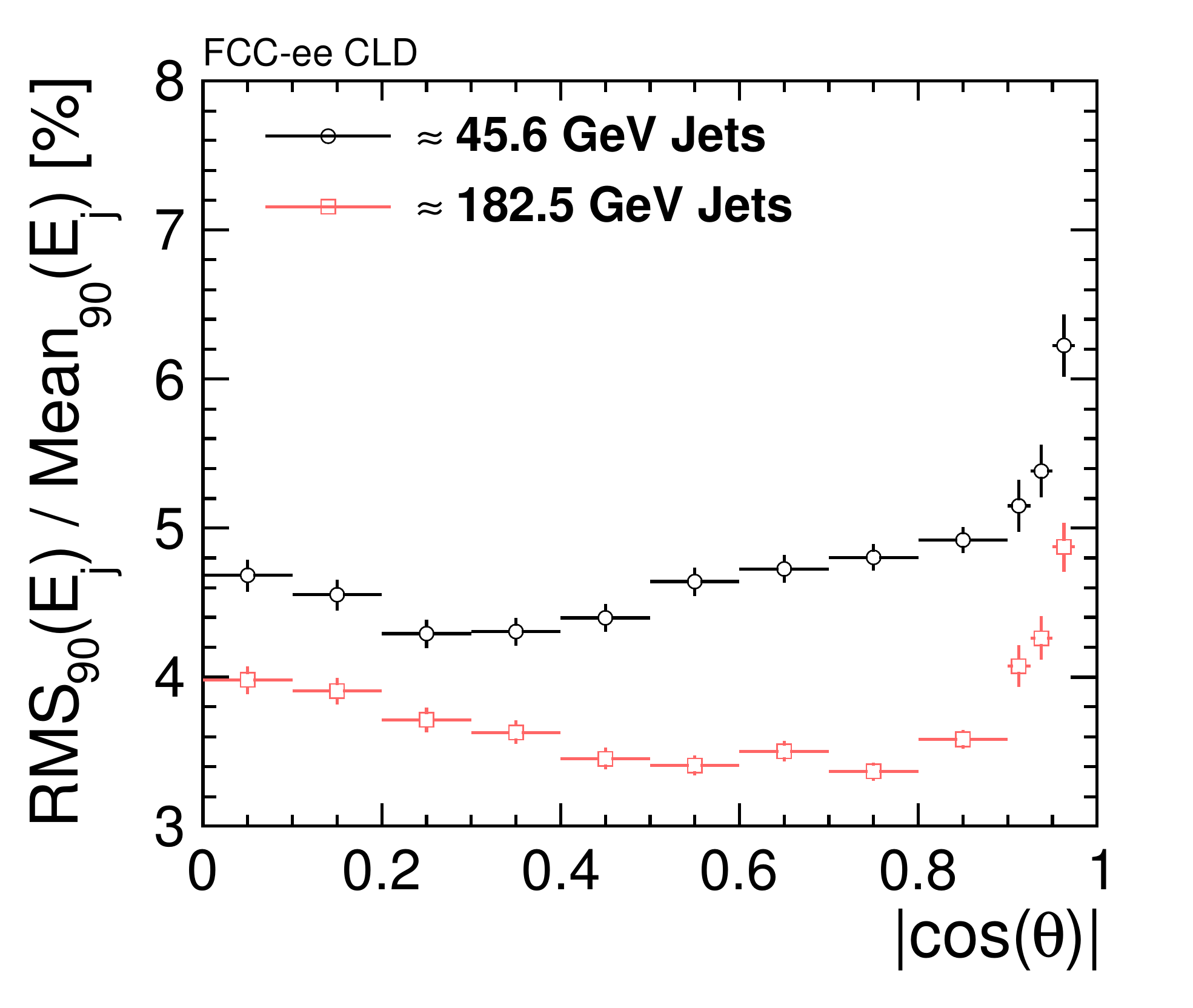}
         \caption{Jet energy resolution as a function of $|\cos\theta|$ for 45.5 and 182.5 GeV jets from \PZgstarToqq (u, d, s) events. PFO reconstruction was used.
}
   \label{fig:JER_vs_E}
\end{figure}

The jet energy resolution  is studied as a function of the quark $|\cos\theta|$. 
As shown in Figure~\ref{fig:JER_vs_E}, for lower energies the jet energy resolution is  4.5\%--5\%, while for higher energy jets the resolution is better than 4\%.
The reason for the $\theta$-dependence, observed in particular at the higher energies, will be the subject of further investigations.

\begin{figure}[htbp]
  \centering
  \includegraphics[width=0.5\linewidth]{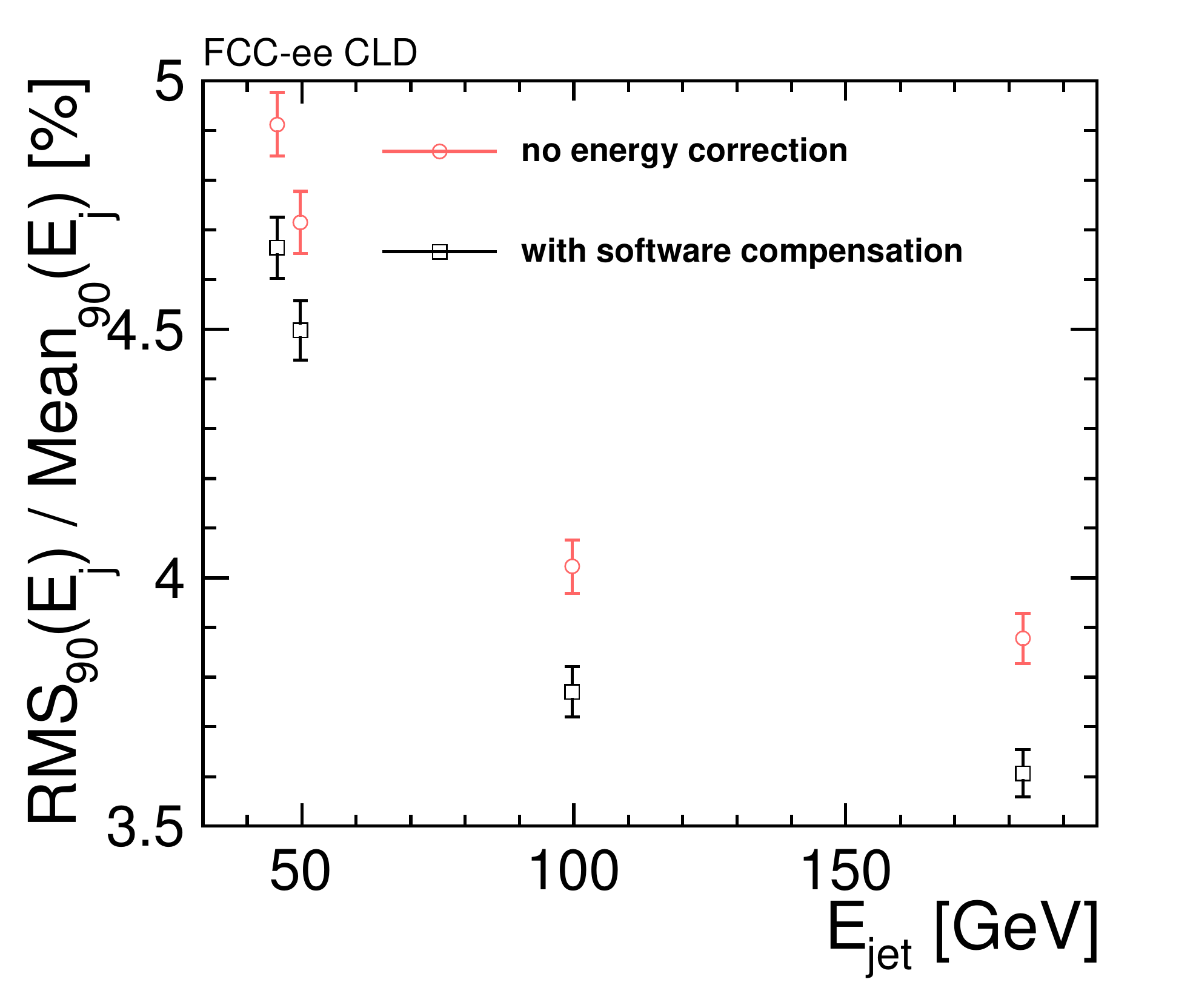}
         \caption{Jet energy resolution for central jets with $|\cos\theta|<0.7$ as a function of jet energy for \PZgstarToqq events at different energies.
PFO reconstruction without energy correction (red) is compared to PFO reconstruction applying software compensation (black).}
   \label{fig:JER_vs_E_SC_vs_noSC}
\end{figure}

To demonstrate the effect of the software compensation, the jet energy resolution was studied without applying software compensation weights.
As shown in Figure~\ref{fig:JER_vs_E_SC_vs_noSC}, the relative improvement of the jet energy resolution due to software compensation is between 5\,to\,7\%, depending on the jet energy. 

\paragraph{Impact of Beam-Induced Background}
To investigate the effect of beam-induced background on the jet performance, incoherent pair events have been overlaid to the physics events.
For this study, an integration time window of 400 ns both for 91.2 GeV and 365 GeV centre-of-mass energies has been chosen.
This corresponds to overlaying 20 and 1 bunch crossings to one physics event  at 91.2 GeV and 365 GeV, respectively.  
Choosing the same integration window allows to use the same calorimeter calibration constants for both centre-of-mass energies.

\begin{figure}[htbp]
  \centering
  \begin{subfigure}{.5\textwidth}
    \centering
    \includegraphics[width=\linewidth]{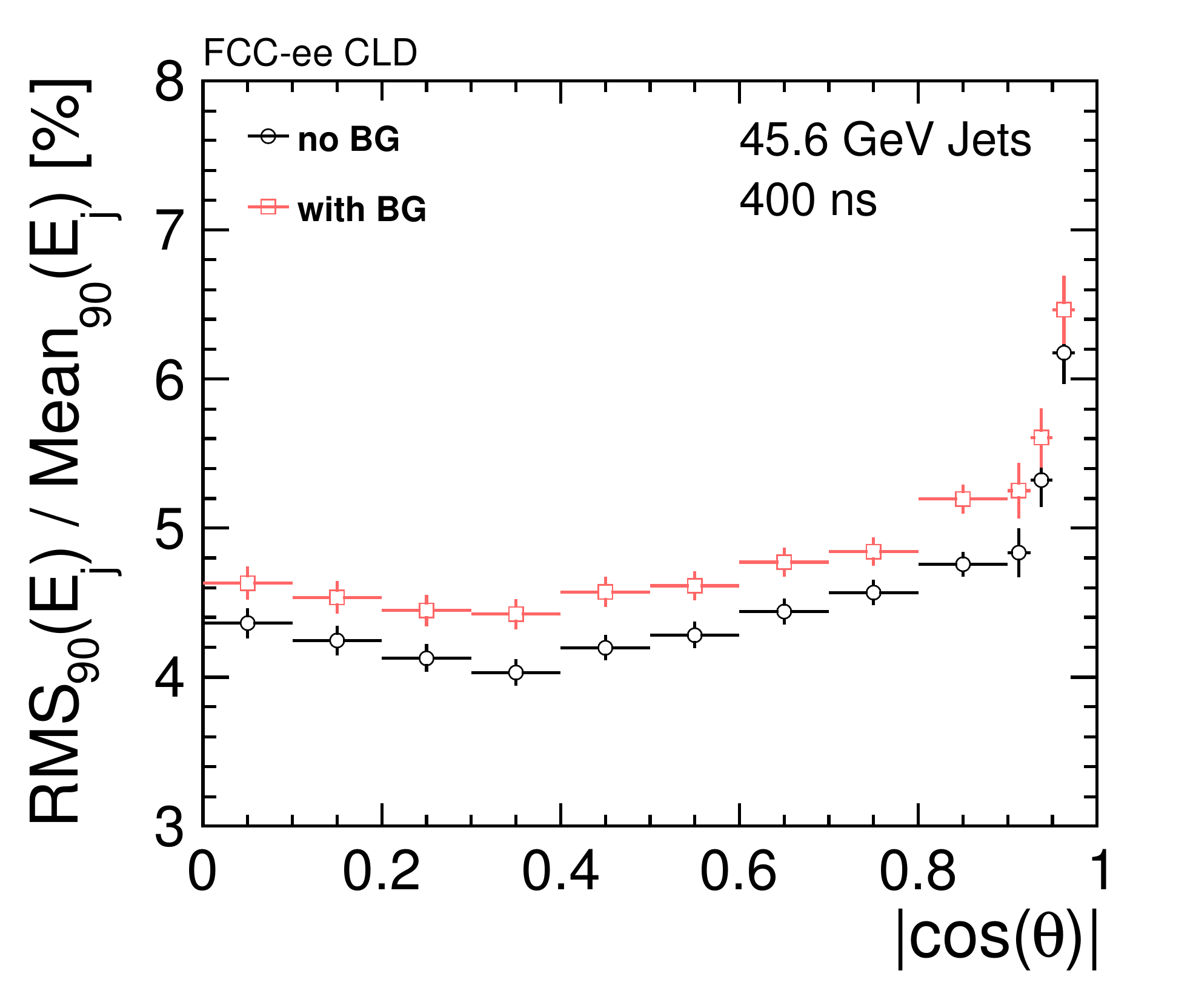}
    \caption{}
    \label{subfig:JER_totalE_400ns_w_wo_bkg_91GeV}
  \end{subfigure}%
  \begin{subfigure}{.5\textwidth}
    \centering
    \includegraphics[width=\linewidth]{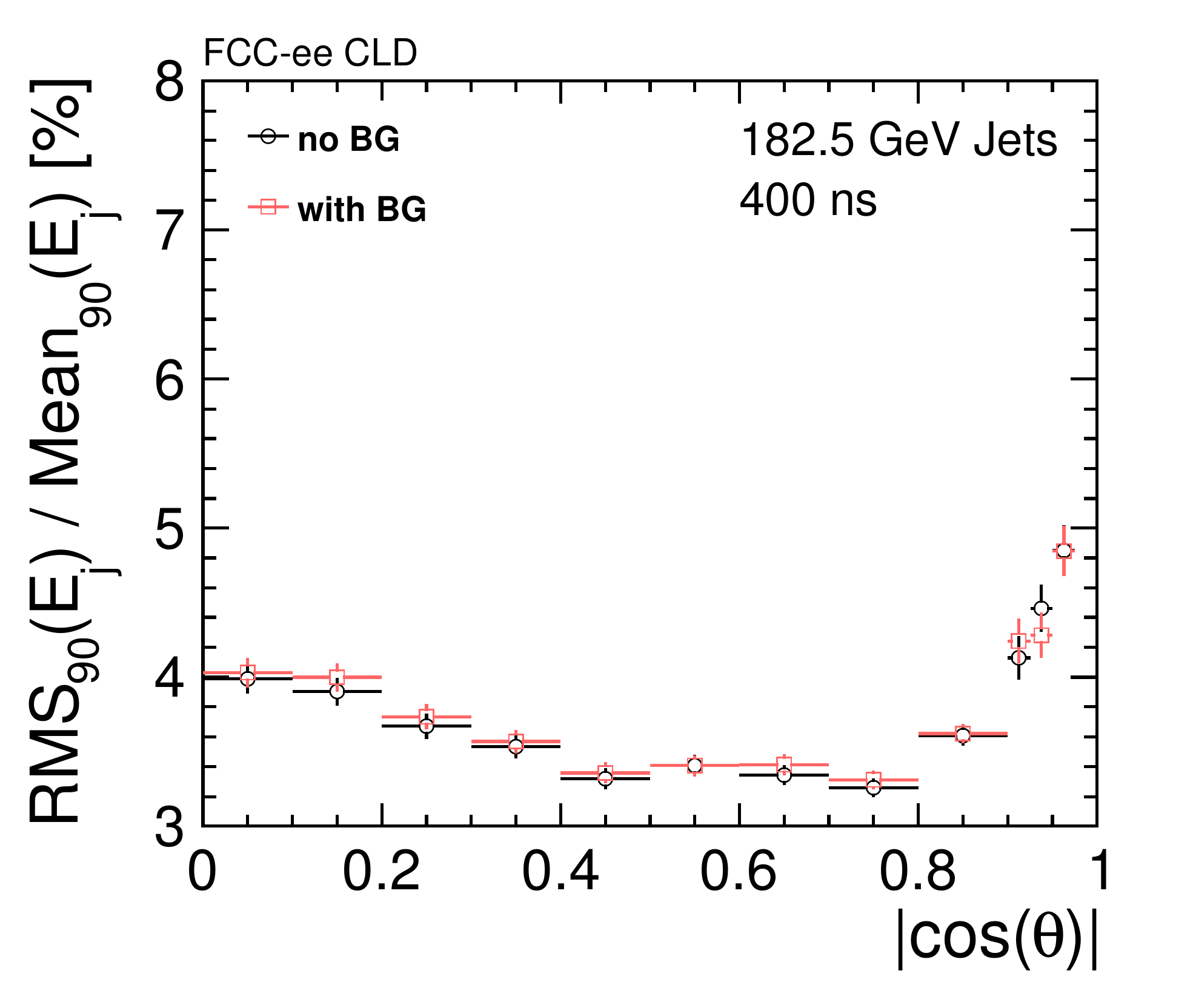}
    \caption{}
    \label{subfig:JER_totalE_400ns_w_wo_bkg_365GeV}
  \end{subfigure}
  \caption{Jet energy resolution as a function of the quark polar angle with and without incoherent pair background overlaid on the di-jet events, (a) at 91.2 GeV and (b) at 365 GeV centre-of-mass energies. The background is integrated over 400 ns.}
  \label{fig:JER_totalE_400ns_w_wo_bkg_91GeV_365GeV}
\end{figure}

The comparison of the jet energy resolution obtained with and without background overlay is shown in Figure~\ref{fig:JER_totalE_400ns_w_wo_bkg_91GeV_365GeV}. 
At 91.2 GeV, one observes a degradation of the jet energy resolution when including the beam-induced background.
Since the energy resolution is calculated using the sum of all reconstructed particles, 
all calorimeter hits originating from background particles are added to the total energy sum, independently of their direction w.r.t. the axis of the jet. 
At 365 GeV, the effect of the background is significantly smaller. 
This is understood, since, as shown in Subsection~\ref{calorimeter_bg}, the total deposited energy from background particles is significantly larger for 20 bunch crossings at 91.2 GeV than for 1 bunch crossing at 365 GeV. 
Additionally, the relative weight of the  energy deposited by background particles is smaller at higher centre-of-mass energies. 

\paragraph{Jet Clustering}
As it has been shown in the previous section, comparing the energy sum of all reconstructed particles with the sum of
all stable particles to estimate jet energy resolution is not a robust method against beam backgrounds.
Thus, a proper jet reconstruction is needed.
In the following, the jet reconstruction is performed with the VLC algorithm~\cite{Boronat:2016tgdVLC} in the two-jets exclusive mode. 
The VLC algorithm parameters $\gamma$ and $\beta$ are fixed to 1.0, while the R parameter is set to 1.1.
The algorithm is run over all reconstructed Particle Flow Objects in the event, in order to build two `reconstructed jets'.
Also, the same algorithm is run over all stable MC particles (excluding neutrinos) to reconstruct two `particle level jets'. 
The two reconstructed jets have to be matched to each of the particle level jets within an angle of 10$^{\circ}$.
The jet energy resolution is determined by the width of the energy response ratio of the reconstructed jets ($E_{j}^{R}$) divided by the particle level jets ($E_{j}^{G}$), normalised
by the mean value of that ratio, $\mathrm{Mean}_{90}$(E$^R_j$/E$^G_j$).

\begin{figure}[htbp]
  \centering
  \begin{subfigure}{.5\textwidth}
    \centering
    \includegraphics[width=\linewidth]{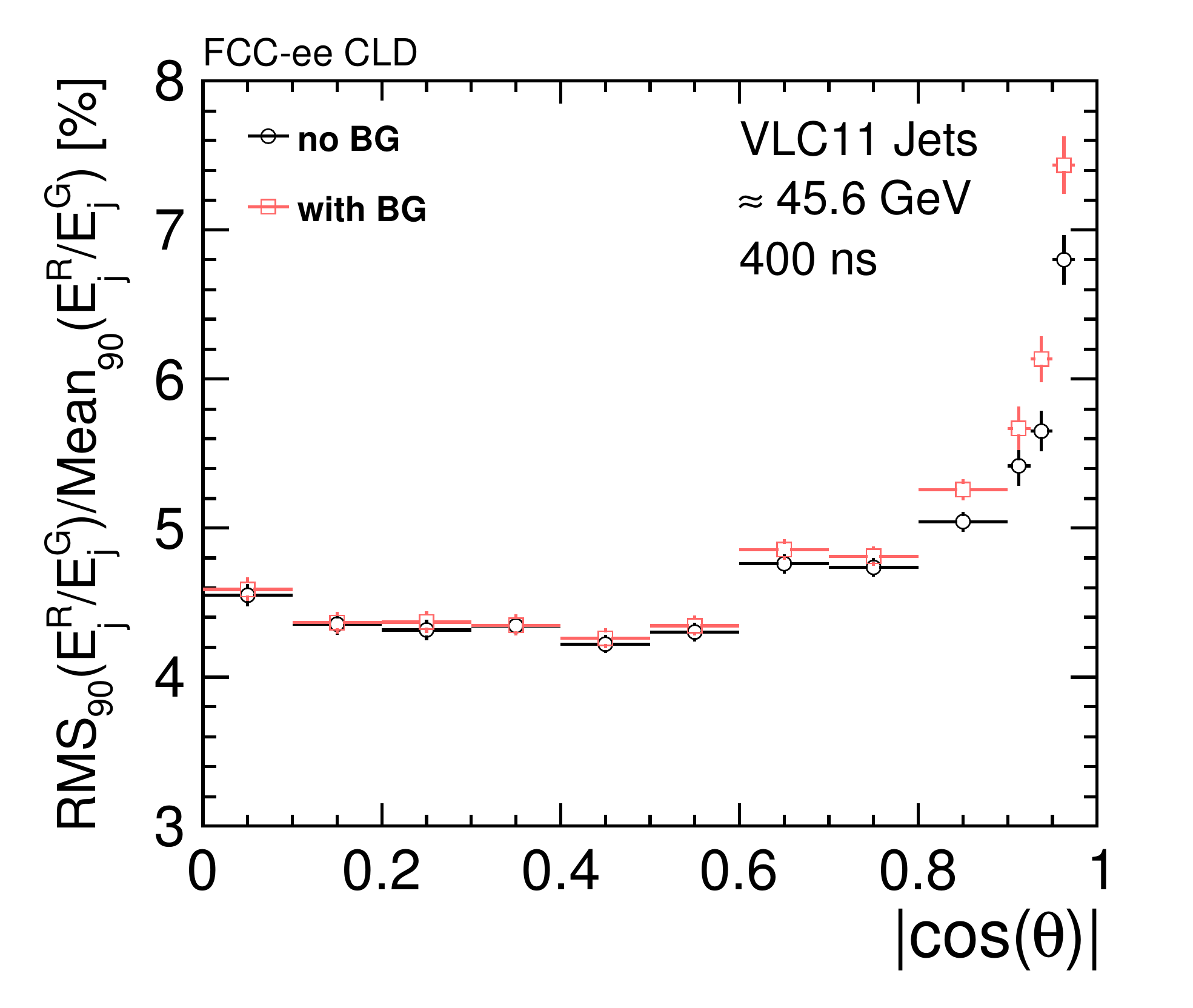}
    \caption{}
    \label{subfig:JER_jetClust_400ns_w_wo_bkg_91GeV}
  \end{subfigure}%
  \begin{subfigure}{.5\textwidth}
    \centering
    \includegraphics[width=\linewidth]{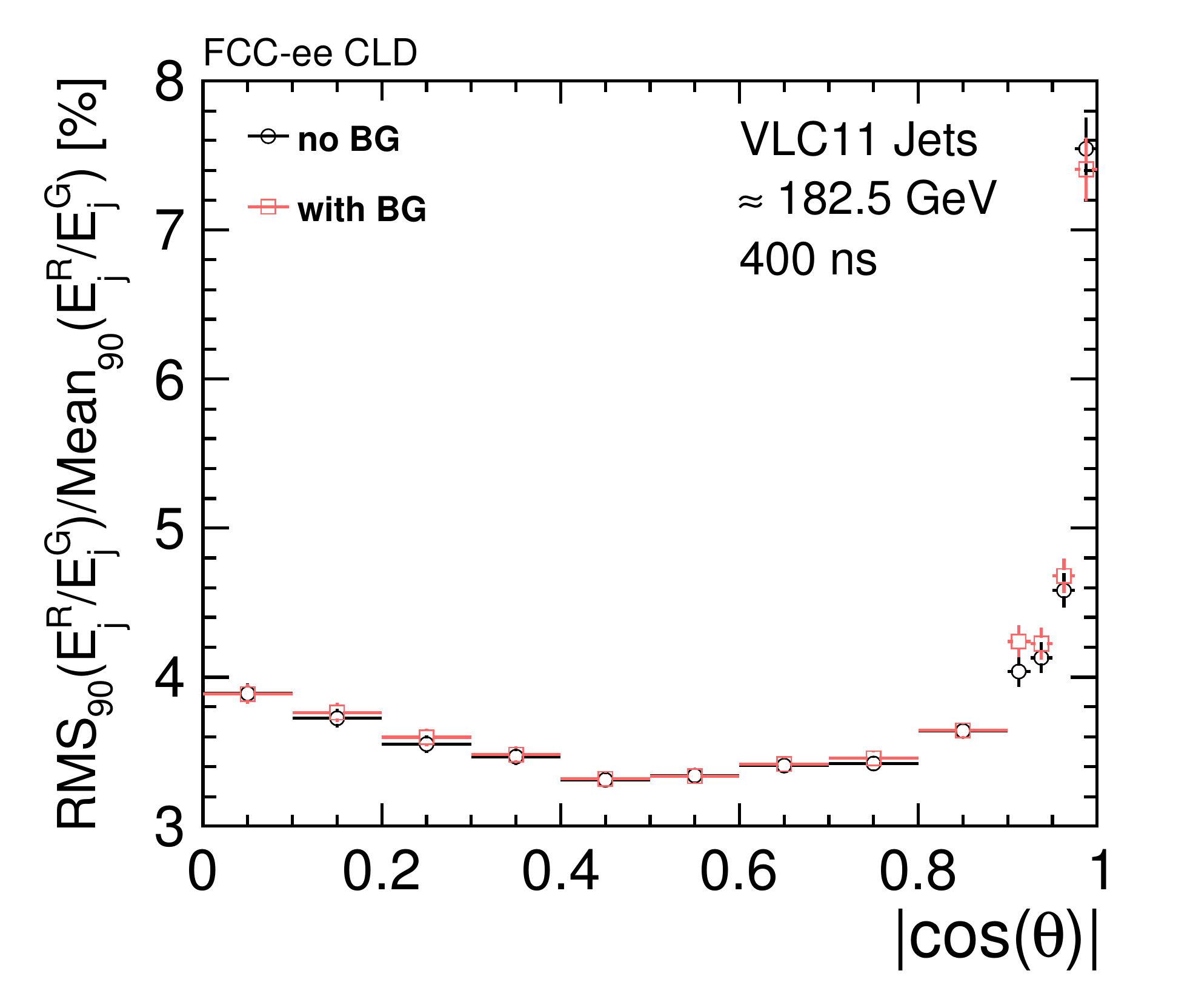}
    \caption{}
    \label{subfig:JER_jetClust_400ns_w_wo_bkg_365GeV}
  \end{subfigure}
  \caption{Jet energy resolution as a function of the quark polar angle, with and without incoherent pair background overlaid on the physics di-jet events, (a) at 91.2 GeV and (b) at 365 GeV centre-of-mass energy. The background is integrated over 400 ns.
The energy resolution is obtained by comparing the energy of reconstructed jets using the VLC algorithm with the energy of particle level jets.}
  \label{fig:JER_jetClust_400ns_w_wo_bkg_91GeV_365GeV}
\end{figure}

Figure~\ref{fig:JER_jetClust_400ns_w_wo_bkg_91GeV_365GeV} shows the jet energy resolution obtained with the jet reconstruction method. 
For both centre-of-mass energies the impact of the background is negligible, except in the forward region at 91.2 GeV, where a significant amount of energy from background particles is deposited. 
Note that these results are obtained without applying any additional timing or $p_T$ cuts (as done e.g. in CLICdet studies) - applying such cuts might suppress the effect of the background in the forward region.

\begin{figure}[htbp]
  \centering
  \begin{subfigure}{.5\textwidth}
    \centering
    \includegraphics[width=\linewidth]{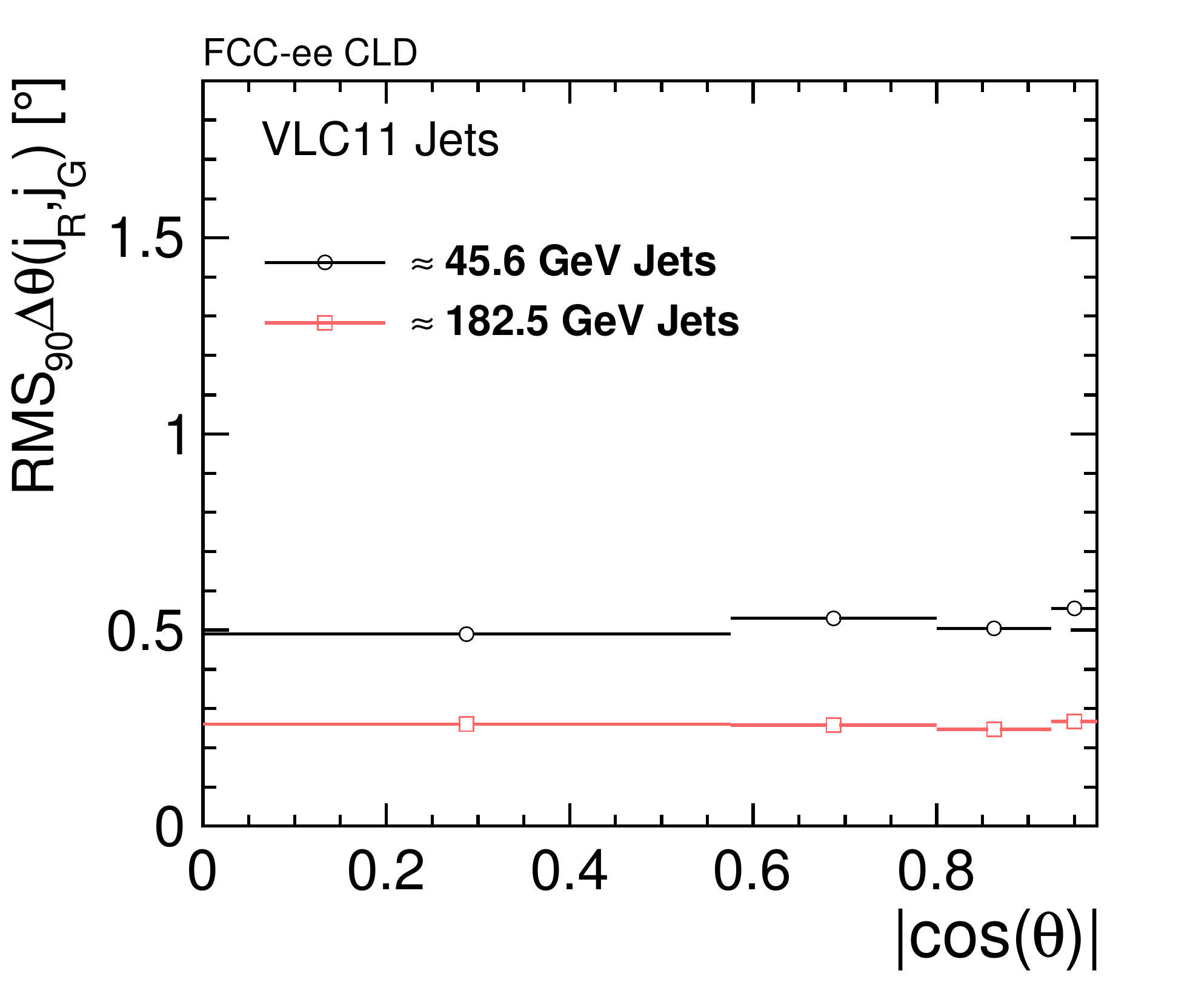}
    \caption{}
    \label{subfig:angulRes_jetClust_400ns_w_wo_bkg_91GeV}
  \end{subfigure}%
  \begin{subfigure}{.5\textwidth}
    \centering
    \includegraphics[width=\linewidth]{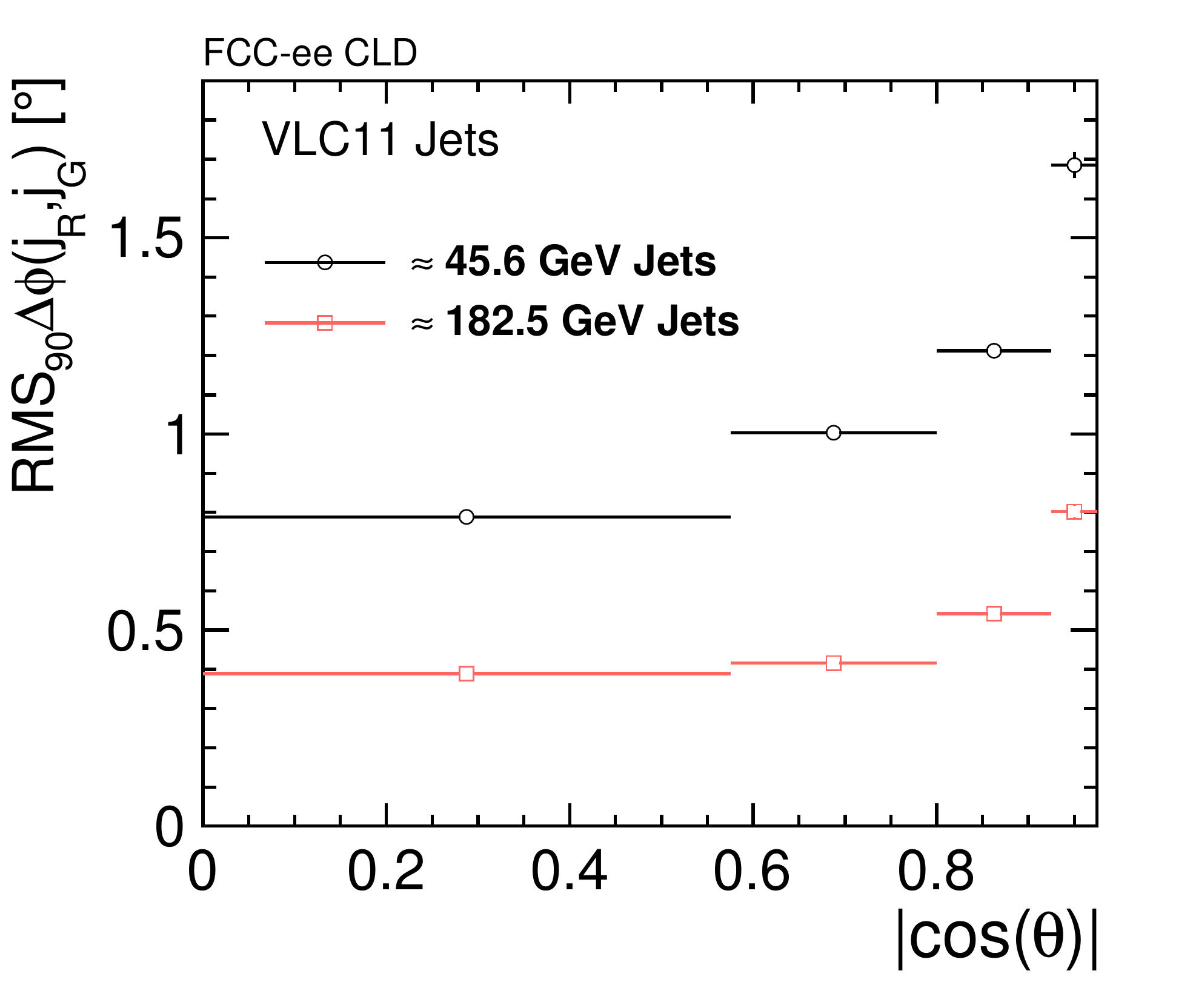}
    \caption{}
    \label{subfig:angulRes_jetClust_400ns_w_wo_bkg_365GeV}
  \end{subfigure}
    \caption{Jet angular resolutions (a) in $\theta$ and (b) in $\phi$  at 91.2 GeV and 365 GeV centre-of-mass energies.}
  \label{fig:angulRes_jetClust_400ns_w_wo_bkg_91GeV_365GeV}
\end{figure}

The angular resolution of jets, in $\theta$ and $\phi$, have been studied by comparing azimuthal $\phi$ and polar $\theta$ angles of reconstructed and particle level jets. 
The resolutions are defined as RMS$_{90}$ of the distributions $\Delta\theta$(j$_{\mathrm{R}}$,j$_{\mathrm{G}}$) and $\Delta\phi$(j$_{\mathrm{R}}$,j$_{\mathrm{G}}$), respectively. 
The results are shown in Figure~\ref{fig:angulRes_jetClust_400ns_w_wo_bkg_91GeV_365GeV}. The $\phi$ resolution for jets is found to be somewhat worse than the $\theta$ resolution, 
which can be explained by the effect of the magnetic field on the jet reconstruction. 
Investigating the angular resolutions with different detector magnetic field, it was found that the $\phi$ resolution strongly depends on the field, while the $\theta$ resolution does not.

\begin{figure}[htbp]
  \centering
  \begin{subfigure}{.5\textwidth}
    \centering
    \includegraphics[width=\linewidth]{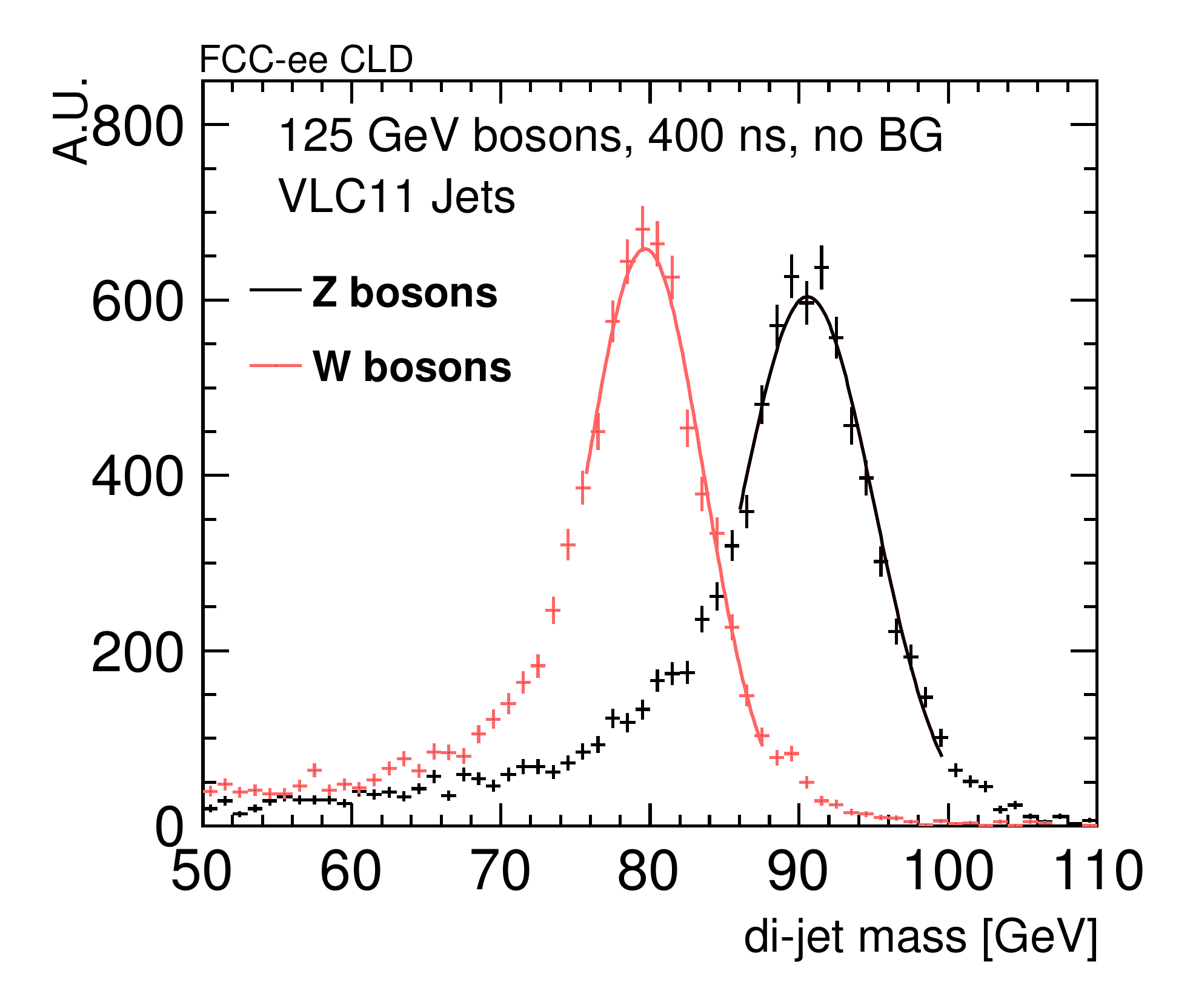}
    \caption{}
    \label{subfig:WZMassPeakSeparation_noBG}
  \end{subfigure}%
  \begin{subfigure}{.5\textwidth}
    \centering
    \includegraphics[width=\linewidth]{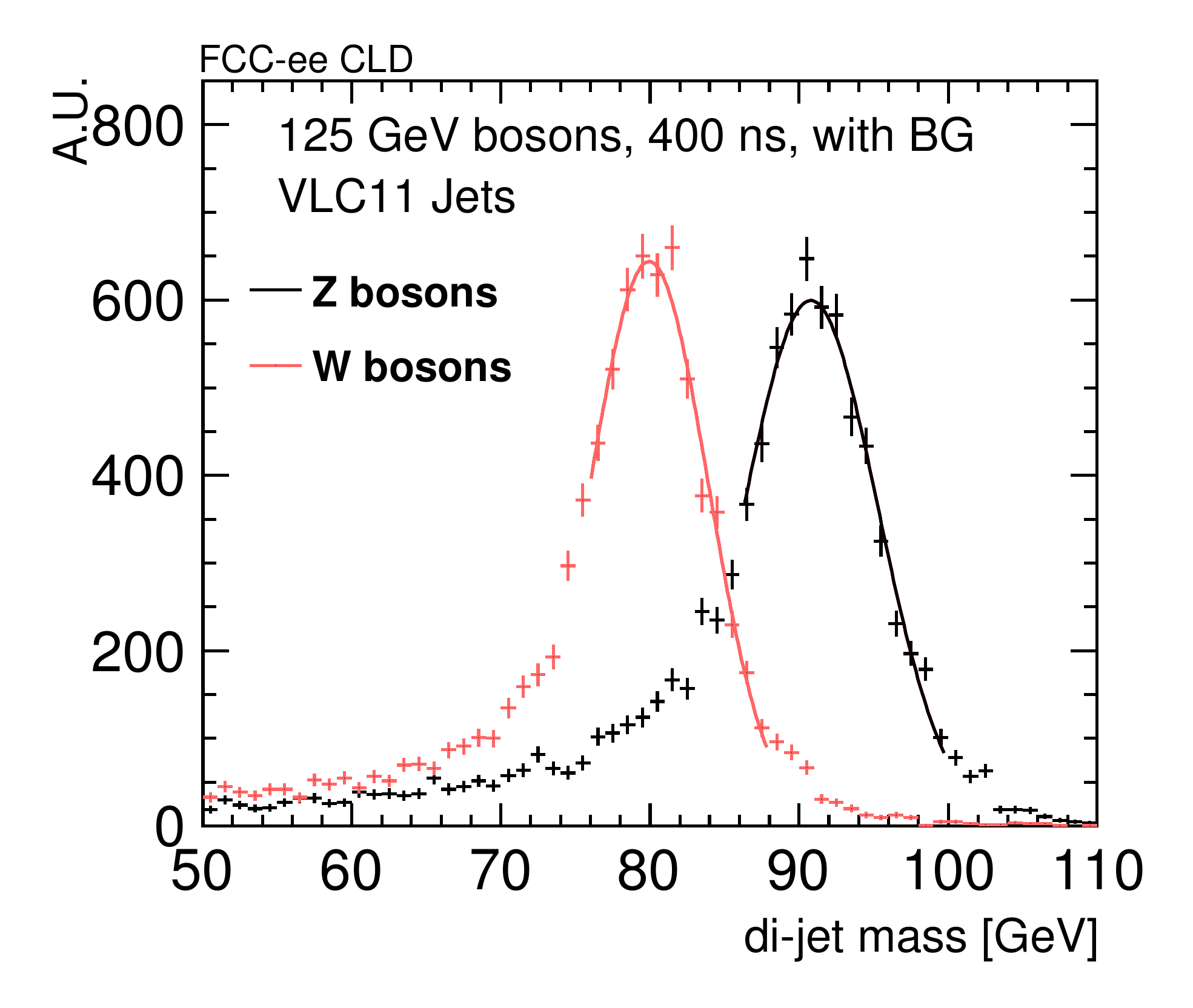}
    \caption{}
    \label{subfig:WZMassPeakSeparation_withBG}
  \end{subfigure}
    \caption{ Invariant di-jet mass distributions of W- and Z-bosons of 125 GeV energy (a) without beam-induced background and (b) with 365 GeV incoherent pair background overlaid. 
The solid lines represent Gaussian fits of the distributions done iteratively within a range of [$\mu - \sigma$, $\mu + 2\sigma$]. }
  \label{fig:WZMassPeakSeparation}
\end{figure}

\paragraph{W--Z Separation}
One of the requirements for the calorimeter system of the CLD detector is the ability to distinguish hadronic decays of W- and Z-bosons. 
To study the W- and Z-boson mass peak separation power, two processes are used: $WW \to \mu\nu_{\mu}qq$ and $ZZ \to \nu\bar{\nu} q\bar{q}$. 
All particles, except decay products from leptonic decays of bosons, are used for the jet reconstruction. 
The invariant masses of W- and Z-bosons are shown in Figure~\ref{fig:WZMassPeakSeparation}. 
In order to estimate the separation power of the two peaks an iterative Gaussian fit is performed in the range [$\mu - \sigma$, $\mu + 2\sigma$] 
around the peak $\mu$, until $\sigma$ of the fit stabilises within $\pm$5$\%$. 
Restricting the fit range is motivated by the presence of significant non-Gaussian tails on the low-mass side of the distributions. 
The separation power is estimated from the fit parameters as: $(m_{Z} - m_{W}) / \sigma_{average}$, 
where $\sigma_{average} = (\sigma_{Z} + \sigma_{W}) / 2$ and $m_{Z}$,  $m_{W}$ are the mean values of the fitted distributions.

The mass resolution and separation power calculated for different values of the VLC parameter R are shown in Table~\ref{tab:WZMassSeparation}.
The separation power is calculated using two different methods: first, the mass of W- and Z-boson is obtained 
as the mean of the Gaussian fit, and second, the mass distributions are scaled such that the mean of the fit becomes equal to the PDG values of the W- and Z-boson mass. 
The results for both methods are displayed in Table~\ref{tab:WZMassSeparation}. 
The separation power for 125 GeV bosons is found to lie within the range 2.1--2.6.  
The impact of the 365\,GeV incoherent pair background on the separation power is found to be negligible.

\begin{table}[!hbt]
  \begin{center}
    \caption{ W- and Z-boson mass peak resolution and separation power calculated with different values of R of the VLC jet clustering algorithm. The energy of the bosons is 125 GeV.  }
      \centering
         \begin{tabular}{|c|c|c|c|c|c|}
             \hline
             background & $R$ & $\sigma_{m(W)}/m(W)$ & $\sigma_{m(Z)}/m(Z)$ & Separation & Separation (fixed mean) \\
             overlay & & [\%] & [\%] & $[\sigma]$ & $[\sigma]$ \\
             \hline
             no BG   & 0.7 & 5.94 & 5.75 & 2.19 & 2.16\\
             with BG & 0.7 & 5.95 & 5.90  & 2.13 & 2.13\\
             \hline
             \hline
             no BG   & 0.9 & 5.26 & 5.11 & 2.46 & 2.43\\
             with BG & 0.9 & 5.18 & 5.19 & 2.43 & 2.43\\
             \hline
             \hline
             no BG   & 1.1 & 4.99 & 4.94 & 2.58 & 2.54\\
             with BG & 1.1 & 5.36 & 4.96 & 2.50  & 2.45\\
             \hline
         \end{tabular}
\label{tab:WZMassSeparation}
\end{center}
\end{table}

\clearpage
\subsubsection{Flavour Tagging}
\label{flavour_tag}

Flavour tagging studies were initially performed for the CLIC\_SiD detector model and described in the CLIC CDR~\cite{cdrvol2}.
These studies were later extended to more realistic vertex detector geometries, with particular emphasis on the material budget~\cite{Alipour_Roloff_2014}.
Recently, further studies were performed for the CLICdet model, using the software chain described in Subsection~\ref{sec:simreco} and the flavour tagging package LCFIPlus~\cite{Suehara:2015ura}. First results are presented in~\cite{CLICdet_performance}, where a path to further improvements of the performance is also described.

New studies on flavour tagging performance in CLD, using the same tools as for CLICdet, improved by an optimized BDT, have been performed.
Samples of di-jet events ($\epem\rightarrow\bb$,$\epem\rightarrow\cc$,$\epem\rightarrow\uu,\dd,\ssbar$) at two centre-of-mass energies (\SI{91}{GeV} and \SI{365}{GeV}) have been simulated and reconstructed. Moreover, the quark pairs have been simulated in different directions in phase space, in order to be able to study the performance dependence on the polar angle. 
For each sample of fixed centre-of-mass energy and quark direction, results are shown in terms of misidentification efficiency as a function of quark tagging efficiency.
Two sources of misidentification are distinguished in all cases: contamination from the other heavy quark (charm for beauty misidentification, beauty for charm misidentification), and contamination from light-flavour quarks.
For each case, two curves are always shown, corresponding to track reconstruction performed with the conformal pattern recognition (Conformal Tracking) and 
with the true (Monte Carlo) pattern recognition (Truth Tracking). The performance ratio between the two reconstruction methods is shown at the bottom of each plot. 
The results obtained with the Truth Tracking allow to evaluate the detector performance, without the effect of the specific pattern recognition algorithm. 
The aim of the comparison with the Conformal Tracking results is to assess how far off the latter is with respect to the true performance.

Flavour tagging is performed by training boosted decision trees (BDTs) as multivariate classifier in the TMVA package in ROOT.
The jets are classified based on the number of reconstructed secondary vertices. Four categories are identified as follows: 2 secondary vertices (A), 1 secondary vertex and 1 pseudo-vertex\footnote{A pseudo-vertex is defined when a track is present, whose trajectory is collinear to and within some distance from the line connecting primary and secondary vertex.} (B), 1 secondary vertex and 0 pseudo-vertices (C), and 0 secondary vertices (D).
The light-flavour jets are confined in category (D), while most of the b-jets fall into categories (A) and (B). More details on the vertex reconstruction, jet clustering and tagging strategy can be found in~\cite{Suehara:2015ura}.

\begin{figure}[htp]
\centering
\includegraphics[width=.3\textwidth]{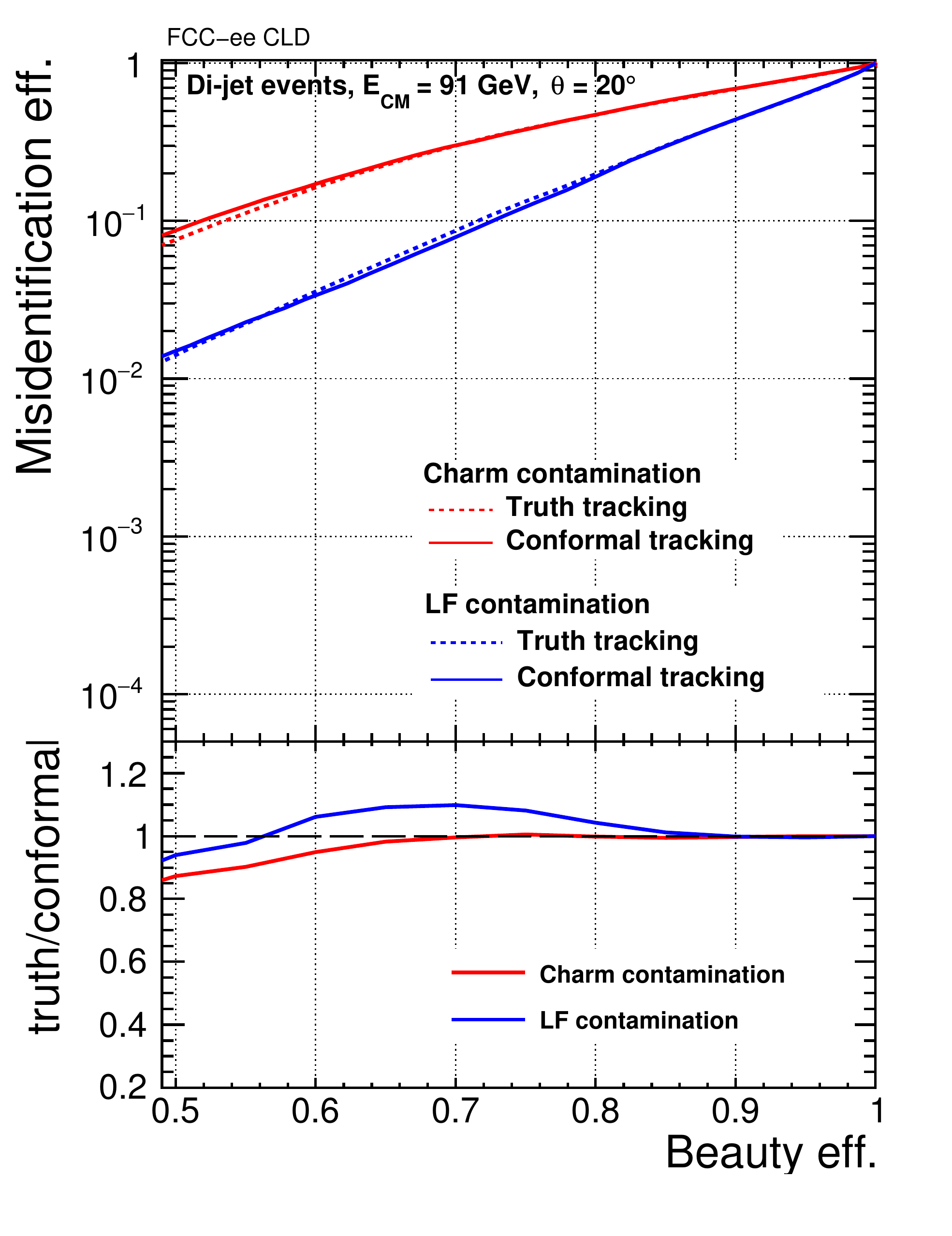}\quad
\includegraphics[width=.3\textwidth]{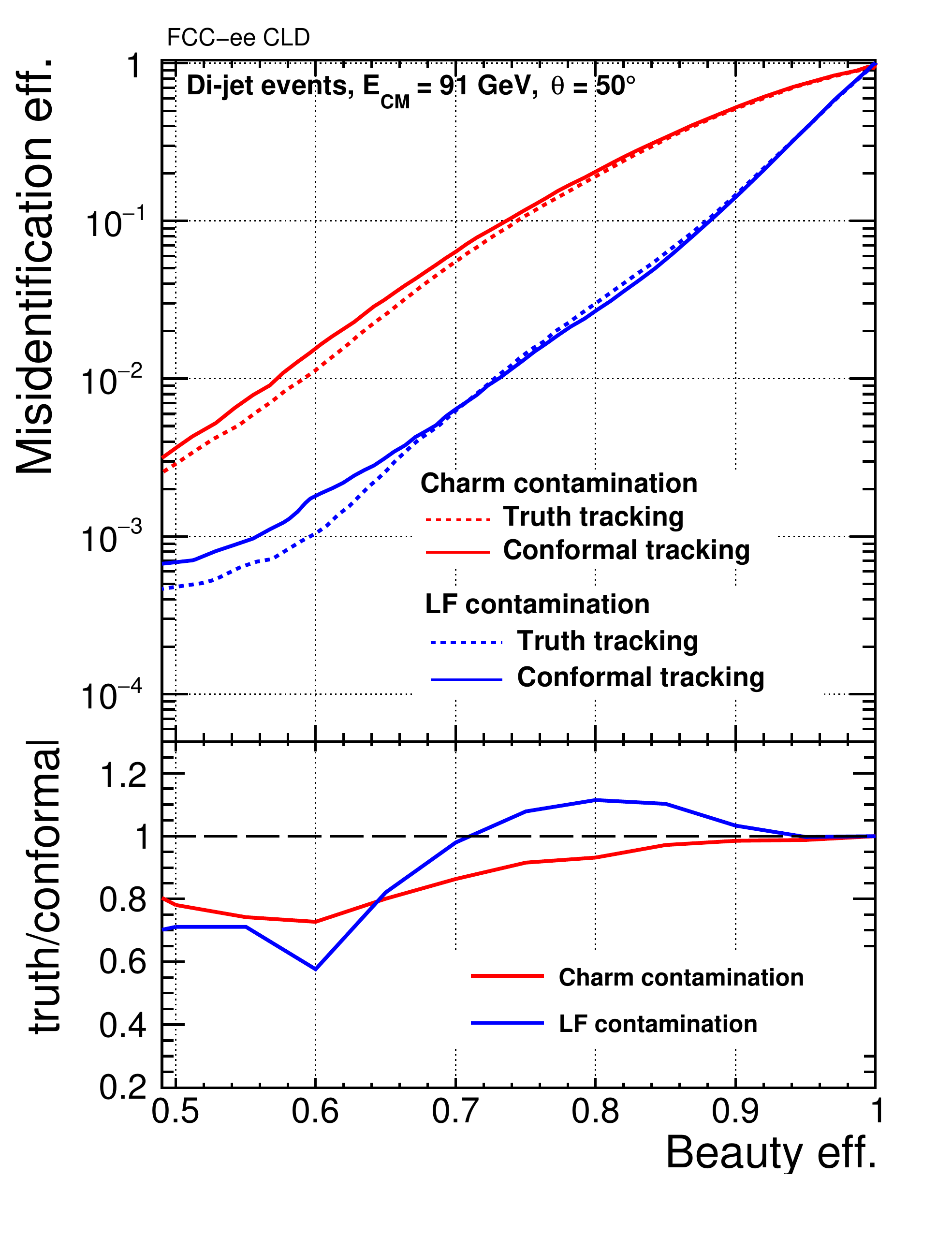}\quad
\includegraphics[width=.3\textwidth]{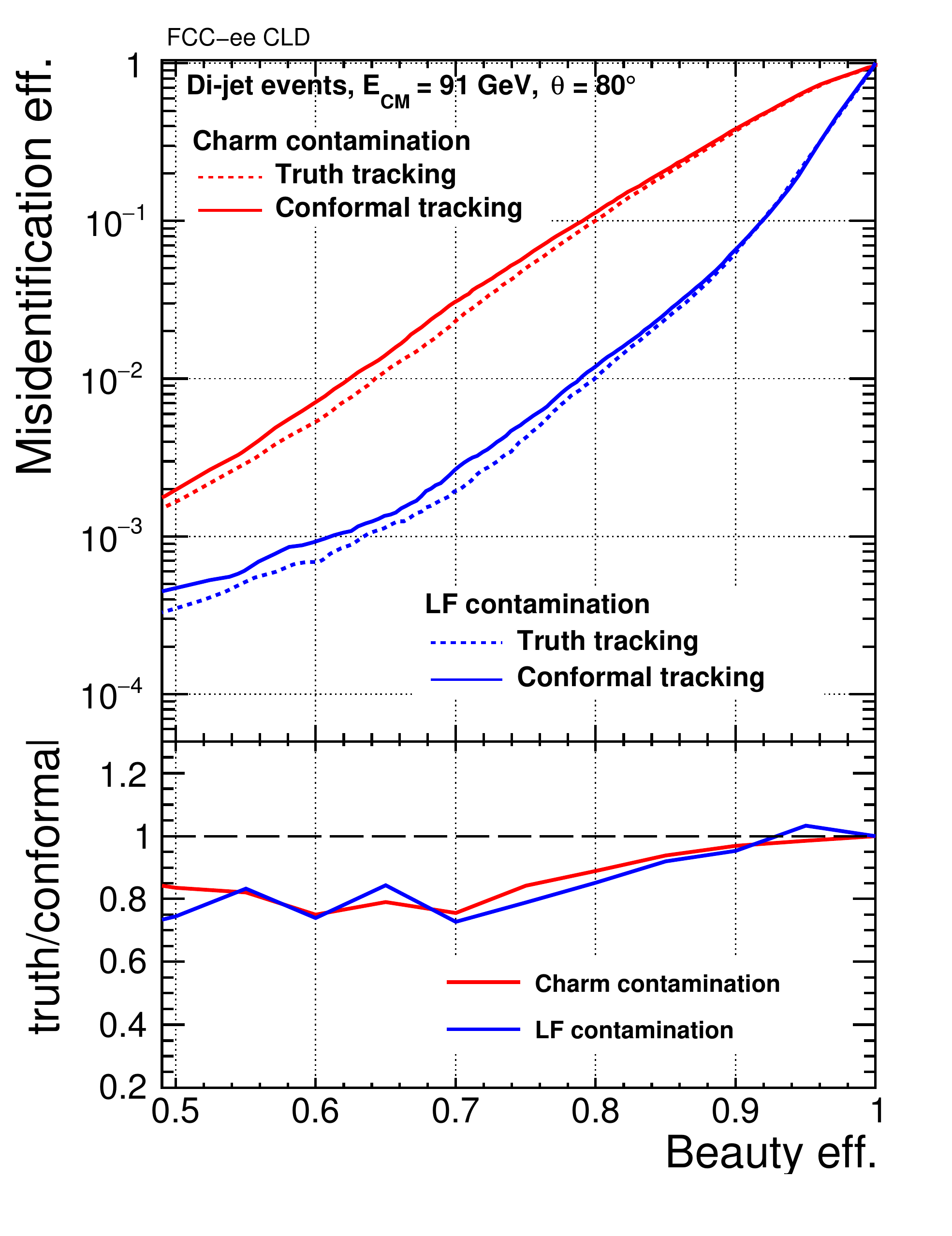}

\medskip
\includegraphics[width=.3\textwidth]{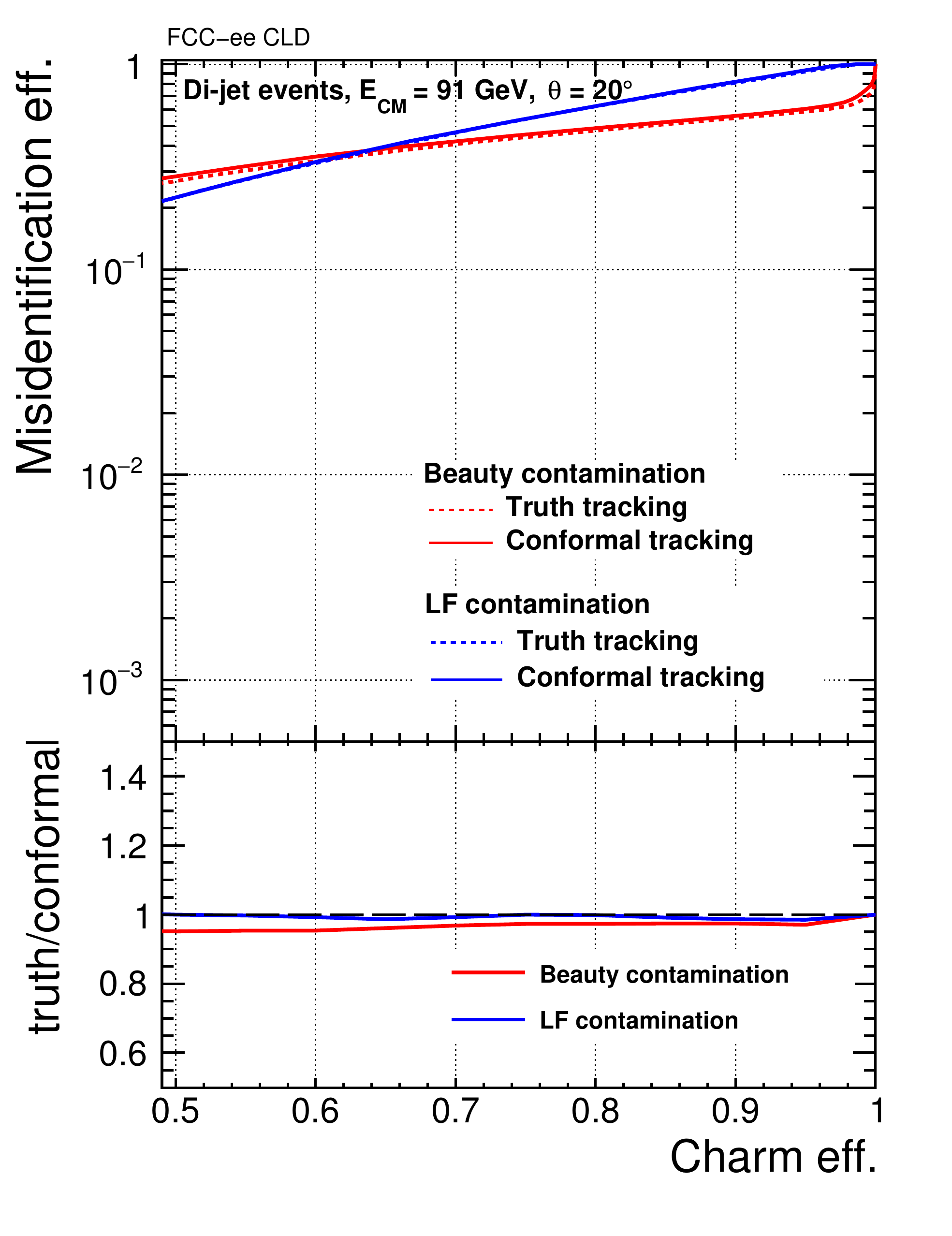}\quad
\includegraphics[width=.3\textwidth]{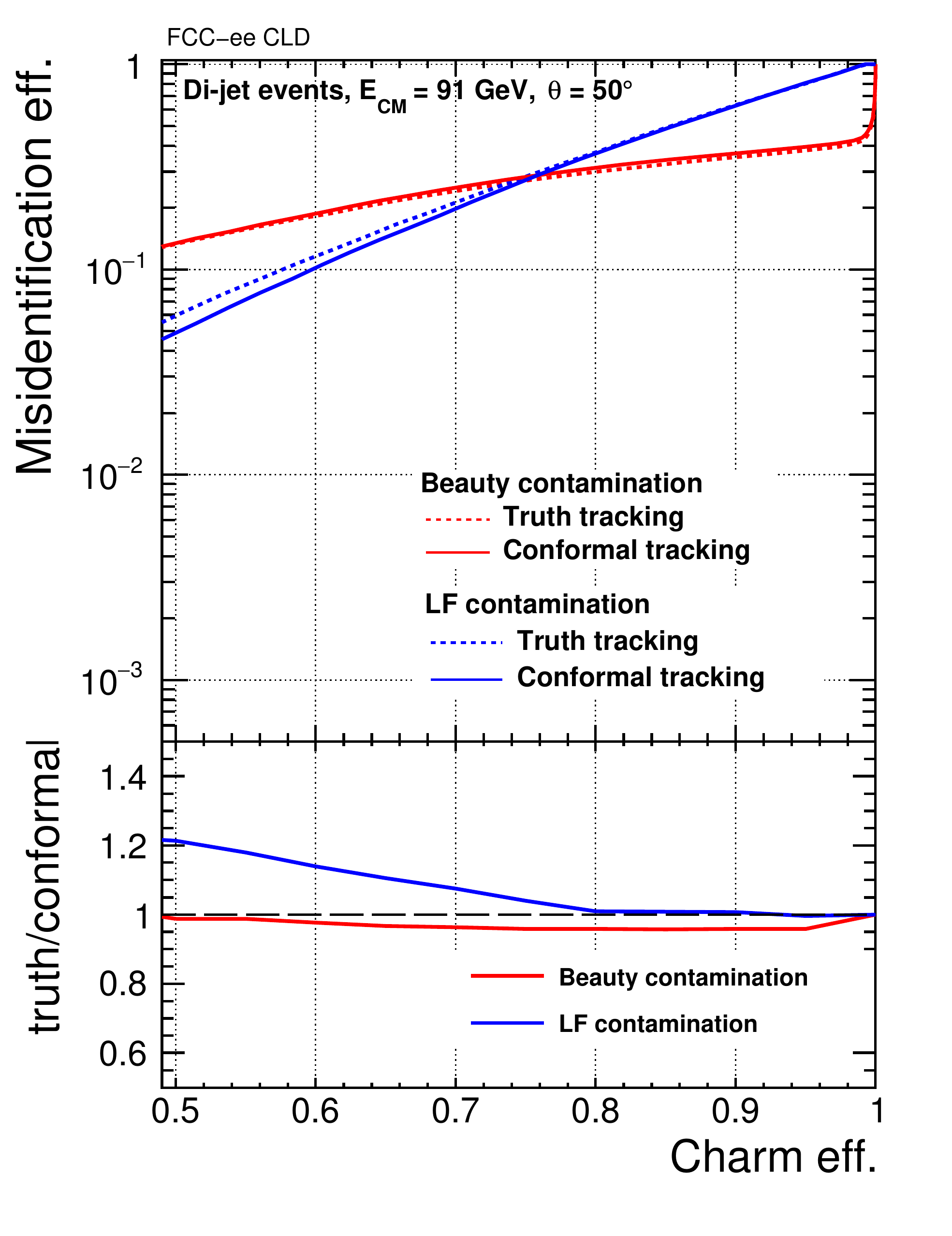}\quad
\includegraphics[width=.3\textwidth]{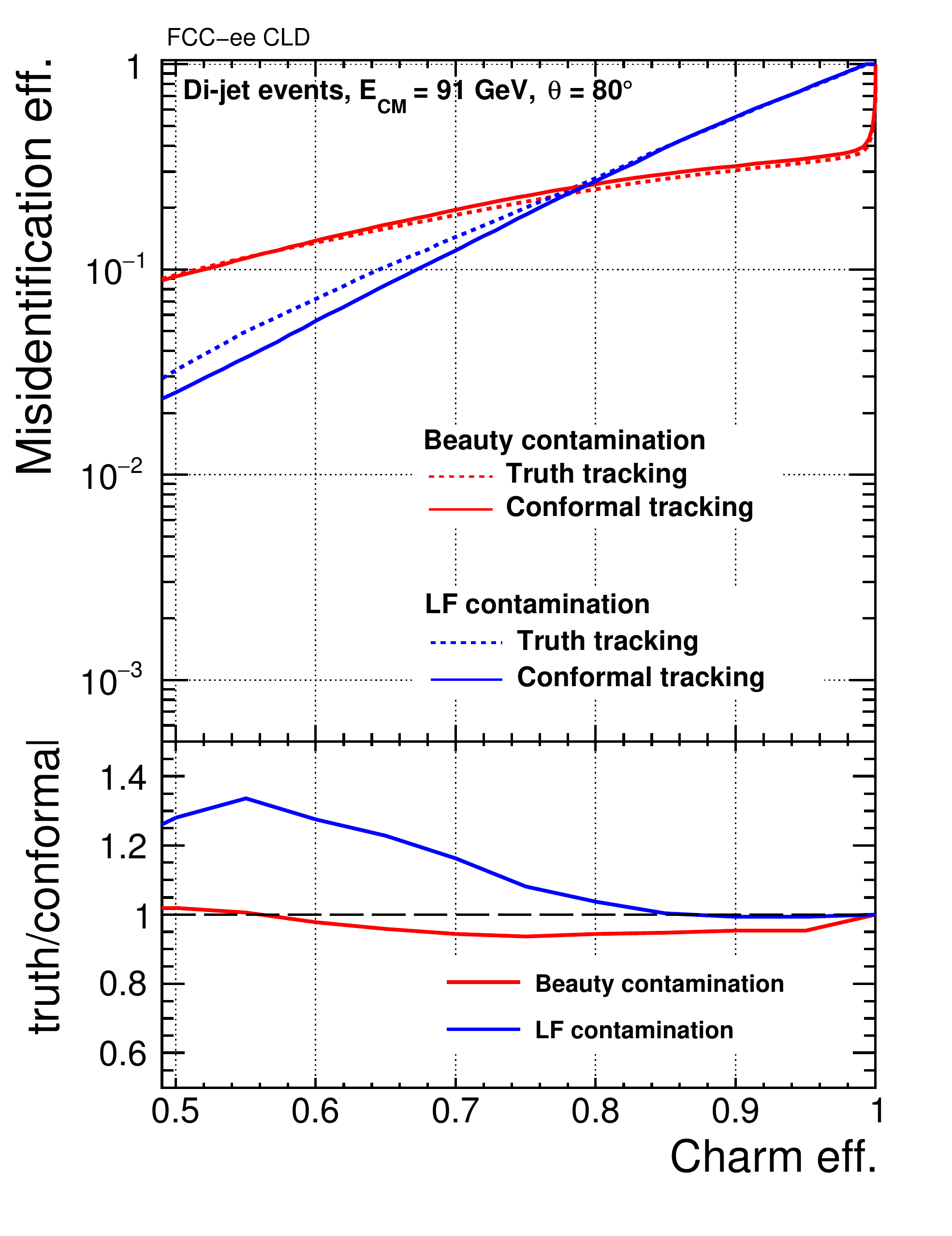}

\caption{B-tagging (top) and C-tagging (bottom) performance for di-jet events at \SI{91}{GeV} centre-of-mass energy and $\theta$ = 20\degrees{}(left), 50\degrees{}(middle) and 80\degrees{}(right). Per contamination source, results obtained with truth and conformal tracking are shown, with their ratio on the bottom canvas. }
  \label{fig:tagging_91GeV}
\end{figure}

\begin{figure}[htp]
\centering
\includegraphics[width=.3\textwidth]{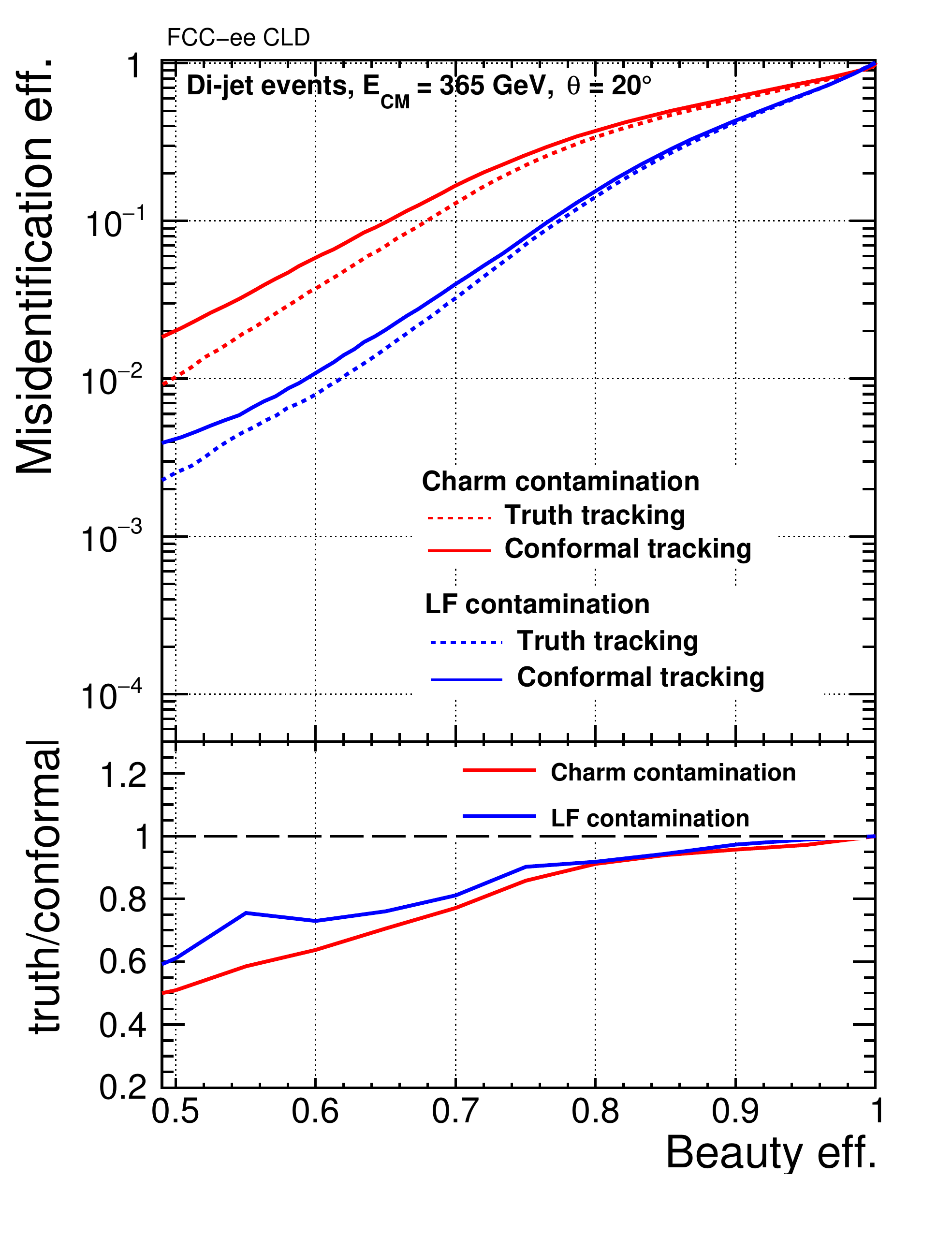}\quad
\includegraphics[width=.3\textwidth]{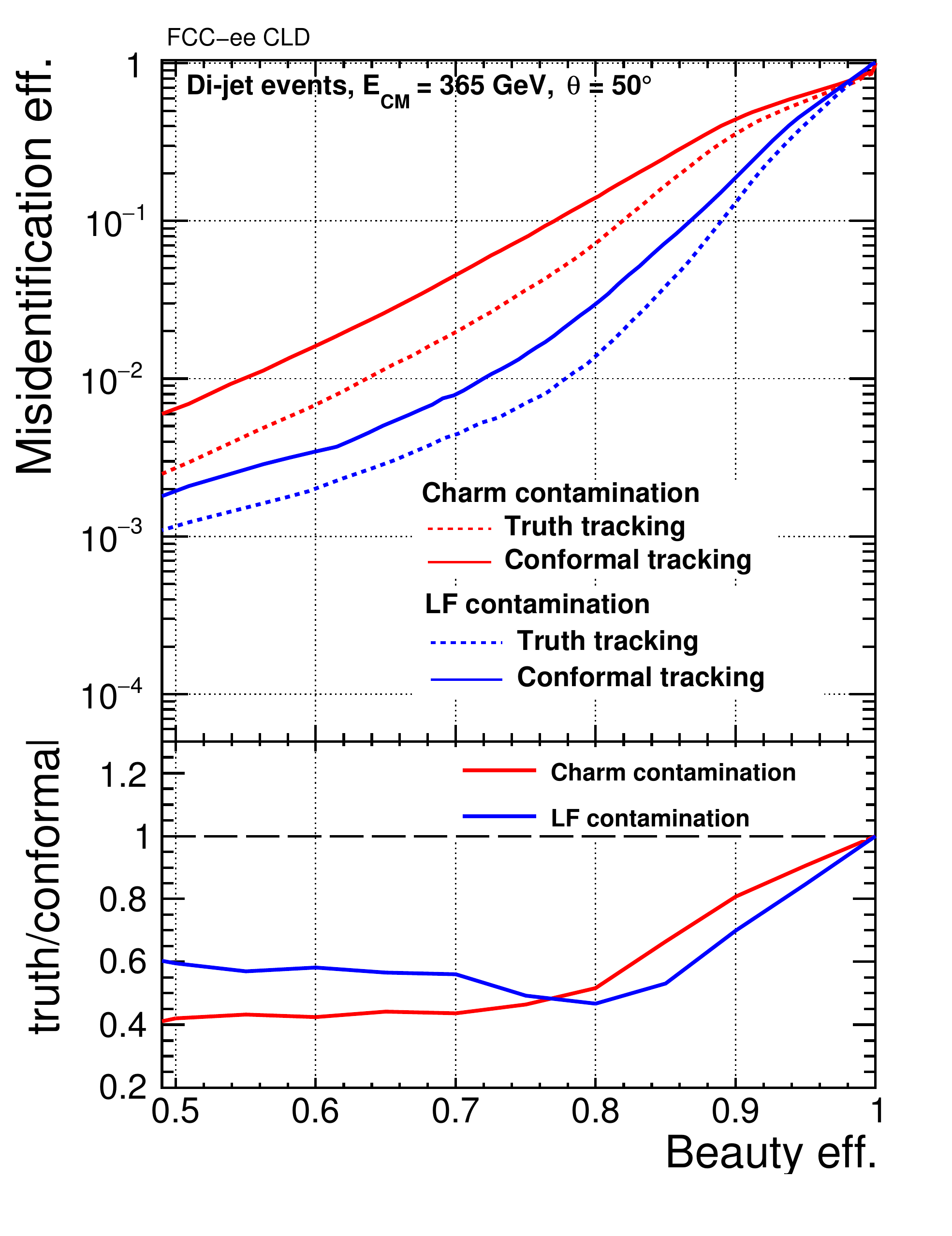}\quad
\includegraphics[width=.3\textwidth]{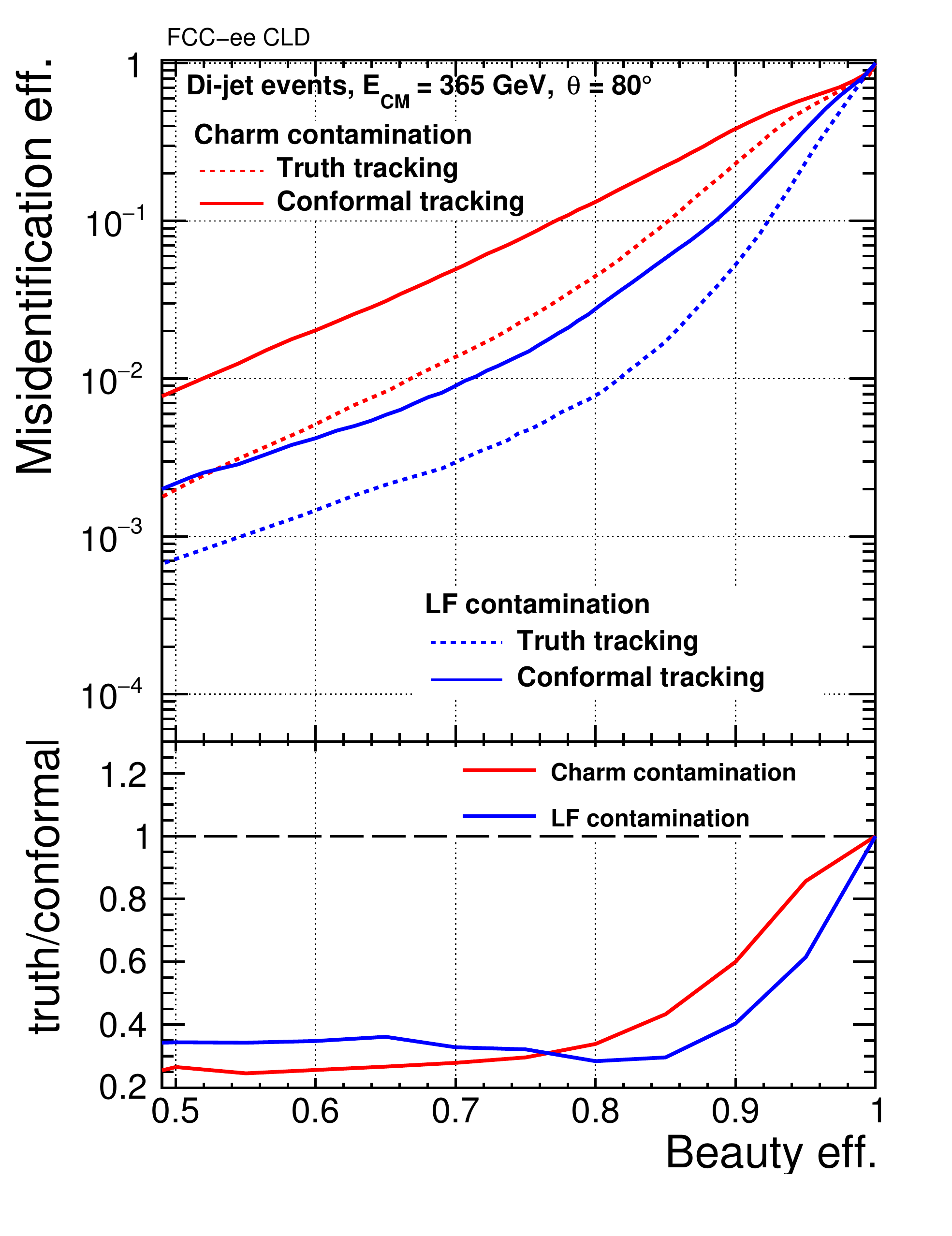}

\medskip
\includegraphics[width=.3\textwidth]{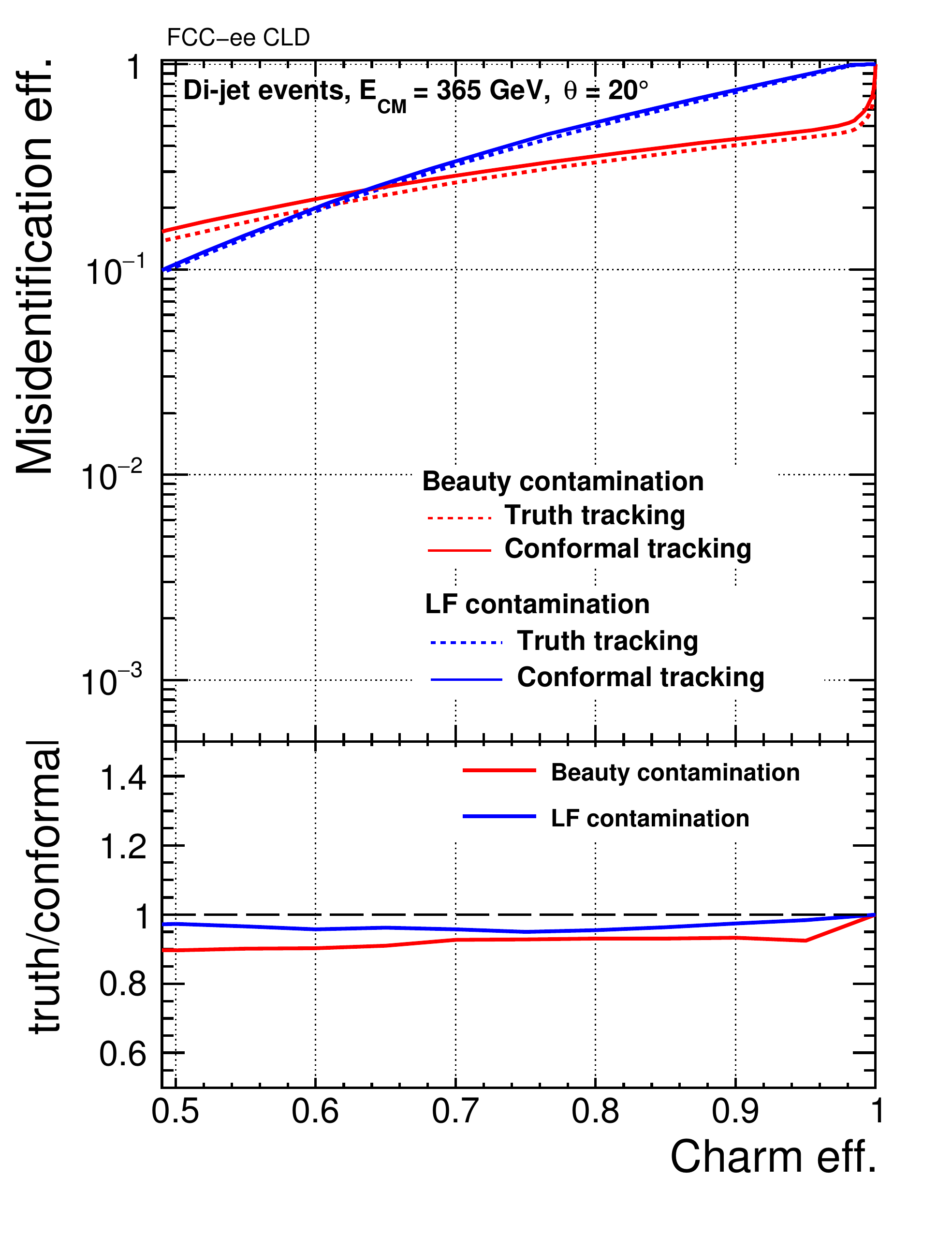}\quad
\includegraphics[width=.3\textwidth]{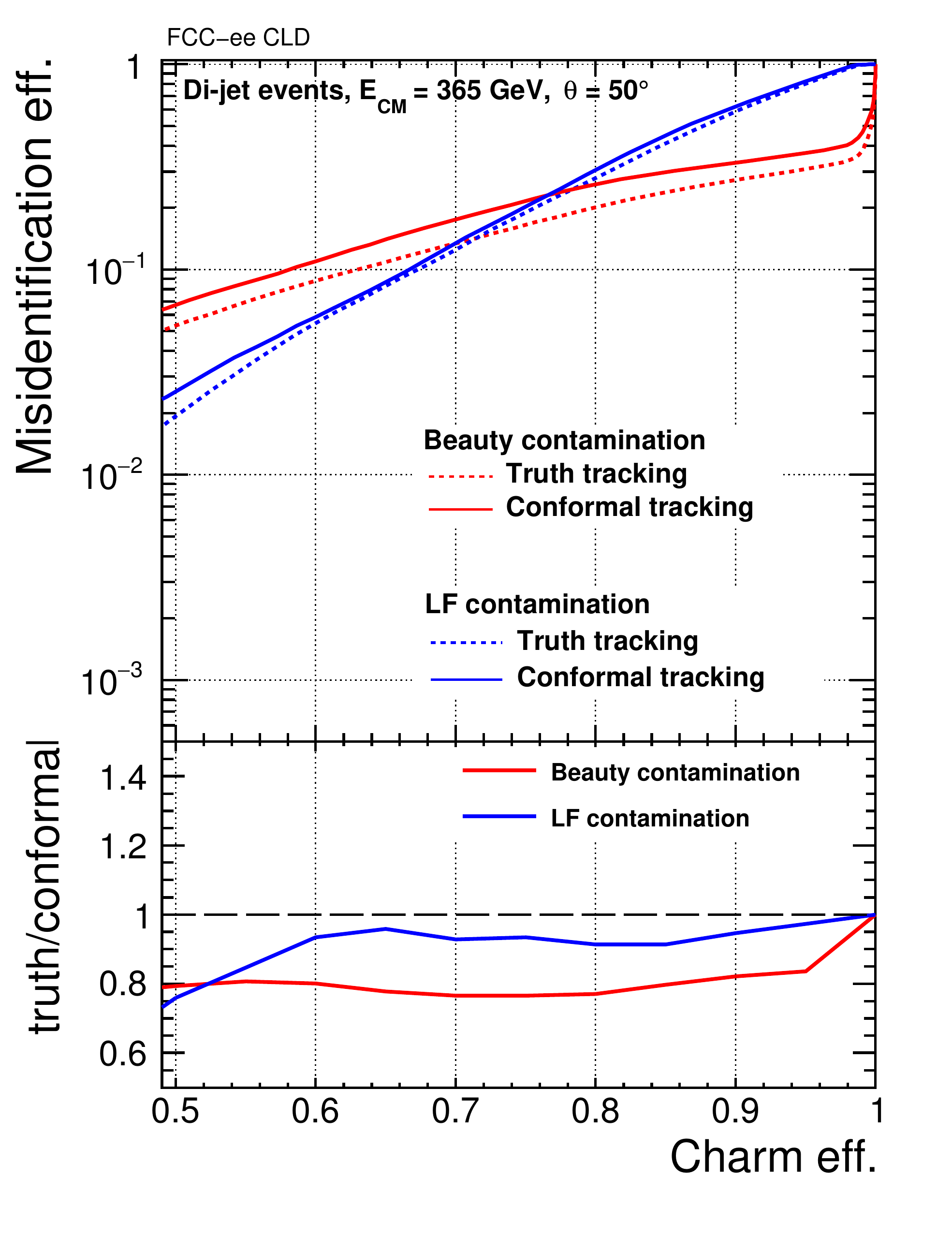}\quad
\includegraphics[width=.3\textwidth]{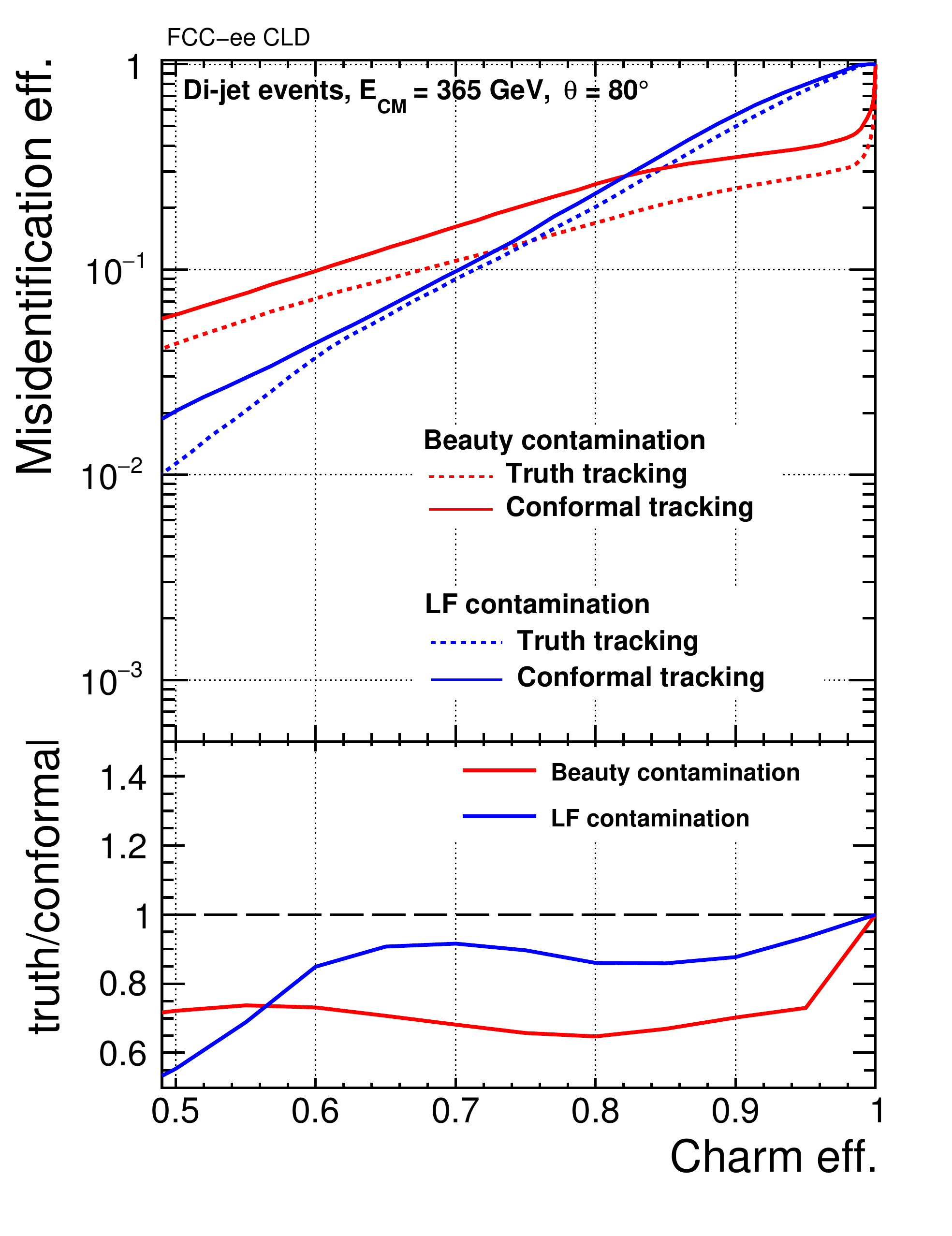}

\caption{B-tagging (top) and C-tagging (bottom) performance for di-jet events at \SI{365}{GeV} centre-of-mass energy and $\theta$ = 20\degrees{}(left), 50\degrees{}(middle) and 80\degrees{}(right). Per contamination source, results obtained with truth and conformal tracking are shown, with their ratio on the bottom canvas. }
  \label{fig:tagging_365GeV}
\end{figure}

Figures~\ref{fig:tagging_91GeV} and~\ref{fig:tagging_365GeV} summarise the b-tagging (top) and c-tagging (bottom) performance for di-jets at \SI{91}{GeV} and \SI{365}{GeV} centre-of-mass energies, for quarks at $\theta$ = 20\degrees{}(left), 50\degrees{}(middle), 80\degrees{}(right).

In all cases, the misidentification for b-jets is smaller than for c-jets, as the latter can be more easily misidentified either as b-jets or light-jets.
At \SI{91}{GeV} centre-of-mass energy, both the b- and c-tagging misidentification is reduced in central jets with respect to forward jets.
The same trend is observed for c-tagging at \SI{365}{GeV}, while b-tagging in the central jets shows a slightly worse misidentification. 
Moreover, in terms of absolute values of misidentification, the c-tagging results are systematically better for \SI{365}{GeV} than for \SI{91}{GeV} centre-of-mass energy jets. This is consistent with the fact that c quarks have a short decay length, and the higher boost at \SI{365}{GeV} is helpful to discriminate between primary and secondary vertices.
This is true also for b-tagging, except for central jets, which is explained by a combined effect of jet kinematics and detector geometry: b quarks in the highest centre-of-mass scenario decay \textit{after} the innermost vertex layer when produced in the central region. Therefore, the misidentification of b-jets increases in this case.

In general, the agreement between the results with truth and conformal tracking is worse at \SI{365}{GeV} centre-of-mass energy and for central jets. This is consistent with the finding that the reconstruction performs worse when no hits are available in the innermost detector layers. Work is in progress to improve the secondary vertex finding with tracks reconstructed with conformal pattern recognition. Further studies foresee to evaluate the flavour tagging performance in presence of background.

\begin{figure}[htp]
\centering
\includegraphics[width=.7\textwidth]{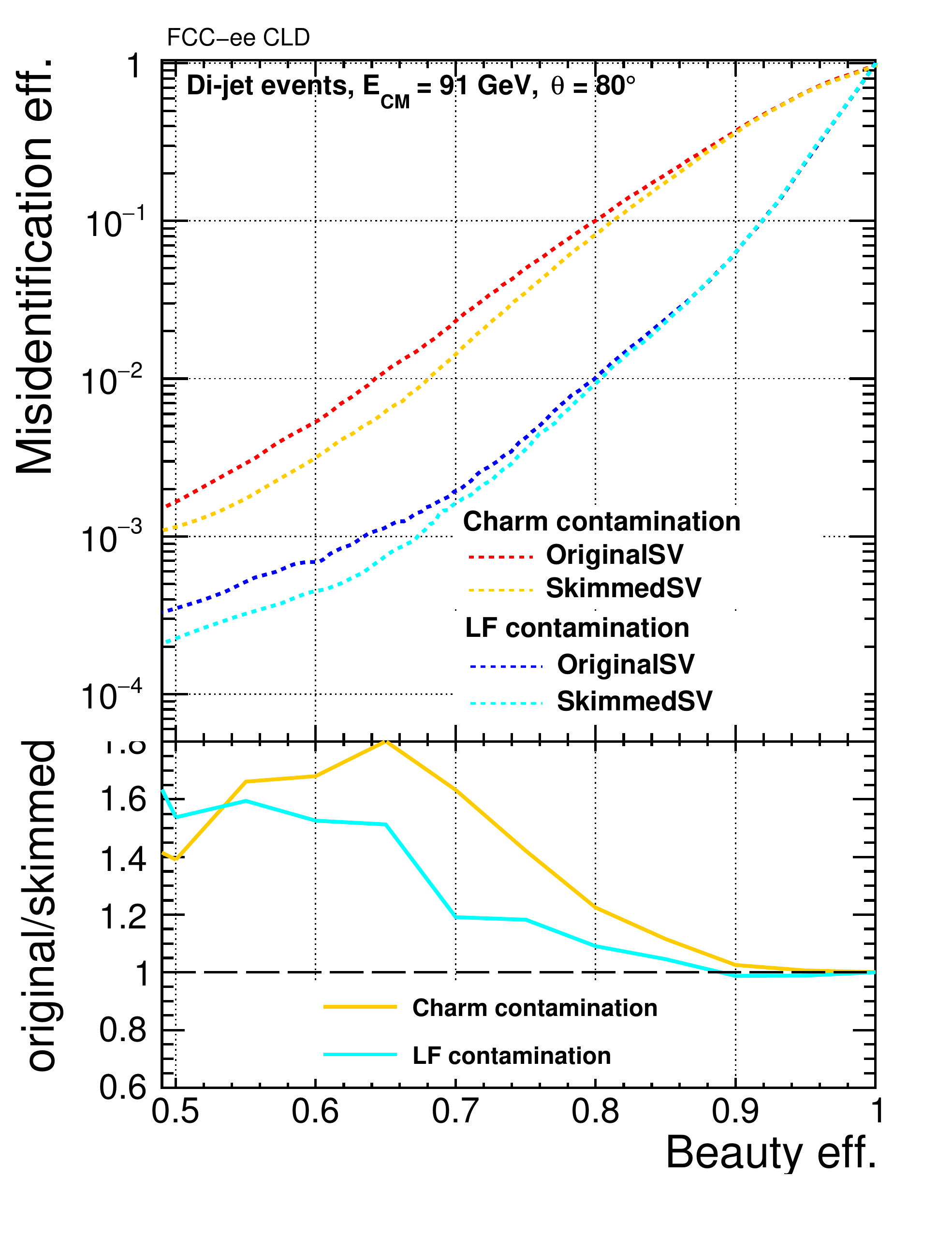}
\caption{B-tagging performance for di-jet events at \SI{91}{GeV} centre-of-mass energy and $\theta$ = 80\degrees{}. Results are shown with and without removing (`skimming') of secondary vertices produced in the interaction with the material of the vertex detector layers.}
\label{fig:tagging_skim}
\end{figure}

Further improvements are currently being obtained with a newly developed strategy that consists in removing from the BDT training the secondary vertices produced in the interaction with the material of the vertex detector layers. Results show that the tagging in central jets is more affected by the presence of such secondary vertices, and the b-tagging in particular is more affected than the c-tagging, since such vertices mimic the beauty quark decay cascade. The gain in removing those additional secondary vertices is particularly high for di-jet events at the lowest centre-of-mass energy. As an example,  Figure~\ref{fig:tagging_skim} shows the case where the largest improvement was found. These results were obtained using truth tracking only. 

\clearpage

\section{Summary and Outlook}
\label{sec:summary}

A detector concept for FCC-ee, CLD,  has been developed, based on the design of the CLIC detector model (CLICdet)
and adapted for the experimental conditions and physics requirements of FCC-ee.
The detector features an all-silicon vertex and tracking system, followed by a silicon-tungsten ECAL and a scintillator-steel HCAL. The detailed layout of the CLD concept, as implemented in the simulation model, has been presented.
Full simulation studies of CLD show promising performance results.

One notable difference between CLICdet and CLD is the absence of power pulsing for the latter. Future steps in the design of CLD will have to include detailed studies of the cooling, cabling and supports of all sub-detectors, not least the calorimeters. This implies detector optimisation by simulation studies, but also a considerable engineering effort.

Additional engineering effort will have to be devoted to the integration of the detector parts, together with the elements of
the machine-detector interface (solenoids, quadrupoles, LumiCal, vacuum system, etc.). Detector installation, opening and maintenance scenarios also need to be studied.

Options for changes in a next version of the CLD design have been described: a detector layout with a smaller beam pipe radius, an option with a smaller tracker radius and an ECAL with less layers in longitudinal segmentation. Once more engineering details on the machine-detector interface are available, an improved forward coverage in the electromagnetic calorimetry can also be envisaged.

\section*{Acknowledgements}

This work benefited from services provided by the ILC Virtual Organisation, supported by the national resource providers of the EGI Federation. This research was done using resources provided by the Open Science Grid, which is supported by the National Science Foundation and the U.S. Department of Energy's Office of Science.

\begin{appendices}

\clearpage
\section{Overview of CLD sub-detector parameters and read-out channel numbers}
\label{sec:Appendix_I}

In the following, sub-detector parameters for all CLD elements are given, together with the expected number of read-out channels.
For the vertex and tracker discs, the numbers shown reflect the total (left plus right discs).
Numbers for two scenarios are provided: In the first one, a tracker with strip geometry is assumed - this scenario appears now to be disfavoured in the CLICdet vertex and tracker studies. The second scenario
assumes (elongated) pixel geometry for the tracker, and is closer to the single point resolution assumed in the simulation model of CLD.

 \begin{table}[hbtp]
\caption{\label{tab:areas} Overview of CLD sub-detector parameters - a tracker with \textbf{strip geometry} is assumed.}
\vspace{5mm}
\centering
\begin{tabular}{l c c l l}
    \toprule
    \textbf{Sub-detector} & \textbf{Sensor area [m$^2$]} & \textbf{Cell size [mm$^2$]}  & \textbf{Number of} \\
 & & \textbf& \textbf{channels [10$^6$]} \\
    \midrule
VTX barrel & 0.358 & 0.025$\times$0.025  &    570 \\    
VTX petal discs & 0.172 & 0.025$\times$0.025 & 270 \\
\midrule
Inner Tracker Discs ITD1 &1.27 & 0.025$\times$0.025 & 2032\\
Inner Tracker Discs ITD2 &  2.66 &   0.05$\times$1 &54 \\
Inner Tracker Discs ITD3 &  2.59 &   0.05$\times$1 & 53\\
Inner Tracker Discs ITD4 &  2.47 &   0.05$\times$1 & 50\\
Inner Tracker Discs ITD5 &  2.32 &   0.05$\times$1 &47 \\
Inner Tracker Discs ITD6 &  2.03 &   0.05$\times$1 &41 \\
Inner Tracker Discs ITD7 &  1.95 &   0.05$\times$1 & 40\\
Outer Tracker Discs OTD1& 23.94 &  0.05$\times$10& 49\\
Outer Tracker Discs OTD2& 23.94 &  0.05$\times$10&49 \\
Outer Tracker Discs OTD3&23.94 &  0.05$\times$10&49 \\
Outer Tracker Discs OTD4& 23.94 &  0.05$\times$10& 49\\
\midrule
Inner Tracker Barrel ITB1 & 0.77 & 0.05$\times$1 &16   \\
Inner Tracker Barrel ITB2 & 2.42 & 0.05$\times$1 &  48 \\
Inner Tracker Barrel ITB3 & 5.83 & 0.05$\times$5 &  24 \\
Outer Tracker Barrel OTB1& 15.88  & 0.05$\times$10   & 32 \\
Outer Tracker Barrel OTB2& 24.91  & 0.05$\times$10   & 50 \\
Outer Tracker Barrel OTB3& 33.93  & 0.05$\times$10   &  68\\
\midrule
ECAL barrel & 2498 &  5$\times$5 &99 \\
ECAL endcaps (including ECAL plugs) & 1486 & 5$\times$5  & 59 \\
\midrule
HCAL barrel &   3629 &  30$\times$30 &4.0 \\
HCAL endcaps (including HCAL rings) & 4750  & 30$\times$30  &5.2 \\
\midrule
MUON barrel &1916 & 30$\times$30& 2.1\\
MUON endcaps &1351 & 30$\times$30& 1.5\\
\midrule
Total & & & 3762\\
    \bottomrule
\end{tabular}
\label{table.overall}
\end{table}

 \begin{table}[hbtp]
\caption{\label{tab:areas} Overview of CLD sub-detector parameters - a tracker with \textbf{pixel geometry} is assumed.}
\vspace{5mm}
\centering
\begin{tabular}{l c c l l}
    \toprule
     \textbf{Sub-detector} & \textbf{Sensor area [m$^2$]} & \textbf{Cell size [mm$^2$]}  & \textbf{Number of} \\
 & & \textbf& \textbf{channels [10$^6$]} \\
    \midrule
VTX barrel & 0.358 & 0.025$\times$0.025  &    570 \\    
VTX petal discs & 0.172 & 0.025$\times$0.025 & 270 \\
\midrule
Inner Tracker Discs ITD1 &1.27 & 0.025$\times$0.025 & 2032\\
Inner Tracker Discs ITD2 &  2.66 &   0.03$\times$0.3 &296 \\
Inner Tracker Discs ITD3 &  2.59 &   0.03$\times$0.3 & 294\\
Inner Tracker Discs ITD4 &  2.47 &  0.03$\times$0.3 & 271\\
Inner Tracker Discs ITD5 &  2.32 &   0.03$\times$0.3 &256 \\
Inner Tracker Discs ITD6 &  2.03 &   0.03$\times$0.3 &224\\
Inner Tracker Discs ITD7 &  1.95 &   0.03$\times$0.3 & 219\\
Outer Tracker Discs OTD1& 23.94 &  0.03$\times$0.3& 2725\\
Outer Tracker Discs OTD2& 23.94 &  0.03$\times$0.3&2725 \\
Outer Tracker Discs OTD3&23.94 &  0.03$\times$0.3&2725 \\
Outer Tracker Discs OTD4& 23.94 & 0.03$\times$0.3& 2725\\
\midrule
Inner Tracker Barrel ITB1 & 0.77 & 0.03$\times$0.3 &88   \\
Inner Tracker Barrel ITB2 & 2.42 & 0.03$\times$0.3 &  266 \\
Inner Tracker Barrel ITB3 & 5.83 & 0.03$\times$0.3&  663\\
Outer Tracker Barrel OTB1& 15.88  & 0.03$\times$0.3   & 1753 \\
Outer Tracker Barrel OTB2& 24.91  &0.03$\times$0.3 & 2754 \\
Outer Tracker Barrel OTB3& 33.93  & 0.03$\times$0.3  &  3783\\
\midrule
ECAL barrel & 2498 &  5$\times$5 &99 \\
ECAL endcaps (including ECAL plugs) & 1486 & 5$\times$5  & 59 \\
\midrule
HCAL barrel &   3629 &  30$\times$30 &4.0 \\
HCAL endcaps (including HCAL rings) & 4750  & 30$\times$30  &5.2 \\
\midrule
MUON barrel &1916 & 30$\times$30& 2.1\\
MUON endcaps &1351 & 30$\times$30& 1.5\\
\midrule
Total & & & 24810\\
    \bottomrule
\end{tabular}
\label{table.overall}
\end{table}

\clearpage
\section{A fast simulation study for a potentially smaller CLD tracker}
\label{sec:Appendix_II}

The CLD detector has been devised by rescaling the baseline detector CLICdet designed for CLIC~\cite{CLICdet_note_2017}.
The size of the tracking system in CLICdet was chosen in order to be able to reconstruct high-momentum tracks in the central barrel with a transverse 
momentum resolution $\sigma(\Delta p_\mathrm{{T}}/p_\mathrm{{T}}^2)$ of the order of ~2$\cdot$10$^{-5}$ GeV$^{-1}$  at momenta above 100~GeV, as required by its physics programme goals~\cite{CLICdet_performance}.
To be able to achieve the same \pT{} resolution for FCC-ee, which features a magnetic field of 2 T (half the magnitude of the CLICdet magnetic field), the outermost layer of the tracker has been moved out from a radius of 1.5 m to a radius of 2.1 m.
However, the requirements on the transverse momentum resolution for a detector at FCC-ee are less stringent than for CLICdet.
For measuring the point-to-point energy error in Z boson width measurements in the scan at the Z-pole run, an accuracy of 50-200 MeV for a muon with 45 GeV momentum would be necessary~\cite{ABlondel}. At normal incidence, 200 MeV accuracy would correspond to ~10$\cdot$10$^{-5}$ GeV$^{-1}$.
With the default design for CLD, the \pT{} resolution achieved for muons of 45 GeV momentum at normal incidence is ~7$\cdot$10$^{-5}$ GeV$^{-1}$.

A study on the variation of the transverse momentum resolution with different positions of the tracking layers has been performed with the fast simulation tool "Lic Detector Toy"~\cite{Lictoy}.
Four different positions of the outermost tracker layer have been tested, with the innermost layer of the Outer Tracker fixed and the intermediate layer rescaled to be equidistant from innermost and outermost ones. The positions of the Outer Tracker layers for the four configurations are summarised in Table~\ref{tab:trackerConfigs}. The positions of the other layers of the tracking system were left unchanged.

\begin{table}[H]
\caption{Outer tracker layer radii for four configurations tested with fast simulation.}
\vspace{0.6em}
\centering
  \begin{minipage}{.4\textwidth}
  \centering
   \caption*{(a) Default}
     \begin{tabular}{cc}
     \toprule Layer & Radius [mm]\\
     \midrule 
     3 & 2136\\
     2 & 1568\\
     1 & 1000\\
     \bottomrule
   \end{tabular}
  \end{minipage}
  \begin{minipage}{.4\textwidth}
  \centering
    \caption*{(b) Variant 1}
    \begin{tabular}{cc} 
    \toprule Layer & Radius [mm]\\
    \midrule 
     3 & 2000\\
     2 & 1500\\
     1 & 1000\\
     \bottomrule
   \end{tabular}
  \end{minipage}
\vspace{5em}
  \begin{minipage}{.4\textwidth}
  \centering
   \caption*{(c) Variant 2}
    \begin{tabular}{cc} 
    \toprule Layer & Radius [mm]\\
    \midrule 
    3 & 1900\\
    2 & 1450\\
    1 & 1000\\
    \bottomrule
   \end{tabular}
  \end{minipage}
  \begin{minipage}{.4\textwidth}
  \centering
   \caption*{(d) Variant 3}
    \begin{tabular}{cc} 
    \toprule Layer & Radius [mm]\\
    \midrule 
    3 & 1800\\
    2 & 1400\\
    1 & 1000\\
    \bottomrule
   \end{tabular}
\end{minipage}
\label{tab:trackerConfigs}
\end{table}

\begin{figure}[ht] 
  \begin{subfigure}{0.5\linewidth}
    \centering
    \includegraphics[width=\linewidth]{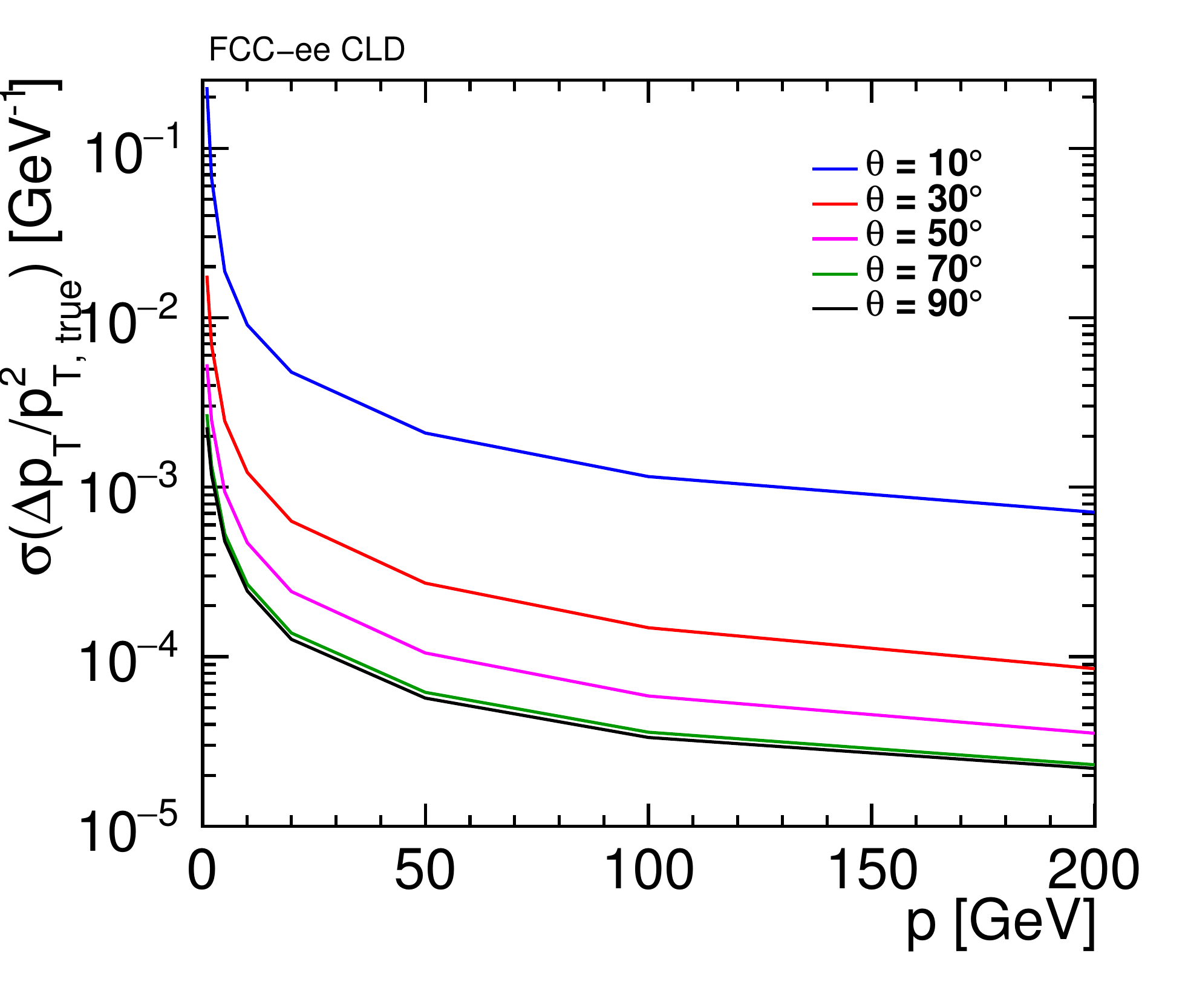} 
    \caption{Default configuration.} 
  \end{subfigure}
  \begin{subfigure}{0.5\linewidth}
    \centering
    \includegraphics[width=\linewidth]{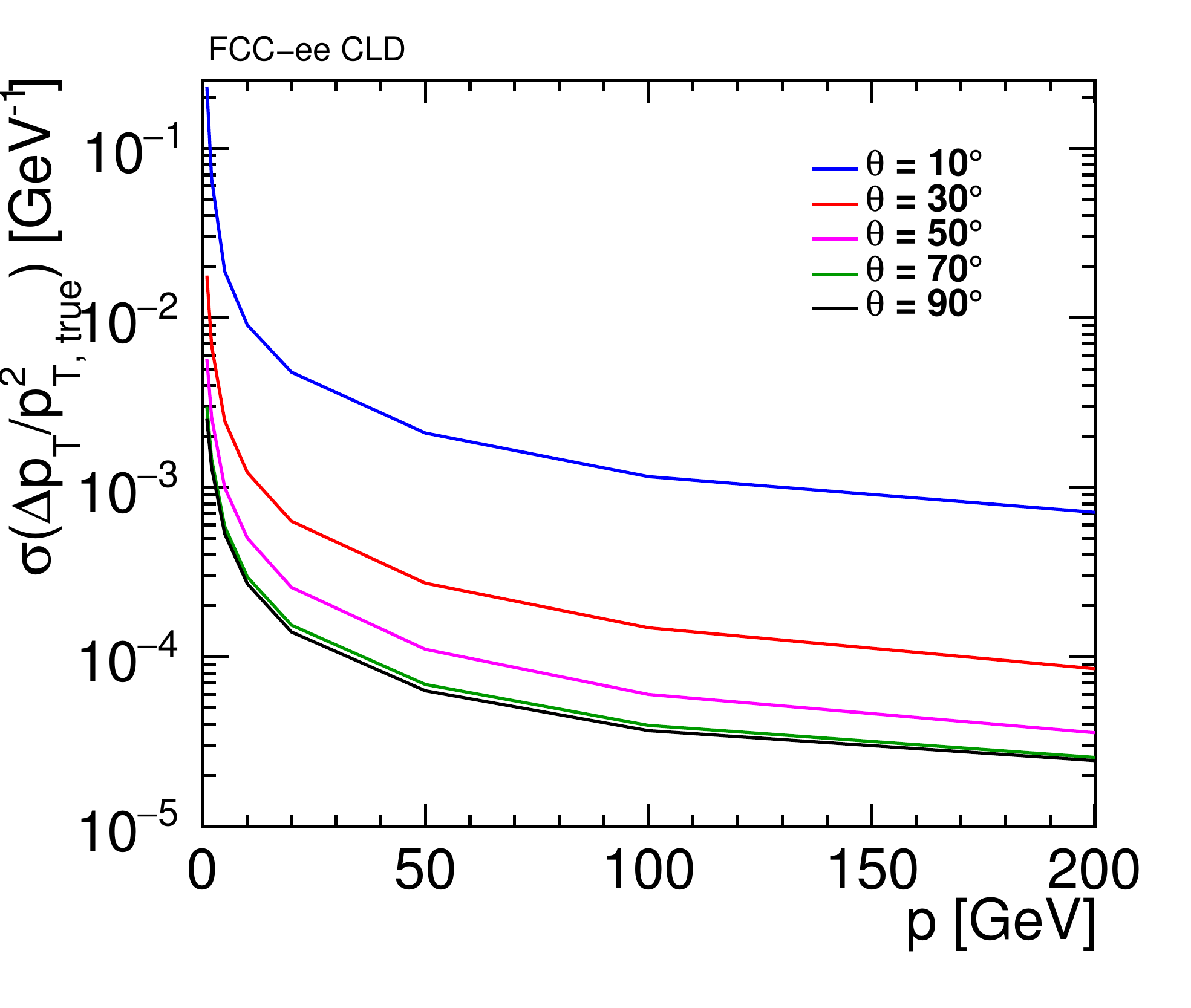} 
    \caption{Variant 1.} 
  \end{subfigure} 
  \begin{subfigure}{0.5\linewidth}
    \centering
    \includegraphics[width=\linewidth]{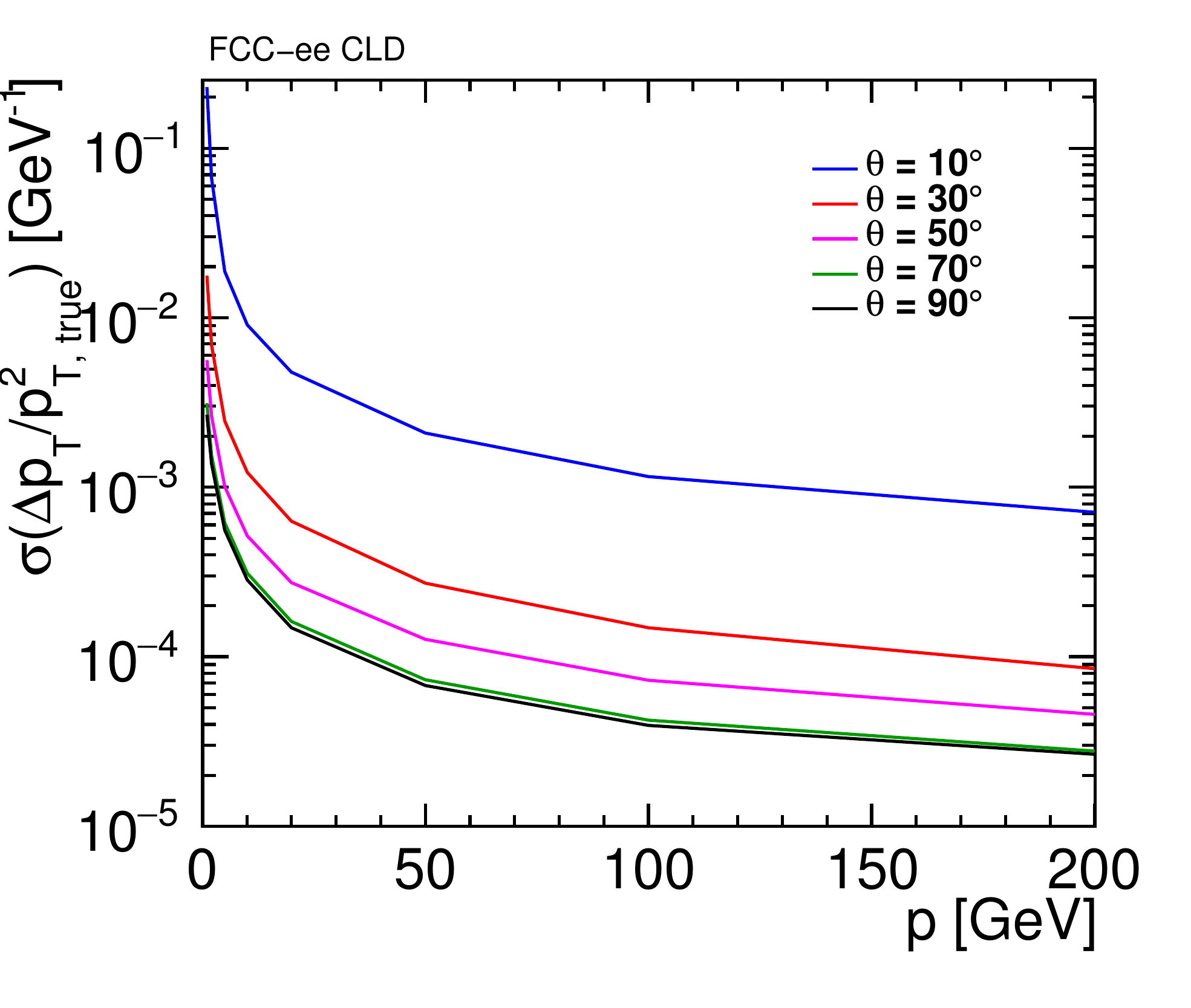} 
    \caption{Variant 2.} 
  \end{subfigure}
  \begin{subfigure}{0.5\linewidth}
    \centering
    \includegraphics[width=\linewidth]{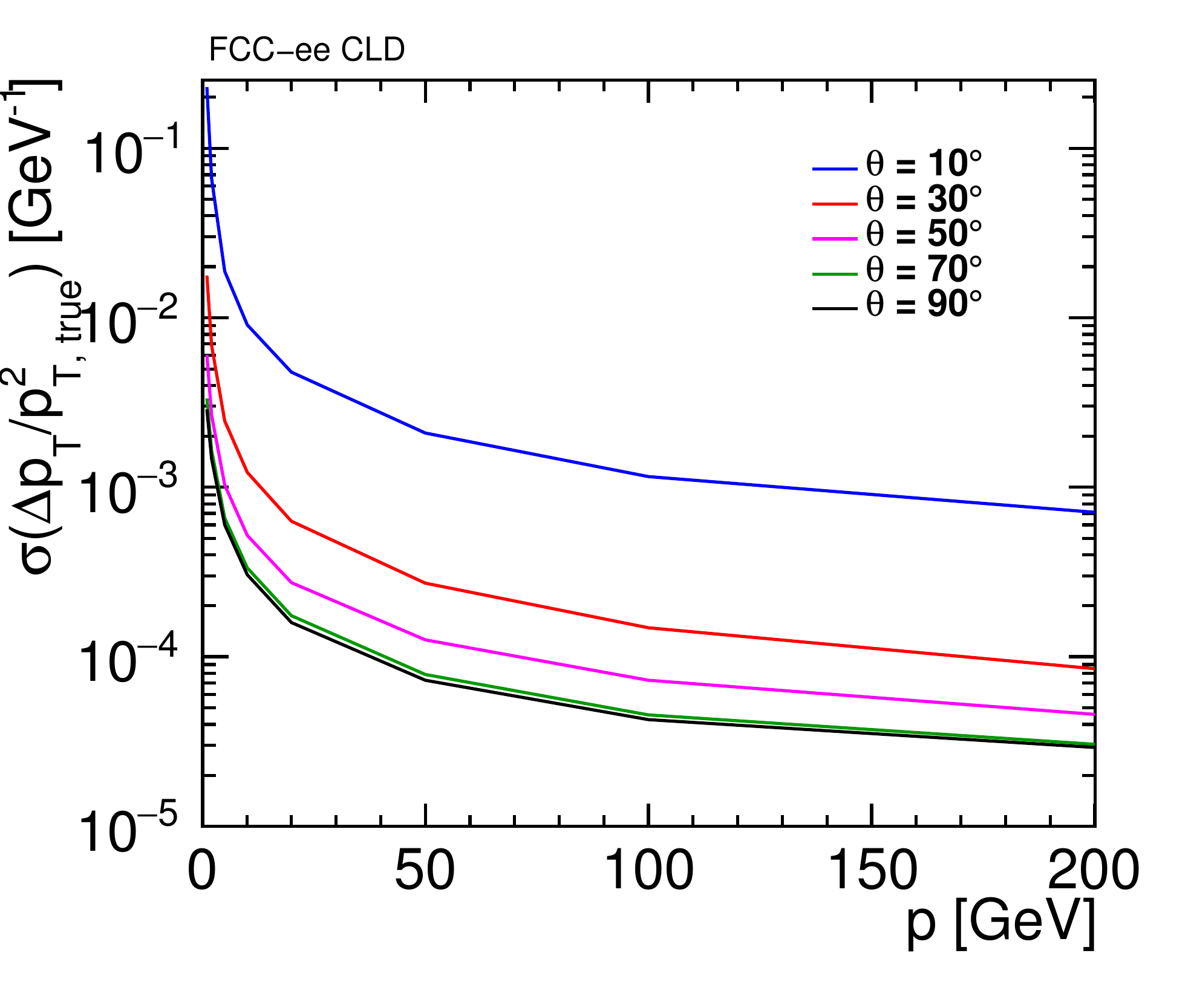} 
    \caption{Variant 3.} 
  \end{subfigure} 
  \caption{Transverse momentum resolution for single muons at polar angles $\theta$ = 10\degrees{}, 30\degrees{}, 50\degrees{}, 70\degrees{} and 90\degrees{} with four tracker configurations: default (a), outermost radius = 2.0 m (b), 1.9 m (c), 1.8 m (d).}
  \label{fig:ptresConfigs} 
\end{figure}

The results are shown in Figure~\ref{fig:ptresConfigs} and the resolutions obtained for 45 GeV muons at 90\degrees{} are summarized in~\cref{tab:res45GeV}. They suggest that a tracker with reduced outermost radius, down to 1.8 m, would provide a \pT{} resolution of ~8$\cdot$10$^{-5}$ GeV$^{-1}$ for muons with 45 GeV momentum at normal incidence. Thus, such a smaller tracker would be a valid option for a forthcoming next version of the CLD design. Consequenly, the innermost radius of the ECal could be reduced, therefore reducing drastically the cost of the most expensive sub-detector.
 
\begin{table}[!hbt]
  \begin{center}
    \caption{ Transverse momentum resolutions obtained for 45 GeV muons in the 4 detector variants studied.  }
      \centering
         \begin{tabular}{|c|c|}
             \hline
             Default & 7$\cdot$10$^{-5}$ GeV$^{-1}$ \\
             Variant 1 & 7.3$\cdot$10$^{-5}$ GeV$^{-1}$ \\
             Variant 2 & 7.75$\cdot$10$^{-5}$ GeV$^{-1}$ \\
             Variant 3 & 8$\cdot$10$^{-5}$ GeV$^{-1}$ \\
             \hline
         \end{tabular}
\label{tab:res45GeV}
\end{center}
\end{table}

\clearpage
\section{Optimization of the ECAL for CLD}
\label{sec:Appendix_III}

As described in Subsection~\ref{ecal}, the ECAL consists of 40 layers which allows reaching an excellent photon resolution. However, such a big number of layers makes the ECAL the most expensive system of CLD. In this appendix, a study on ECAL layout optimisation is presented.

Four options of the ECAL with  different numbers of layers are considered. Three options consist of 40, 30 or 20 uniformly distributed layers. The last option consists of a combination of 20 thin and 10 thick layers, similarly to the ILD detector concept. Each of the options has a different thickness of the absorber layers (tungsten plates) in order to keep the same total ECAL thickness of $\approx$ 22 X$_0$. A detailed description of each option is presented in Table~\ref{table:ecal_options}. All other parameters of the ECAL segmentation remain unchanged (cf. Table~\ref{tab:ECAL}).

\begin{table}[hbtp]
\caption{ECAL configurations considered for the optimisation study.}
\vspace{5mm}
\centering
\begin{tabular}{lccc}
\toprule
Layer structure & Thickness         & Total thickness \\
                & tungsten alloy    &  per layer \\
                & [mm]              &  [mm] \\[0.1cm]
\midrule
40 uniform          &  1.9          & 5.05 \\[0.1cm]
30 uniform          &  2.62         & 5.77 \\[0.1cm]
20 uniform          &  3.15         & 7.19 \\[0.1cm]
20 thin + 10 thick  & 1.9 + 3.8     & 5.05 + 6.95  \\
\bottomrule
\end{tabular}%
\label{table:ecal_options}
\end{table}

\begin{figure}[htbp]
  \centering
  \begin{subfigure}{.5\textwidth}
    \centering
      \includegraphics[width=\linewidth]{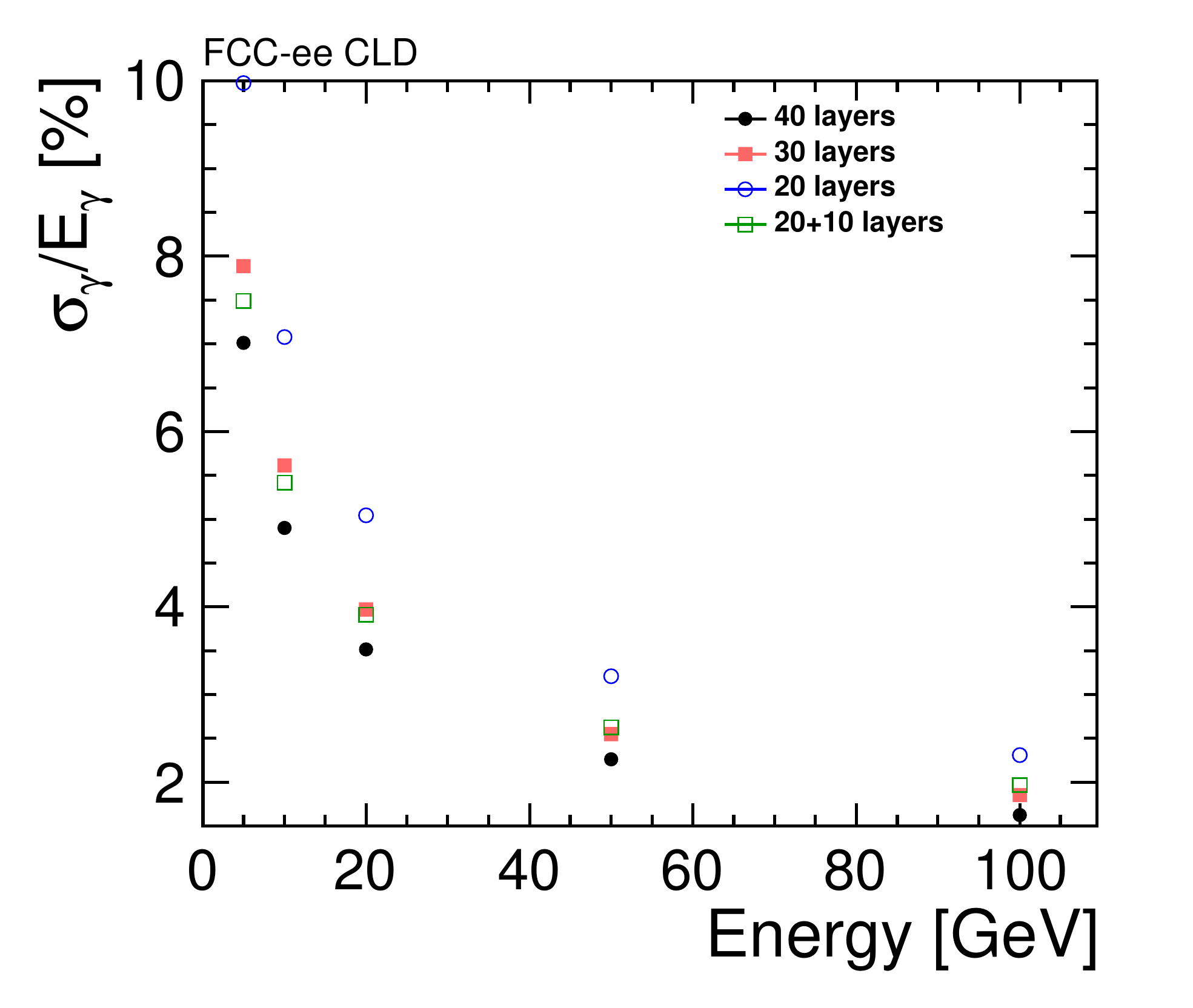}
    \caption{}
  \end{subfigure}%
  \begin{subfigure}{.5\textwidth}
    \centering
      \includegraphics[width=\linewidth]{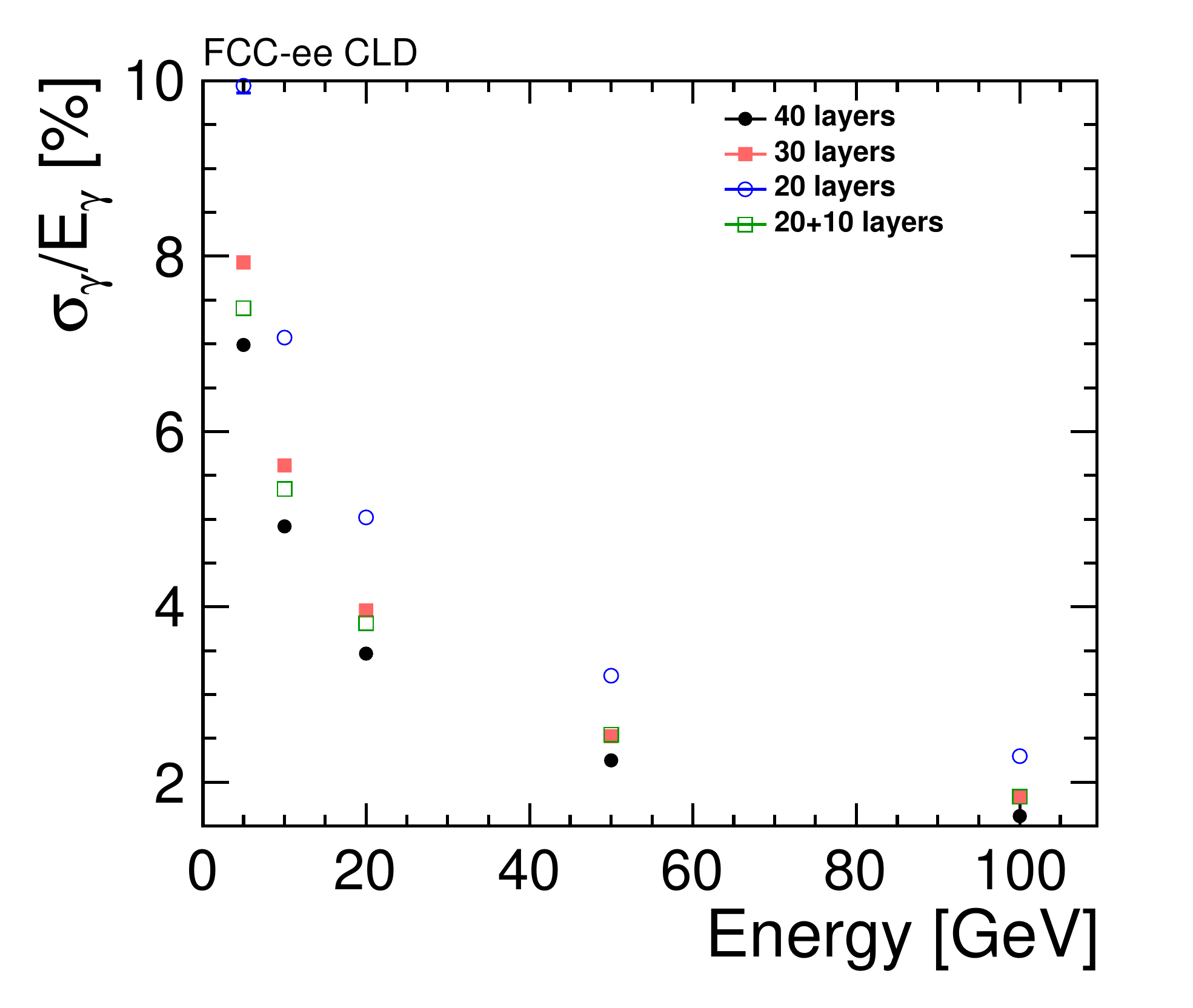}
    \caption{}
  \end{subfigure}
  \caption{Photon energy resolution as a function of energy in the (a) barrel and (b) endcap regions. Results are shown for different ECAL layer configurations.}
  \label{fig:photonResolutionVsEnergyVsNlayers}
\end{figure}

To study the photon energy resolution, the calorimeter calibration has been performed for each detector option. Energy resolutions have been studied for photons with energies of 5, 10, 20, 50 and 100 GeV and are shown in Figure~\ref{fig:photonResolutionVsEnergyVsNlayers}. As expected, the best photon resolution at the whole energy range is achieved with the 40 layer configuration, while the worst resolution is achieved with the 20 layer configuration. For the 20+10 layer mixed configuration the resolution is better than the 30 layer option for low photon energies, while it is worse for higher energies.

The ECAL configuration with only 20 layers results in a significant degradation of the photon energy resolution, while the 20+10 and the 30 layer options could be good alternatives to the default 40 layers - both would reduce the cost of the detector significantly, while not compromising too much on the resolution.

Additionally, the jet energy resolution with \PZgstarToqq events of 91.2 and 365 GeV masses have been studied with the four ECAL configurations. The resulting resolutions are show in Table~\ref{table:jer_vs_ecalLayers}. Configurations with 30 and 20+10 layers provide comparable resolutions to the base-line configuration with 40 layers. A slight degradation in jet energy resolution can be observed for the 20 layer option.

\begin{table}[hbtp]
    \caption{Jet energy resolution for central jets (|cos($\theta$)|<0.6) for different ECAL configurations.}
\vspace{5mm}
\centering
\begin{tabular}{lccc}
\toprule
    Layer structure & JER [$\%$]          & JER [$\%$] \\
                    & $\sqrt{s}=$ 365 GeV & $\sqrt{s}=$ 91.2 GeV \\[0.1cm]
\midrule
40 uniform          & 3.62 $\pm$ 0.05 & 4.52 $\pm$ 0.06 \\[0.1cm]
30 uniform          & 3.72 $\pm$ 0.05 & 4.45 $\pm$ 0.06 \\[0.1cm]
20 uniform          & 3.78 $\pm$ 0.05 & 4.82 $\pm$ 0.07 \\[0.1cm]
20 thin + 10 thick  & 3.67 $\pm$ 0.05 & 4.56 $\pm$ 0.06 \\
\bottomrule
\end{tabular}
\label{table:jer_vs_ecalLayers}
\end{table}

\clearpage
\section{Comparison of RMS90 and double-sided Crystal Ball fit methods}
\label{sec:Appendix_IV}

In this appendix, a comparison of two methods of extracting the jet energy resolution is presented. 
The first method, referred to as $\mathrm{RMS}_{90}$ and used in the linear collider community,
uses the RMS in the smallest range of the jet energy response distribution which contains 90\% of the events. This method is used, unless indicated, throughout this note.
The second method, inspired by what is done in CMS calorimetry studies, is based on  fitting of the jet energy response and extracting the energy resolution from the fit.
As shown in this appendix, both methods lead to similar results.

\begin{figure}[htbp]
  \centering
  \begin{subfigure}{.45\textwidth}
    \centering
    \includegraphics[width=\linewidth]{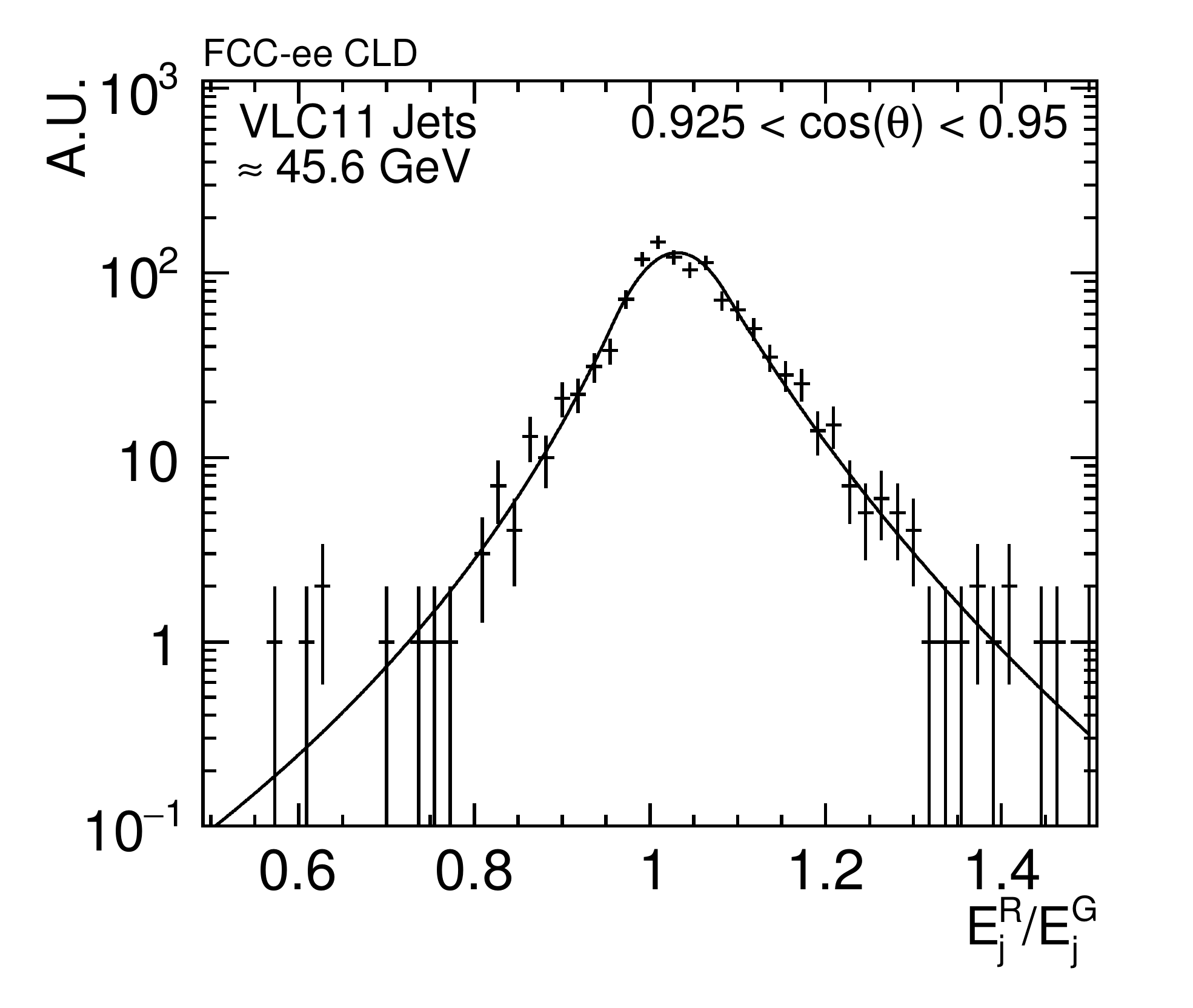}
    \caption{}
    \label{subfig:JER_RMS90_vs_CBfit_91_2}
  \end{subfigure}
  \begin{subfigure}{.45\textwidth}
    \centering
    \includegraphics[width=\linewidth]{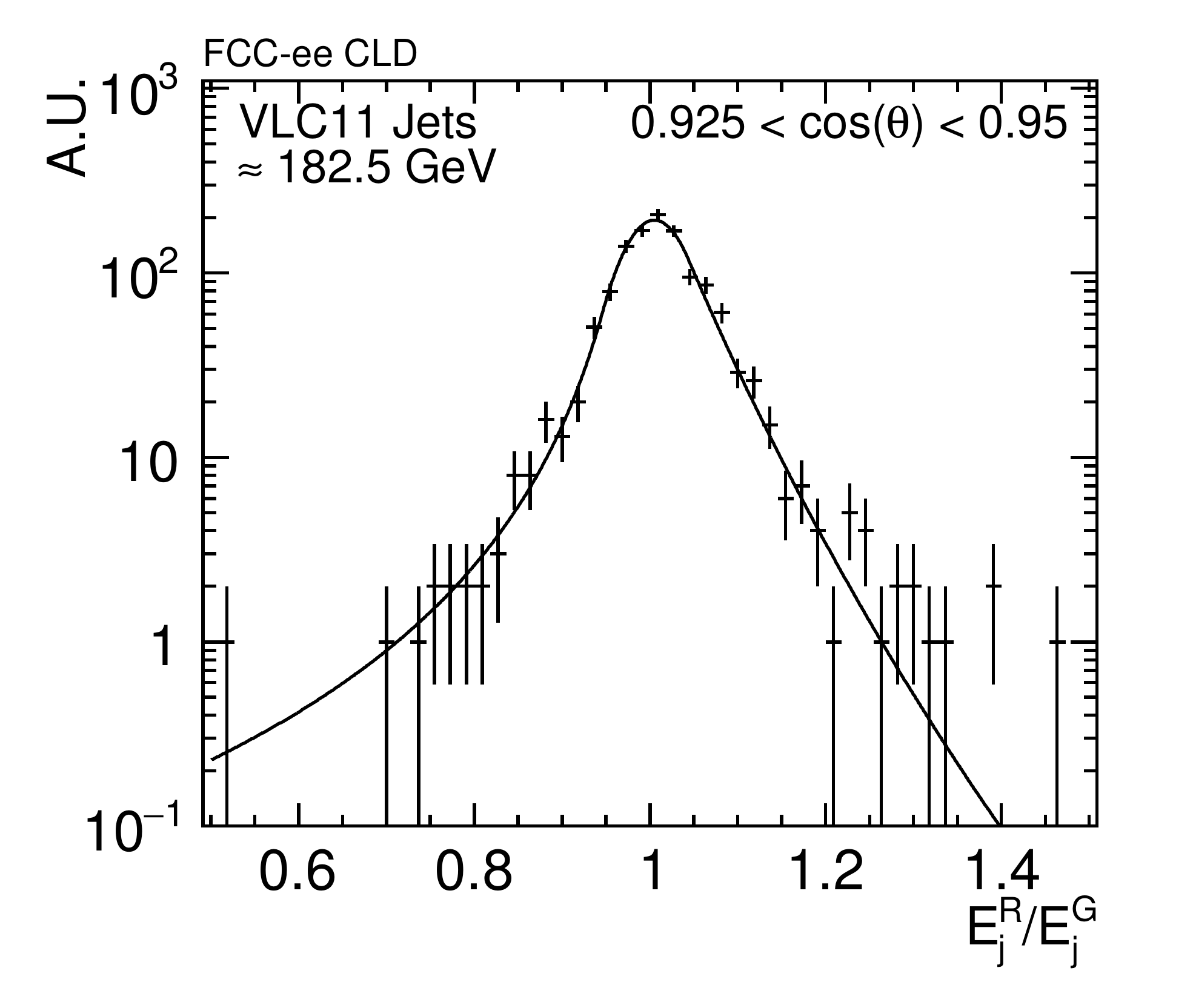}
    \caption{}
    \label{subfig:JER_RMS90_vs_CBfit_365}
  \end{subfigure}
  \caption{Jet energy response of forwards jets (0.925 $< |\cos(\theta)| <$ 0.95) with incoherent pair background overlaid 
on the physics di-jet events (a) at 91.2 GeV and (b) at 365 GeV centre-of-mass energies. The double-sided Crystal Ball fit is shown as black solid line. 
  }
  \label{fig:JER_fitting_91GeV_365GeV}
\end{figure}

As described in Subsection~\ref{jet_energy_res}, the jet energy response is defined as $E_j^R/E_j^G$, where $E_j^R$ is the measured energy 
of a reconstructed jet and $E_j^G$ is the energy of the MC truth particle jet. In this study, VLC jet clustering is used.
To fit these distributions a set of different functions has been studied: 
a Gaussian with 2$\sigma$ and 3$\sigma$ fit ranges, a one-sided Crystal Ball and a double-sided Crystal Ball 
function~\cite{Oreglia:1980cs}, using the Minuit2 library~\cite{James:1994vla} as implemented in \ROOT~6.08.00~\cite{Antcheva:2009zz}.
Due to the presence of asymmetric non-Gaussian tails in the distributions, the best fit was obtained with a double-sided Crystal Ball function, 
which consists of a Gaussian core and two exponential tails.
An example of the fit to the jet energy response of forward jets (0.925 $< |\cos(\theta)| <$ 0.95) with beam-induced background (incoherent pairs) overlaid 
on the physics di-jet event at 91.2 GeV and 365 GeV centre-of-mass energies is shown in Figure~\ref{fig:JER_fitting_91GeV_365GeV}.

Fitting is performed by the following procedure. First, the distribution is fitted with a Gaussian function over the full range. 
Second, an iterative Gaussian fit is done in the range of $\pm$2$\sigma$ until the value of $\sigma$ stabilises within $\pm$5$\%$. 
The fitted parameters of the Gaussian are used as initial parameters for the double-sided Crystal Ball function. 
Finally, an iterative fit with the double-sided Crystal Ball function is done until the $\sigma_{CB}$ parameter stabilises within $\pm$5$\%$. 
The jet energy resolution is defined as the $\sigma$ parameter of the Gaussian core of the Crystal Ball function.

In order to make sure that a large fraction of the events are described by the Gaussian core (which is used to define the energy resolution) 
rather than by the tails, an additional restriction is introduced. Note that, in the double-sided Crystal ball function, the points of transition
between Gaussian core and exponential tails, on both sides of the Gaussian core, are labelled a$_1$ and a$_2$.  Using the condition
$\mid$a$_{1,2}$$\mid$\,$>$\,1$\sigma_{CB}$ ensures that the tails of the fit will not dominate over the Gaussian core. 

\begin{figure}[htbp]
  \centering
  \begin{subfigure}{.5\textwidth}
    \centering
    \includegraphics[width=\linewidth]{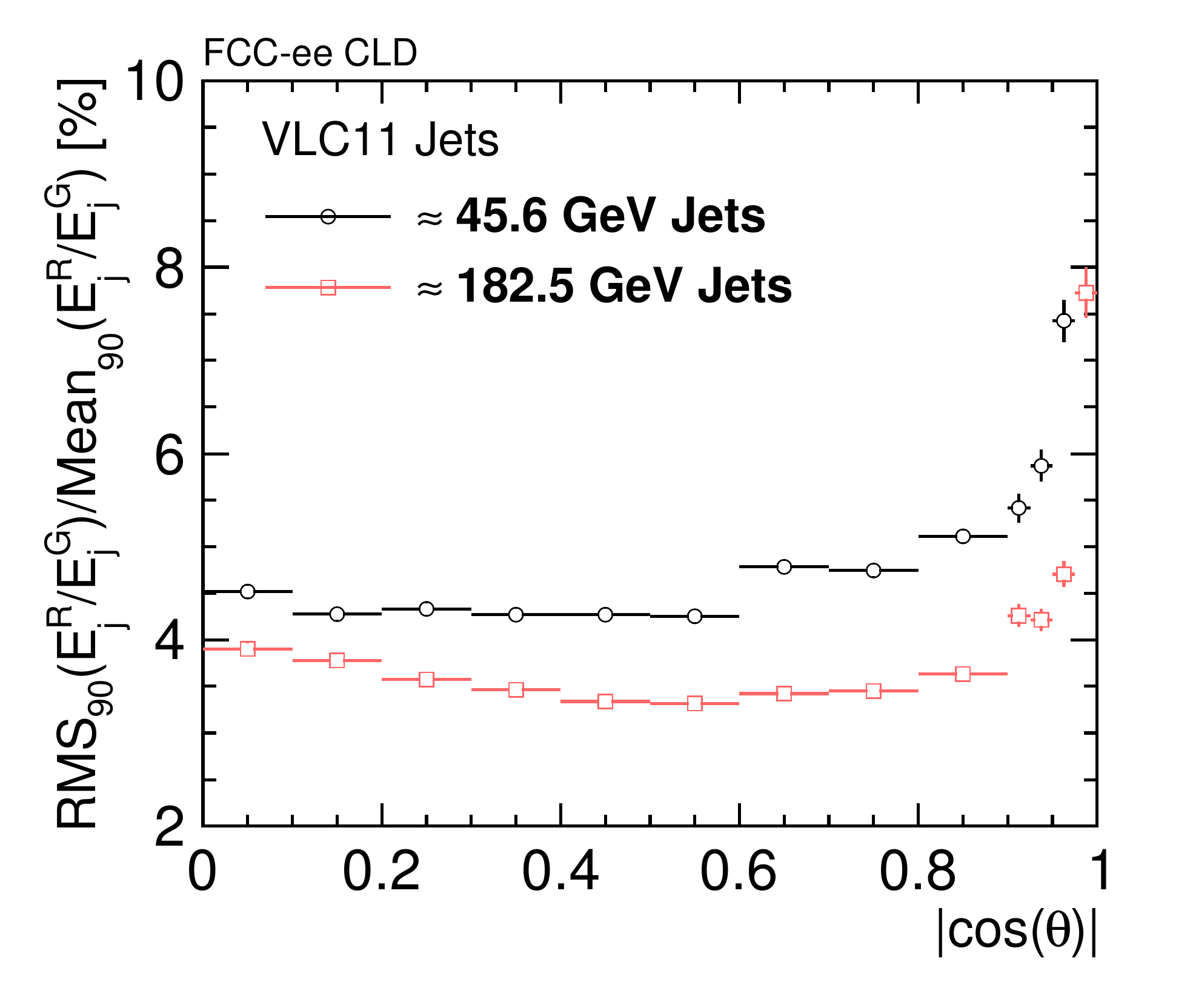}
    \caption{}
    \label{subfig:JER_RMS90_vs_CBfit_91_2}
  \end{subfigure}%
  \begin{subfigure}{.5\textwidth}
    \centering
    \includegraphics[width=\linewidth]{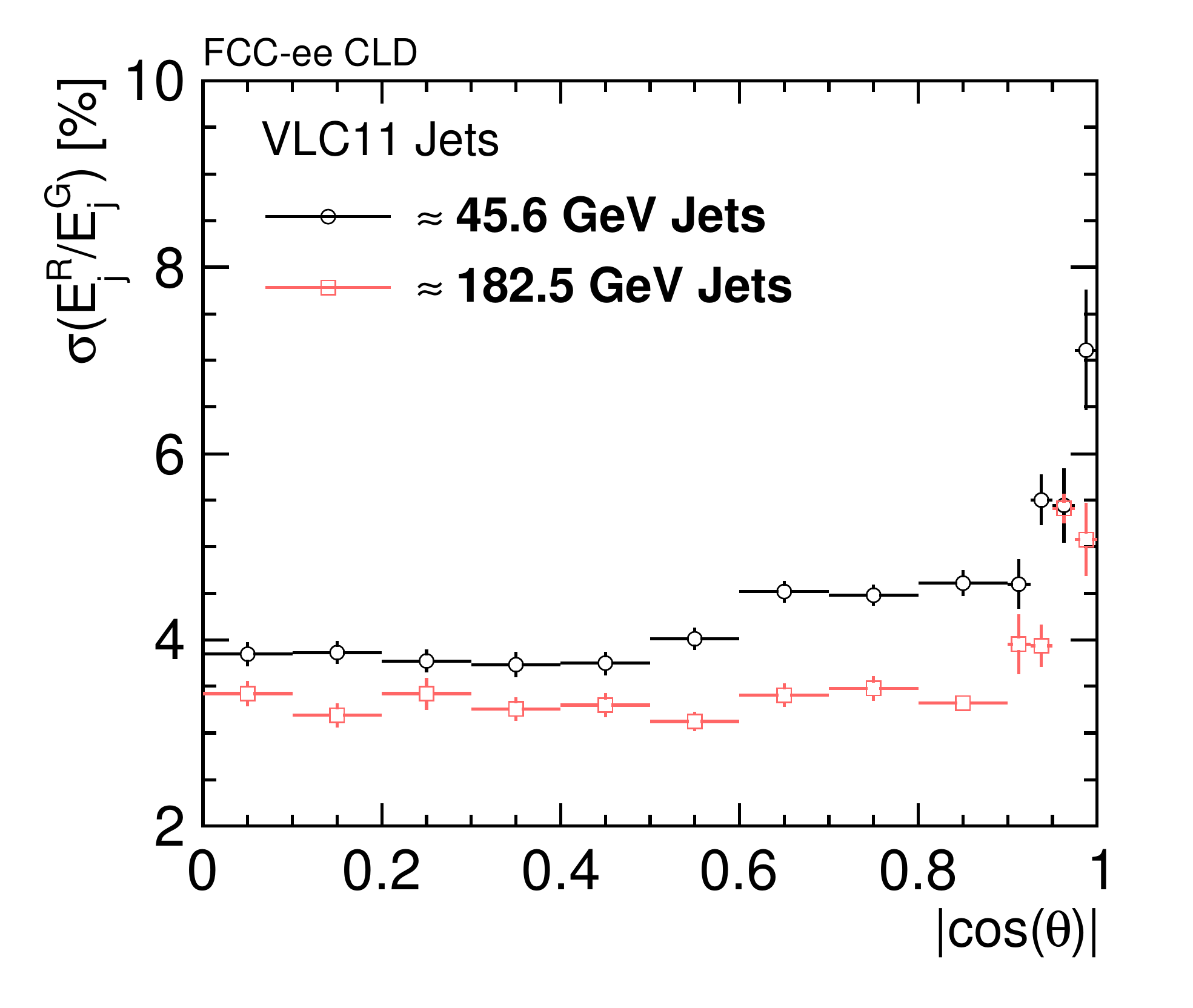}
    \caption{}
    \label{subfig:JER_RMS90_vs_CBfit_365}
  \end{subfigure}
    \caption{Jet energy resolution as function of the quark $|\mathrm{cos}(\theta)|$, with incoherent pair background overlaid on the physics di-jet events, at 91.2 GeV and 365 GeV centre-of-mass energies, (a) using $\mathrm{RMS}_{90}$ and (b) using the double-sided Crystal ball function.}
  \label{fig:JER_RMS90_vs_CBfit}
\end{figure}

A comparison of the jet energy resolution as function of the quark $|\mathrm{cos}(\theta)|$ obtained using $\mathrm{RMS}_{90}$ and the fitting procedure, both at 91.2 GeV 
and 365 GeV centre-of-mass energies, is shown in Figure~\ref{fig:JER_RMS90_vs_CBfit}. 
Overall, the two methods give comparable results, with the $\mathrm{RMS}_{90}$ method yielding slightly more conservative values at low energy. 
This can be understood by the fact that the fraction of events described by the Gaussian core of the double-sided Crystal Ball function in the fitting procedure
 is usually lower than 90$\%$ (as used in the $\mathrm{RMS}_{90}$ method).

\clearpage
\section{Flavour tagging performance for a detector adapted to a smaller beam pipe}
\label{sec:Appendix_V}

Following a proposal for an alternative design of the FCC-ee interaction region with a smaller beam pipe radius at the interaction point, an adapted version of the CLD detector has been implemented in the simulation software.
With the beam pipe radius reduced from \SI{15}{mm} to \SI{10}{mm}, the modified detector features the innermost double layer of the vertex barrel detector at a smaller radius. 
The position of the third double layer is left unchanged, while the second double layer has been placed to be equidistant to the innermost and outermost. 
The new radii are summarised and compared to the default model in Table~\ref{tab:newCLD}. Note that the layout of the vertex detector discs is not changed.

\begin{table}[h!]
\caption{Positions of the double barrel layer of the vertex detector with the default detector model and the modified model with updated beam pipe radius.}
\label{tab:newCLD}
\centering
\begin{tabular}{c c c}
\toprule
Vertex barrel layer & Radius for the default model [mm] & Radius for the new model [mm]\\
\midrule
Layer 1 & 17.5 & 12.5 \\
Layer 2 & 18.5 & 13.5 \\
Layer 3 & 37 & 35 \\
Layer 4 & 38 & 36 \\
Layer 5 & 57 & 57 \\
Layer 6 & 58 & 58\\
\bottomrule
\end{tabular}
\end{table}

\begin{figure}[htp]
\centering
\includegraphics[width=.4\textwidth]{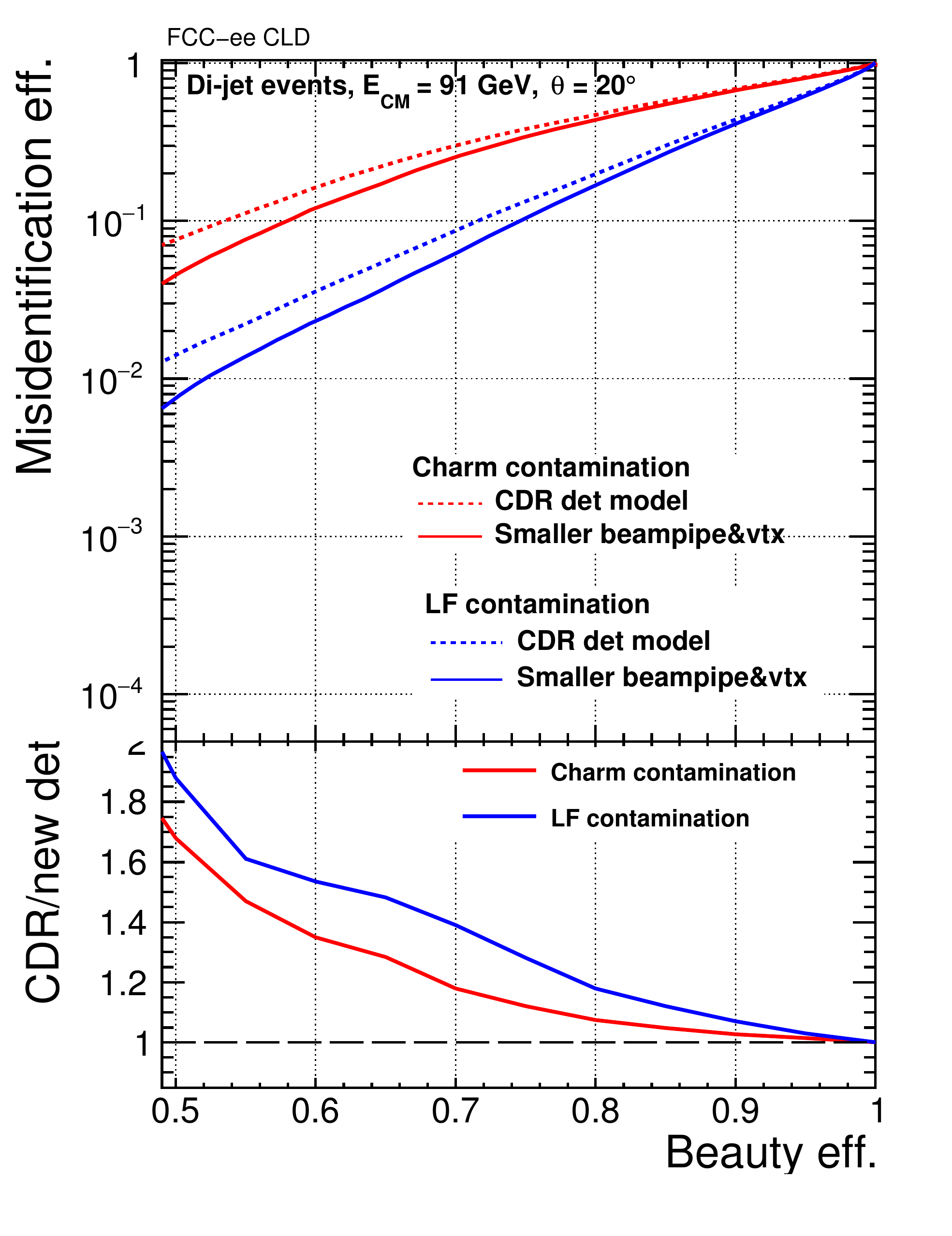}\quad
\includegraphics[width=.4\textwidth]{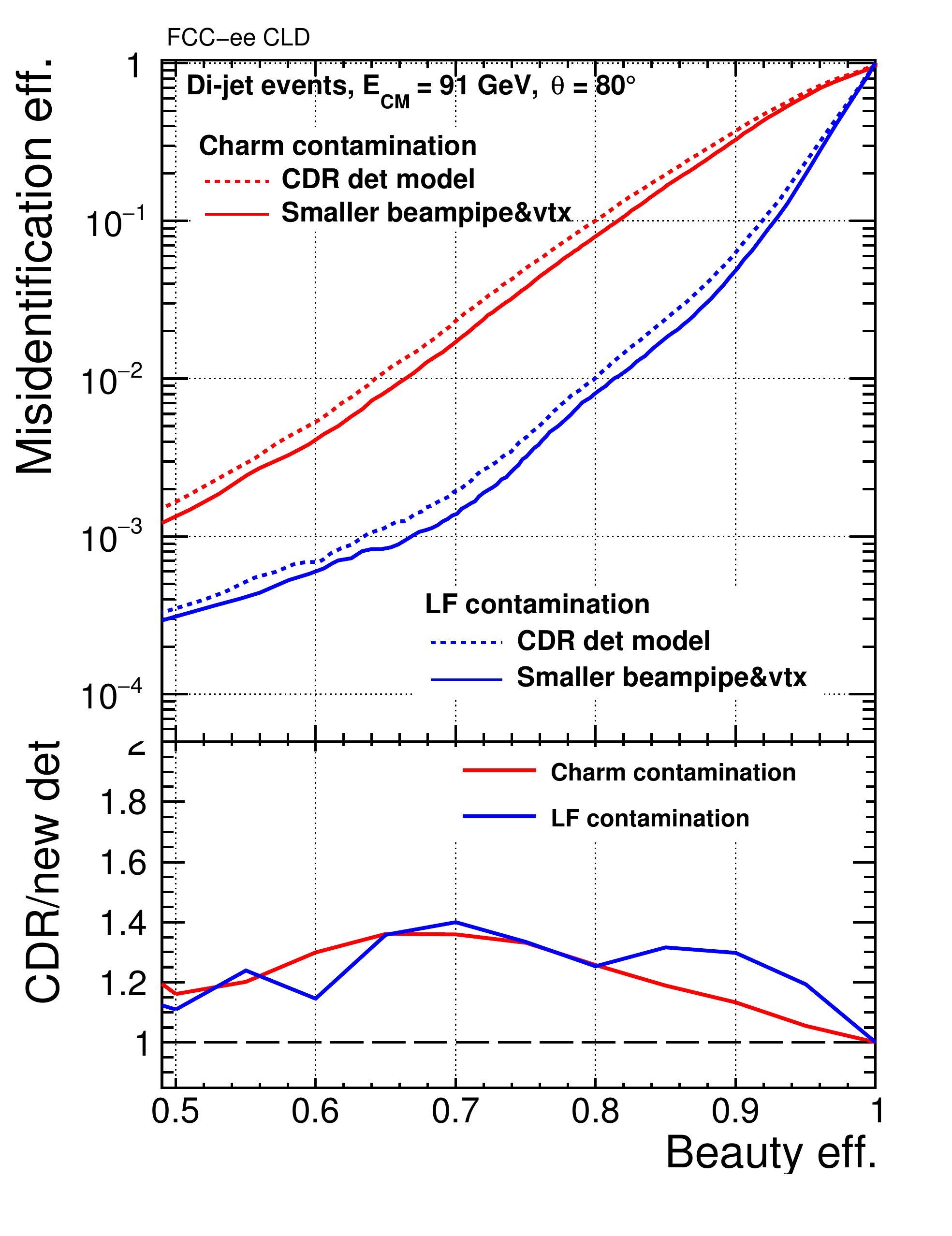}

\medskip
\includegraphics[width=.4\textwidth]{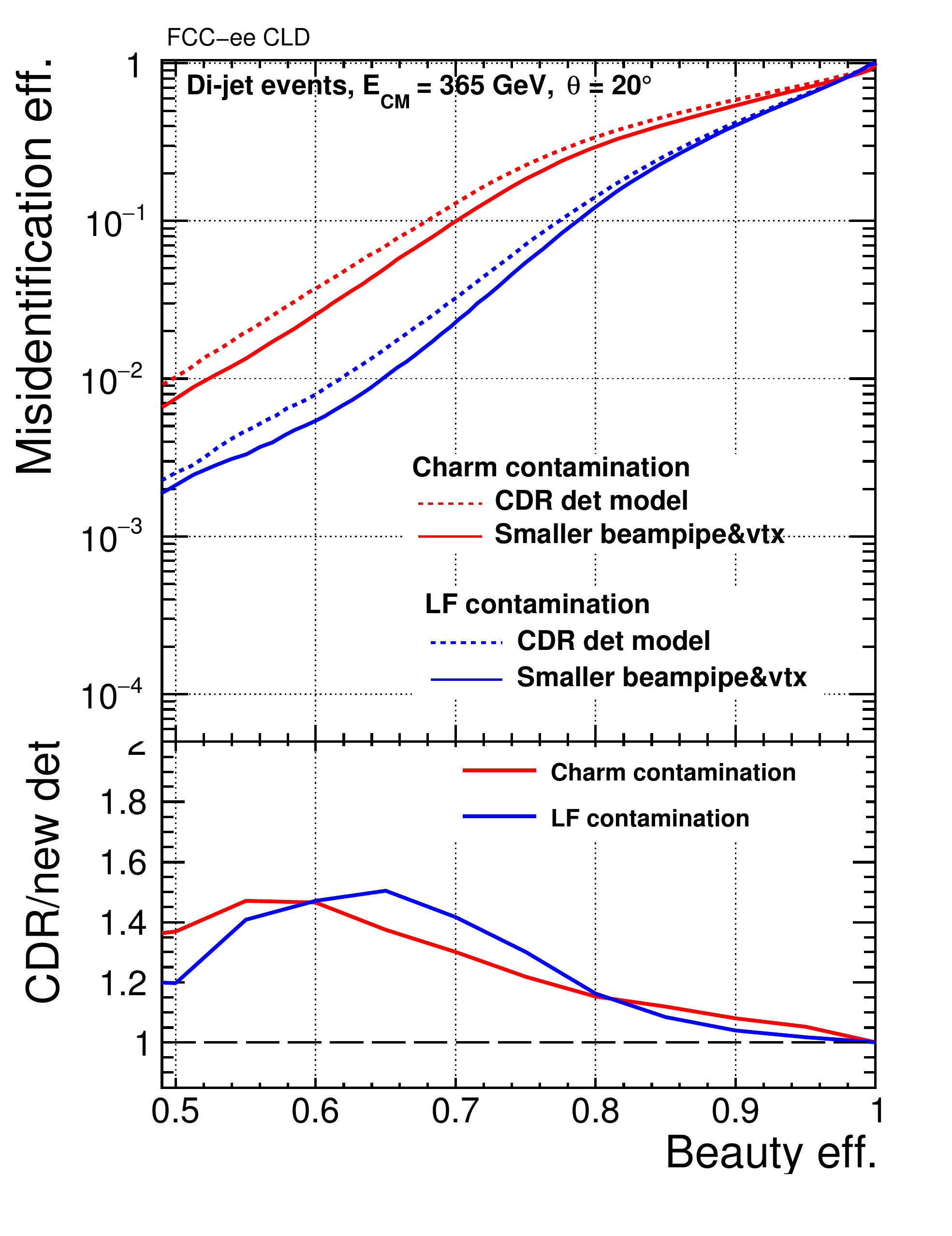}\quad
\includegraphics[width=.4\textwidth]{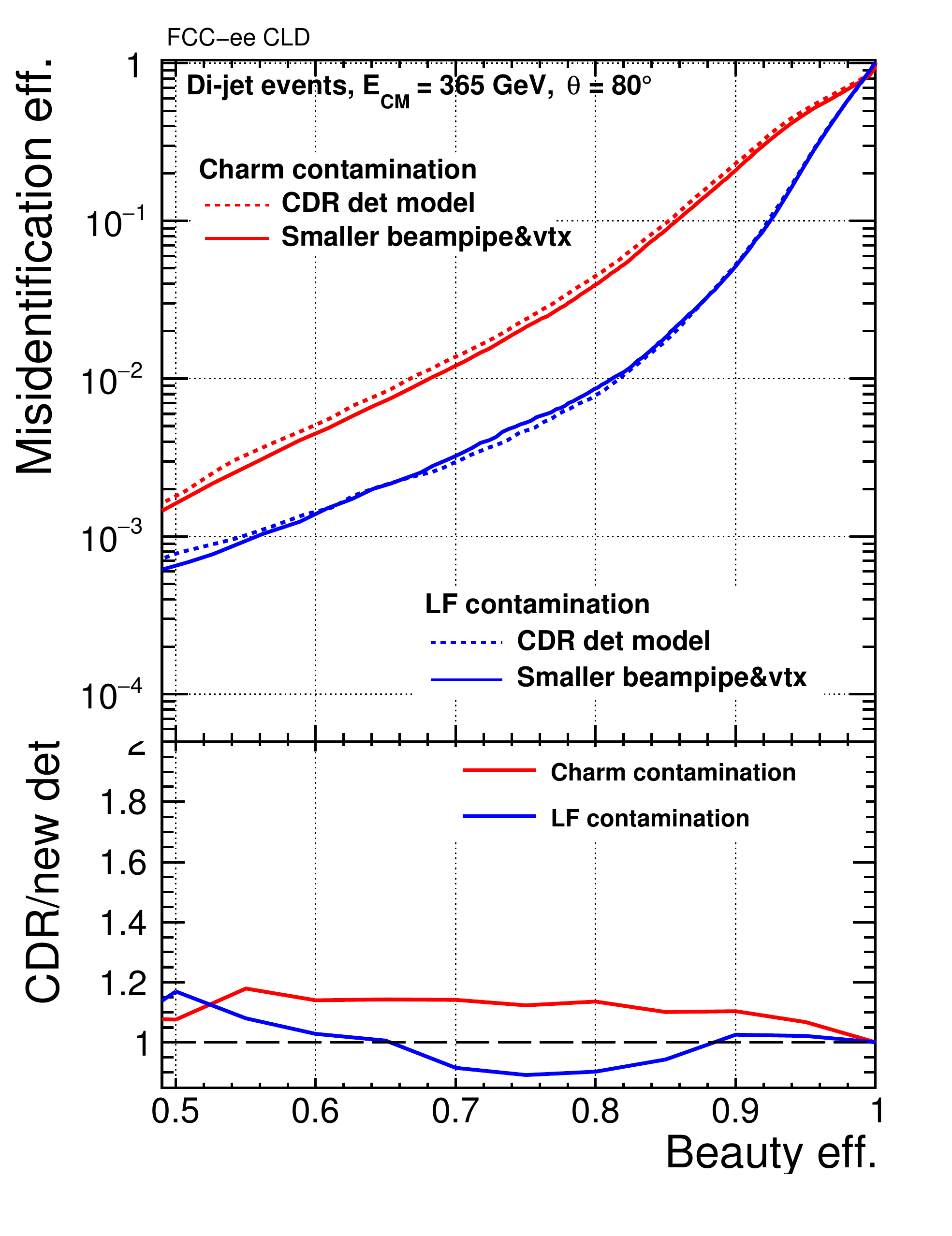}

\caption{B-tagging performance for di-jet events at \SI{91}{GeV} (top) and \SI{365}{GeV} (bottom) centre-of-mass energy and $\theta$ = 20\degrees{}(left) and 80\degrees{}(right). Per contamination source, results obtained with the default and the modified CLD detector model are shown, with their ratio on the bottom canvas. }
  \label{fig:Btag_newCLD}
\end{figure}

\begin{figure}[htp]
\centering
\includegraphics[width=.4\textwidth]{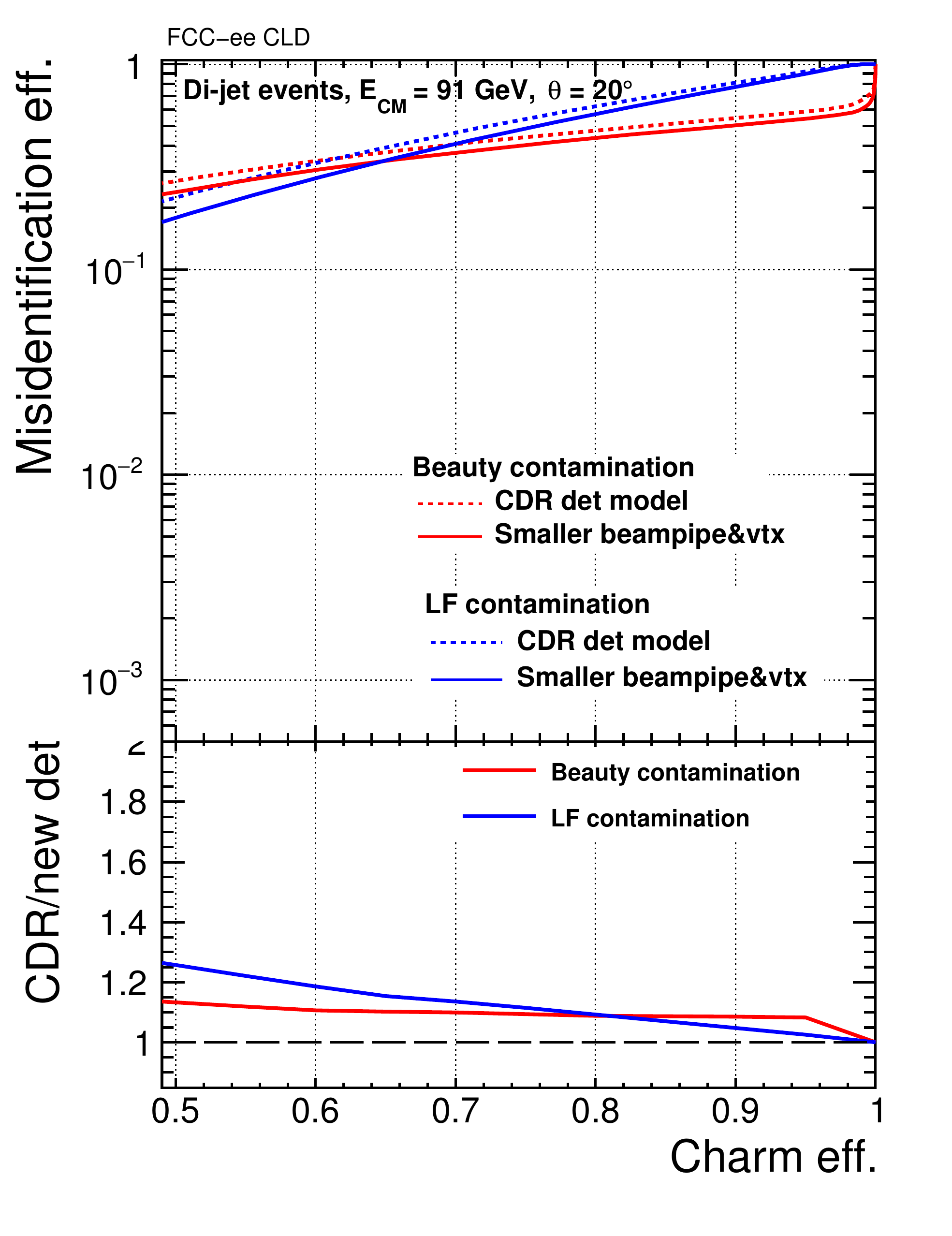}\quad
\includegraphics[width=.4\textwidth]{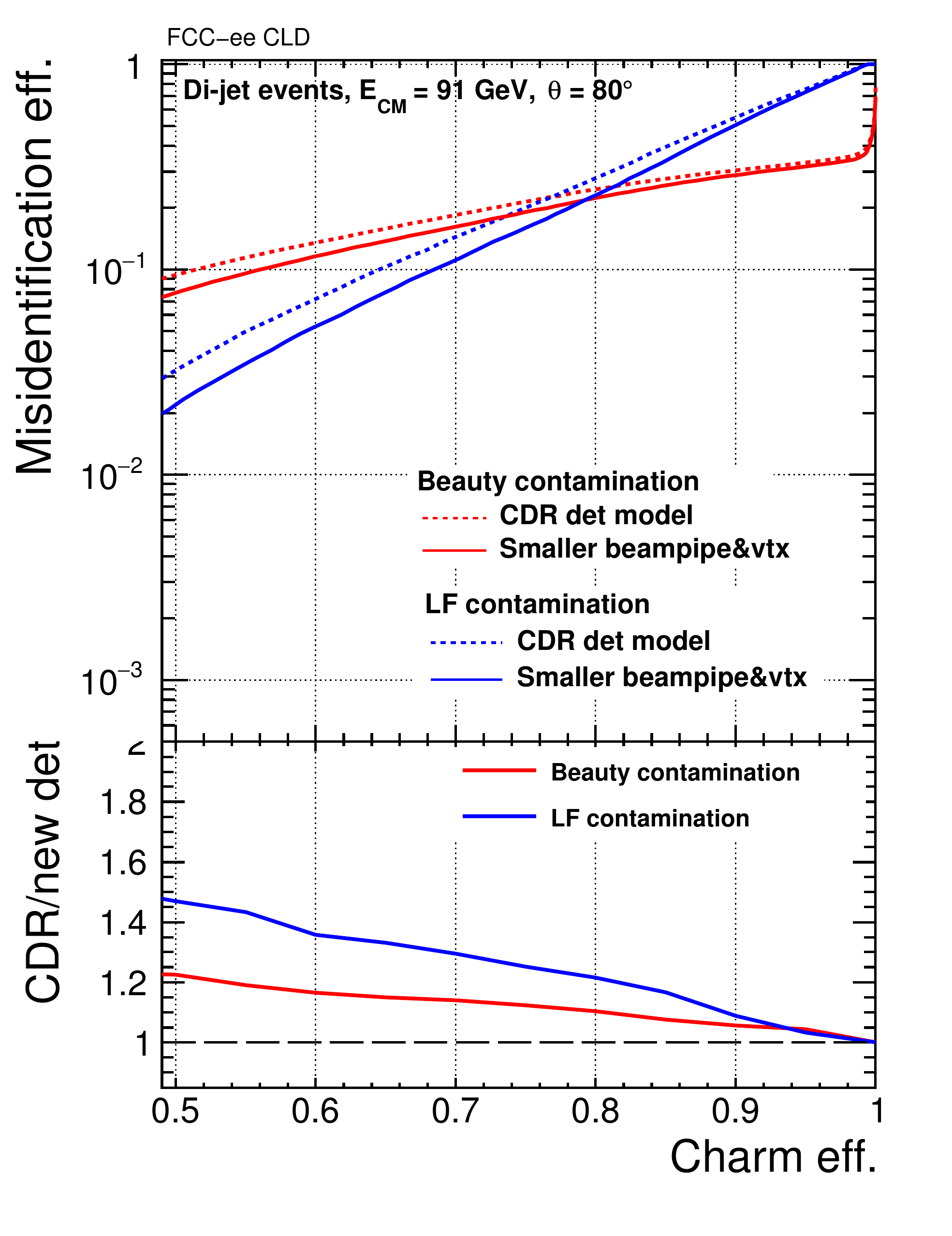}

\medskip
\includegraphics[width=.4\textwidth]{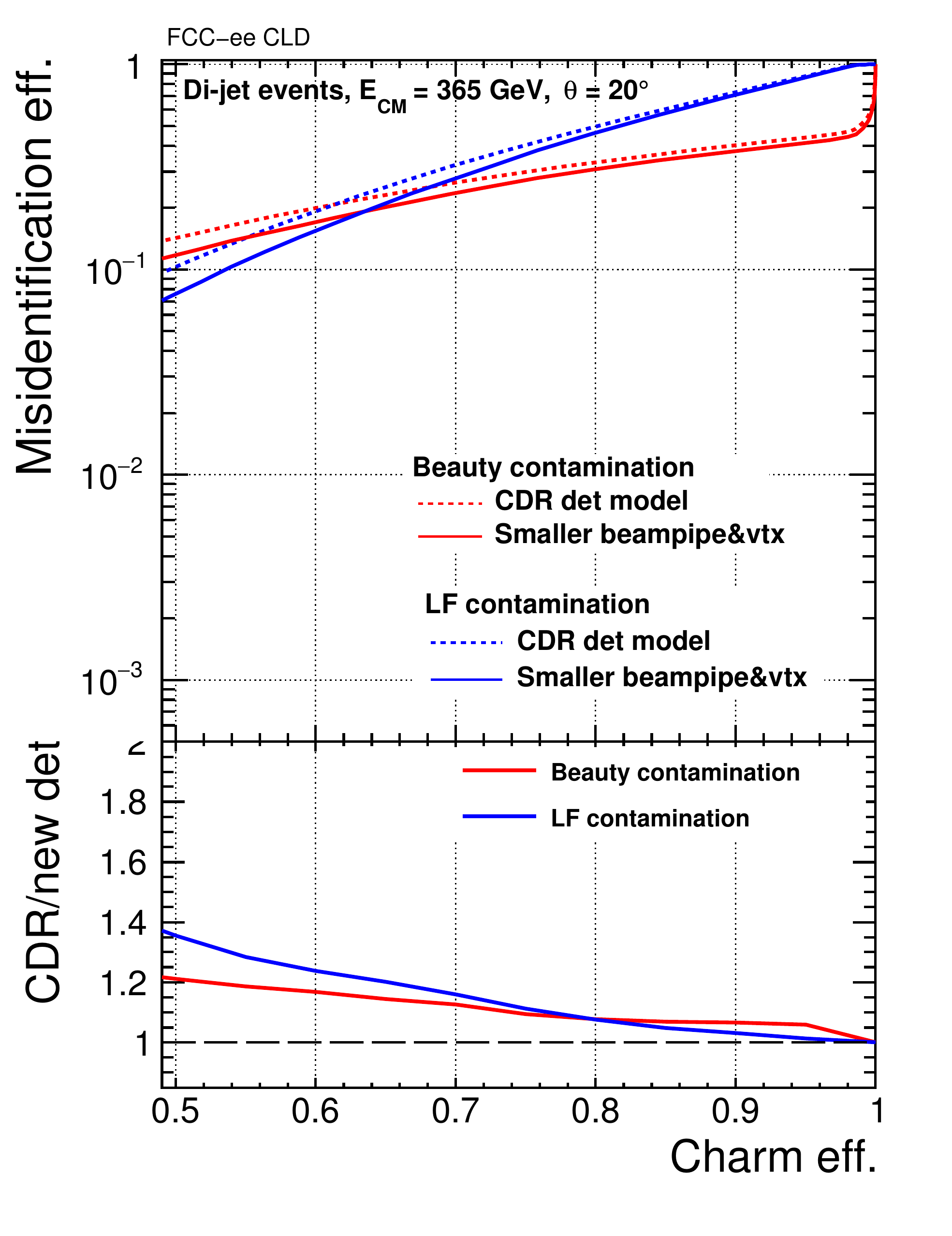}\quad
\includegraphics[width=.4\textwidth]{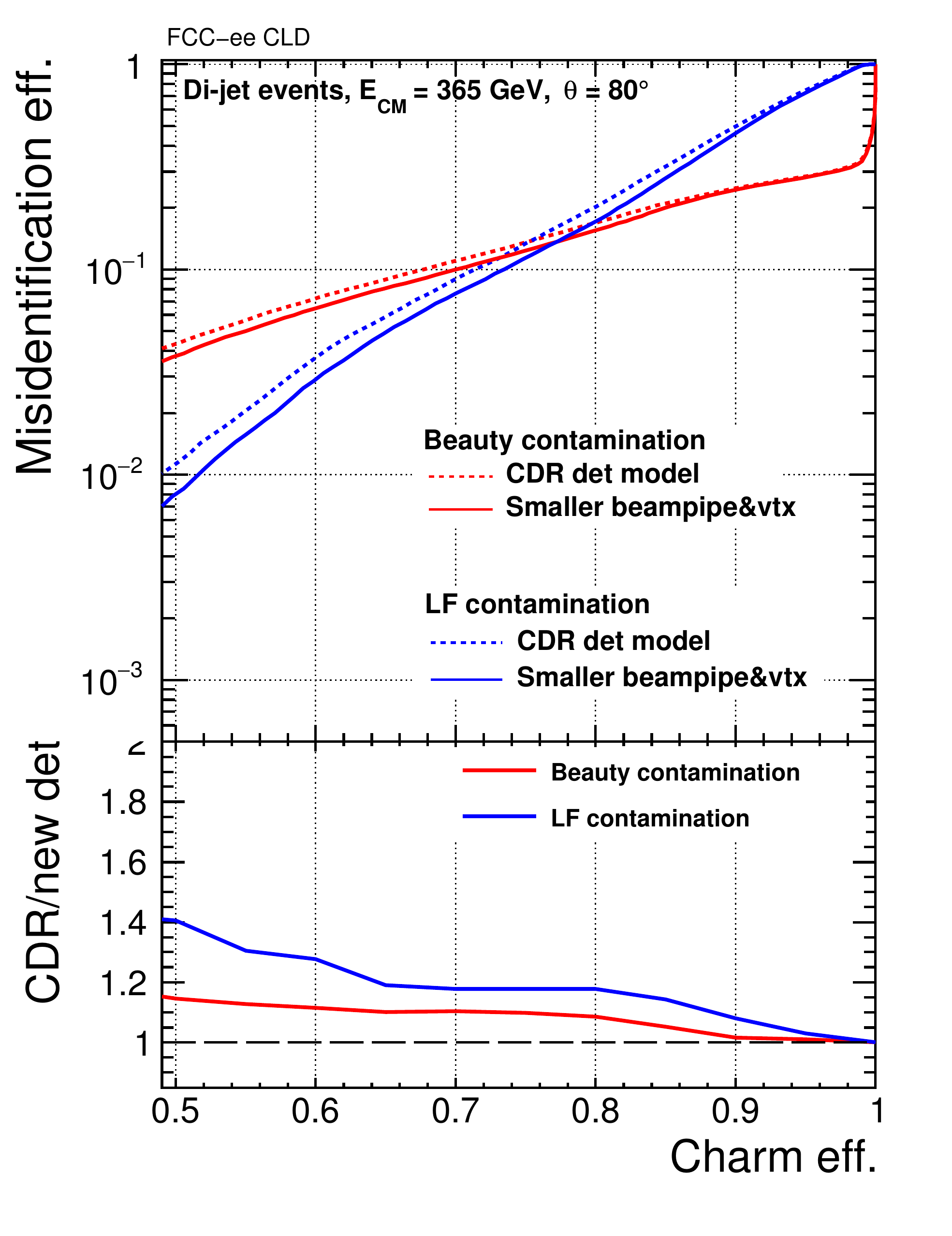}

\caption{C-tagging performance for di-jet events at \SI{91}{GeV} (top) and \SI{365}{GeV} (bottom) centre-of-mass energy and $\theta$ = 20\degrees{}(left) and 80\degrees{}(right). Per contamination source, results obtained with the default and the modified CLD detector model are shown, with their ratio on the bottom canvas. }
  \label{fig:Ctag_newCLD}
\end{figure}

The flavour tagging performance for the new detector has been studied using the truth tracking. The results for B-tagging and C-tagging are shown in Figures~\ref{fig:Btag_newCLD} and~\ref{fig:Ctag_newCLD}, respectively, 
in comparison with the default detector model, for di-jet events at \SI{91}{GeV} (top) and \SI{365}{GeV} (bottom) centre-of-mass energy, at $\theta$ = 20\degrees{}(left) and 80\degrees{}(right). 
The bottom canvas shows the performance ratio between the CLD default and the modified detector model.

As expected, an overall improvement is obtained. The most striking improvement is observed for b-tagging, in particular for forward jets.

\end{appendices}

\clearpage

\printbibliography[title=References]

\end{document}